%% file: main.tex
\documentclass[fontsize=11pt, paper=b5, twoside, open=right, BCOR=7mm, DIV=13, headinclude=true,
					chapterprefix,numbers=noenddot]{scrbook}

\input{preamble}

\begin{document}

\renewcommand{\thepage}{\roman{page}}

\include{frontBack/titlepage}
\include{frontBack/abstract}
\include{frontBack/epigraph}
\include{frontBack/contents}

\renewcommand{\thepage}{\arabic{page}}
\setcounter{page}{1}

\include{main/introduction}		
\include{main/strongInteractions}
\include{main/dispersionRelations}
\include{main/pipi}		
\include{main/eta3pi}
\include{main/eta3piDispersionRelations}
\include{main/eta3piSolution}
\include{main/numerics}

\include{main/results}

\include{main/conclusion}

\include{frontBack/acknowledgements}
\include{frontBack/appendix}
\include{frontBack/bibliography}

\end{document}

%% file: preamble.tex
\usepackage[british]{babel}
\usepackage[utf8]{inputenc}
\usepackage[T1]{fontenc}
\usepackage{lmodern}

\usepackage{textcomp}		
\usepackage{bbm}				
\usepackage[titletoc]{appendix}		
\usepackage{enumitem}		
\usepackage[font=small,labelfont={sf,bf},format=plain]{caption}	
\usepackage{placeins}
\usepackage{pstricks,pst-plot}
\usepackage{xcolor, graphicx, epsfig, psfrag}
\usepackage{axodraw4j}
\usepackage[numbers,sort&compress]{natbib}
\usepackage{amsfonts,amsmath,amssymb}
\usepackage{mathrsfs}			
\usepackage{subfigure}
\usepackage[openbib,NoDate]{currvita}	
\usepackage{ifthen}
\usepackage{array}
\usepackage[ps2pdf,bookmarks,colorlinks,bookmarksnumbered]{hyperref}
\usepackage{breakurl}

\hypersetup{	pdftitle={Determination of the quark mass ratio Q from eta to 3 pions},
					pdfauthor={Stefan Lanz},
					pdfsubject={},
					pdfkeywords={}
				}

\newcommand{\vect}[2][c]{\left( \hspace{-3pt}
\begin{array}{#1}
	#2	
\end{array}
\hspace{-2pt} \right)}
\newcommand{\mat}[3][c]{\left( \hspace{-2pt}
\begin{array}{*{#2}#1}
#3
\end{array} 
\hspace{-1pt} \right)}

\newcommand{\inv}[1]{\frac{1}{#1}}
\newcommand{\Mbar}{M\hspace{-0.85em}\overline{\rule[0.65em]{0.8em}{0pt}}}
\newcommand{\Kbar}{K\hspace{-.7em}\overline{\rule[0.65em]{0.65em}{0em}}}

\newcommand{\bra}[1]{\langle #1 |}
\newcommand{\ket}[1]{| #1 \rangle}
\newcommand{\braket}[2]{\langle #1 | #2 \rangle}
\newcommand{\ins}{, \text{in}}
\newcommand{\outs}{, \text{out}}
\newcommand{\aav}[1]{\langle #1 \rangle}
\newcommand{\psgn}{\phantom{$+$}}
\newcommand{\ee}[1]{\!\times\! 10^{#1}}
\renewcommand{\th}{_\text{th}}
\newcommand{\msbar}{\overline{\text{MS}}}

\newcommand{\etapi}{{\eta \to 3 \pi}}
\newcommand{\chpt}{\texorpdfstring{$\chi$PT}{ChPT}}
\newcommand{\lqcd}{\L_\text{QCD}}
\newcommand{\ldqcd}{\Lambda_\text{QCD}}
\newcommand{\lib}{\L_\text{IB}}
\newcommand{\intlib}{\int \! d^4 x\; \lib}

\renewcommand{\Im}{\text{Im}\,}
\renewcommand{\Re}{\text{Re}\,}
\newcommand{\disc}{\text{disc}\,}
\newcommand{\Res}{\text{Res}\,}
\newcommand{\tr}{\text{Tr}\,}		
\newcommand{\dega}{^\dagger}		
\newcommand{\dcos}{d\!\cos}
\newcommand{\tree}{^\text{tree}}

\newcommand{\mpi}{m_\pi}
\newcommand{\mpip}{m_{\pi^+}}

\newcommand{\mpiz}{m_{\pi^0}}
\newcommand{\mK}{m_K}
\newcommand{\mKp}{m_{K^+}}
\newcommand{\mKz}{m_{K^0}}
\newcommand{\meta}{m_\eta}
\newcommand{\Fpi}{F_\pi}
\newcommand{\FK}{F_K}

\newcommand{\GeV}{\text{GeV}}
\newcommand{\MeV}{\text{MeV}}

\newcommand{\eV}{\text{eV}}

\renewcommand{\L}{\mathscr{L}}
\renewcommand{\O}{\mathcal{O}}
\newcommand{\A}{\mathcal{A}}
\newcommand{\T}{\mathcal{T}}
\newcommand{\G}{\mathcal{G}}
\newcommand{\M}{\mathcal{M}}
\newcommand{\N}{\mathcal{N}}

\newcommand{\eolp}{\,.}
\newcommand{\eolc}{\,,}

\newcommand{\pos}[1]{#1}

\newlength{\dotsize}
\newcommand{\entryTextPos}{}
\newcounter{entrynum}
\newcommand{\entryNum}{0}
\newcommand{\entryFont}{\small}
\newcommand{\entry}[4]{	\addtocounter{entrynum}{-1}
								\ifthenelse{\equal{#2}{\empty}}{
									\psline{|-|}(#3,\theentrynum)(#4,\theentrynum)
								}{
									\pscircle*(#2,\theentrynum){2.5pt}
									\ifthenelse{\equal{#3}{\empty}}{}{
										\SpecialCoor
										\psline{|-|}(!#2 #3 sub \theentrynum)(!#2 #4 add \theentrynum)
										\NormalCoor
									}
								}
								\rput[l](\entryTextPos,\theentrynum){\entryFont #1}
							}

\newcommand{\dotybars}[3]{\psdot[dotsize=5pt](#1,#2)
		\SpecialCoor
		\psline{|-|}(!#1 #2 #3 sub)(!#1 #2 #3 add)
		\NormalCoor
}
\newcommand{\boxybars}[3]{\psdot[dotsize=5pt,dotstyle=square*](#1,#2)
		\SpecialCoor
		\psline{|-|}(!#1 #2 #3 sub)(!#1 #2 #3 add)
		\NormalCoor
}

\newlength{\mylw}
\newlength{\mydotlw}
\colorlet{fillblue}{blue!30}
\colorlet{fillgreen}{green!25}
\newgray{shadegray}{0.85}

\makeatletter
\def\cleardoublepage{\clearpage\if@twoside \ifodd\c@page\else
        \hbox{}
        \thispagestyle{empty}              
        \newpage%
        \if@twocolumn\hbox{}\newpage\fi\fi\fi}
\makeatother

\makeatletter
\DeclareRobustCommand*{\bfseries}{%
  \not@math@alphabet\bfseries\mathbf
  \fontseries\bfdefault\selectfont
  \boldmath
}
\makeatother



%% file: frontBack/titlepage.tex
\begin{titlepage}

\enlargethispage{5 \baselineskip}

   \begin{center}
		\sffamily
       \Huge{\textbf{Determination of the quark mass ratio $Q$ from  $\eta \to 3 \pi$}} \\[2cm]

		\linespread{1.2}

       \large Inauguraldissertation

       der Philosophisch-naturwissenschaftlichen Fakult\"at

       der Universit\"at Bern\\[4cm]

				vorgelegt von\\[.5cm]

       \textbf{\Large Stefan Lanz} \\[.4cm]

		von Huttwil/BE

		\vfill

       Leiter der Arbeit:\\[.5cm]
       Prof. Dr. Gilberto Colangelo\\[.4cm]

		Albert Einstein Zentrum f\"ur fundamentale Physik

       Institut f\"ur theoretische Physik, Universit\"at Bern

   \end{center}

\end{titlepage}

\cleardoublepage

\begin{titlepage}
\setcounter{page}{3}

\enlargethispage{6 \baselineskip}

   \begin{center}
		\sffamily
       \Huge{\textbf{Determination of the quark mass ratio $Q$ from  $\eta \to 3 \pi$}} \\[1.5cm]

		\linespread{1.2}

       \large Inauguraldissertation

       der Philosophisch-naturwissenschaftlichen Fakult\"at

       der Universit\"at Bern\\[1.5cm]

				vorgelegt von\\[.4cm]

       \textbf{\Large Stefan Lanz} \\[.4cm]

		von Huttwil/BE

		\vfill

       Leiter der Arbeit:\\[.4cm]
       Prof. Dr. Gilberto Colangelo\\[.4cm]

		Albert Einstein Zentrum f\"ur fundamentale Physik

       Institut f\"ur theoretische Physik, Universit\"at Bern\\[1cm]

		Von der Philosophisch-naturwissenschaftlichen Fakult\"at angenommen.\\[.5cm]

		Bern, 12.05.2011 \hfill \begin{minipage}[t]{3.9cm} Der Dekan:\\[1.6cm]Prof. Dr. S. Decurtins \end{minipage}

   \end{center}

\end{titlepage}

\cleardoublepage

%% file: frontBack/abstract.tex
\thispagestyle{empty}

\vspace*{3cm}

\begin{center} \textbf{\sffamily Abstract} \end{center}

\noindent
The decay $\etapi$ proceeds exclusively through isospin violating operators and is therefore an excellent probe to
examine the strength of isospin breaking in the strong interaction. The latter can be expressed through the quark mass
ratio $Q^2 = (m_s^2 - \hat{m}^2)/(m_d^2 - m_u^2)$. The main object of this thesis is to analyse the decay $\etapi$ using
dispersion relations in order to extract $Q$ as well as other physical quantities. The dispersion relations are a set of
integral equations that lead to a representation of the decay amplitude in terms of four unknown subtraction constants.
We apply two different methods to determine these. Following the procedure of Anisovich and Leutwyler, we match our
dispersive representation to the one-loop result from chiral perturbation theory, thus updating their old analysis. In
addition, we also make use of the recent experimental determination of the Dalitz distribution by the KLOE collaboration
to obtain information on the subtraction constants. In this way we find $Q = 21.31^{+0.59}_{-0.50}$.

\cleardoublepage

%% file: frontBack/epigraph.tex
\thispagestyle{empty}

\vspace*{3cm}

\begin{center}
	\Large \textit{for Mirjam, Ruben \& Liv}
\end{center}

\cleardoublepage






%% file: frontBack/contents.tex
\tableofcontents

\cleardoublepage

%% file: main/introduction.tex
\addtocontents{toc}{\protect\vspace{2.5ex}}
\phantomsection
\addcontentsline{toc}{chapter}{Introduction}
\markboth{Introduction}{Introduction}

\chapter*{Introduction}

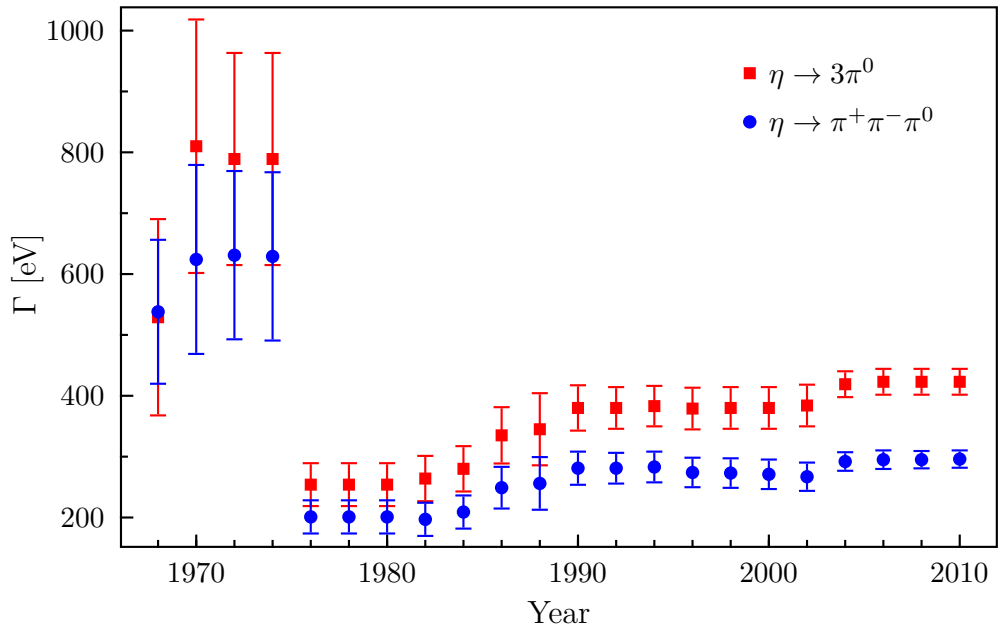
\begin{figure}[t]
	\psset{xunit=.25,yunit=0.008}
	\begin{pspicture*}(1960,0)(2012,1040)
		\psframe(1966,150)(2012,1040)
		\multips(1970,150)(10,0){5}{\psline(0,5pt)}
		\multips(1968,150)(2,0){21}{\psline(0,3pt)}
		\multido{\n=1970+10}{5}{\uput{.2}[270](\n,150){\n}}
		\multips(1966,200)(0,200){5}{\psline(5pt,0)}
		\multips(1966,300)(0,200){4}{\psline(3pt,0)}
		\multido{\n=200+200}{5}{\uput{0.2}[180](1966,\n){\n}}
		\uput{0.7}[270](1989,150){\large Year}
		\uput{1}[180]{90}(1966,600){\large $\Gamma$ [\eV]}

		\psdot[dotsize=5pt,dotstyle=square*,linecolor=red](1999,930) \rput[l](2000,930){$\eta \to 3 \pi^0$}
		\psdot[dotsize=5pt,linecolor=blue](1999,850) \rput[l](2000,850){$\eta \to \pi^+ \pi^- \pi^0$}

		\psset{linecolor=red}
		\boxybars{1968}{529}{163}
		\boxybars{1970}{810}{210}
		\boxybars{1972}{789}{176}
		\boxybars{1974}{789}{176}
		\boxybars{1976}{254}{37}
		\boxybars{1978}{254}{37}
		\boxybars{1980}{254}{37}
		\boxybars{1982}{264}{39}
		\boxybars{1984}{280}{39}
		\boxybars{1986}{335}{48}
		\boxybars{1988}{345}{61}
		\boxybars{1990}{380}{39}
		\boxybars{1992}{380}{36}
		\boxybars{1994}{383}{35}
		\boxybars{1996}{379}{36}
		\boxybars{1998}{380}{36}
		\boxybars{2000}{380}{36}
		\boxybars{2002}{384}{36}
		\boxybars{2004}{419}{23}
		\boxybars{2006}{423}{23}
		\boxybars{2008}{423}{23}
		\boxybars{2010}{423}{23}

		\psset{linecolor=blue}
		\dotybars{1968}{538}{120}
		\dotybars{1970}{624}{157}
		\dotybars{1972}{631}{140}
		\dotybars{1974}{629}{140}
		\dotybars{1976}{201}{29}
		\dotybars{1978}{201}{29}
		\dotybars{1980}{201}{29}
		\dotybars{1982}{197}{29}
		\dotybars{1984}{209}{29}
		\dotybars{1986}{249}{36}
		\dotybars{1988}{256}{45}
		\dotybars{1990}{281}{29}
		\dotybars{1992}{281}{27}
		\dotybars{1994}{283}{27}
		\dotybars{1996}{274}{26}
		\dotybars{1998}{273}{26}
		\dotybars{2000}{271}{26}
		\dotybars{2002}{267}{25}
		\dotybars{2004}{292}{17}
		\dotybars{2006}{295}{17}
		\dotybars{2008}{295}{16}
		\dotybars{2010}{296}{16}

	\end{pspicture*}
	\caption{The historical development of the decay width for the $\eta$ decay channels $\eta \to \pi^+ \pi^- \pi^0$
and $\eta \to 3 \pi^0$ from 1968 until 2010. Before, only an upper limit for the total decay width was measured. The
large fluctuation is almost entirely due to the total decay width, which is fixed via the process $\eta \to \gamma
\gamma$; the branching ratios have changed only moderately since 1970. All the values stem from the PDG report of the
corresponding year.}
\label{fig:introHistory}
\end{figure}

Over the last almost three decades, chiral perturbation theory (\chpt)~\cite{Weinberg1979,Gasser+1985,Gasser+1984} in
its many variants has helped to considerably increase our understanding of the strong interactions at low energies.
While for some processes---for instance $\pi \pi$ scattering---the theoretical description has reached astonishing
precision, for others the situation seems less favourable. A prominent example of the latter class is the decay
$\etapi$. Both, experimental and theoretical results for its decay width have varied a lot over the years and often
been in flat disagreement. Figure~\ref{fig:introHistory} shows the historical development of the decay width for both
$\etapi$ channels since 1968. A brief account of attempts to calculate the same quantities is given in the following.

\looseness=-1			
Since three pions cannot couple to vanishing isospin and angular momentum at the same time, the
decay must proceed through isospin breaking operators. The electromagnetic interactions do violate isospin symmetry and
could thus be responsible for the transition, but already in the late sixties it was shown that their contribution is
suppressed due to chiral symmetry and cannot account for the large decay width~\cite{Bell+1968,Sutherland1966}. From
current algebra~\cite{Cronin1967,Osborn+1970}, or equivalently from \chpt{} at leading order, follows a decay width of
merely $66~\eV$. Several authors examined the effects from final-state rescattering~\cite{Neveu+1970,Roiesnel+1981}.
In Ref.~\cite{Roiesnel+1981}, it was found that final-state interactions lead to a decay width of $200~\eV$ in
excellent agreement with the experimental value at that time. Only a few years later,
\chpt{} at one loop was formulated which allowed for a systematic extension of the current algebra result. Gasser and
Leutwyler~\cite{Gasser+1985a} calculated a decay width of $160 \pm 50~\eV$, where the error is due to higher order
corrections. This value was not only in excellent agreement with experiment but also confirmed that indeed final-state
rescattering effects could fully account for the discrepancy between current algebra and experiment. But they also
mentioned that new experiments hinted at the possibility of a considerably larger decay width, which would ``resurrect
the puzzle'' they ``claim[ed] to solve'' (p.~548). Later, it was confirmed that the decay width had been underestimated
so far which motivated two dispersive analyses~\cite{Anisovich+1996,Kambor+1996} since in this way, more than one
final-state rescattering process could be taken into account.

The second theoretical puzzle is related to the neutral channel $\eta \to 3 \pi^0$. It is usually parametrised in terms
of a single slope parameter $\alpha$, which many experiments have consistently found to be
negative~\cite{Alde+1984,Abele+1998,Tippens+2001,Achasov+2001,Bashkanov+2007,Adolph+2009,Unverzagt+2009,Prakhov+2009,
Ambrosino+2010} and which is known very precisely today. The current PDG average~\cite{PDG2010} is
\begin{equation}
	\alpha = -0.0317 \pm 0.0016 \eolp \nonumber
\end{equation}
Chiral perturbation theory, on the other hand, leads to a positive value at the one-loop~\cite{Gasser+1985a} and the
two-loop level~\cite{Bijnens+2007}. The error bars on the latter are, however, quite large
and also encompass negative values. Kambor et al. were the first to find a negative sign with their dispersive
analysis~\cite{Kambor+1996} but with too small a magnitude. 

The particular interest in this process is, however, not only due to the problems involved in its theoretical
understanding. Since it violates isospin, but not dominantly through electromagnetic interactions, it is ideally suited
to probe the strength of isospin breaking terms in the strong interactions. It offers access to the quark mass
difference $m_u - m_d$ that measures isospin breaking, or to the quark mass double ratio
\begin{equation}
	Q^2 = \frac{m_s^2 - \hat{m}^2}{m_d^2 - m_u^2} \eolc \qquad \text{with}\quad \hat{m} = \inv{2} (m_u + m_d) \eolc
	\nonumber
\end{equation}
that compares isospin breaking with chiral symmetry breaking. The ratio is useful, e.g., in conjunction with lattice
results for the light quark masses. Lattice simulations are always performed in the isospin limit, i.e., the up and
down quark masses are set equal, and can accordingly only measure $m_s$ and $\hat{m}$. The missing information that is
needed in order to extract also $m_u$ and $m_d$ is provided by the ratio $Q$. The determination of this parameter is one
of the main goals of this thesis.

Despite the long history of research on $\etapi$, the decay is still not fully understood and arouses a lot of
interest even today, as can be seen from several recent publications. In Ref.~\cite{Borasoy+2005}, the decay amplitude
was computed using unitarised chiral perturbation theory in good agreement with experiment. Also the aforementioned
two-loop calculation has been published only recently~\cite{Bijnens+2007}. An analytical dispersive analysis following a
different approach than ours was presented in Refs.~\cite{Kampf+2009,Kampf+2011}. One of its advantages is that it will
allow to include isospin breaking effects due to the mass difference $\mpiz - \mpip$~\cite{Kampf+tobe}. Furthermore, in
Ref.~\cite{Schneider+2011} the decay was examined by means of the non-relativistic effective field theory (NREFT)
approach. Also this method allows to include higher-order isospin breaking effects and in addition is particularly
suited to analyse the reasons for the failure of \chpt{} to reproduce the amplitude for the neutral decay. All these
articles give a negative value for the slope parameters $\alpha$. Furthermore, the full electromagnetic corrections up
to $\O(e^2 q)$ have been calculated~\cite{Ditsche+2009} and an analysis using resummed chiral perturbation theory
is under way~\cite{Kolesar+2010}.

The goal of this thesis is to compute a dispersive representation of the $\etapi$ decay amplitude following the
approach of Anisovich and Leutwy\-ler~\cite{Anisovich+1996,Walker1998}. From the amplitude we can obtain the quark
mass ratio $Q$, but also the slope parameter $\alpha$ and other parameters. There are several reasons to redo this old
analysis. Foremost, since final-state rescattering seems to be the main source of trouble on the theoretical side,
dispersion relations are indeed an excellent tool to tackle this process. Many of the inputs to the calculation have
been improved considerably over the years, most prominently the $\pi \pi$ phase shifts and are now used to update the
old analysis. The Dalitz distribution has been measured with unprecedented accuracy by the KLOE
collaboration~\cite{Ambrosino+2008} and this data can be used as an input to the dispersive analysis. In this way we can
go beyond the analysis of Anisovich and Leutwyler who did not have suitable experimental data at their hand. But also
the new theoretical results that have been mentioned above are a valuable possibility for comparison and may also
provide inputs for a future extension of our analysis.

This thesis is structured as follows. The first four chapters introduce a number of basic topics that will be crucial
for the later discussion of the dispersive analysis. We start with a brief overview of the theory of the strong
interactions, which provides the physical framework we will work in. Quark masses and quark mass ratios, in
particular also $Q$, are discussed at the end of the first chapter. This is followed by an introduction to the
mathematical concept of dispersion relations. By means of the dispersion relations, we want to account for the large
contributions from final-state rescattering to the $\etapi$ decay amplitude and therefore dedicate a chapter to a
detailed discussion of $\pi \pi$ scattering. Afterwards, we also review the decay $\etapi$ in some detail, with a
special focus on the low-energy approximation of the decay amplitude and on an overview of experimental and theoretical
results. This concludes the introductory part as we have now all the ingredients ready for the dispersive analysis. In
Chapter~\ref{chp:etaDisp}, we derive the dispersion relations for $\eta \to \pi^+ \pi^- \pi^0$ and discuss how they are
solved in the next two chapters. First, we describe the solution algorithm and then address the implementation of the
numerical solution in a computer program. The thesis is closed with a detailed account of our results in
Chapter~\ref{chp:Results}.

%% file: main/strongInteractions.tex
\chapter{Theory of the strong interaction} 

After the great success of quantum electrodynamics (QED) as a relativistic quantum theory for electrons, muons and
photons, it was generally believed that all the other phenomena of particle physics could be described in terms of a 
quantum field theory involving only few degrees of freedom. In the framework of QED, predictions for physical
quantities are obtained using perturbation theory. The interaction terms in the Lagrangian are proportional to a small
coupling constant $\alpha_\text{\tiny QED}$, such that an expansion in $\alpha_\text{\tiny QED}$ around the free
theory converges rather rapidly. In order to deal with the infinities that appear when going beyond the leading order
approximation, the technique of renormalisation was devised, where the divergences are absorbed in the non-physical
parameters in the Lagrangian. Renormalisability became one of the main guiding principle in the construction of quantum
field theories.

At the time when QED was developed, the number of known elementary particles was quite modest. The electron, the muon
and their antiparticles had been observed, as well as a small number of hadrons, including the proton, the neutron and
the pion. But many discoveries increased this number to more than 100. A quantum field theory with that many elementary
degrees of freedom was not very appealing and soon the hope that the whole of particle physics could be formulated
following this paradigm was given up. A textbook from 1970 entitled ``Elementary Particle Theory'' (\cite{Martin+1970},
p.~63) remarks concerning quantum field theory:
\begin{quote}
	This method has completely failed to account for the dynamics of strongly interacting particles, the reason for this
	failure presumably being the strength of the strong interactions which makes it meaningless to treat the interaction
	part of the Hamiltonian as a perturbation.

	Since the Hamiltonian approach has failed to provide a theory of strongly interacting particles we shall abandon it
	and look for an alternative method of attack.
\end{quote}
One of the alternative methods that, while not being a full success, lead to many interesting and fruitful results is
$S$-matrix theory. The general idea was to find enough mathematical guidelines (e.g.\ unitarity) and symmetry
principles (e.g.\ baryon number conservation or parity) to constrain physical amplitudes to a unique expression
compatible with all the requirements. This goal was never reached, but nevertheless many useful results on the
properties of scattering and decay amplitudes have been gained. Also the technique of dispersion relations has grown
out of $S$-matrix theory and some knowledge on the latter is therefore indispensable for the understanding of this
thesis.

During the heydays of $S$-matrix theory, Gell-Mann proposed that all the observed hadrons could be composite states
built from elementary fermions he called quarks~\cite{Gell-Mann1961}, but since these particles failed to show up in
\nocite{Gell-Mann+1964} 
experiments they were generally considered to be mathematical tools rather than physical objects. The decisive step
towards a quantum field theory of the strong interactions came with the discovery of asymptotic freedom
\cite{Gross+1973,Gross+1973a,Politzer1973,Gross1999}. Not only was it rigorously proved that the coupling
constant of some non-Abelian gauge theories becomes small at very large energies, allowing thus for dispersive
calculations, but also indicated that at low-energies, the quarks could be bound together with such strength that it is
simply not possible to separate them. This mechanism of confinement was seen as a possible explanation for the apparent
absence of quarks in nature. In the words of Gross and Wilczek (\cite{Gross+1973a}, p.~3650):
\begin{quote}
	This is an exciting possibility which might provide a mechanism for having a theory of quarks without real quark
	states. Whether this can be realized $[\ldots]$ deserves much attention.
\end{quote}
Not long afterwards, a quantum field theory with an $SU(3)$ gauge group was proposed as the fundamental theory of the
strong interaction~\cite{Fritzsch+1973}. It is asymptotically free such that at high energies perturbation theory can
be used successfully. But in the low-energy region, the coupling constant is too large for the perturbative series to
converge. Following a folk theorem by Weinberg~\cite{Weinberg1979}, Gasser and Leutwyler~\cite{Gasser+1985,Gasser+1984}
constructed the framework of chiral perturbation theory that successfully describes the low-energy dynamics of
QCD. They had, however, to give up the principle of renormalisability that had been thought to be indispensable so far.
A large number of accurate results have been obtained from \chpt{} until today, including amplitudes for many decay and
scattering processes and expressions for quark mass ratios in terms of meson masses.

Besides \chpt, also lattice QCD has been producing results in the low-energy region of the strong interaction for many
years now. For a long time, however, these simulations had to be performed in the quenched approximation and 
with large pion masses. The corresponding systematic uncertainties are very difficult to estimate. Only in recent
years, simulations under realistic conditions have become possible. The relation between the lattice and \chpt\ has been
one of mutual benefit. With lattice calculations, one can among other things determine the quark masses and the
low-energy constants from first principles. Chiral perturbation theory, on the other hand, has been used in lattice
calculations in order to control lattice artefacts. In particular, it can help in the chiral extrapolation, where the
pion masses are brought to their physical value, and take care of finite volume effects.

The intent of this chapter is not to discuss the strong interaction in its entirety, but rather to introduce the
topics relevant to the rest of this thesis. It is hence kept short with details added where they are required for later
use. We start with a brief presentation of $S$-matrix theory that culminates in the discussion of the analytic
properties of the scattering amplitude. Especially this last topic plays a central role in the construction of
dispersion relations, as we will see in the next chapter.
This is followed by a brief introduction to the quantum field theory of the strong interaction, quantum chromodynamics
(QCD). We focus mainly on the global symmetries of QCD, since they are the essential ingredients to the construction
of its effective low-energy theory, chiral perturbation theory, that is presented afterwards. The low-energy
approximation of an amplitude that is obtained in this framework serves as ideal theoretical input to the dispersion
relations. The last pages of this chapter are dedicated to the masses of quarks and mesons. In particular, also the
quark mass double ratio $Q$ will be introduced.

\section{\texorpdfstring{$S$}{S}-matrix theory} \label{sec:strongSmatrix}

Particle physics experiments usually consist in the investigation of scattering or decay processes. In a scattering
experiment, two particles are brought to collision and two or more particles emerge. If the final state contains the
same particles as the initial state, the process is said to be elastic. Otherwise, new particles are produced
and the process is called inelastic. A decay process is inelastic by definition. In both cases, scattering and decay
experiments, one tries to determine the species and the momenta of the outgoing particles. This is, of course, a highly
non-trivial task, but the discussion of the problems involved in particle detection is beyond the scope of this thesis.
This section is rather intended to discuss how such processes are treated theoretically. In the following, we will
focus on scattering processes, but the same observations can be applied to decay processes as well.

\subsection{Asymptotic states, \texorpdfstring{$S$}{S}- and \texorpdfstring{$T$}{T}-matrix}

The initial state for a scattering experiment contains two particles that are on a collision course. A long time before
the scattering takes place, that is as $t\to-\infty$, these particles are well separated and do not interact with each
other, so that they can be regarded as free particles. Such an initial state is called an asymptotic state and
denoted by $\ket{i \ins}$, where $i$ stands for initial, or by $\ket{p_1,p_2 \ins}$, where $p_1$ and $p_2$ are the
momenta of the incoming particles. The label ``in'' declares the state as an asymptotic state of incoming particles. The
``in'' states are postulated to form a complete set of states, called the Fock space.

Similarly, a long time after the scattering, as $t \to \infty$, the particles in the final state form again free states
that are denoted by $\ket{f \outs}$, where $f$ stands for final, or $\ket{p_3,p_4, \ldots \outs}$. There is no reason to
restrict the number of particles to two, hence the ellipsis. Also the ``out'' states are postulated to form a complete
set.

The probability amplitude for an initial state $\ket{i\ins}$ to be found as the final state $\ket{f\outs}$ is given by
\begin{equation}
	S_{fi} \equiv \braket{f\outs}{i\ins} = \bra{f\ins}S\ket{i\ins} \eolc
\end{equation}
where the second equality defines the scattering operator $S$. The matrix composed of all its matrix elements $S_{fi}$
is called the $S$-matrix. From the above definition it is immediately clear that
\begin{equation}
	\bra{p_1,p_2,\ldots \outs} = \bra{p_1,p_2,\ldots \ins} S \eolc
	\label{eq:strongRelInOut1}
\end{equation}
and $S$ is thus the operator that evolves the ``in'' state in time from $t=-\infty$ to $t=\infty$. Furthermore, as both,
``in'' and ``out'' states, form a complete set of states, $S$ cannot be singular and its inverse must exist. Therefore,
\begin{equation}
	\bra{p_1,p_2,\ldots \ins} = \bra{p_1,p_2,\ldots \outs} S^{-1} \eolp
	\label{eq:strongRelInOut2}
\end{equation}

Symmetry principles impose constraints on the $S$-matrix. If the system under consideration has a conserved
quantity $A$, that is $[A,S] = 0$, and if the initial and final states are eigenstates of $A$ with eigenvalues $a_i$ 
and $a_f$, respectively, then
\begin{equation}
	0 = \bra{f\ins}[A,S]\ket{i\ins} = (a_f - a_i) \bra{f\ins}S\ket{i\ins} \eolp
\end{equation}
This implies that the transition amplitude must vanish unless $a_f = a_i$. In other words, a conserved quantity indeed
remains unchanged in a scattering process, as the scattering can only take place between states with the same
eigenvalue. The eigenvalue of such an operator is called a good quantum number, as it can be used to label the states.
In particular, the four momentum $P_\mu$ is a conserved quantity and according to the above rules we must have
\begin{equation}
	\bra{f\ins}S\ket{i\ins} \propto \delta^4(p_f - p_i) \eolc
\end{equation}
where $p_i$ and $p_f$ denote the total momentum of the incoming and outgoing particles, respectively.

It is customary to separate unity from the $S$-matrix,
\begin{equation}
	S = 1 + i T \eolc
	\label{eq:strongTmatrix}
\end{equation}
thus defining the $T$-matrix. The $S$-matrix element is then split as
\begin{equation}
	\bra{f\ins}S\ket{i\ins} = \braket{f\ins}{i\ins} + i \bra{f\ins}T\ket{i\ins} \eolc
	\label{eq:strongTmatrixElement}
\end{equation}
and it becomes clear that the $T$-matrix contains the non-trivial part of the $S$-matrix, while the first term on the
right-hand side is the amplitude for the case, where the particles in the initial state do not interact at all. From the
$T$-matrix, the Lorentz invariant scattering amplitude $\T_{fi}$ is defined as
\begin{equation}
	i \bra{f\ins}T\ket{i\ins} = i (2\pi)^4 \delta^4(p_f - p_i) \T_{fi} \eolp
	\label{eq:strongAmplitude}
\end{equation}

\subsection{Unitarity of the \texorpdfstring{$S$}{S}-matrix}

This section introduces the fundamental physical postulate often referred to as probability conservation.
By this we mean that a system that starts out in an initial state $\ket{i\ins}$ must end in \emph{some} final state
$\ket{f\outs}$ after it has evolved in time. Consequently, if the probability to find the system in the final state
$\ket{f\outs}$ is $P_{i\to f}$ and if the final states form a complete set of states (which we postulated above), we
must have $\sum_f P_{i\to f} = 1$. The transition probability $P_{i\to f}$ is nothing else than the square of the
corresponding $S$-matrix element and so,
\begin{equation}
	1 = \sum_f |\bra{f\ins}S\ket{i\ins}|^2
		= \sum_f \bra{i\ins}S\dega\ket{f\ins} \bra{f\ins}S\ket{i\ins} = \bra{i\ins}S\dega S\ket{i\ins} \eolp
\label{eq:strongProbSum}
\end{equation}
On the other hand, we can choose the initial state
\begin{equation}
	\ket{i\ins} = \inv{\sqrt{|\alpha|^2 + |\beta|^2}} \big( \alpha \ket{a} + \beta \ket{b} \big) \eolc
\end{equation}
where $\ket{a}$ and $\ket{b}$ are two orthonormal states and $\alpha$ and $\beta$ are arbitrary complex numbers. Using
this state in Eq.~\eqref{eq:strongProbSum}, we find
\begin{equation}\begin{split}
	1 &= \inv{|\alpha|^2 + |\beta|^2} \Big( 
	\begin{aligned}[t]
		&|\alpha|^2 \bra{a} S\dega S \ket{a} + |\beta|^2 \bra{b} S\dega S\ket{b} \\[1mm]
		& + \alpha^\ast \beta \bra{a} S\dega S \ket{b} + \alpha \beta^\ast \bra{b} S\dega S \ket{a} \Big)
	\end{aligned}
	\\[2mm]
		&= 1 + \inv{|\alpha|^2 + |\beta|^2} 
				\left( \alpha^\ast \beta \bra{a} S\dega S \ket{b} + \alpha \beta^\ast \bra{b} S\dega S \ket{a} \right) \eolp
\end{split}\end{equation}
Because $\alpha$ and $\beta$ are arbitrary complex numbers, both terms inside the brackets must vanish separately:
\begin{equation}
	\bra{a} S\dega S \ket{b} = \bra{b} S\dega S \ket{a} = 0 \eolp
	\label{eq:strongSdegaS0}
\end{equation}
Equations~\eqref{eq:strongProbSum} and \eqref{eq:strongSdegaS0} together with the fact that the inverse of $S$ exists
lead
to
\begin{equation}
	S\dega S = S S\dega = 1 \eolp
	\label{eq:strongSunitarity}
\end{equation}
In other words, the $S$-matrix is unitary. For this reason, probability conservation is usually called
\emph{unitarity}. This result allows to extend some of the formul\ae\ presented above. The relations between in and out
states in Eqs.~\eqref{eq:strongRelInOut1} and \eqref{eq:strongRelInOut2} can be replaced by
\begin{equation}\begin{split}
	\bra{p_1,p_2,\ldots \outs} &= \bra{p_1,p_2,\ldots \ins} S \eolc \quad
	\bra{p_1,p_2,\ldots \ins} = \bra{p_1,p_2,\ldots \outs} S\dega \eolc \\[1mm]
	\ket{p_1,p_2,\ldots \outs} &= S\dega \ket{p_1,p_2,\ldots \ins} \eolc \quad
	\ket{p_1,p_2,\ldots \ins} = S \ket{p_1,p_2,\ldots \outs} \eolp
	\label{eq:strongRelInOut3}
\end{split}\end{equation}
From these equations follows immediately that the $S$-matrix element can be written in two ways, namely
\begin{equation}
	S_{fi} \equiv \braket{f\outs}{i\ins} = \bra{f\ins}S\ket{i\ins} = \bra{f\outs}S\ket{i\outs} \eolc
\end{equation}
such that we can safely omit the labels ``in'' and ``out'' in $S$-matrix elements.
For the $T$-matrix, unitarity implies
\begin{equation}
	T - T\dega = i T T\dega \eolc
	\label{eq:strongSTunitarity}
\end{equation}
as can be shown by inserting Eq.~\eqref{eq:strongTmatrix} into Eq.~\eqref{eq:strongSunitarity}. Using the definition of
the amplitude in
Eq.~\eqref{eq:strongAmplitude}, the left hand side can be rewritten as
\begin{equation}
	\bra{f} T \ket{i} - \bra{f} T\dega \ket{i} = (2\pi)^4 \delta^4(p_f-p_i) \left( \T_{fi} - \T_{if}^\ast \right) \eolp
\end{equation}
On the right hand side we insert a complete set of intermediate states and find
\begin{equation}
	\bra{f} i T T\dega \ket{i} = i \sum_n \bra{f}T\ket{n}\bra{n}T\dega\ket{i} 
			= i (2\pi)^8 \sum_n \delta^4(p_f-p_n) \delta^4(p_n-p_i) \T_{fn} \T_{in} ^\ast \eolp
\end{equation}
Putting everything together leads to the unitarity condition
\begin{equation}
	\T_{fi} - \T_{if}^\ast = i (2\pi)^4 \sum_n \delta^4(p_n-p_i) \T_{fn} \T_{in} ^\ast \eolp
	\label{eq:strongUnitarityCondition1}
\end{equation}
If the interaction respects time reversal symmetry, as the strong interaction does, the unitarity condition can be
brought to a more appealing form. For ``in'' and ``out'' states made up of scalar particles, time reversal invariance
implies
\begin{equation}
	\braket{f\outs}{i\ins} = \braket{i\outs}{f\ins} \eolc
\end{equation}
and the $S$- and $T$-matrix elements then satisfy
\begin{equation}
	\bra{f}S\ket{i} = \bra{i}S\ket{f} \quad \Longrightarrow \quad 
	\braket{f}{i} + \bra{f}T\ket{i} = \braket{i}{f} + \bra{i}T\ket{f} \eolp
\end{equation}
Because the amplitude in the case of no interaction is time reversal invariant, that is $\braket{f}{i} =
\braket{i}{f}$, the $T$-matrix elements and the amplitude are symmetric, $\T_{fi} = \T_{if}$. The left-hand side of
Eq.~\eqref{eq:strongUnitarityCondition1} is then basically the imaginary part of the amplitude and the unitarity
condition becomes
\begin{equation}
	\Im\,\T_{fi} = \frac{(2\pi)^4}{2} \sum_n \delta^4(p_n-p_i) \T_{fn} \T_{in} ^\ast \eolp
	\label{eq:strongUnitarityCondition2}
\end{equation}
As this is a statement on the imaginary part of an amplitude, it is not surprising that it has deep consequences on the
amplitude's analytic properties. We will come back to this point below.

\subsection{Kinematics of scattering processes and crossing} \label{sec:strongScattering}

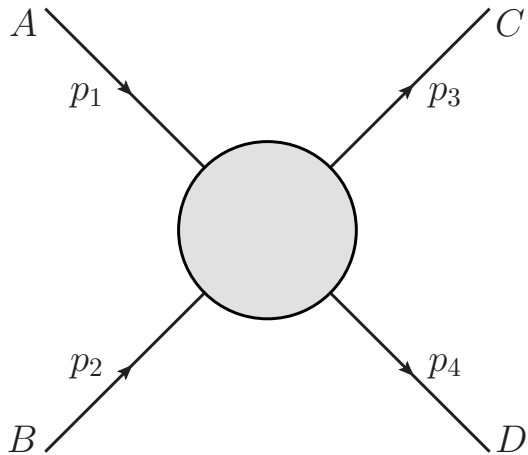
\begin{figure}[tb]
	\SetScale{0.55}
	\setlength{\unitlength}{0.55pt}
	\begin{center}
	\begin{picture}(368,338) (17,-29)
		\SetWidth{2.0}
		\SetColor{Black}
		\Line[arrow,arrowpos=0.5,arrowlength=8.333,arrowwidth=3.333,arrowinset=0.2](240,178)(350,288)
		\Line[arrow,arrowpos=0.5,arrowlength=8.333,arrowwidth=3.333,arrowinset=0.2](50,-12)(160,98)
		\Line[arrow,arrowpos=0.5,arrowlength=8.333,arrowwidth=3.333,arrowinset=0.2](50,288)(160,178)
		\Line[arrow,arrowpos=0.5,arrowlength=8.333,arrowwidth=3.333,arrowinset=0.2](240,98)(350,-12)
		\Text(90,238)[rt]{\Large{\Black{$p_1$}}}
		\Text(90,38)[rb]{\Large{\Black{$p_2$}}}
		\Text(310,238)[lt]{\Large{\Black{$p_3$}}}
		\Text(310,38)[lb]{\Large{\Black{$p_4$}}}
		\Text(45,288)[rt]{\Large{\Black{$A$}}}
		\Text(45,-12)[rb]{\Large{\Black{$B$}}}
		\Text(355,288)[lt]{\Large{\Black{$C$}}}
		\Text(355,-12)[lb]{\Large{\Black{$D$}}}
		\GOval(200,138)(60,60)(0){0.882}
	\end{picture}
	\end{center}
\caption{The scattering process $A(p_1) B(p_2) \to C(p_3) D(p_4)$.}
\label{fig:strong22scattering}
\end{figure}

We discuss the kinematics of a generic $2 \to 2$ scattering process $A(p_1) B(p_2) \to C(p_3) D(p_4)$ for scalar
particles as depicted in Fig.~\ref{fig:strong22scattering}. The $S$-matrix element for this process is given by
\begin{equation}
	\braket{C(p_3) D(p_4)\outs}{A(p_1) B(p_2)\ins} 
				= \bra{C(p_3) D(p_4)} S \ket{A(p_1) B(p_2)} \eolc
\end{equation}
and the four-momenta are $p_i = (E_i,\vec{p}_i)$ with $E_i = \sqrt{|\vec{p}_i|^2 + m_i^2}$, where $m_i$ is the mass.
From the $T$-matrix element, the Lorentz invariant scattering amplitude $\A$ is defined by
\begin{equation}
	\bra{C(p_3) D(p_4)} T \ket{A(p_1) B(p_2)} = (2 \pi)^4 \, \delta^4(p_1 + p_2 - p_3 - p_4)\, \A(p_1,p_2,p_3,p_4) \eolp
\end{equation}
It is convenient to express the amplitude in terms of the Lorentz invariant Mandelstam variables
\begin{equation}\begin{split}
	s &= (p_1 + p_2)^2 = (p_3 + p_4)^2\eolc\\
	t &= (p_1 - p_3)^2 = (p_2 - p_4)^2\eolc\\
	u &= (p_1 - p_4)^2 = (p_2 - p_3)^2 \eolp
	\label{eq:strongScatMandelstam}
\end{split}\end{equation}
No other Lorentz invariant quantities that are independent from these can be built from the momenta and the Lorentz
invariant amplitude can thus only be a function of $s$, $t$ and $u$. Since the Mandelstam variables are related by
\begin{equation}
	s + t + u = \sum_{i = 1}^4 m_i^2 \eolc
	\label{eq:strongScatMandelstamSum}
\end{equation}
the amplitude only depends on two independent kinematic variables. This becomes also clear from counting the
degrees of freedom: the four-momenta of the particles contain 16 degrees of freedom, but 6 can be fixed due to Lorentz
invariance, four by the on-shell condition and another four by energy-momentum conservation.

Because they are Lorentz invariant quantities, the Mandelstam variables can be evaluated in an arbitrary coordinate
system. The simplest choice is the centre of mass frame, where $\vec{p}_1 = -\vec{p}_2$ and $\vec{p}_3 = -\vec{p}_4$.
We can then re-express the Mandelstam variables in terms of the masses, energies and the scattering angle 
$\theta_s = \angle (\vec{p}_1,\vec{p}_3)$%
\footnote{Note that one could define the scattering angle to be the angle between any pair of an in- and an outgoing
momentum and our choice is thus purely conventional. All four possible definitions lead to the same absolute value for 
$\cos \theta_s$, but the sign may differ.},
\begin{equation}\begin{split}
	s &= (E_1 + E_2)^2 = (E_3 + E_4)^2 \eolc\\[1mm]
	t &= m_1^2 + m_3^2 - 2 E_1 E_3 + 2 |\vec{p}_1| |\vec{p}_3| \cos \theta_s \\
		&= m_2^2 + m_4^2 - 2 E_2 E_4 + 2 |\vec{p}_1| |\vec{p}_3| \cos \theta_s\eolc\\[1mm]
	u &= m_1^2 + m_4^2 - 2 E_1 E_4 - 2 |\vec{p}_1| |\vec{p}_3| \cos \theta_s \\
		&= m_2^2 + m_3^2 - 2 E_2 E_3 - 2 |\vec{p}_1| |\vec{p}_3| \cos \theta_s\eolp
	\label{eq:strongScatMandelstamCOM}
\end{split}\end{equation}
After replacing $E_i$ by $\sqrt{|\vec{p}_i|^2 + m_i^2}$, the expressions for $s$ can be solved for $|\vec{p}_1|$ and
$|\vec{p}_3|$, respectively, yielding
\begin{equation}
	|\vec{p}_1|^2 = \frac{\lambda(s,m_1^2,m_2^2)}{4s}\eolc\qquad |\vec{p}_3|^2 = \frac{\lambda(s,m_3^2,m_4^2)}{4s}\eolc
	\label{eq:strongScat3Momentum}
\end{equation}
where $\lambda$ is the K\"all\'en triangle function defined as
\begin{equation}\begin{split}
	\lambda(a,b,c) &= a^2 + b^2 + c^2 -2(ab+bc+ca)\\& = \left( a - \big(\sqrt{b} + \sqrt{c} \, \big)^2 \right) 
																	\left( a - \big(\sqrt{b} - \sqrt{c} \, \big)^2 \right) \eolp
\end{split}\end{equation}
The second form is sometimes convenient if, as in Eq.~\eqref{eq:strongScat3Momentum}, $b$ and $c$ are squares.
Plugging the above expressions for $t$ and $u$ into Eq.~\eqref{eq:strongScatMandelstamSum} leads to
\begin{equation}
	E_{1} = \frac{s + \Delta_{12}}{2 \sqrt{s}}, \quad E_{2} = \frac{s - \Delta_{12}}{2 \sqrt{s}}, \quad
	E_{3} = \frac{s + \Delta_{34}}{2 \sqrt{s}}, \quad E_{4} = \frac{s - \Delta_{34}}{2 \sqrt{s}}\eolc
	\label{eq:strongScatEnergies}
\end{equation}
where we introduced $\Delta_{ij} = m_i^2 - m_j^2$. Inserting Eqs.~\eqref{eq:strongScat3Momentum} and
\eqref{eq:strongScatEnergies} in Eq.~\eqref{eq:strongScatMandelstamCOM}, we finally obtain
\begin{equation}\begin{split}
	t = \frac{\sum m_i^2 - s}{2} - \frac{\Delta_{12}\Delta_{34}}{2s} 
						+ \frac{\lambda^{1/2}(s,m_1^2,m_2^2)\, \lambda^{1/2}(s,m_3^2,m_4^2)}{2s} \cos \theta_s\eolc\\[2mm]
	u = \frac{\sum m_i^2 - s}{2} + \frac{\Delta_{12}\Delta_{34}}{2s} 
						- \frac{\lambda^{1/2}(s,m_1^2,m_2^2)\, \lambda^{1/2}(s,m_3^2,m_4^2)}{2s} \cos \theta_s\eolp
	\label{eq:strongScattu}
\end{split}\end{equation}
As an immediate consequence, we get for the scattering angle
\begin{equation}
	\cos \theta_s = \frac{s(t-u) + \Delta_{12}\Delta_{34}}
									{\lambda^{1/2}(s,m_1^2,m_2^2)\, \lambda^{1/2}(s,m_3^2,m_4^2)}\eolp
	\label{eq:strongScatsChannelScatteringAngle}
\end{equation}

With the Mandelstam variables in the form of Eqs.~\eqref{eq:strongScatMandelstamCOM} and \eqref{eq:strongScattu}, it
is now straightforward to find the physically allowed values. The centre-of-mass momentum $s$ must satisfy both of the
relations
\begin{align}\begin{split}
	s &= (E_1 + E_2)^2 = \left( \sqrt{m_1 + |\vec{p}_1|^2} + \sqrt{m_2 + |\vec{p}_1|^2} \right)^2 
											\geq \left( m_1 + m_2 \right)^2 \eolc \\
	s &= (E_3 + E_4)^2 = \left( \sqrt{m_3 + |\vec{p}_3|^2} + \sqrt{m_4 + |\vec{p}_3|^2} \right)^2 
											\geq \left( m_3 + m_4 \right)^2 \eolc
\end{split}\end{align}
that is
\begin{equation}
	s \geq \max \left( (m_1+m_2)^2, (m_3+m_4)^2 \right) \eolp
\end{equation}
For each value of $s$ in this range we can then find the range of allowed values for $t$ and $u$ from the requirement
that $-1 \leq \cos \theta_s \leq 1$. This range of values for the kinematic variables is called the
physical region.

The amplitude possesses a very simple symmetry: it is unaffected by swapping the two incoming particles. This means that
the amplitude is invariant under the simultaneous interchange of $p_1$ with $p_2$ as well as of all the quantum numbers
that characterize the particle $A$ with their counterparts for $B$. In terms of the Mandelstam variables, this
corresponds to interchanging $t$ and $u$ and the amplitude must thus satisfy
\begin{equation}
	\A(s,t,u) = \A(s,u,t) \eolp
\end{equation}

So far we have only considered the so-called $s$-channel process, where the invariant total momentum of the incoming 
particles is given by $s$. But other processes can also be described by Fig.~\ref{fig:strong22scattering}, namely
\begin{alignat}{2}
	A(p_1) \bar{C}(-p_3) &\to \bar{B}(-p_2) D(p_4),&\hspace{3em}&\text{($t$-channel)}, \label{eq:strongScattChannel} \\
	A(p_1) \bar{D}(-p_4) &\to \bar{B}(-p_2) C(p_3),&&\text{($u$-channel)}, \label{eq:strongScatuChannel}
\end{alignat}
where the invariant total momentum of the incoming particles is given by $t$ and $u$, respectively. Their kinematics
can be discussed in complete analogy to the $s$-channel. In the centre of mass frame of the incoming (or
outgoing) particles, we can express the Mandelstam variables in terms of the centre-of-mass energy (which is, of course,
itself one of the Mandelstam variables) and the corresponding scattering angle. The latter is defined as $\theta_t =
\angle(\vec{p}_1,-\vec{p}_2)$ or $\theta_u = \angle(\vec{p}_1,\vec{p}_3)$, respectively%
\footnote{Again, we must choose from four equivalent definitions for the scattering angle. This choice is convenient
because the scattering angles in the $t$- and $u$-channel are related to $\theta_s$ by $\vec{p}_2 \leftrightarrow
-\vec{p}_3$ and $\vec{p}_2 \leftrightarrow -\vec{p}_4$, respectively.}%
. But it is not even necessary to redo the calculations. One can get the formul\ae{} for $t$- and $u$-channel from the
$s$-channel expressions by applying the appropriate variable exchanges. We obtain in the $t$-channel:
\begin{equation}\begin{split}
	s = \frac{\sum m_i^2 - t}{2} - \frac{\Delta_{13}\Delta_{24}}{2t} + \frac{\lambda^{1/2}(t,m_1^2,m_3^2)\,
\lambda^{1/2}(t,m_2^2,m_4^2)}{2t} \cos \theta_t\eolc\\[2mm]
	u = \frac{\sum m_i^2 - t}{2} + \frac{\Delta_{13}\Delta_{24}}{2t} - \frac{\lambda^{1/2}(t,m_1^2,m_3^2)\,
\lambda^{1/2}(t,m_2^2,m_4^2)}{2t} \cos \theta_t\eolc
	\label{eq:strongScatsu}
\end{split}\end{equation}
and in the $u$-channel:
\begin{equation}\begin{split}
	s = \frac{\sum m_i^2 - u}{2} - \frac{\Delta_{14}\Delta_{23}}{2u} - \frac{\lambda^{1/2}(u,m_1^2,m_4^2)\,
\lambda^{1/2}(u,m_2^2,m_3^2)}{2u} \cos \theta_u\eolc\\[2mm]
	t = \frac{\sum m_i^2 - u}{2} + \frac{\Delta_{14}\Delta_{23}}{2u} + \frac{\lambda^{1/2}(u,m_1^2,m_4^2)\,
\lambda^{1/2}(u,m_2^2,m_3^2)}{2u} \cos \theta_u\eolp
	\label{eq:strongScatst}
\end{split}\end{equation}
The physical regions for the three related processes are distinct. The principle of crossing, sometimes also called the
substitution law, now states that the invariant amplitudes of all three processes are described by a single analytic
function $\A(s,t,u)$. This statement is only useful if an analytic continuation between the physical regions of
the different channels exists.

\begin{figure}[t!]
\begin{center}
	\psset{unit=.86cm,linewidth=\mylw}
	\begin{pspicture*}(-1,-1)(11,11)
		\psline(-1,2)(11,2)
		\psline(0.42,-1)(7.35,11)
		\psline(9.58,-1)(2.65,11)
\pscurve[fillstyle=hlines](5.02,10.07)(4.6,10.15)(4.43,10.23)(4.29,10.31)(4.18,10.39)(4.07,10.47)(3.97,10.55)(3.88,
10.63)(3.8,10.71)(3.72,10.79)(3.64,10.87)(3.56,10.95)(3.49,11.03)(3.42,11.11)(3.35,11.19)(3.28,11.27)(6.75,11.27)(6.68,
11.19)(6.61,11.11)(6.54,11.03)(6.47,10.95)(6.39,10.87)(6.31,10.79)(6.23,10.71)(6.15,10.63)(6.06,10.55)(5.96,10.47)(5.86,
10.39)(5.74,10.31)(5.6,10.23)(5.43,10.15)(5.02,10.07)
\pscurve[fillstyle=hlines](1.88,1.53)(1.7,1.68)(1.58,1.73)(1.47,1.76)(1.37,1.78)(1.26,1.8)(1.16,1.82)(1.06,1.83)(0.96,
1.84)(0.87,1.85)(0.77,1.86)(0.67,1.86)(0.58,1.87)(0.48,1.87)(0.39,1.88)(0.29,1.88)(0.2,1.89)(0.1,1.89)(0.01,1.9)(-0.09,
1.9)(-0.18,1.9)(-0.28,1.91)(-0.37,1.91)(-0.46,1.91)(-0.56,1.91)(-0.65,1.92)(-0.75,1.92)(-0.84,1.92)(-0.93,1.92)(-1.03,
1.93)(-1.12,1.93)(-1.21,1.93)(-1.31,1.93)(-1.4,1.93)(-1.49,1.93)(-1.59,1.94)(-1.68,1.94)(-1.77,1.94)(-1.87,1.94)(-1.96,
1.94)(0.21,-1.83)(0.26,-1.75)(0.31,-1.67)(0.36,-1.59)(0.4,-1.51)(0.45,-1.43)(0.5,-1.35)(0.54,-1.28)(0.59,-1.2)(0.64,
-1.12)(0.69,-1.04)(0.73,-0.96)(0.78,-0.88)(0.83,-0.8)(0.87,-0.73)(0.92,-0.65)(0.97,-0.57)(1.01,-0.49)(1.06,-0.41)(1.11,
-0.33)(1.16,-0.25)(1.2,-0.17)(1.25,-0.09)(1.29,-0.01)(1.34,0.07)(1.39,0.15)(1.43,0.23)(1.48,0.31)(1.52,0.39)(1.57,
0.47)(1.61,0.55)(1.66,0.64)(1.7,0.72)(1.74,0.81)(1.79,0.9)(1.82,0.99)(1.86,1.09)(1.89,1.19)(1.92,1.31)(1.88,1.53)
\pscurve[fillstyle=hlines](7.12,2.07)(7.21,2.07)(7.3,2.06)(7.39,2.06)(7.48,2.06)(7.58,2.06)(7.67,2.06)(7.76,2.06)(7.85,
2.06)(7.94,2.06)(8.04,2.05)(8.13,2.05)(8.22,2.05)(8.31,2.05)(8.4,2.05)(8.5,2.05)(8.59,2.05)(8.68,2.05)(8.77,2.05)(8.86,
2.05)(8.96,2.05)(9.05,2.05)(9.14,2.05)(9.23,2.04)(9.32,2.04)(9.42,2.04)(9.51,2.04)(9.6,2.04)(9.69,2.04)(9.78,2.04)(9.88,
2.04)(9.97,2.04)(10.06,2.04)(10.15,2.04)(10.24,2.04)(10.34,2.04)(10.43,2.04)(10.52,2.04)(10.61,2.04)(10.71,2.04)(10.8,
2.04)(10.89,2.04)(10.98,2.03)(11.07,2.03)(11.17,2.03)(11.26,2.03)(11.35,2.03)(11.44,2.03)(11.54,2.03)(11.63,2.03)(11.72,
2.03)(11.81,2.03)(11.9,2.03)(12.,2.03)(9.74,-1.87)(9.69,-1.8)(9.65,-1.72)(9.6,-1.64)(9.55,-1.57)(9.5,-1.49)(9.45,
-1.41)(9.41,-1.33)(9.36,-1.26)(9.31,-1.18)(9.26,-1.1)(9.21,-1.03)(9.17,-0.95)(9.12,-0.88)(9.07,-0.8)(9.02,-0.72)(8.97,
-0.65)(8.92,-0.57)(8.87,-0.49)(8.83,-0.42)(8.78,-0.34)(8.73,-0.27)(8.68,-0.19)(8.63,-0.12)(8.58,-0.04)(8.53,0.03)(8.48,
0.11)(8.43,0.18)(8.39,0.26)(8.34,0.33)(8.29,0.41)(8.24,0.48)(8.19,0.56)(8.14,0.63)(8.09,0.71)(8.04,0.78)(7.99,0.85)(7.94
,0.93)(7.89,1.)(7.84,1.07)(7.79,1.15)(7.74,1.22)(7.69,1.29)(7.64,1.36)(7.59,1.43)(7.53,1.51)(7.48,1.58)(7.43,1.65)(7.38,
1.72)(7.33,1.79)(7.28,1.86)(7.22,1.93)(7.17,2.)(7.12,2.07)

		\rput(9,2.5){\small $u$-channel}
		\rput(5,9.5){\small $s$-channel}
		\rput(2.95,0.8){\small $t$-channel}
		\rput(0.3,2.3){\footnotesize{$s=0$}}
		\rput{-60}(2.8,10){\footnotesize{$t=0$}}
		\rput{60}(7.2,10){\footnotesize{$u=0$}}
		\psline{->}(4.5,2)(4.5,4)
		\psarc{-}(4.5,2){0.25}{0}{90}  
		\psdot[linewidth=0.1pt](4.6,2.1)
		\psline{->}(6.143,4.949)(4.5,4)
		\psarc{-}(6.143,4.949){0.25}{210}{300}
		\psdot[linewidth=0.1pt](6.11,4.82)
		\psline{->}(3.607,4.516)(4.5,4)
		\psarc{-}(3.607,4.516){0.25}{-120}{-30}
		\psdot[linewidth=0.1pt](3.65,4.38)
		\rput(4.7,3){\small $s$}
		\rput(5.5,4.3){\small $t$}
		\rput(4.1,4.5){\small $u$}
	\end{pspicture*}
\end{center}
\caption{Representation of the Mandelstam variables $s$, $t$ and $u$ in a Mandelstam diagram. The
arrows point in the direction of positive values. The shaded region represents the physical region for a generic
scattering process with $m_1 = 1$, $m_2 = 1.5$, $m_3 = 2.5$ and $m_4 = 3$. Note that the physical regions for the three
channels are distinct.}
\label{fig:strongMandelstamDiagram}
\end{figure}
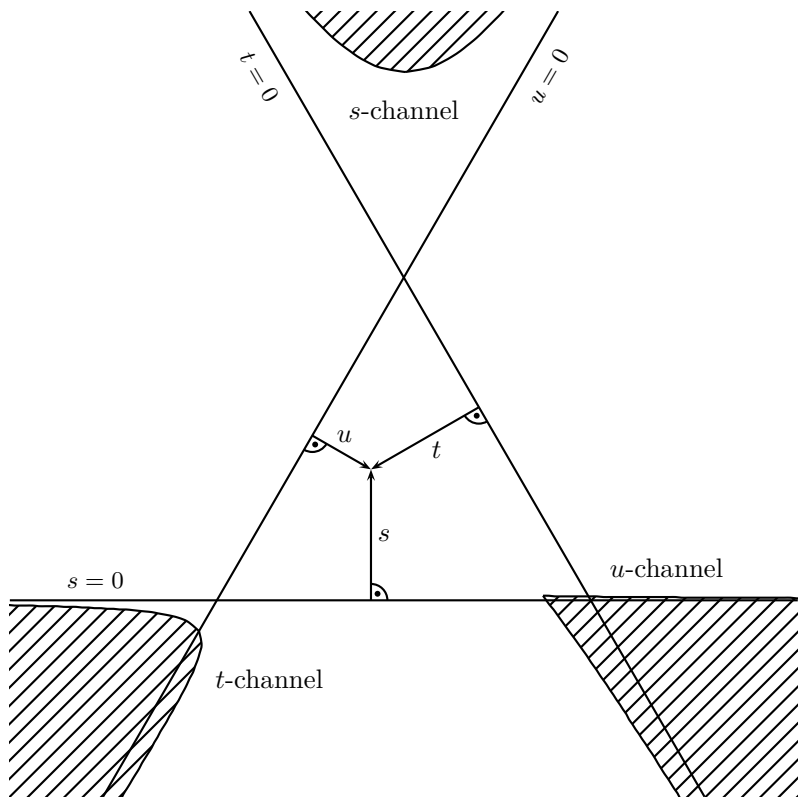

There is a convenient way to represent the $s$-$t$-$u$ plane graphically. For a given point
inside an equilateral triangle, the sum of the perpendicular distances to each side is equal to the height of the
triangle. This remains true for points outside of the triangle, if the distances pointing away from the triangle are
regarded as negative. Choosing the sides of the triangle to be the coordinate axes and the perpendicular distances to be
$s$, $t$ and $u$, the constraint~\eqref{eq:strongScatMandelstamSum} on the sum of the Mandelstam variables is
automatically satisfied, if the height of the triangle is equal to $\sum m_i^2$. Such a Mandelstam diagram is shown in
Fig.~\ref{fig:strongMandelstamDiagram}\footnote{I thank Peter Stoffer, who made his figure from Ref.~\cite{Stoffer2010}
available, which formed the basis for all the Mandelstam diagrams in this thesis.}.

\subsection{Maximal analyticity} \label{sec:strongAnalyticity}

Based on the previous sections, we can now discuss the analytic properties of the scattering amplitude. The importance
of the analytic properties lies in the fact that they are supposed to be fundamental like unitarity or crossing. Later
it will become clear that they form the essential input to the dispersion relations.

\begin{figure}
	\SetScale{0.6}
	\setlength{\unitlength}{0.6pt}
	\begin{center}
	\subfigure[]{
		\begin{picture}(258,130) (15,-15)
			\SetWidth{1.2}
			\SetColor{Black}
			\Line[arrow,arrowpos=0.5,arrowlength=5,arrowwidth=2,arrowinset=0.2](16,114)(80,50)
			\Line[arrow,arrowpos=0.5,arrowlength=5,arrowwidth=2,arrowinset=0.2](16,-14)(80,50)
			\Line[arrow,arrowpos=0.5,arrowlength=5,arrowwidth=2,arrowinset=0.2](80,50)(208,50)
			\Line[arrow,arrowpos=0.5,arrowlength=5,arrowwidth=2,arrowinset=0.2](208,50)(272,114)
			\Line[arrow,arrowpos=0.5,arrowlength=5,arrowwidth=2,arrowinset=0.2](208,50)(272,-14)
		\end{picture}
		\label{fig:strongIntermediateState1}
	}
	\hspace{1em}
	\subfigure[]{
		\begin{picture}(258,130) (15,-15)
			\SetWidth{1.2}
			\SetColor{Black}
			\Line[arrow,arrowpos=0.5,arrowlength=5,arrowwidth=2,arrowinset=0.2](16,114)(80,50)
			\Line[arrow,arrowpos=0.5,arrowlength=5,arrowwidth=2,arrowinset=0.2](16,-14)(80,50)
			\Line[arrow,arrowpos=0.5,arrowlength=5,arrowwidth=2,arrowinset=0.2](208,50)(272,114)
			\Line[arrow,arrowpos=0.5,arrowlength=5,arrowwidth=2,arrowinset=0.2](208,50)(272,-14)
			\Arc[arrow,arrowpos=0.5,arrowlength=5,arrowwidth=2,arrowinset=0.2,clock](144,31.333)(66.667,163.74,16.26)
			\Arc[arrow,arrowpos=0.5,arrowlength=5,arrowwidth=2,arrowinset=0.2](144,68.667)(66.667,-163.74,-16.26)
		\end{picture}
		\label{fig:strongIntermediateState2}
  	}
  	\end{center}
  	\caption{Two examples of Feynman diagrams that may contribute to a $2 \to 2$ scattering process.}
  	\label{fig:strongIntermediateState}
\end{figure}
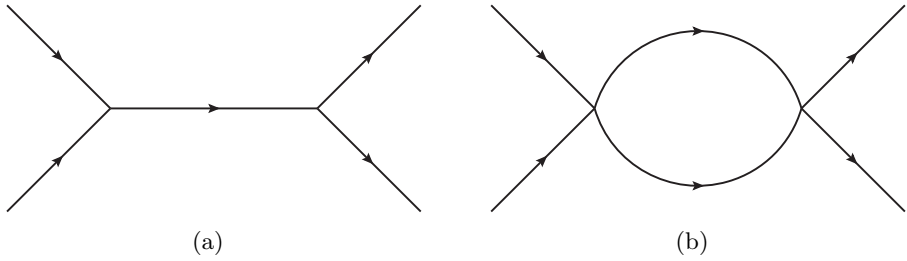

The scattering amplitude contains two kinds of singularities that originate from the presence  of intermediate
states. The first kind is due to single particle intermediate states that contribute through processes of the type
depicted in Fig.~\ref{fig:strongIntermediateState1}. Taking the incoming momentum to be $s$, then, if the exchanged
particle has mass $M$, its propagator is
\begin{equation}
	\inv{s - M^2} \eolc
\end{equation}
such that the intermediate state contributes an isolated pole at $s=M^2$ to the amplitude. On the other hand, diagrams
of the kind shown in Fig.~\ref{fig:strongIntermediateState2} lead to cuts in the amplitude. From the unitarity condition
in Eq.~\eqref{eq:strongUnitarityCondition1} follows that whenever the initial momentum $p_i$ crosses the threshold for
the production of a new multiparticle intermediate state, the transition amplitude $\T_{fi}$ obtains an additional
contribution to the imaginary part. For $s$ below the threshold of the lowest lying state, the amplitude must be real
and can, at a fixed value for $t$, be represented as a polynomial in $s$ with real coefficients that may depend on $t$%
\footnote{Since the Mandelstam variables are related, the amplitude is for fixed $t$ a function of one variable only,
which we choose to be $s$.}.
As a consequence, for complex values of $s$ and real $t$ the amplitude $\A(s,t,u)$ satisfies the Schwarz reflection
principle:
\begin{equation}
	\A(s^*,t,u(s^*)) = \A^*(s,t,u(s)) \eolp
\end{equation}
If the amplitude picks up an imaginary part for some real values of $s$, the Schwarz reflection principle immediately
implies the appearance of a cut, since then
\begin{equation}
	\A(s-i \epsilon,t,u) = \A^*(s+i \epsilon,t,u) \stackrel{\Im \A \neq 0}{\neq} \A(s+i \epsilon,t,u) \eolp
\end{equation}
Taking the limit $\epsilon \to 0$, we find that the amplitude is not uniquely defined on the
real $s$ axis. Each production threshold is thus the starting point of a new cut---a so-called branch point. We have
only discussed the $s$ channel for now, but both kinds of singularities appear, of course, in each of the Mandelstam
variables.

The poles and cuts of the amplitude due to intermediate states are called dynamical singularities. The postulate of
\emph{maximal analyticity} states that the amplitude contains no other singularities than these. As we will see later
this is a powerful constraint on the amplitude.

\section{The strong interaction as a quantum field theory}

In this section, we give a very brief introduction to the quantum field theory of the strong interaction at low
energies. A short overview of quantum chromodynamics is followed by a sketchy derivation and discussion of chiral
perturbation theory, focussing on the aspects that will become relevant later in this thesis. A highly recommended
pedagogical introduction to chiral perturbation theory can be found in Ref.~\cite{Scherer2002}. The following pages 
are largely based on that article.

\subsection{Quantum chromodynamics}

\begin{table}[tb]
\renewcommand{\arraystretch}{1.7}
\begin{center}
\begin{tabular}{lccccc}
			& Mass in MeV 						& $Q$ & $I$ & $I_3$ & $J^P$ \\ \hline
	$u$	& 1.7--3.3							& $+\frac{2}{3}$	& $\inv{2}$ 	& $+\inv{2}$ 	& $\inv{2}^+$ \\
	$d$	& 4.1--5.8							& $-\frac{1}{3}$	& $\inv{2}$ 	& $-\inv{2}$ 	& $\inv{2}^+$ \\
	$s$	& $101^{+29}_{-21}$				& $-\frac{1}{3}$	& 0			 & \phantom{+}0	& $\inv{2}^+$ \\
	$c$	& $1270^{+70}_{-90}$				& $+\frac{2}{3}$	& 0			 & \phantom{+}0	& $\inv{2}^+$ \\
	$b$	& $4190^{+180}_{-60}$			& $-\frac{1}{3}$	& 0			 & \phantom{+}0	& $\inv{2}^+$ \\
	$t$	& $(172 \pm 1.6) \cdot 10^3$	& $+\frac{2}{3}$	& 0			 & \phantom{+}0	& $\inv{2}^+$
\end{tabular}
\end{center}
\caption{The six quark flavours together with their $\msbar$ masses and some quantum numbers~\cite{PDG2010}.
			The isospin quantum numbers $I$ and $I_3$ will be explained at the end of
			Sec.~\ref{sec:strongGlobalSymmetries}.}
\label{tab:strongQuarks}
\end{table}

The theory that successfully describes the dynamics of the strong interaction is quantum chromodynamics (QCD)
\cite{Weinberg1973,Fritzsch+1973}. It is a gauge theory and its gauge group, $SU(3)_c$, is referred to as colour. The
fundamental degrees of freedom are the quarks and gluons and the Lagrangian is given by
\begin{equation}
	\lqcd = \sum_{f} \bar{q}_f \left( i \gamma_\mu D^\mu - m_f \right) q_f 
							- \inv{4} \G_{\mu \nu, a} \G^{\mu \nu}_a \eolc
\end{equation}
where the sum goes over the six quark flavours $f=u,d,s,c,b,t$ (up, down, strange, charm, bottom, top). The masses of
the quarks and some of their quantum numbers are listed in Table~\ref{tab:strongQuarks}.
The quark fields transform under the fundamental representation of the gauge group, thus forming a colour triplet. For a
quark field $q_f$ of flavour $f$ this means that
\begin{equation}
	q_f = \vect{ q_{f,r} \\ q_{f,g} \\ q_{f,b} } \mapsto 
		\vect{ q'_{f,r} \\ q'_{f,g} \\ q'_{f,b} } = U(x) q_f \eolc \qquad U \in SU(3)_c \eolc
\end{equation}
where the the second label on the quark fields is the quantum number for colour that takes the values $r$ for red,
$g$ for green and $b$ for blue. Quarks that belong to the same triplet only differ in their color charge and are
identical in all other properties as, e.g., mass, and electric charge. A colour transformation of the quark
fields, being an $SU(3)$ matrix, is characterised by a set of 8 parameters $\theta_a(x)$, $a=1,\ldots,8$, according to
\begin{equation}
	U(x) = \exp \left( -i \sum_{a=1}^8 \theta_a(x) \frac{\lambda_a}{2} \right) \eolp
\end{equation}
The $\lambda_a$ are the $3 \times 3$ Gell-Mann matrices given by
\begin{equation}
 \begin{array}{lll}
 		\lambda_1 = \mat[r]{3}{0&1&0\\1&0&0\\0&0&0} \eolc
	&  \lambda_2 = \mat[r]{3}{0&-i&0\\i&0&0\\0&0&0} \eolc
	&  \lambda_3 = \mat[r]{3}{1&0&0\\0&-1&0\\0&0&0} \eolc \\
	\vspace{-1mm} &&\\
 		\lambda_4 = \mat[r]{3}{0&0&1\\0&0&0\\1&0&0} \eolc
	&  \lambda_5 = \mat[r]{3}{0&0&-i\\0&0&0\\i&0&0} \eolc
	&  \lambda_6 = \mat[r]{3}{0&0&0\\0&0&1\\0&1&0} \eolc \\
	\vspace{-1mm} &&\\
 		\lambda_7 = \mat[r]{3}{0&0&0\\0&0&-i\\0&i&0} \eolc
	&  \multicolumn{2}{l}{\lambda_8 = \dfrac{1}{\sqrt{3}} \mat[r]{3}{1&0&0\\0&1&0\\0&0&-2} \eolp}
 \end{array}
\end{equation}
They form a basis of the set of all complex, traceless, Hermitian $3 \times 3$ matrices. The matrices $i \lambda_a$ are
antihermitian and thus constitute a basis of the Lie algebra $\textit{su}(3)$ of $SU(3)$. If they are supplemented by
the unit matrix, here for convenience denoted by $\lambda_0 = \sqrt{2/3}\; \text{diag}(1,1,1)$, an arbitrary $3 \times
3$ matrix $M$ can be written as
\begin{equation}
	M = \sum_{a=0}^8 \lambda_a M_a \eolc
\end{equation}
where the complex coefficients $M_a$ are given by
\begin{equation}
	M_a = \inv{2} \tr \left( \lambda_a M \right) \eolp
	\label{prel:Ma}
\end{equation}

The gauge symmetry is a local symmetry such that the kinetic term in $\lqcd$ would not be gauge invariant, if it
contained an ordinary partial derivative $\partial^\mu$. Instead, it must be written with the covariant derivative
$D^\mu$ which is defined by the requirement
\begin{equation}
	D_\mu q_f \mapsto U(x) (D_\mu q_f) \eolc 
\label{prel:covDerivativeTransf}
\end{equation}
that is, the covariant derivative of the quark field transforms in exactly the same way as the quark field itself under
gauge transformations. The covariant derivative with this property is given by
\begin{equation}
	D_\mu q_f = \partial_\mu q_f - i g \sum_{a=1}^8 \frac{\lambda_a}{2}\, \A_{\mu,a}\, q_f \eolc
\label{prel:covDerivative}
\end{equation}
where $g$ is the strong coupling constant and the $\A_{\mu,a}$ are eight independent gauge potentials that describe the
gluons. The interaction between the quarks and gluons is entirely due to the gluon field in the covariant derivative.
Its strength is determined by $g$ and it is independent of the quark flavour. The covariant derivative
transforms according to Eq.~\eqref{prel:covDerivativeTransf} provided that the gauge field transforms as
\begin{equation}
	\frac{\lambda_a}{2}\, \A_{\mu,a} \mapsto U(x) \frac{\lambda_a}{2}\, \A_{\mu,a} U\dega(x)
			-\frac{i}{g} \left( \partial_\mu U(x) \right) U\dega(x) \eolp
\end{equation}
Finally, $\L_\text{QCD}$ also contains the non-Abelian field-strength tensor,
\begin{equation}
	\G_{\mu\nu,a} = \partial_\mu \A_{\nu,a} - \partial_\nu \A_{\mu,a} + g f_{abc} \A_{\mu,b} \A_{\nu,c} \eolc
\end{equation}
where $f_{abc}$ are the structure constants of $SU(3)$. From its square follows the kinetic term for the gluon field as
well as the gluon-gluon interaction terms.

Even though the QCD Lagrangian does, in principle, fully determine the dynamics of the quarks and gluons, it is not
possible to make full use of its power due to a number of reasons. A theory should be able to produce predictions
on scattering cross sections and decay widths, which can be compared with experimental data and hence allow for a
rigorous test of its accuracy. Such measurable quantities are in field theories usually obtained using the method of
perturbation theory. The free theory, where all interactions are turned off, is solved exactly and the interaction term
in the Lagrangian is then added as a perturbation, which is small due to the smallness of the coupling constant.
Results in the interacting theory are then expressed as a power series in the coupling constant. Since the coupling
constant is small, the size of the contributions diminishes rapidly with increasing order, leading to a quick
convergence of the series.

While the procedure sketched above proved extremely successful in quantum electrodynamics, where predictions
of astonishing accuracy have been achieved, the situation is less favourable in QCD. The degrees of freedom of the free
theory---the quarks and gluons---have never been observed in nature, instead one finds a huge number of bound states.
While the free theory of QED already represents a first, if crude, approximation to the real world situation, the free
theory of QCD, describing free quarks and gluons, seems to have nothing in common with nature. It is nowadays
believed that this is a consequence of a mechanism called colour confinement. The strong force does not, as one would
expect from the electromagnetic or gravitational interactions, decrease but grow with distance, such that particles
carrying colour charge can never be separated and the spectrum only contains states with neutral colour charge.
Confinement has not been mathematically rigorously proved. Its proof is, in fact, part of
one of the seven famous Millennium Prize Problems stated by the Clay Mathematics Institute~\cite{Jaffe+2000}. 

Another peculiarity of QCD (and other non-Abelian Yang-Mills theories) saves perturbation theory at least partially:
asymptotic freedom~\cite{Gross+1973,Gross+1973a,Politzer1973,Gross1999}.
In processes that take place at very high energies, the strong interactions become weak and quarks and gluons start to
behave as free particles allowing for a perturbative expansion around the free theory. But while this permits to obtain
theoretical predictions for processes at high energies, low-energy QCD is still not accessible. Based on a proposal by
Weinberg~\cite{Weinberg1979}, Gasser and Leutwyler developed the framework of chiral perturbation theory
\cite{Gasser+1985,Gasser+1984} that makes perturbative calculations possible also at low energies. Rather than on an
expansion in the coupling constant, this method relies on a perturbative series in terms of the small external momenta.
In the following subsections, we will very briefly introduce the basic concepts and equations.

\subsection{Accidental global symmetries of the QCD Lagrangian} \label{sec:strongGlobalSymmetries}

In the quark model, the proton is understood to be a bound state of two up- and one down-quark. Comparing the masses of
these constituent particles with the proton mass reveals the astonishing fact that
\begin{equation}
	2 m_u + m_d \approx 10~\MeV \ll m_p \approx 938~\MeV \eolp
	\label{pipi:protonMass}
\end{equation}
Other than the mass of an atom, which is given by the sum of the proton and electron masses minus a very small
amount of binding energy, the contribution of the quark masses to the proton mass is negligible. Almost the entire mass
of the proton must be due to binding energy, such that the scale that determines the strength of the strong
interactions, $\ldqcd$, is of the order of $m_p \approx 1~\GeV$. Low-energy QCD is concerned with processes at energies
well below $\ldqcd$. The list of the quarks in Table~\ref{tab:strongQuarks} reveals that this scale divides the quarks
in two groups: the $u$, $d$ and $s$ quarks have masses that lie substantially below $\ldqcd$ and are accordingly called
light. The remaining three quark flavours are called heavy.

The lightest particle that contains one of the heavy quarks is the $D^0$, being a bound state of a $c$-quark and a
$\bar{u}$-antiquark, with a mass of about 1865~MeV. A reaction that is energetic enough to create such a particle thus
takes place well outside of the low-energy region of QCD that we are interested in. In the following we will completely
ignore the occurrence of such states and neglect the heavy quarks in the Lagrangian. For this purpose it is convenient
to introduce the vector of the light quark fields,
\begin{equation}
	q = \vect{ u \\ d \\ s } \eolc \qquad \bar{q} = \left( \bar{u}, \bar{d}, \bar{s} \right) \eolc
\end{equation}
and the corresponding quark mass matrix,
\begin{equation}
	\M = \mat{3}{m_u&0&0\\0&m_d&0\\0&0&m_s} \eolp
\end{equation}
The QCD Lagrangian without heavy quarks can then be expressed as
\begin{equation}
	\lqcd = \bar{q} ( i \gamma_\mu D^\mu - \M )q - \inv{4} \G_{\mu \nu, a} \G^{\mu \nu}_a \eolp
	\label{prel:LQCD}
\end{equation}
Equation~\eqref{pipi:protonMass} even suggests that the light quark masses can, in a first approximation, be neglected.
The QCD Lagrangian in this so-called chiral limit---the name will become clear later---is given by 
\begin{equation}
	\lqcd^0 = \bar{q} i \gamma_\mu D^\mu q - \inv{4} \G_{\mu \nu, a} \G^{\mu \nu}_a \eolp
\end{equation}
Since the covariant derivative acts on all flavours in the same way, this Lagrangian has a global $U(3)$ symmetry. It
is invariant under $q \mapsto V e^{-i \theta_V} q$, where $V$ is a $SU(3)$ matrix and $\theta_V$ is a real number. But
there is an even larger global symmetry. We define the projection operators
\begin{equation}
	P_R = \frac{1 + \gamma_5}{2} = P_R\dega \eolc \qquad P_L = \frac{1 - \gamma_5}{2} = P_L\dega \eolc
	\label{eq:strongPLR}
\end{equation}
which satisfy,
\begin{equation}
	P_R+P_L=1 \eolc \qquad P_{R/L}^2 = P_{R/L}, \qquad P_R P_L = P_L P_R = 0 \eolc
\end{equation}
as projection operators should. They project the Dirac field onto its left- and right-handed components, respectively,
i.e., $q_L = P_L q$ and $q_R = P_R q$. Using the anti-commutation relations for the gamma matrices one finds that the
QCD Lagrangian in the chiral limit in terms of the chiral fields reads
\begin{equation}
	\lqcd^0 = \bar{q}_R i \gamma_\mu D^\mu q_R + \bar{q}_L i \gamma_\mu D^\mu q_L 
		- \inv{4} \G_{\mu \nu, a} \G^{\mu \nu}_a \eolp
\end{equation}
Because the left- and right-handed fields do not mix, $\lqcd^0$ is invariant under
\begin{equation}
	q_L \mapsto U_L q_L \eolc \qquad q_R \mapsto U_R q_R \eolc
\end{equation}
where $U_L$ and $U_R$ are independent $U(3)$ matrices that can be parametrised in terms of the Gell-Mann matrices as
\begin{equation}
	U_L = \exp \left( -i \sum_{a=1}^8 \theta^L_a \frac{\lambda_a}{2} \right) \, e^{-i \theta^L} \eolc \qquad
	U_R = \exp \left( -i \sum_{a=1}^8 \theta^R_a \frac{\lambda_a}{2} \right) \, e^{-i \theta^R} \eolp
\end{equation}
$\lqcd^0$ thus has a global $U(3)_L \times U(3)_R$ symmetry. Because the symmetry transformation acts on left- and
right-handed fields independently, this is called the \emph{chiral symmetry}.

If the quark masses are not neglected, the Lagrangian contains a mass term, which can be expressed in terms
of the chiral fields as
\begin{equation}
	-\bar{q} \M q = -\bar{q}_L \M q_R -\bar{q}_R \M q_L \eolp
\end{equation}
This term mixes left- and right-handed fields and consequently breaks chiral symmetry. This is the origin of the term
\emph{chiral limit}: it is in this limit of vanishing light quark masses that the QCD Lagrangian exhibits chiral
symmetry.

According to Noether's theorem, from the symmetry arise a total number of 18 conserved currents,
\begin{equation}
	L^{\mu,a} = \bar{q}_L \gamma^\mu \frac{\lambda_a}{2} q_L \eolc \quad 
		L^\mu = \bar{q}_L \gamma^\mu q_L \eolc \quad
	R^{\mu,a} = \bar{q}_R \gamma^\mu \frac{\lambda_a}{2} q_R \eolc \quad 
		R^\mu = \bar{q}_R \gamma^\mu q_R \eolc
\end{equation}
with
\begin{equation}
	\partial_\mu L^{\mu,a} = \partial_\mu R^{\mu,a} = \partial_\mu L^{\mu} = \partial_\mu R^{\mu} = 0 \eolc
\end{equation}
and the same number of conserved charges that are defined from the currents by
\begin{equation}\begin{split}
	Q_L^a(t) = \int d^3x \, L^{0,a}(x) \eolc \qquad &Q_L(t) = \int d^3x \, L^0(x) \eolc \\
	Q_R^a(t) = \int d^3x \, R^{0,a}(x) \eolc \qquad &Q_R(t) = \int d^3x \, R^0(x) \eolp \\
\end{split}\end{equation}
The charge operators are the generators of their respective symmetry group.

Instead of the chiral currents, one often uses the linear combinations
\begin{equation}\begin{aligned}
	V^{\mu,a} &= R^{\mu,a} + L^{\mu,a} =  \bar{q} \gamma^\mu \frac{\lambda_a}{2} q \eolc \qquad
		&V^\mu &= R^\mu + L^\mu = \bar{q} \gamma^\mu q \eolc \\
	A^{\mu,a} &= R^{\mu,a} - L^{\mu,a} =  \bar{q} \gamma^\mu \gamma_5 \frac{\lambda_a}{2} q \eolc
		&A^\mu &= R^\mu - L^\mu  = \bar{q} \gamma^\mu \gamma_5 q \eolc
\end{aligned}
\end{equation}
and the corresponding charges
\begin{equation}
	Q_V^a = Q_R^a + Q_L^a \eolc \quad Q_V = Q_R + Q_L \eolc \quad
		Q_A^a = Q_R^a - Q_L^a \eolc \quad Q_A = Q_R - Q_L \eolp
\end{equation}
Under parity, $V^{\mu,(a)}$ and $A^{\mu,(a)}$ transform as vector and axial-vectors, respectively. The singlet vector
current $V^\mu$ is due to a transformation of the left- and right-handed fields by the same phase, while the singlet
axial-vector current $A^\mu$ comes from a transformation with opposite phase. All these currents are conserved at the
classical level, but after quantisation it turns out that the axial-vector current is anomalous and thus not conserved
\cite{Adler1969,Adler+1969,Bardeen1969,Bell+1969}:
\begin{equation}
	\partial_\mu A^\mu = \frac{3 g^2}{32 \pi^2} \epsilon_{\mu \nu \rho \sigma} \G^{\mu \nu}_a \G^{\rho \sigma}_a \eolp
\end{equation}
The transformation corresponding to the singlet axial-vector current is thus not a symmetry of QCD. All the other
currents are conserved also at the quantum level and thus QCD has in the chiral limit the global symmetry 
$SU(3)_L \times SU(3)_R \times U(1)_V$. It is generated by the charge operators $Q_R^a$, $Q_L^a$, (or $Q_V^a$, $Q_A^a$)
and $Q_V$. Note that while the spaces associated with $Q_R^a$, $Q_L^a$, and $Q_V^a$ are subgroups of the full global
symmetry group, namely $SU(3)_R$, $SU(3)_L$, and $SU(3)_V$, the space generated by $Q_A^a$ is not a group because the
product of two pseudoscalar operators is not a pseudoscalar itself.

The quark masses explicitly break chiral symmetry. Apart from the fact that the Lagrangian is no longer invariant under
a chiral transformation, this also shows up in the divergences of the currents:
\begin{equation}\begin{aligned}
	\partial_\mu V^{\mu,a} &= i \bar{q} \left[ \M,\frac{\lambda_a}{2} \right] q \eolc \qquad
		&\partial_\mu V^\mu &=  0 \eolc \\[1mm]
	\partial_\mu A^{\mu,a} &= i \bar{q} \left\{ \M,\frac{\lambda_a}{2} \right\} \gamma_5 q \eolc
		&\partial_\mu A^\mu &= 2 i \bar{q} \M \gamma_5 q 
		+ \frac{3 g^2}{32 \pi^2} \epsilon_{\mu \nu \rho \sigma} \G^{\mu \nu}_a \G^{\rho \sigma}_a \eolp
		\label{eq:strongCurrentDivergences}
\end{aligned}\end{equation}
In the presence of quark masses, only the singlet vector current is still conserved, implying that also the massive
Lagrangian possesses the corresponding symmetry. That this is indeed true can be directly verified in
Eq.~\eqref{prel:LQCD}. But, as it turns out, not all is lost. Interpreting the small quark masses as a perturbation on
$\lqcd^0$ and thus expanding physical quantities in the quark masses leads to a successful description of the QCD
dynamics at low energies.

Apart from chiral symmetry, the QCD Lagrangian has another accidental global symmetry that is of great importance. In
the limit where the $u$- and $d$-quark have the same mass, $\lqcd$ is invariant under the global transformation
\begin{equation}
	\vect{u\\d} \mapsto \vect{u'\\d'} = V \vect{u\\d} \eolc \qquad V \in SU(2) \eolp
\end{equation}
Since the mass difference $m_u-m_d$ is very small compared to $\ldqcd$, this approximate symmetry is rather
accurately realised. Because of the similarity to spin, which is also a $SU(2)$ symmetry, it is called \emph{isospin},
sometimes also isotopic spin or isobaric spin. Note, however, that it is an internal symmetry and has nothing to do with
space.

The symmetry was originally proposed by Heisenberg~\cite{Heisenberg1932} in order to explain the similarity of the
proton and the newly discovered neutron. He suggested that they are two instances of the same particle that simply
differ in the value of some quantum number $\tau$. Only later, the name isotopic spin was introduced by Wigner
\cite{Wigner1937}. 

According to our modern understanding of isospin, the two lightest quarks transform under the fundamental representation
of $SU(2)$, thus forming a doublet. They are assigned a total isospin $I = \inv{2}$; the $u$-quark then has third
component $I_3 = +\inv{2}$, while the $d$-quark has $I_3 = -\inv{2}$. All the other quark flavours are isospin singlets
with $I = I_3 = 0$.

Since the $u$- and the $d$-quark have different electrical charges, the electromagnetic interactions violate isospin
symmetry. Electromagnetic corrections destroy the degeneracy of isospin multiplets as they lead to a
shift in the mass of charged, but not of electrically neutral particles. The Standard Model as a whole becomes symmetric
under isospin transformations if $m_u = m_d$ and if the electromagnetic interactions are turned off, i.e., $e = 0$. This
is called the isospin limit. In this limit, the proton and the neutron indeed have equal masses. In the real world the
degeneracy is lifted due to two effects: Because of the quark mass difference, the neutron is heavier than the proton,
but the mass of the latter is increased by electromagnetic corrections. The two effects almost cancel, resulting in a
neutron mass that is only slightly larger than the proton mass.

\subsection{Spontaneous symmetry breaking and Goldstone bosons} \label{sec:strongGoldstoneBosons}

One would expect that the hadrons organise themselves into approximately degenerate multiplets corresponding to the
irreducible representation of the symmetry group of the QCD Lagrangian, \mbox{$SU(3)_L \times SU(3)_R \times U(1)_V$}.
The $U(1)_V$ symmetry indeed induces an ordering principle: its consequence is baryon number conservation and the
hadrons can accordingly be grouped into mesons (baryon number $B=0$) and baryons ($B=1$). If the axial charges $Q_A^a$
are the generators of a symmetry of QCD, that is if $[Q_A^a,H_{QCD}] = 0$, then one expects that for each positive
parity state in the spectrum, one finds a degenerate negative parity state. This parity doubling is, however, not
observed in nature: a negative parity octet of baryons does not exist. In fact, the hadrons are ordered in multiplets of
$SU(3)$~\cite{Gell-Mann1961}, suggesting that the actual global symmetry group of QCD is $SU(3)_V$. Furthermore, the
lightest pseudoscalar mesons form an octet and have masses much smaller than the corresponding vector mesons. These
observations are strong evidence for the occurrence of spontaneous symmetry breaking
\cite{Nambu1960,Nambu+1961,Nambu+1961a}, as will be argued in the following. 

Let us fist explain what spontaneous symmetry breaking means. A common example that can be visualised rather easily is
the magnetisation of a ferromagnet. The laws of solid state physics that describe the magnet are invariant under spatial
rotations. Above the Curie temperature (or critical temperature) $T_C$, the elementary magnets inside the material are
unordered and point in arbitrary directions such that the magnet is symmetric under the full group of spatial rotations
as well. But if the temperature drops below $T_C$, neighbouring elementary magnets tend to be oriented in the
same direction and a spontaneous magnetisation occurs. The system is now no longer invariant under arbitrary spatial
rotations, but only under rotations around the magnetisation vector. The symmetry is said to be spontaneously broken
from $SO(3)$ to the subgroup $SO(2)$. On the other hand, if the magnet is put in an external magnetic field, the
elementary magnets are aligned along its direction leading to a net magnetisation even if the temperature is above
$T_C$. This phenomenon is called induced magnetisation and the symmetry breaking is said to be explicit. It is caused by
the external field that selects one particular direction. By contrast, the spontaneous magnetisation is oriented
randomly and is independent of any external influence.

In general terms, spontaneous symmetry breaking occurs if a Lagrangian is symmetric under a group $G$, but the ground
state is found to be symmetric under a subgroup $H \subset G$ only. The generators of the coset space $G/H$ do not
annihilate the vacuum. With each one of them is associated a massless state and the theory thus contains a total number
of $n_G - n_H$ massless particles, $n_G$ and $n_H$ being the respective number of generators of the groups $G$ and $H$.
This is the famous Goldstone theorem~\cite{Goldstone1961,Goldstone+1962} and the massless particles are accordingly
called Goldstone bosons. If the group $G$ is not an exact symmetry of the theory, i.e., if some \emph{small} parameter
breaks the symmetry explicitly, $G$ is said to be an approximate symmetry. In that case, the Goldstone bosons pick up a
small mass proportional to the symmetry breaking parameter.

\begin{table}[tb]
\renewcommand{\arraystretch}{1.6}
\begin{center}
\begin{tabular}{lr@{.}lcccccc}
	& \multicolumn{2}{c}{Mass in MeV} & $Q$ & $I$ & $I_3$ & $S$ & $J^P$ & Quark content \\ \hline
	$\pi^+$		& 139&57018	& $+1$ 	& 1 			& $+1$ 		& \psgn0	& $0^-$ & $u\bar{d}$ \\
	$\pi^0$		& 134&9766	& \psgn0	& 1 			& \psgn0		& \psgn0 & $0^-$ & $(u\bar{u} - d\bar{d}\,)/\sqrt{2}$ \\
	$\pi^-$		& 139&57018	& $-1$ 	& 1 			& $-1$ 		& \psgn0	& $0^-$ & $d\bar{u}$ \\
	$K^+$			& 493&677	& $+1$	& $\inv{2}$	& $+\inv{2}$& $+1$	& $0^-$ & $u\bar{s}$ \\
	$K^-$			& 493&677	& $-1$	& $\inv{2}$	& $-\inv{2}$& $-1$	& $0^-$ & $s\bar{u}$ \\
	$K^0$			& 497&614	& \psgn0	& $\inv{2}$	& $-\inv{2}$& $+1$	& $0^-$ & $d\bar{s}$ \\
	$\bar{K}^0$	& 497&614	& \psgn0	& $\inv{2}$	& $+\inv{2}$& $-1$	& $0^-$ & $s\bar{d}$ \\
	$\eta$		& 547&853	& \psgn0	& 0 			& \psgn0		& \psgn0	& $0^-$ & 
		$(u\bar{u} + d\bar{d} - 2 s\bar{s})/\sqrt{6}$
\end{tabular}
\end{center}
\caption{The mesons in the light pseudoscalar octet together with their masses and some quantum numbers~\cite{PDG2010}.}
\label{tab:strongMesons}
\end{table}

Let us now apply the concept of spontaneous symmetry breaking to QCD. The symmetry group of the massless QCD Lagrangian,
$G=SU(3)_L \times SU(3)_R$, is spontaneously broken to the subgroup $H=SU(3)_V$%
\footnote{The symmetry $U(1)_V$ is not broken at all. It is not important for the following arguments and thus
omitted.}%
. There are a number of reasons why the symmetry
breaking must proceed in this way. The absence of parity doubling in nature and the fact that hadrons are organised in
$SU(3)$ multiplets indicate that the actual symmetry of QCD is $SU(3)_V$. According to Goldstone's theorem, with each
broken generator is associated a Goldstone boson with a small mass and with symmetry properties closely related
to those of the generators. The Goldstone bosons must thus be light, pseudoscalar and transform as an octet under
$SU(3)_V$. Indeed, the lightest mesons (see Table~\ref{tab:strongMesons}) have all these properties, making
them excellent candidates for being the Goldstone bosons of the spontaneous symmetry breaking.

It can be shown that there exists an isomorphic mapping between the quotient space $G/H$ and the Goldstone boson
fields. This allows to parametrise the latter in terms of a $3 \times 3$ $SU(3)$ matrix as
\begin{equation}
	U(x) = \exp \left( \frac{i \phi(x)}{F_0} \right) \eolc
\end{equation}
where $F_0$ is the pion decay constant in the chiral limit, defined from the matrix element
\begin{equation}
	\bra{0} A_\mu^a(0) \ket{\phi^b(p)} = i p_\mu F_0 \delta^{ab} 
	\label{eq:strongF0}
\end{equation}
for some single Goldstone boson state $\ket{\phi^b(p)}$ and
\begin{equation}\begin{split}
	\phi(x) = \sum_{a=1}^8 \lambda_a \phi_a(x) 
	&= \mat{3}{%
	\phi_3 + \inv{\sqrt{3}} \phi_8	& \phi_1 - i \phi_2							& \phi_4 - i \phi_5	\\[1mm]
	\phi_1 + i \phi_2						& -\phi_3 + \inv{\sqrt{3}} \phi_8		& \phi_6 - i \phi_7 \\[1mm]
	\phi_4 + i \phi_5						& \phi_6 + i \phi_7							& -\frac{2}{\sqrt{3}} \phi_8
	}\\[3mm]
	&= \mat{3}{%
	\phi_3 + \inv{\sqrt{3}} \phi_8	& \sqrt{2} \pi^+								& \sqrt{2} K^+	\\[3mm]
	\sqrt{2} \pi^-							& -\phi_3 + \inv{\sqrt{3}} \phi_8		& \sqrt{2} K^0 \\[3mm]
	\sqrt{2} K^-							& \sqrt{2} \bar{K}^0							& -\frac{2}{\sqrt{3}} \phi_8
	} \eolp
\end{split}\end{equation}
In the second line we have not yet introduced the fields for the $\pi^0$ and the $\eta$. The reason for doing so is
that the fields $\phi_3$ and $\phi_8$ mix and the physical mesons are thus linear combinations thereof. This is
discussed in detail in Sec.~\ref{sec:strongMesonsMasses}. Under \mbox{$SU(3)_L \times SU(3)_R$} the matrix $U$
transforms as
\begin{equation}
	U(x) \mapsto R U(x) L\dega \eolc \qquad R \in SU(3)_R \eolc \ L \in SU(3)_L \eolp 
\end{equation}
The space of matrices $U$ is no vector space and the mapping does thus not define a representation but rather a
nonlinear realisation of the group. The ground state of the system is given by $\phi(x) = 0$, which corresponds to 
$U(x) = U_0 = 1$, and as expected, it is invariant under the subgroup $SU(3)_V$, where right- and left handed quarks are
rotated by the same matrix $V$:
\begin{equation}
	U_0 \mapsto V U_0 V\dega = V V\dega = 1 = U_0 \eolp
\end{equation}
On the other hand, rotating the left-handed quarks by $A$ and the right-handed quarks by $A\dega$ indeed does change
the ground state:
\begin{equation}
	U_0 \mapsto A\dega U_0 A \dega = A\dega A\dega \neq U_0 \eolp
\end{equation}

\subsection{Chiral perturbation theory}

The Lagrangian for chiral perturbation theory is now formulated in terms of the matrix field $U(x)$, its derivatives
and the quark mass matrix $\M$. In contrast to the QCD Lagrangian, it is expressed in terms of degrees of freedom that
are observed in nature, such that the free theory of non-interacting, massless mesons is indeed a valid first
approximation. Its construction is based on a folk theorem by Weinberg~\cite{Weinberg1979} stating that from the
effective Lagrangian containing all the terms compatible with the symmetries of some underlying theory, one can
calculate perturbatively the most general $S$-matrix consistent with the fundamental principles of quantum field theory
and the underlying theory. The theorem has later been proved in Refs.~\cite{Leutwyler1994,D'Hoker+1994}. From the
symmetry only follows the structure of the terms in the Lagrangian, but not the strength of the respective couplings.
These are encoded in a number of low-energy constants (LECs) that are unconstrained by the symmetry and must be
determined otherwise. They can be obtained, for example, by comparison of theoretical predictions with experimental data
or by direct calculation on the lattice.

The number of terms that can be constructed along these lines is infinite because once one has found a single term
allowed by the symmetry restrictions, any power of this term can be added to the Lagrangian as well. It is therefore
indispensable to find some ordering principle that permits to decide on the importance of the contributions that come
from a given term. Furthermore, even from a Lagrangian with a finite number of terms, an infinite number of Feynman
diagrams is generated and the ordering principle should hence also allow to compare the importance of
two diagrams. This is achieved by Weinberg's power counting scheme~\cite{Weinberg1979} that orders terms and Feynman
diagrams according to their scaling with $p/\ldqcd$. The momentum thus takes the role of the expansion parameter in the
perturbative series. The building blocks of the Lagrangian are counted as
\begin{equation}
	U = \O(p^0) \eolc \quad \partial_\mu U = \O(p) \eolc \quad \partial_\mu \partial_\nu U = \O(p^2) \eolc \quad \ldots
 \eolc \quad \M = \O(p^2) \eolp
\end{equation}
With these simple rules one can determine the chiral order of any term built from the meson fields and the quark
masses. The effective Lagrangian of \chpt\ is a series of terms with an increasing number of derivatives and quark mass
terms,
\begin{equation}
	\L_{\chi \text{PT}} = \L_2 + \L_4 + \L_6 + \ldots \eolc
\end{equation}
where the indices stand for the order in the chiral expansion. Since Lorentz invariance requires derivatives to appear
pairwise, only even chiral orders are possible.

The power counting scheme then also states that a Feynman diagram with $N_L$ loops and $N_{2n}$ vertices from the
Lagrangian $\L_{2n}$ contributes at $\O(p^D)$ with
\begin{equation}
	D = 2 + 2 N_L + \sum_{n=1}^\infty 2(n-1) N_{2n} \eolp
\end{equation}
Clearly, also the chiral order of Feynman diagrams is always an even number. The implication of this formula is that at
$\O(p^2)$, only tree level graphs containing exclusively vertices from $\L_2$ contribute. At $\O(p^4)$ one must add
contributions from tree-level graphs with exactly one $\L_4$-vertex and from one-loop graphs composed only from
$\L_2$-vertices.

The lowest-order Lagrangian is of chiral order $p^2$ and, in the absence of external fields, it reads
\begin{equation}
	\L_2 = \frac{F_0^2}{4} \tr \left( \partial_\mu U \partial^\mu U\dega \right)
			+ \frac{F_0^2 B_0}{2} \tr \left( \M U\dega + U \M \dega \right) \eolc
\end{equation}
where $F_0$ and $B_0$ are two low-energy constants. $F_0$ is the pion decay constant in the chiral limit as defined in
Eq.~\eqref{eq:strongF0} and $B_0$ is related to the chiral quark condensate by
\begin{equation}
	B_0 = -\frac{\langle \bar{q} q \rangle}{3 F_0^2} \eolp
	\label{eq:strongB0}
\end{equation}
The overall normalisation of $\L_2$ is chosen such that the kinetic term for the Goldstone bosons takes standard form,
that is
\begin{equation}
	\inv{2} \partial_\mu \phi_i \, \partial^\mu \phi_i \eolc
\end{equation}
for $i = 1, \ldots, 8$.

The Lagrangians $\L_4$ at $\O(p^4)$~\cite{Gasser+1985} and $\L_6$ at $\O(p^6)$
\cite{Akhoury+1991,Bijnens+1999,Bijnens+2002a,Ebertshauser+2002,Fearing1994} are also known. In principle, the
derivation of these proceeds along the same lines as for the lowest-order Lagrangian, even though the complications
involved increase considerably with each order. The number of independent invariant terms, and thus the number of
low-energy constants, grows rapidly: $\L_4$ contains 12 LECs $(L_1, \ldots, L_{10}, H_1, H_2)$ and $\L_6$ more than
100. In view of the large number of terms involved, two main difficulties appear. One has, on the one hand, to make
sure that none of the allowed terms are forgotten, on the other hand there exists no general algorithm that permits to
decide whether a given set of operators is independent or not. The latter is also reflected in the fact that the number
of independent terms contained in $\L_6$ has decreased with time. The reason for the ever growing number of coupling
constants with increasing chiral order is the non-renormalisability of the theory. At each order, new counter terms must
be introduced in order to cancel the additional divergences that arise. Despite these difficulties, a large number of
processes have been calculated up to $\O(p^6)$; for a review see Ref.~\cite{Bijnens2007}.

\subsection{Lowest-order meson masses and mixing} \label{sec:strongMesonsMasses}

Expanding the matrix $U(x)$ up to quadratic order,
\begin{equation}
	U(x) = 1 + i \frac{\phi(x)}{F_0} - \inv{2} \frac{\phi^2(x)}{F_0^2} \eolc
\end{equation}
in the second term of the leading-order chiral Lagrangian leads to
\begin{equation}
	\frac{F_0^2 B_0}{2} \tr \left( \M U\dega + U \M \dega \right)
		= F_0^2 B_0 ( m_u + m_d + m_s ) - \frac{B_0}{2} \tr ( \phi^2 \M ) + \O(\phi^3) \eolc
\end{equation}
where we used the fact that the trace is cyclic and $\M\dega = \M$. The first term is only a trivial shift in the
energy, but the second term is quadratic in the fields and hence contains the mass terms for the mesons. With
$\tr (\lambda_a \lambda_b ) = 2 \delta_{ab}$, we find
\begin{equation}\begin{split}
	- \frac{B_0}{2} \tr ( \phi^2 \M ) =
	&-B_0(m_u + m_d) \pi^+ \pi^- - B_0(m_u + m_s) K^+ K^-\\[2mm]
	&- B_0(m_d + m_s) K^0 \bar{K}^0 -\frac{B_0}{2} (m_u + m_d) (\phi^3)^2\\[2mm]
	&- \frac{B_0}{\sqrt{3}} (m_u - m_d) \phi^3 \phi^8 - \frac{B_0}{6} (m_u + m_d + 4 m_s) (\phi^8)^2 \eolp
	\label{eq:strongMassTerms1}
\end{split}\end{equation}
The term proportional to $\phi^3 \phi^8$ is responsible for isospin symmetry breaking since it mixes the
triplet $\phi^3$ with the singlet $\phi^8$. Accordingly, it vanishes in the limit $m_u=m_d$. The meson mass matrix can
be diagonalised by a rotation:
\begin{equation}
	\vect[l]{\pi^0\\ \eta }
			= \mat[r]{2}{\cos \epsilon&\sin \epsilon \\ -\sin \epsilon&\cos \epsilon} \vect{\phi^3 \\ \phi^8}
			= \vect{ \phi^3 + \epsilon\, \phi^8 \\ \phi^8 - \epsilon\, \phi^3 } + \O(\epsilon^2) \eolp
\end{equation}
If $\epsilon$ is set to
\begin{equation}
	\epsilon = \frac{\sqrt{3}}{4} \, \frac{m_d - m_u}{m_s - \hat{m}} \eolc \qquad 
			\hat{m} = \frac{m_u + m_d}{2}  \eolc
	\label{eq:strongEpsilon}
\end{equation}
the mixing term vanishes and the terms containing $\phi^3$ and $\phi^8$ in Eq.~\eqref{eq:strongMassTerms1} are
replaced by
\begin{multline}
	-\frac{B_0}{2} \left\{ (m_u + m_d) + 2 \epsilon\, \frac{m_u-m_d}{\sqrt{3}} \right\} (\pi^0)^2\\[2mm]
	-\frac{B_0}{2} \left\{ \frac{m_u + m_d + 4 m_s}{3} - 2 \epsilon\, \frac{m_u-m_d}{\sqrt{3}} \right\} \eta^2
	+ \O(\epsilon^2) \eolp
	\label{eq:strongMassTerms2}
\end{multline}
The inverse of the quark mass ratio appearing in Eq.~\eqref{eq:strongEpsilon} is usually called $R$:
\begin{equation}
	R = \frac{m_s - \hat{m}}{m_d - m_u} \eolp
	\label{eq:strongRDefinition}
\end{equation}
From the mass terms in Eqs.~\eqref{eq:strongMassTerms1} and \eqref{eq:strongMassTerms2} we can now readily read off the
lowest-order expressions for the mesons masses:
\begin{equation}\begin{split}
	\mpip^2 &= B_0 (m_u + m_d)\\[1mm]
	\mpiz^2 &=B_0 (m_u + m_d) + \frac{2 \epsilon}{\sqrt{3}} B_0 (m_u - m_d) + \O(\epsilon^2)\\[1mm]
	\mKp^2 &=B_0 (m_u + m_s)\\[2mm]
	\mKz^2 &=B_0 (m_d + m_s)\\[1mm]
	\meta^2 &= B_0 \frac{m_u + m_d+4 m_s}{3} - \frac{2 \epsilon}{\sqrt{3}} B_0 (m_u - m_d) + \O(\epsilon^2)
	\label{eq:strongLOMesonMasses}
\end{split}\end{equation}
These relations have first been derived based on pure symmetry arguments by Gell-Mann, Oakes, and
Renner~\cite{Gell-Mann+1968}. They are valid at lowest order in the quark mass expansion and in the absence of
electromagnetic interactions, that is they obtain corrections of $\O (\M^2,e^2)$. At this order, the mass difference
between the $\pi^\pm$ and the $\pi^0$ is entirely due to $\eta \pi^0$ mixing and is very small. We can use $\delta =
(m_d-m_u)$ as an additional expansion parameter and expand around the
isospin limit, where $m_u = m_d = \hat{m}$. To leading order in $\delta$, the pions and the kaons
are degenerate:
\begin{equation}\begin{split}
	\mpi^2 &= 2 B_0 \hat{m} + \O(\delta^2) \eolc \hspace{4em} \mK^2 = B_0 (\hat{m} + m_s) + \O(\delta) \eolc \\[2mm]
	\meta^2 &= 2 B_0 \frac{\hat{m} + 2 m_s}{3} + \O(\delta^2) \eolp
\end{split}\end{equation}
The kaon mass in the isospin limit is related to the masses of the $K^+$ and the $K^0$ by
\begin{equation}
	m^2_{K,\text{QCD}} = B_0 \left( \frac{m_u+m_s}{2} + \frac{m_d + m_s}{2} \right)
			= \inv{2} ( \mKp^2 + \mKz^2)_\text{QCD} \eolc
	\label{eq:strongmKisoemlimit}
\end{equation}
where the subscript QCD indicates that only the strong interactions have been taken into account. In the isospin limit,
at leading order in the quark-mass expansion, the masses satisfy the Gell-Mann--Okubo relation
\cite{Gell-Mann1961,Okubo1962},
\begin{equation}
	4 \mK^2 = 3 \meta^2 + \mpi^2 \eolp
	\label{eq:strongGellMannOkubo}
\end{equation}

So far, we have only discussed the masses within QCD. But the masses of the charged particles are shifted due to
electromagnetic contributions to the self-energy. These break isospin symmetry, such that they must be
subtracted, if we want to calculate the degenerate masses from the physical ones. The pion mass difference contains a
strongly suppressed QCD contribution of $\O(\delta^2)$ and is hence dominated by electromagnetic effects:
\begin{equation}
	( \mpip^2 - \mpiz^2 )_\text{QCD} \ll ( \mpip^2 - \mpiz^2 )_\text{QED} \eolp
\end{equation}
The pion mass in the isospin limit is thus given by $\mpiz$ up to a correction of $\O(\delta^2)$. 
According to Dashen's theorem~\cite{Dashen1969}, the electromagnetic corrections to the pion and the kaon mass are the
same up to corrections of $\O(\M e^2)$, allowing us to write
\begin{equation}
	( \mKp^2 - \mKz^2 )_\text{QCD} = (\mKp^2 - \mKz^2) - ( \mpip^2 - \mpiz^2 ) + \O(\M e^2, \delta^2) \eolc
	\label{eq:strongDashen}
\end{equation}
which implies for the electromagnetic kaon mass splitting
\begin{equation}
	\Delta m_K^\text{em} = ( \mKp^2 - \mKz^2 )_\text{QED} = \frac{\mpip^2 - \mpiz^2}{\mKz + \mKp} \approx 1.3~\MeV
\eolp
\end{equation}
We can now remove the electromagnetic contribution from the $K^+$ mass in Eq.~\eqref{eq:strongmKisoemlimit} and thus
arrive at the following expressions for the meson masses in the isospin limit:
\begin{equation}\begin{split}
	\mpi^2 &= \mpiz^2 + \O(\delta^2) \eolc\\[1mm]
	\mK^2  &= \inv{2} \left( \mKp^2 + \mKz^2 - \mpip^2 + \mpiz^2 \right) + \O(\M e^2, \delta^2)
	\eolp \label{eq:strongIsospinMesonMasses}
\end{split}\end{equation}
From the physical meson masses in Table~\ref{tab:strongMesons} we obtain $\mK = 495.013~\MeV$. The Gell-Mann--Okubo
relation can now be explicitly rewritten in terms of the physical meson masses as
\begin{equation}
	2 \mKp^2 + 2 \mKz^2 = 3 \meta^2 - \mpiz^2 + 2 \mpip^2 \eolp
\end{equation}
Even though this is a leading order relation in $m_s/\ldqcd$, it holds to an accuracy of about 7\%. Generally, at
next-to-leading order, one expects corrections of the order of
\begin{equation}
	\frac{m_s}{\ldqcd} \sim \frac{m_K^2}{16 \pi^2 \Fpi^2} \approx 20 \% \eolp
\end{equation}

\subsection{Quark mass ratios}

The quark masses are fundamental parameters of QCD and for a complete understanding of the strong interaction their
values must be known. They are, however, not directly measurable due to confinement. In order to estimate
them, one must therefore turn to indirect methods. The relations for the leading-order meson masses allow one to
determine ratios of the lightest quark masses. 

From the leading-order meson masses, Eq.~\eqref{eq:strongLOMesonMasses}, and Dashen's theorem,
Eq.~\eqref{eq:strongDashen}, follows
\begin{equation}
	B_0 (m_u - m_d) = (\mKp^2 - \mKz^2) - ( \mpip^2 - \mpiz^2 ) + \O(\M e^2, \M^2, \delta^2) \eolc
	\label{eq:strongB0mumd}
\end{equation}
from which one obtains $2 B_0 m_u$ and $-2 B_0 m_d$ by adding or subtracting $\mpiz^2$, respectively. In a similar
way, we also find
\begin{equation}
	2 B_0 m_s = (\mKp^2 + \mKz^2) - \mpip^2 + \O(\M e^2, \M^2) \eolp
\end{equation}
From these expressions we can immediately calculate the quark ratios that Weinberg originally derived using 
$SU(3) \times SU(3)$ current algebra~\cite{Weinberg1977}:
\begin{equation}\begin{split}
	\frac{m_d}{m_u} &\approx \frac{\mKz^2 - \mKp^2 + \mpip^2}{\mKp^2 - \mKz^2 - \mpip^2 + 2 \mpiz^2} \approx 1.79 \eolc
\\[2mm]
	\frac{m_s}{m_d} &\approx \frac{\mKp^2 + \mKz^2 - \mpip^2}{\mKz^2 - \mKp^2 + \mpip^2} \approx 20.2 \eolp
	\label{eq:strongWeinbergRatios}
\end{split}\end{equation}
According to the Particle Data Group~\cite{PDG2010}, $m_d/m_u$ lies between 1.6 and 2.9, and $m_s/m_d$ between 17 and
22. In the same way one can also derive an approximation for $R$ (see Eq.~\eqref{eq:strongRDefinition}), namely
\begin{equation}
	R = \inv{2} \frac{\mKp^2+\mKz^2-\mpip^2-\mpiz^2}{(\mKz^2-\mKp^2)-(\mpiz^2-\mpip^2)} 
			= \frac{\mK^2 - \mpi^2}{(\mKz^2-\mKp^2)_{QCD}} \eolp
\end{equation}
The meson masses receive corrections from higher orders that shift the value of the ratio to
\begin{equation}
	\frac{\mK^2 - \mpi^2}{(\mKz^2-\mKp^2)_{QCD}} 
				= \frac{m_s - \hat{m}}{m_d - m_u} \left\{ 1 - \Delta_M + \O(\M^2) \right\} \eolp
	\label{eq:strongRmesonMasses}
\end{equation}
The correction term $\Delta_M$ is $\O(\M)$ and is given by~\cite{Gasser+1984}
\begin{equation}
	\Delta_M = \inv{32 \pi^2 F_0^2} \left( \meta^2 \log \frac{\meta^2}{\mu^2} - \mpi^2 \log \frac{\mpi^2}{\mu^2} \right)
						+ \frac{8}{F_0^2} (\mK^2 - \mpi^2) (2 L_8^r - L_5^r) \eolc
\end{equation}
where $L_8^r$ and $L_5^r$ are two renormalised LECs from $\L_4$ and $\mu$ is the renormalisation scale.
The same term appears also in another ratio of meson masses,
\begin{equation}
	\frac{\mK^2}{\mpi^2} = \frac{m_s + \hat{m}}{m_d + m_u} \left\{ 1 + \Delta_M + \O(\M^2) \right\} \eolc
\end{equation}
such that the terms of $\O(\M)$ cancel in the product of the two relations:
\begin{equation}
	\frac{\mpi^2}{\mK^2} \frac{(\mKz^2-\mKp^2)_{QCD}}{\mK^2 - \mpi^2} 
			= \frac{m_d^2 - m_u^2}{m_s^2 - \hat{m}^2} \left\{ 1 + \O(\M^2) \right\} \eolp
	\label{eq:strongQmesonMasses}
\end{equation}

\begin{figure}[tb]
\psset{xunit=2.6cm,yunit=.9cm}
\begin{center}\begin{pspicture*}(0.45,19)(5.4,26.5)

\pscurve[linewidth=1pt](0.8, 25.51)(0.85, 25.38)
(0.9,25.24)(0.95,25.11)(1.,24.98)(1.05,24.85)(1.1,24.73)(1.15,24.6)(1.2,24.48)(1.25,24.36)(1.3,24.24)
(1.35,24.12)(1.4,24.01)(1.45,23.89)(1.5,23.78)(1.55,23.67)(1.6,23.56)(1.65,23.45)(1.7,23.34)(1.75,23.24)(1.8,23.14)
(1.85,23.03)(1.9,22.93)(1.95,22.83)(2.,22.73)(2.05,22.64)(2.1,22.54)(2.15,22.44)(2.2,22.35)(2.25,22.26)(2.3,22.17)
(2.35,22.08)(2.4,21.99)(2.45,21.9)(2.5,21.81)(2.55,21.72)(2.6,21.64)(2.65,21.55)(2.7,21.47)(2.75,21.39)(2.8,21.31)
(2.85,21.22)(2.9,21.14)(2.95,21.07)(3.,20.99)(3.05,20.91)(3.1,20.83)(3.15,20.76)(3.2,20.68)(3.25,20.61)(3.3,20.53)
(3.35,20.46)(3.4,20.39)(3.45,20.32)(3.5,20.25)(3.55,20.18)(3.6,20.11)

\pscircle*(1.272,24.30){0.08}	
\pscircle*(1.9,22.93){0.08}	
\pscircle*(2.3,22.17){0.08}	
\pscircle*(2.6,21.64){0.08}	
\pscircle*(3.2,20.68){0.08}	

\uput{0.2}[45](1.272,24.30){Dashen}
\uput{0.2}[45](1.9,22.93){Duncan et al.}
\uput{0.2}[45](2.3,22.17){Bijnens \& Prades}
\uput{0.2}[45](2.6,21.64){Donoghue \& Perez}
\uput{0.2}[45](3.2,20.68){Ananthanarayan \& Moussallam}

\psset{linewidth=.5pt}
\psline(0.7,20)(3.7,20)
\psline(0.7,21)(2.99,21)
\psline(0.7,22)(2.39,22)
\psline(0.7,23)(1.87,23)
\psline(0.7,24)(1.40,24)
\psline(0.7,25)(0.99,25)
\multips(1,20)(1,0){3}{\psline(0,5pt)}
\multips(1.5,20)(1,0){2}{\psline(0,3pt)}
\multido{\n=1+1}{3}{\rput(\n,19.6){\n}}
\multido{\n=20+1}{6}{\rput(0.6,\n){\n}}
\rput(0.65,26.1){\Large $Q$}
\rput(3.8,19.5){$\Delta m_K^\text{em}$ in MeV}

\end{pspicture*}\end{center}
\caption{$Q$ as calculated from a meson ratio with different values for the electromagnetic kaon mass splitting. The
left-most point has been calculated in the absence of Dashen violation and thus agrees with $Q_D$. The other points,
from left to right, have been taken from Refs.~\cite{Duncan+1996}, \cite{Bijnens+1997a}, \cite{Donoghue+1997},
and~\cite{Ananthanarayan+2004}. The figure has been inspired by a similar picture in Ref.~\cite{Leutwyler1996}. }
\label{fig:strongQFromKaon}
\end{figure}
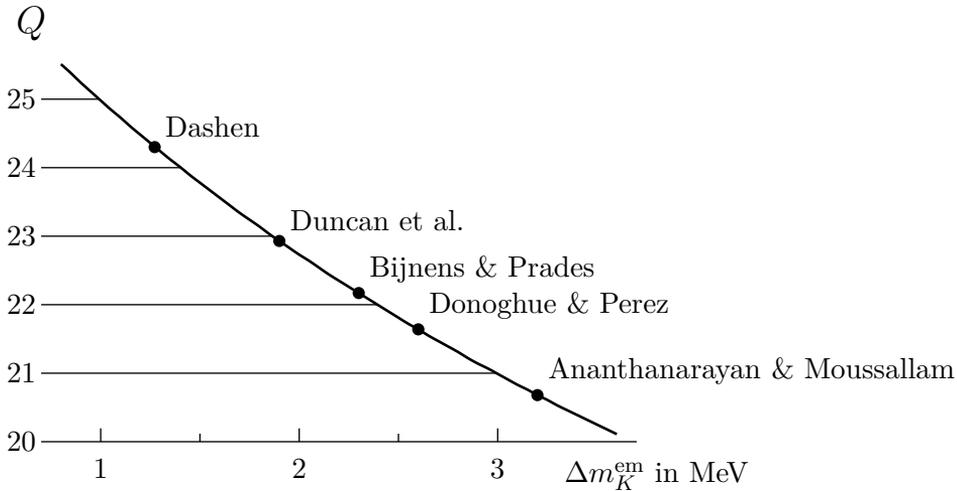

\noindent On the right hand side appears the quark mass double ratio
\begin{equation}
	Q^2 = \frac{m_s^2 - \hat{m}^2}{m_d^2 - m_u^2} \eolc
	\label{eq:strongQDefinition}
\end{equation}
and we have found that it is given by a ratio of meson masses up to corrections of $\O(\M^2)$. In order to
calculate $Q$ from the physical meson masses, we must take the electromagnetic corrections into account. This is
achieved by using Eq.~\eqref{eq:strongIsospinMesonMasses}, leading to
\begin{equation}
	Q_D^2 = \frac{(\mKp^2 + \mKz^2 - \mpip^2 + \mpiz^2)(\mKp^2 + \mKz^2 - \mpip^2 - \mpiz^2)}
			{4 \mpiz^2 (\mKz^2 - \mKp^2 - \mpiz^2 + \mpip^2)} \approx (24.3)^2 \eolp
\end{equation}
This is the value of $Q$ in the absence of Dashen violation. 
The electromagnetic kaon mass splitting $\Delta m_K^\text{em}$ is substantially changed by higher order effects.
Several authors have calculated Dashen violating contributions, e.g., Refs.~\cite{Duncan+1996,Bijnens+1997a,
Donoghue+1997,Ananthanarayan+2004}, and found deviations from Dashen's theorem that range from 50 up to 150 per cent.
Figure~\ref{fig:strongQFromKaon} shows their values for $\Delta m_K^\text{em}$ together with the corresponding
results for $Q$.

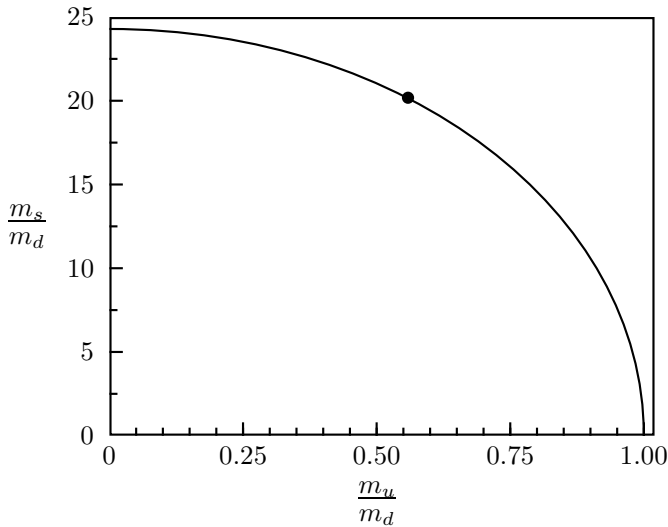
\begin{figure}[tb]
\psset{xunit=7cm,yunit=0.22cm}
\begin{center}\begin{pspicture*}(-.2,-5.4)(1.1,26)
\psset{linewidth=\mylw}
\psframe(0,0)(1.02,25)
\multips(0.25,0)(0.25,0){4}{\psline(0,5pt)}
\multips(0.05,0)(0.05,0){20}{\psline(0,3pt)}
\multido{\n=0+0.25}{5}{\uput{.15}[270](\n,0){\small \n}}
\multips(0,5)(0,5){4}{\psline(5pt,0)}
\multips(0,2.5)(0,5){5}{\psline(3pt,0)}
\multido{\n=0+5}{6}{\uput{0.2}[180](0,\n){\small \n}}

\uput{0.6}[270](0.501,0){\Large $\frac{m_u}{m_d}$}
\uput{.8}[180](0,12.5){\Large $\frac{m_s}{m_d}$}
\psellipticarc(0,0)(1.002,24.35){0}{90}
\pscircle*(0.5587,20.18){0.08}
\end{pspicture*}\end{center}
\caption{The figure shows the Leutwyler ellipse for $Q = Q_D = 24.3$. The dot marks the ratios given in
		Eq.~\eqref{eq:strongWeinbergRatios}.}
\label{fig:strongLeutwylerEllipse}
\end{figure}

Kaplan and Manohar~\cite{Kaplan+1986} have shown that a change in the quark masses of the form $m_u \mapsto m_u +
\alpha m_d m_s$ (and cyclic) can be absorbed into $\O(p^4)$ operators by shifting the low-energy constants $L_6$, $L_7$,
and $L_8$ accordingly. The quark mass ratios $m_u/m_d$, $m_s/m_d$, and $R$ are not invariant under this transformation
which implies that corrections from $\L_4$ can, in principle, change them to any value that can be reached by the
aforementioned shift of the quark masses. The double ratio $Q$, on the other hand, is invariant not affected by the
transformation up to corrections of $\O(\M^2)$. The transformation of the quark masses depends on a single parameter
$\alpha$, such that the ratios $m_u/m_d$ and $m_s/m_d$ are not independent. They are rather constrained to lie on an
elliptic curve, whose semi-major axis is given by $Q$,
\begin{equation}
	\left( \frac{m_u}{m_d} \right)^2 + \inv{Q^2} \left( \frac{m_s}{m_d} \right)^2 = 1 \eolc
\end{equation}
where a term of the order of $(\hat{m}/m_s)^2$ has been neglected. The curve is plotted in
Fig.~\ref{fig:strongLeutwylerEllipse} together with the lowest-order estimates for the quark mass ratios from
Eq.~\eqref{eq:strongWeinbergRatios}.

In Sec.~\ref{sec:eta3piIsoStructure} we will show that the decay amplitude for the process $\etapi$ is proportional to
$B_0 (m_u - m_d)$. This factor is not directly measurable, but it can be expressed in terms of measurable quantities and
the quark mass ratio $Q$ or $R$, as we will show in the following. We recall from Eq.~\eqref{eq:strongB0mumd} that
\begin{equation}
	B_0 (m_u - m_d) = (\mKp^2 - \mKz^2)_\text{QCD} + \O(\M^2) \eolp
\end{equation}
Equation~\eqref{eq:strongRmesonMasses} relates the strong kaon mass difference to the quark mass ratio $R$ and it
follows that
\begin{equation}\begin{split}
	B_0 (m_u - m_d) &= -\inv{R} (\mK^2 - \mpi^2) + \O(\M^2)\\
						 &= -\frac{3}{4 R} (\meta^2 - \mpi^2)  + \O(\M^2) \eolc
	\label{eq:B0mumdR}
\end{split}\end{equation}
where we have used the Gell-Mann--Okubo mass formula \eqref{eq:strongGellMannOkubo} in the second equality. Similarly,
from Eq.~\eqref{eq:strongQmesonMasses} one obtains
\begin{equation}
	B_0 (m_u - m_d) = -\inv{Q^2} \frac{\mK^2(\mK^2 - \mpi^2)}{\mpi^2} + \O(\M^3) \eolp
	\label{eq:B0mumdQ}
\end{equation}
Of the two expressions for $B_0 (m_u - m_d)$ in Eqs.~\eqref{eq:B0mumdR} and \eqref{eq:B0mumdQ}, the latter has the
advantages that it is valid up to $\O(p^6)$ and is expressed in terms of $Q$, which is invariant under the
Kaplan-Manohar transformation.

%% file: main/dispersionRelations.tex
\chapter{Introduction to dispersion relations}

We introduce the concept of dispersion relations with a number of examples, focused mainly on mathematical
aspects. In the first part on dispersion relations for a function of a single variable, the subject is introduced as a
purely mathematical problem, and the contact to physical applications is only made in a brief final comment. This is
followed by the construction of the dispersion relations for four-particle amplitudes, which are functions of two
variables. The physics relevant for this discussion has been introduced in Sec.~\ref{sec:strongSmatrix}. The
chapter ends with the derivation of the solution to the Omn\`es problem, which plays an important role in the
dispersive analysis of $\etapi$.

For a more detailed discussion of this chapter's subject we suggest the
textbooks~\cite{Alfaro+1973,Barton1965,Queen+1974}.

\section{Dispersion relation for a function of one variable} \label{sec:dispRelOneVar}

\begin{figure}[tb]
\begin{center}
	\psset{linewidth=\mylw,unit=.6cm}

	\begin{pspicture*}(-6,-6)(7,6.5)
		\psline{-}(-6,0)(5,0)
		\psline{->}(0,-6)(0,6)
		\uput{.4}[0](0,5.9){$\Im s'$}
		\uput{.4}[90](6.2,0){$\Re s'$}
		\psline[linewidth=.1]{->}(2,0)(6,0)
		\pscircle*(2,0){.1}
		\uput{.4}[270](2,0){$M^2$}
		\psarc(2,0){.2}{90}{-90}
		\psline(2,.2)(5.4,.2)
		\psline(2,-.2)(5.4,-.2)
		\psarc(0,0){5.4}{2}{-2}
		\uput{.3}[25](3.82,3.82){$\gamma$}
		\psarc[linewidth=3\mylw]{->}(0,0){5.4}{43}{47}
		\psarc[linewidth=3\mylw]{->}(0,0){5.4}{223}{227}
		\psline{->}(0,0)(-3.82,3.82)
		\uput{.1}[45](-1.91,1.91){$\Lambda^2$}	
		\pscircle*(3,1.2){.05}
		\uput{.1}[45](3,1.2){$s$}
	\end{pspicture*}

\end{center}

\caption{The integration contour $\gamma$ that is used for the representation of $f(s)$ in Eq.~\eqref{eq:cauchy}.}
\label{fig:contour}
\end{figure}
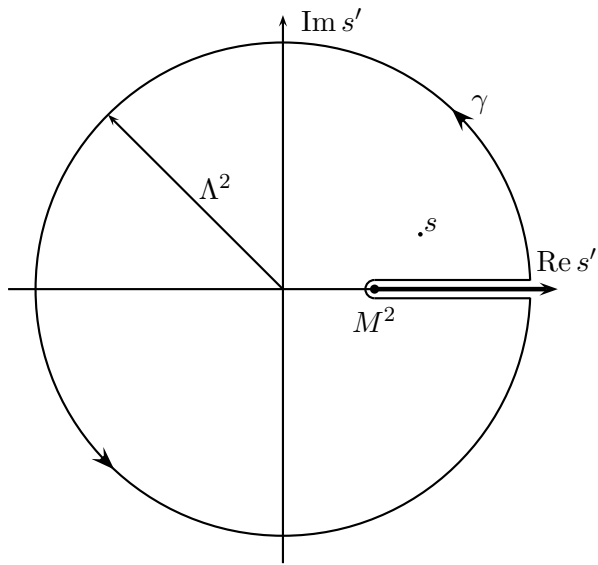

Consider a complex valued function $f(s)$ that is analytic in the entire complex plane up to a branch cut for $s \geq
M^2$ and real on the real axis below the cut. This implies that it satisfies the Schwarz reflection principle,
\begin{equation}
	f(s^*) = f^*(s) \eolp
	\label{eq:Schwarz}
\end{equation}
We can now invoke Cauchy's integral formula,
\begin{equation}
	f(s) = \inv{2 \pi i} \oint_\gamma \frac{f(s')}{s' - s} \; ds' \eolc
	\label{eq:cauchy}
\end{equation}
where the integration along the closed contour $\gamma$ has to be taken in anticlockwise sense. It is valid for any $s$
that lies within the contour, provided that $f(s)$ is analytic in the entire region enclosed by $\gamma$. This means
that we have to choose an integration path that avoids the cut as shown in Fig.~\ref{fig:contour} and, in order to have
a representation that is valid for any $s$ away from the cut, that we must send the radius $\Lambda^2$ to infinity.
Using the definition of the discontinuity,
\begin{equation}
	\disc f(s) \equiv \frac{ f(s + i \epsilon) - f(s - i \epsilon) }{2 i} \eolc
\end{equation}
we can rewrite Eq.~\eqref{eq:cauchy} as
\begin{equation}
	f(s \pm i \epsilon) = \inv{\pi} \int\limits_{M^2}^{\ \ \Lambda^2} \frac{\disc f(s')}{s' - s \mp i \epsilon} ds'
				+ \inv{2 \pi i} \oint\limits_{|s'| = \Lambda^2} \frac{f(s')}{s' - s} \; ds' \eolc
\end{equation}
where the limit $\epsilon \to 0^+$ is implied. For values of $s$ away from the cut this limit can be taken trivially,
but if $s$ lies on the cut, $\epsilon$ ensures that the integration path avoids the pole at $s' = s$. Choosing the
negative (positive) sign in the denominator corresponds to evaluating $f$ on the upper (lower) rim of the cut. In the
following, we will simplify the notation by omitting $+ i \epsilon$ in the argument of a function, whenever the function
has to be evaluated on the upper rim of the cut, i.e., $f(s) \equiv \lim_{\epsilon \to 0^+} f(s + i \epsilon)$. If, on
the other hand, the function must be evaluated on the lower rim of the cut, we will write explicitly $f(s - i
\epsilon)$, where, of course, the limit is still implied.

Next, we want to take the limit $\Lambda^2 \to \infty$. Let us assume for now that in this limit, the integral along the
circle vanishes and the integral along the cut converges. We then find
\begin{equation}
	f(s) = \inv{\pi} \int\limits_{M^2}^{\ \infty} \frac{\disc f(s')}{s' - s} ds' \eolp
	\label{eq:disprelUnsubtr}
\end{equation}
This is an example of an unsubtracted dispersion relation. In the present case, it can be rewritten due to the
Schwarz reflection principle~\eqref{eq:Schwarz}, which implies
\begin{equation}
	\disc f(s) = \frac{f(s+i\epsilon) - f^\ast(s+i\epsilon)}{2i} = \Im f(s + i\epsilon) = \Im f(s)\eolc
\end{equation}
leading to
\begin{equation}
	f(s) = \inv{\pi} \int\limits_{M^2}^{\ \infty} \frac{\Im f(s')}{s' - s} ds' \eolp
	\label{eq:disprelUnsubtrIm}
\end{equation}
Often, dispersion relations are stated in this form. But one should not forget that the integral is always to be taken
over the discontinuity and only if the Schwarz reflection principle is satisfied, the discontinuity is identical to the
imaginary part.

So far we have assumed the integral along $|s'| = \Lambda^2$ to vanish and the integral along the cut to converge in the
limit $\Lambda^2 \to \infty$. We show now that both assumptions are true if $f(s)$ tends to zero for $|s| \to \infty$.
In this case, we have for the integral along the arc
\begin{equation}
\begin{split}
	\left| \oint \frac{f(s')}{s' - s} \; ds' \right|  &\leq \oint \left| \frac{f(s')}{s' - s} \right| \left| ds' \right| 
		\leq \; \max_{|s'| = \Lambda^2} \; \left| f(s') \right| \; \oint  \inv{ | |s'| - |s| |} \left| ds' \right|\\[2mm]
	&= \; \max_{|s'| = \Lambda^2} \; \left| f(s') \right| \; \frac{2 \pi \Lambda^2}{\Lambda^2 - |s|} \;
			\stackrel{\Lambda^2 \to \infty}{\longrightarrow} \;
			2 \pi \lim_{\Lambda^2 \to \infty} \; \max_{|s'| = \Lambda^2} \; \left| f(s') \right| \eolc
\end{split}
\end{equation}
which vanishes if and only if $f(s)$ goes to zero for $|s| \to \infty$. If this is true, it is immediately clear
that the integral along the cut converges because the integrand vanishes asymptotically faster than $1/s'$.

Owing to a theorem of Sugawara and Kanazawa~\cite{Sugawara+1961}, this condition for the validity of the unsubtracted
dispersion relation can even be relaxed. The theorem is valid for a much more general situation than the one at hand,
as it also allows for poles and a left-hand cut on the real axis. Both of these situations will be treated later such
that it is useful to state the theorem in its general form already here. Let $h(s)$ be a function with the following
properties
\begin{enumerate}[label=(\roman*)]
	\item It is analytic everywhere in the complex plane except for poles and cuts on the real axis.
			The cuts in general extend to $\pm \infty$.
	\item It is bounded by some finite power of $|s|$ as $|s| \to \infty$ in any direction.
	\item It has finite limits $h(\infty \pm i \epsilon)$ as $s \to \infty \pm i \epsilon$ along the right-hand cut and
			definite (not necessarily finite) limits $h(-\infty \pm i \epsilon)$ as $s \to -\infty \pm i \epsilon$ along
			the left-hand cut.
\end{enumerate}
Then the limits of $h(s)$ when $s$ tends to infinity in any other direction are
\begin{equation}
\lim_{|s| \to \infty} h(s) = \left\{ \begin{array}{l} h(\pm\infty + i \epsilon), \qquad \Im s > 0 \\[2mm]
																		h(\pm\infty - i \epsilon), \qquad \Im s < 0 \end{array} \right.
	\eolp
\end{equation}
If only one cut, say the right-hand one, extends to infinity, then the limits in the upper and lower half-plane are
the same because they coincide at $-\infty$. This also implies that $\disc~h(+\infty) = 0$.

Owing to the theorem, we can now state that the dispersion relation in Eq.~\eqref{eq:disprelUnsubtrIm} is valid for any
function $f(s)$ that is bounded by some finite power of $|s|$ as $|s| \to \infty$, provided it vanishes for $s \to
\infty$ along the cut. In this case the integral along the arc is zero, and the integral along the cut converges.

If these conditions are not satisfied and, for example, $f(s)$ goes to a constant, we have to subtract the dispersion
relation. We make use of the identity
\begin{equation}
	\frac{1}{s' - s} = \inv{s' - s_1} + \frac{s - s_1}{(s' - s_1) (s' - s)}
	\label{eq:subtraction}
\end{equation}
in Eq.~\eqref{eq:cauchy} and find
\begin{equation}
	f(s) = \inv{2 \pi i} \oint_\gamma \frac{f(s')}{s' - s_1} \; ds' + \frac{s - s_1}{2 \pi i} \oint_\gamma 
				\frac{f(s')}{(s'-s_1)(s' - s)} \; ds' \eolp
\end{equation}
The first term is nothing else than $f(s_1)$ and is called a subtraction constant. $s_1$ is called the subtraction point
and can be chosen arbitrarily with $s_1 < M^2$. The integrand in the second term now falls off fast enough such that the
integral along the arc vanishes and thus we find the once subtracted dispersion relation
\begin{equation}
	f(s) = f(s_1) + \frac{s - s_1}{\pi} \int\limits_{M^2}^{\ \infty} \frac{\Im f(s')}{(s'-s_1)(s' - s - i\epsilon)} ds' \eolp
\end{equation}

In the general case, where $f(s)$ grows asymptotically with some power of $s$, $f(s) \asymp s^N$, we have to perform
$N+1$ subtractions. For each one of them we can choose a different subtraction point, such that subtracting in the way
described above becomes rather tedious. It is easier to consider the function
\begin{equation}
	g(s) \equiv \frac{f(s)}{P_{N+1}(s)} = \frac{f(s)}{(s-s_1)^{N_1} (s-s_2)^{N_2} \cdots} \eolc
\end{equation}
where $P_{N+1}(s)$ is an arbitrary polynomial of order $N+1$ in $s$, $s_j < M^2$ are the subtraction points and, of course, $\sum N_j = N+1$.
$g(s)$ vanishes for $|s| \to \infty$ such that the integral along the arc in Cauchy's integral formula is zero. Furthermore, it has the same
analytic structure as $f(s)$ up to the poles at $s = s_j$ generated by the zeros of the polynomial. They are situated within the contour
$\gamma$ and therefore Cauchy's integral formula as formulated in Eq.~\eqref{eq:cauchy} does not apply any longer.

The required extension of Cauchy's integral formula comes from the residue theorem.
Let $f(s)$ be a function that is analytic within a contour $\gamma$ up to a number of isolated poles $s_j$. Then, the residue theorem states that
\begin{equation}
	\inv{2 \pi i} \oint_{\gamma} f(s') \; ds' = \sum_j \Res(f,s_j) \eolc \label{eq:residueTheorem}
\end{equation}
where the integration along $\gamma$ is taken in anticlockwise sense. If $f$ has a pole of order k at $s_j$, the residue is given by
\begin{equation}
	\Res(f,s_j) =  \inv{(k-1)!} \left( \frac{d}{ds} \right)^{k-1} (s - s_j)^k \; f(s) \; \bigg|_{s = s_j} \eolp
	\label{eq:ResDef}
\end{equation}
This allows us now to calculate
\begin{equation}
	\inv{2 \pi i} \oint_\gamma \frac{g(s')}{(s' - s)} \; ds' = \Res \left(\frac{g(s')}{(s' - s)},s \right)
			+ \sum_j \Res \left(\frac{g(s')}{(s' - s)},s_j \right) \eolc
	\label{eq:residueG}
\end{equation}
where we have assumed $s$ to be distinct from all the $s_j$ for now. 
The sum goes over the zeros of $P_{N+1}$. The pole at $s' = s$ is of order 1 and we can calculate its residue
from Eq.~\eqref{eq:ResDef},
\begin{equation}
	\Res \left(\frac{g(s')}{(s' - s)},s \right) = \lim_{s' \to s}\ (s'-s) \, \frac{g(s')}{s'-s} = g(s) \eolp
\end{equation}
As the next step, we want to simplify the sum over the residues at the subtraction points. First, we assume that the
polynomial has $N+1$ distinct zeros, i.e., all the $N_j$ are equal to 1, such that the corresponding poles are simple.
Then, because of $s \neq s_j$, the residues in the sum can be rewritten as
\begin{equation}
	\Res \left(\frac{g(s')}{(s' - s)},s_j \right) = \frac{\Res \left(g(s'),s_j \right)}{s_j-s} \eolc
\end{equation}
and, after solving for $g(s)$ we get from Eq.~\eqref{eq:residueG}
\begin{equation}
	g(s) = \inv{2 \pi i} \oint_\gamma \frac{g(s')}{(s' - s)} \; ds' 
							- \sum_j \frac{\Res \left(g(s'),s_j \right)}{s_j-s} \eolp
\end{equation}
Note that in the absence of poles the last term vanishes and the equation is reduced to Cauchy's integral formula.
Finally, we reintroduce $f(s)$ and remove the contribution of the circle at infinity to find
\begin{equation}
	f(s) = Q_N(s) + \frac{P_{N+1}(s)}{2 \pi i} \int\limits_{M^2}^{\ \infty} \frac{\Im f(s')}{P_{N+1}(s')(s' - s)} \; ds' \eolc
	\label{eq:dispRelfFinal}
\end{equation}
where
\begin{equation}
	Q_N(s) = P_{N+1}(s) \sum_j \frac{\Res \left(f(s')/P_{N+1}(s'),s_j \right)}{s-s_j}
	\label{eq:subPolynomial}
\end{equation}
is a polynomial of order $N$ because for all $j$, $(s - s_j)$ is contained in the factorisation of $P_{N+1}(s)$.
It is appropriately called the subtraction polynomial, since its coefficients are the $N+1$ subtraction constants. For a
given polynomial $P_{N+1}(s)$, the latter are fully determined by the values of $f$ at the subtraction points.

Next, we consider the case where several subtractions are performed at the same point, say at $s_1$. In this situation,
the integrand of the contour integral has a pole of order $N_1$ at $s_1$. We define
\begin{equation}
	\tilde{P}_{N+1}(s) = \frac{P_{N+1}(s)}{(s-s_1)^{N_1}} \quad \Rightarrow \quad g(s) = \frac{f(s)}{(s-s_1)^{N_1} \tilde{P}_{N+1}(s)} \eolp
\end{equation}
$\tilde{P}_{N+1}(s)$ is a polynomial of order $N+1-N_1$ that is nonzero at $s_1$. The residue at $s_1$ can then be
written as
\begin{equation}\begin{split}
	\Res \left( \frac{g(s')}{(s'-s)},s_1 \right) &=  \inv{(N_1-1)!} \left( \frac{d}{ds'} \right)^{N_1-1} 
																\frac{f(s')}{(s'-s) \tilde{P}_{N+1}(s')}\, \bigg|_{s' = s_1}\\[1mm]
												&= \sum_{i = 0}^{N_1-1} a_i^{(1)}(s) \, f^{(i)}(s_1) \eolp
\end{split}\end{equation}
This shows that the residue depends on $f(s_1)$ as well as the first $N_1-1$ derivatives of $f$ at $s_1$. We refrain
from giving the full expression for the coefficients $a_i^{(1)}(s)$, but only consider their $s$ dependence,
\begin{equation}
	a_i^{(1)}(s) \propto  \left( \frac{d}{ds'} \right)^{N_1-1-i} 
							\inv{(s'-s) \tilde{P}_{N+1}(s')}\, \bigg|_{s' = s_1}
					= \sum_{j=1}^{N_1-i} \frac{b_j^{(1)}}{(s_1-s)^j} \eolp
\end{equation}
The coefficients $b_j^{(1)}$ are independent of $s$. It is important to note that the highest inverse power of $(s_1-s)$
that appears in the residue is $N_1$. The same procedure applies, if several zeros of $P_{N+1}(s)$ have multiplicity
larger than one. Each residue is then determined by the values $f(s_j), f'(s_j), \ldots, f^{(N_j-1)}(s_j)$, which
amounts to a total number of $N_j$ free parameters. The sum over the residues then contains $\sum N_j = N+1$ free
parameters. The dispersion relation for $f(s)$ is again given by Eq.~\eqref{eq:dispRelfFinal}, but the subtraction
polynomial $Q_N$ cannot be written as in Eq.~\eqref{eq:subPolynomial}, because the dependence on $s_1$ is more
complicated, as we have seen. Instead, it is given by
\begin{equation}
	Q_N(s) = -P_{N+1}(s) \sum_j \sum_{k=0}^{N_j-1} a_k^{(j)}(s) f^{(k)}(s_j) \eolc
\end{equation}
which is true for arbitrary values of the $N_j$. The singular factors $(s_j-s)^{-n}$, $n \leq N_j$ contained in the
$a_k^{(j)}(s)$ are cancelled by the corresponding factors in $P_{N+1}(s)$, such that $Q_N$ is indeed a polynomial. It is
of order $N$, because each term in the sum contains at least one factor $(s_j-s)^{-1}$. As we argued before, it contains
a total number of $N+1$ free parameters. As the $a_k^{(j)}(s)$ that multiply these free parameters are all linearly
independent, $Q_N$ is an arbitrary polynomial of order $N$.

There is one last step left: we have assumed $s$ to be distinct from all the subtraction points. But if we now evaluate
Eq.~\eqref{eq:dispRelfFinal} at one of the subtraction points, the dispersion integral is multiplied by zero and only
the subtraction polynomial remains. The same is true for the first $N_j - 1$ derivatives at the points $s_j$. So we see
that the dispersion relation is regular and well defined also at the subtraction points and Eq.~\eqref{eq:dispRelfFinal}
is the most general form of a $N+1$ times subtracted dispersion relation for $f(s)$.

At this point, a few comments are in order. This example demonstrates very well the strength of the concept of
dispersion relations: if we know the discontinuity of $f(s)$ up to infinity, we can calculate real and imaginary part of
$f(s)$ everywhere in the complex plane up to a polynomial. This is a consequence of only the analytic properties of
the function and no other assumptions are needed. Using dispersion relations, a physical amplitude can thus be
parametrised in terms of a few subtraction constants. Unfortunately, no further information on these constants can
be drawn from the dispersion relation itself, which is its most obvious limitation. External input, be it from
experiment or theory, is necessarily required to complete the dispersive representation of the amplitude.

In practice, information on the discontinuity of a physical amplitude is usually not available up to arbitrarily large
energies as is required by the dispersion integrals. As a consequence, one either has to introduce a cut-off in the
integration or make assumptions on the behaviour of the discontinuity at high energies. In order to suppress the
contributions from this high energy region and thus to reduce the resulting uncertainty in the dispersive representation
of the amplitude, one can perform additional subtractions. The subtraction constants introduced in this way can then be
understood to parametrise the lack of knowledge on the discontinuity at high energies. In that sense, they are very
similar to the low-energy constants in effective field theories.

\section{Fixed \texorpdfstring{$t$}{t} dispersion relations and the Mandelstam representation}

\begin{figure}[tb]
\begin{center}
	\psset{unit=.9cm}
	\begin{pspicture*}(-0.7,-0.5)(10.7,9)
		\psset{linewidth=\mylw}
		\psline(-0.7,2)(10.7,2)
		\psline(0.71,-0.5)(6.77,10)
		\psline(9.29,-0.5)(3.23,10)
		\psline[linestyle=dashed](-0.7,6.93)(10.7,6.93)
		\psline[linestyle=dashed](5,-2.93)(-0.7,6.93)
		\psline[linestyle=dashed](5,-2.93)(10.7,6.93)
		\pspolygon[fillstyle=hlines*](0,2)(2.15,2)(0.71,-0.5)(0.71,-2)(-2,-2)(-2,2)
		\pspolygon[fillstyle=hlines*](10,2)(7.85,2)(9.29,-0.5)(9.29,-2)(12,-2)(12,2)
		\pspolygon[fillstyle=hlines*](3.23,10)(5,6.93)(6.77,10)
		\rput(0.4,2.25){\footnotesize{$s=0$}}
		\rput{-60}(3.7,8.6){\footnotesize{$t=0$}}
		\rput{60}(6.25,8.55){\footnotesize{$u=0$}}
		\psline{->}(4.5,2)(4.5,4)
		\psarc{-}(4.5,2){0.25}{0}{90}  
		\psdot[linewidth=0.1pt](4.6,2.1)
		\psline{->}(6.143,4.949)(4.5,4)
		\psarc{-}(6.143,4.949){0.25}{210}{300}
		\psdot[linewidth=0.1pt](6.11,4.82)
		\psline{->}(3.607,4.516)(4.5,4)
		\psarc{-}(3.607,4.516){0.25}{-120}{-30}
		\psdot[linewidth=0.1pt](3.65,4.38)
		\rput(4.7,3){$s$}
		\rput(5.5,4.3){$t$}
		\rput(4.1,4.5){$u$}
		\psline[linewidth=2\mylw](10.29,-0.5)(4.23,10)
		\pscircle*(8.35,2.866){0.09}
		\uput{.3}[0](8.35,2.866){\small $s\th^-$}
		\pscircle*(6,6.93){0.09}
		\uput{.3}[45](6,6.93){\small $s\th^+$}
	\end{pspicture*}
\end{center}
\caption{The Mandelstam diagram for the scattering of two particles with equal mass. Outside the dashed triangle, the
amplitude $\A(s,t,u)$ has cuts. A line of constant $t$ is shown together with the branch points that mark the start of
the cuts when one is moving along that line.}
\label{fig:dispMandelstamDiagram}
\end{figure}
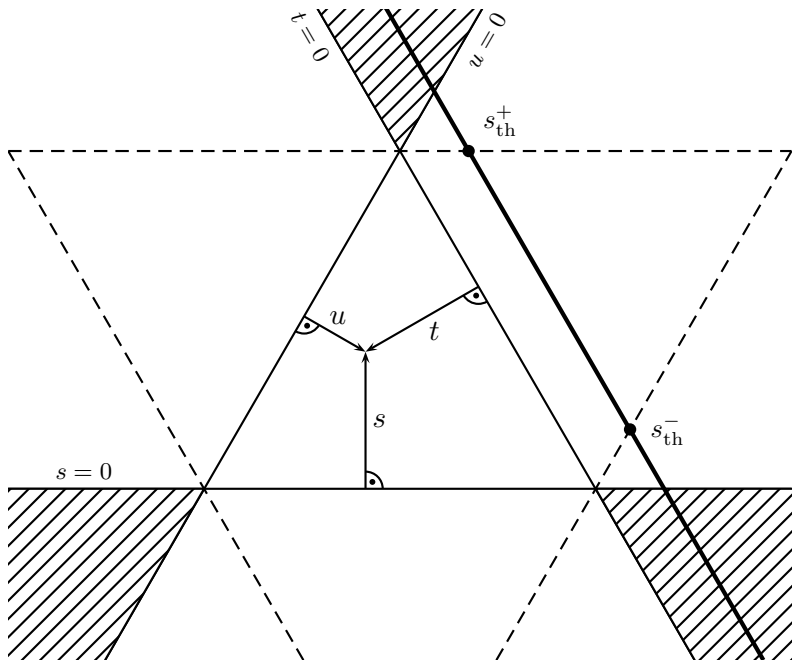

So far we have only constructed the dispersive representation for a function of a single variable. Scattering amplitudes
are, however, in general functions of several variables and do not allow for a dispersion relation of the form of
Eq.~\eqref{eq:dispRelfFinal}. In this section we derive a dispersion relation for the \mbox{$2\to2$} scattering process
in Fig.~\ref{fig:strong22scattering}. In order to avoid unnecessary complications, we restrict
ourselves to a theory that contains only one type of scalar particle with mass $m$. The kinematic relations for this
situation can be found from the general expressions that we derived in Sec.~\ref{sec:strongScattering}. In the
centre-of-mass frame of the $s$-channel, for example, we have  $|\vec{p}_1| = |\vec{p}_2| = |\vec{p}_3| = |\vec{p}_4|
\equiv p$ and $E_1 = E_2 = E_3 = E_4 = \sqrt{m^2 + p^2} \equiv E_{CM}$. We then obtain from
Eqs.~\eqref{eq:strongScatMandelstamCOM} and \eqref{eq:strongScattu}
\begin{equation}\begin{split}
	s &= 4 E_{CM}^2 = 4 (m^2 + p^2) \eolc\\[1mm]
	t &= -2 p^2 (1 - \cos \theta) \eolc\\[1mm]
	u &= -2 p^2 (1 + \cos \theta) \eolp
	\label{eq:dispScatMandelstam4m}
\end{split}\end{equation}
and the physical region is given by $s \geq 4 m^2$ and $t, u \leq0$. Note that $u$ and $t$ also must be larger than
$-4 p^2$, but as this is automatically satisfied if $t$ and $u$ are both negative, there is no need to use this as an
additional constraint. Similarly, one can show that for the $t$ channel process, the physical region is $t \geq 4 m^2$
and $s, u \leq0$, while it is $u \geq 4 m^2$ and $t, u \leq0$ for the $u$-channel. The Mandelstam diagram for this
process with the physical region for each channel shaded is shown in Fig.~\ref{fig:dispMandelstamDiagram}.

Unitarity requires $\A(s,t,u)$ to have cuts for certain values of the Mandelstam variables. Since the lightest
two-particle intermediate state in each channel starts at a centre-of-mass energy of $s^+\th = t^+\th = u^+\th = 4
m^2$, the amplitude has a cut in each Mandelstam variable starting at this point. The Mandelstam diagram in the
figure helps to visualise this situation. In each channel a dashed line marks the starting points of the cuts, where the
centre-of-mass energy is equal to $4 m^2$. Inside the triangle formed by the dashed lines, the amplitude is real, and as
a consequence the Schwarz reflection principle is valid in each Mandelstam variable, e.g.
\begin{equation}
	\frac{\A(s+i \epsilon,t,u) - \A(s-i \epsilon,t,u)}{2 i} = \frac{\A(s,t,u) - \A^*(s,t,u)}{2 i} = \Im \A(s,t,u) \eolc
	\label{eq:fixedtImPart}
\end{equation}
and similarly in the other channels.

\begin{figure}[tb]
\begin{center}
	\psset{linewidth=\mylw,unit=.61cm}

	\begin{pspicture*}(-6,-6)(7,6.5)
		\psline{-}(-5,0)(5,0)
		\psline{->}(0,-6)(0,6)
		\uput{.4}[0](0,5.9){$\Im s'$}
		\uput{.4}[90](6.2,0){$\Re s'$}
		\psline[linewidth=.1]{->}(2,0)(6,0)
		\pscircle*(2,0){.1}
		\uput{.4}[270](2,0){$s\th^+$}
		\psline[linewidth=.1]{<-}(-6,0)(-2,0)
		\pscircle*(-2,0){.1}
		\uput{.4}[270](-2,0){$s\th^-$}
		\psarc(2,0){.2}{90}{-90}
		\psline(2,.2)(5.4,.2)
		\psline(2,-.2)(5.4,-.2)
		\psarc(-2,0){.2}{-90}{90}
		\psline(-2,.2)(-5.4,.2)
		\psline(-2,-.2)(-5.4,-.2)
		\psarc(0,0){5.4}{2}{178}
		\psarc(0,0){5.4}{-178}{-2}
		\uput{.3}[25](3.82,3.82){$\gamma$}
		\psarc[linewidth=3\mylw]{->}(0,0){5.4}{43}{47}
		\psarc[linewidth=3\mylw]{->}(0,0){5.4}{223}{227}
		\psline{->}(0,0)(-3.82,3.82)
		\uput{.1}[45](-1.91,1.91){$\Lambda^2$}	
		\pscircle*(3,1.2){.05}
		\uput{.1}[45](3,1.2){$s$}
	\end{pspicture*}

\end{center}

\caption{The integration contour $\gamma$ in the complex $s$ plane that is used for the fixed $t$ dispersion relation
			for $\A(s,t,u)$ in Eq.~\eqref{eq:fixedtDispRel2}.}
\label{fig:contour2}
\end{figure}
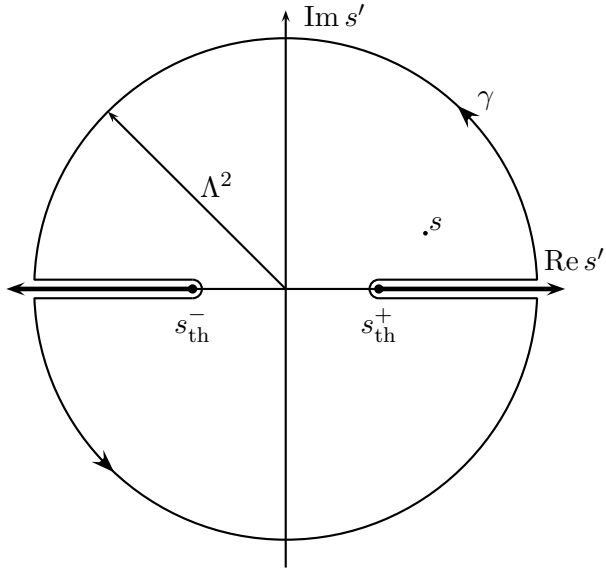

Since the Mandelstam variables are related as $s+t+u = 4 m^2$, the amplitude is in fact a function of two variables.
If one of the Mandelstam variables, say $t$, is kept at a fixed negative value, then $\A(s,t,u)$ is a function of one
variable only. It is defined along a straight line of constant negative $t$ in the Mandelstam plane as indicated in the
figure. One can see immediately that it has two cuts on this line, the branch points being marked with two dots. The cut
stretching to negative infinity in $s$ is due to the cut in the $u$-channel and it starts at
\begin{equation}
	s^-\th = 4 m^2 - t - u^+\th = -t \eolp
\end{equation}
The general procedure for the derivation of the dispersion relation is the same also in the presence of two cuts. We
decide to express the dispersion integrals in $s$. In the complex $s$ plane, the cuts of the amplitude are
\begin{equation}
	s \in (-\infty,s\th^-) = (-\infty,-t) \eolc \qquad 
	s \in (s\th^+,\infty) = (4 m^2, \infty) \eolp
\end{equation}
The integration path must avoid both cuts, as is shown in Fig.~\ref{fig:contour2} and, assuming the integral along the
arc at infinity to vanish, the unsubtracted fixed-$t$ dispersion relation reads
\begin{equation}
	\A(s,t,u) = \inv{\pi} \int\limits_{-\infty\ }^{\ \ s\th^-} \frac{\Im \A(s',t,u)}{s'-s} ds' + 
					\inv{\pi} \int\limits_{s\th^+}^{\ \infty} \frac{\Im \A(s',t,u)}{s'-s} ds' \eolp
	\label{eq:fixedtDispRel2}
\end{equation}
As in the case of the dispersion relation for a function of one variable, the imaginary part must be evaluated on the
upper rim of the cut. $u$ is to be understood as a function of $s$ and $t$. In a similar way, one can derive the
dispersion relation in terms of integrals over $u$. It is of the same form, containing integrals over the left- and
right-hand cut in $u$ with $u\th^- = -t$ and $u\th^+ = 4 m^2$. From Fig.~\ref{fig:dispMandelstamDiagram} it is
immediately clear that the left-hand cut in $u$ is identical to the right-hand cut in $s$
and vice versa. In the following, we rewrite the integral over the left-hand cut in $s$ in an integral over the
right-hand cut in $u$. To this end, we make use of $s^{(\prime)} = 4 m^2 - t - u^{(\prime)}$ and obtain
\begin{equation}
	\inv{\pi} \int\limits_{\infty}^{u\th^+} (-du')\; \frac{\Im \A(4 m^2 - t - u',t,u')}{u-u'}
		= -\inv{\pi} \int\limits_{u\th^+}^{\infty} du'\; \frac{\Im \A(s,t,u')}{u'-u} \eolp
	\label{eq:fixedtEquiv}
\end{equation}
The imaginary part must be evaluated on the upper rim of the cut in $s$, as this was originally an integral
over $s'$. For a value of $s'$ on the upper rim of the cut we have
\begin{equation}
	s' + i \epsilon = \left(4 m^2 - t - u'\right) + i \epsilon = 4 m^2 - t - (u' - i \epsilon) \eolc
\end{equation}
and the corresponding value for $u'$ lies thus on the lower rim of the cut. For the imaginary part of the amplitude this
implies
\begin{equation}
	\Im \A(s'+i\epsilon,t,u) = \Im \A(s,t,u'-i\epsilon) = -\Im \A(s,t,u'+i\epsilon) \eolc
\end{equation}
where we made use of the Schwarz reflection principle in the last equality. Inserting this result into
Eq.~\eqref{eq:fixedtEquiv} yields the alternative formulation of the unsubtracted fixed-$t$ dispersion relation,
\begin{equation}
	\A(s,t,u) = \inv{\pi} \int\limits_{s\th}^\infty ds'\; \frac{\Im \A(s',t,u)}{s'-s} 
					+ \inv{\pi} \int\limits_{u\th}^\infty du'\; \frac{\Im \A(s,t,u')}{u'-u} \eolc
	\label{eq:fixedtDispRel1}
\end{equation}
where in each integral the imaginary part must be evaluated on the upper rim of the cut that is integrated over.

So far we have assumed that the scattering amplitude does not contain any isolated poles. These can be generated
by intermediate single particle states of which, depending on the process under consideration, an arbitrary number can
be present. As we have discussed in the previous section, their contribution can be calculated using the residue theorem
in Eq.~\eqref{eq:residueTheorem}. The exchange of a particle of mass $M$ in the $s$-channel leads to a simple pole at 
$s= M^2$ in the amplitude. It must be included in the dispersion relation by adding
\begin{equation}
	\frac{\Res\left(\A(s,t,u),s=M^2\right)}{s-M^2} \eolp
\end{equation}
Each pole leads to a similar contribution and these must simply be added up.

It is not compulsory to keep $t$ fixed in order to write a dispersion relation. Indeed, Mandelstam~\cite{Mandelstam1958}
argued\footnote{A formal proof of the conjecture does not exist.} that, apart from subtractions, the amplitude can be
written as
\begin{equation}
\begin{split}
	\A(s,t,u) = &\inv{\pi^2} \int\limits_{s\th}^\infty ds' \int\limits_{t\th}^\infty dt'
\frac{\rho_{st}(s',t')}{(s'-s)(t'-t)}
			+ \inv{\pi^2} \int\limits_{t\th}^\infty dt' \int\limits_{u\th}^\infty du'
\frac{\rho_{tu}(t',u')}{(t'-t)(u'-u)}\\[3mm]
			&+ \inv{\pi^2} \int\limits_{s\th}^\infty ds' \int\limits_{u\th}^\infty du'
\frac{\rho_{su}(s',u')}{(s'-s)(u'-u)}
			+ \text{pole contributions} \eolc
	\label{eq:MandelstamRepresentation}
\end{split}
\end{equation}
where the spectral functions $\rho_{st}$, $\rho_{tu}$ and $\rho_{su}$ are real. This is usually called the Mandelstam
representation or the double dispersion relation. It gives the analytic structure of $\A(s,t,u)$ for all complex $s$,
$t$ and $u$ and implies maximal analyticity in each channel, with cuts only for real values of the Mandelstam variables.

We show now that the fixed $t$ dispersion relation in Eq.~\eqref{eq:fixedtDispRel1} can be derived from the more general
Mandelstam representation.
As a first step, we split the third integral in Eq.~\eqref{eq:MandelstamRepresentation} into two by means of the
identities
\begin{equation}
	\inv{(s'-s)(u'-u)} = \left( \inv{s'-s} + \inv{u'-u} \right) \left( \inv{s'-s + u'-u} \right)
\end{equation}
and
\begin{equation}
	(s'-s) + (u'-u) = s' - \left( 4 m^2 - t - u'\right) = u' - \left( 4 m^2 - t - s'\right) \eolp
\end{equation}
The Mandelstam representation then becomes
\begin{equation}\begin{split}
	\A(s,t,u) &= \inv{\pi} \int\limits_{s\th}^\infty ds' \inv{s'-s} \left[
							\inv{\pi} \int\limits_{t\th}^\infty dt'\; \frac{\rho_{st}(s',t')}{t'-t} 
							+ \inv{\pi} \int\limits_{u\th}^\infty du'
							\frac{\rho_{su}(s',u')}{u'-\left(4 m^2 - t - s'\right)} \right] \\[3mm]
					&+ \inv{\pi} \int\limits_{u\th}^\infty du' \inv{u'-u} \left[
							\inv{\pi} \int\limits_{t\th}^\infty dt'\; \frac{\rho_{tu}(t',u')}{t'-t} 
							+ \inv{\pi} \int\limits_{s\th}^\infty ds'
							\frac{\rho_{su}(s',u')}{s'-\left(4 m^2 - t - u'\right)} \right] \\[3mm]
			&+ \text{pole contributions} \eolp
		\label{eq:MandelstamRepresentation2}
\end{split}\end{equation}
Using the Sokhotski-Plemelij formula,
\begin{equation}
	\inv{x \pm i \epsilon} = \mp i \pi \delta(x) + \mathcal{P} \inv{x} \eolc
	\label{eq:dispS-P}
\end{equation}
we can evaluate the imaginary part of $\A$ at $s+i\epsilon$ along the $s$-channel cut and find it to be the first square
bracket in Eq.~\eqref{eq:MandelstamRepresentation2} evaluated at $s' = s$. Similarly, the imaginary part on the upper
rim of the $u$-channel cut is given by the second square bracket evaluated at $u' = u$. This then leads to the fixed $t$
dispersion relation
\begin{equation}\begin{split}
	\A(s,t,u) = \inv{\pi} \int\limits_{s\th}^\infty ds'\; \frac{\Im \A(s',t,u)}{s'-s} 
					+ \inv{\pi} \int\limits_{u\th}^\infty du'\; \frac{\Im \A(s,t,u')}{u'-u}\\
					+ \text{pole contributions} \eolp
\end{split}\end{equation}

The Mandelstam representation involves double integrals and, if required by the amplitude's asymptotic behaviour,
subtractions must be performed in both integration variables. After one subtraction, it takes the form
\begin{equation}\begin{split}
	\A(s,t,u) = &\A(s_1, t_1, u_1) \\&+ \frac{(s-s_1)(t-t_1)}{\pi^2}
					\int\limits_{s\th}^\infty ds' \int\limits_{t\th}^\infty dt'
\frac{\rho_{st}(s',t')}{(s'-s_1)(s'-s)(t'-t_1)(t'-t)}\\[3mm]
					&+ \text{two analogous integrals} + \text{pole contributions} \eolp
\end{split}\end{equation}
From this follows the usual once-subtracted form of the fixed $t$ dispersion relation in the same way as demonstrated
above for the unsubtracted form.

\section{The Omn\`es problem} \label{sec:dispOmnes}

We are looking for functions $\varphi(s)$ that are analytic in the entire complex plane up to a cut running from $s\th$
to $\infty$ and real on the real axis below the cut. Their discontinuity is given by
\begin{equation}
	\disc \varphi(s) = \theta(s-s\th)\, e^{-i \delta(s)} \sin \delta(s) \varphi(s) \eolc
	\label{eq:discPhi}
\end{equation}
where $\delta(s)$ is a bounded real-valued function. This problem was solved by Omn\`es~\cite{Omnes1958}%
\footnote{Omn\`es studied integral equations in this article and Eq.~\eqref{eq:discPhi}, which we take as starting
point, appears as an intermediate step.}.

Using the definition of the discontinuity, Eq.~\eqref{eq:discPhi} can be rewritten as
\begin{equation}
	\varphi(s+i \epsilon) = e^{2i\delta(s)} \varphi(s-i \epsilon) \eolc
\end{equation}
and for the discontinuity of the logarithm of $\varphi(s)$ we thus obtain
\begin{equation}
	\disc \log\, \varphi(s) = \frac{\log\, \varphi(s+i\epsilon) - \log\, \varphi(s-i\epsilon)}{2i} = \theta(s-s\th)\,
\delta(s) \eolp
\end{equation}
Because $\delta(s)$ goes to a constant at infinity, we can write a once subtracted dispersion relation,
\begin{equation}
	\log\, \varphi(s) = \log\, \varphi(0) + \frac{s}{\pi} \int\limits_{s\th}^\infty ds'\, \frac{\disc \log\,
\varphi(s')}{s'(s'-s)}
							= \log\, \varphi(0) + \frac{s}{\pi} \int\limits_{s\th}^\infty ds'\, \frac{\delta(s')}{s'(s'-s)}
\eolp
\end{equation}
The solution, normalised to 1 at $s=0$, is called the Omn\`es function and given by
\begin{equation}
	\Omega(s) = \frac{\varphi(s)}{\varphi(0)} = \exp \left\{ \frac{s}{\pi} \int\limits_{s\th}^\infty ds'\,
\frac{\delta(s')}{s'(s'-s)} \right\} \eolp
	\label{eq:dispOmnes}
\end{equation}
On the rims of the cut, it can be written as
\begin{equation}
	\Omega(s \pm i \epsilon) = \exp \left\{ \frac{s}{\pi}\; \mathcal{P}\hspace{-1.05em} \int\limits_{s\th}^\infty ds'\,
\frac{\delta(s')}{s'(s'-s)} \right\}
										e^{\pm i \delta(s)} \eolc
	\label{eq:dispOmnesCut}
\end{equation}
and consequently, $|\Omega(s)|$ is continuous and the phase on the upper and lower rim of the cut is simply given by
$e^{i \delta}$ and $e^{-i \delta}$, respectively.
Furthermore, $\Omega(s)$ has no zeros. Even though it seems now that the general solution of Eq.~\eqref{eq:discPhi} is
simply $\varphi(s) = c\, \Omega(s)$, this
is not true: the Omn\`es function can be multiplied by any entire function $P(s)$ and the equation still holds because
of
\begin{equation}\begin{split}
	\disc P(s) \Omega(s) &= \frac{P(s+i\epsilon) \Omega(s+i\epsilon) - P(s-i\epsilon) \Omega(s-i\epsilon)}{2i}\\[2mm]
									&= \frac{P(s) \Omega(s+i\epsilon) - P(s) \Omega(s-i\epsilon)}{2i} = P(s)\, \disc \Omega(s)
	\eolp
\end{split}\end{equation}
Thus, $\varphi(s) = P(s) \Omega(s)$ for an arbitrary polynomial $P(s)$ is the general solution of
Eq.~\eqref{eq:discPhi}. If the asymptotic growth of $\varphi(s)$ is limited to some power $s^N$, then the polynomial is
also restricted. In order to find the constraint on the polynomial, we first have to discuss the asymptotic behaviour of
the Omn\`es function.

Assume that the phase shift $\delta(s)$ reaches $k \pi$ at $s=\Lambda^2$ and stays at that value for larger $s$. If $k$
is an integer, the cut ends at $\Lambda^2$ because we then have $\sin\,(k \pi) = 0$ and the discontinuity vanishes. For
$s \geq \Lambda^2$ we can now evaluate the integral by splitting up the integration path. We find for the first part
\begin{equation}
	\frac{s}{\pi} \int\limits_{s\th}^{\Lambda^2} ds' \frac{\delta(s')}{s'(s'-s)}
		\, \stackrel{s \to \infty}{=}  \, -\inv{\pi} \int\limits_{s\th}^{\Lambda^2} ds' \frac{\delta(s')}{s'} \eolc
\end{equation}
which is finite and independent of $s$ and for the second part
\begin{equation}
	\frac{s}{\pi} \int\limits_{\Lambda^2}^{\infty} ds' \frac{\delta(s')}{s'(s'-s)}
		= k \int\limits_{\Lambda^2}^{\infty} ds' \left( \inv{s'-s}-\inv{s'} \right)
		= -k \log \left( \frac{\Lambda^2-s}{\Lambda^2} \right) \eolp 
\end{equation}
We can thus conclude for the asymptotic behaviour of the Omn\`es function:
\begin{equation}
	\Omega(s) \asymp \left( \frac{\Lambda^2-s}{\Lambda^2} \right)^{-k} \eolp
	\label{eq:dispOmnesAsymptotic}
\end{equation}
We stress that this is valid for arbitrary $k$ and not only for integers. This result implies that the general solution 
for $\varphi(s)$ is of the form
\begin{equation}
	\varphi(s) = \Omega(s) \sum_{i = 0}^{N+\lfloor k \rfloor} c_i s^i \eolp
\end{equation}
The Omn\`es function will play an essential role in the dispersive treatment of $\eta \to 3 \pi$.

%% file: main/pipi.tex
\chapter{\texorpdfstring{${\pi \pi}$}{Pion pion} scattering} \label{chp:pipi}

Rescattering effects in the final state account for a very substantial contribution to the $\etapi$ decay amplitude. 
Indeed, the failure of chiral perturbation theory at one-loop level to reproduce the measured decay width is understood
to be caused by the large size of these contributions \cite{Neveu+1970,Roiesnel+1981}. The dispersive method allows one
to treat, in principle, arbitrary many rescattering processes of the pions in the final state. In this chapter, we
introduce those aspects of $\pi \pi$ scattering that are relevant to the dispersive analysis. After the discussion of
the kinematics, we derive two useful representations of the scattering amplitude. Expressed in terms of the so-called
phase shifts---three real functions that are briefly discussed at the end of this chapter---$\pi\pi$ scattering will
later enter the dispersion relations.

\section{One-pion states} \label{sec:1pionStates}

In Sec.~\ref{sec:strongGoldstoneBosons}, we have discussed the Goldstone bosons of QCD, the light pseudoscalar
octet, in detail and listed these particles with their masses and some quantum numbers in Table~\ref{tab:strongMesons}.
The lightest of these mesons are the pions, since they are the Goldstone bosons of $SU(2)_L \times SU(2)_R$, which is
only weakly broken due to the smallness of the up and down quark masses. The pions have total isospin $I = 1$, hence
forming an isospin triplet, and are also eigenstates of the third component $I_3$. Accordingly, the one-pion states are
of the form $\ket{I,I_3}$ with
\begin{equation}
	\ket{1,+1} = \ket{\pi^+}, \qquad \ket{1,0} = \ket{\pi^0}, \qquad \ket{1,-1} = \ket{\pi^-} \eolp
	\label{eq:1pionStates}
\end{equation}
It is sometimes more convenient to work with the pion fields $\pi^i$ instead. They are related to the physical pion
fields by

\begin{equation}
	\pi^\pm \equiv \inv{\sqrt{2}} \left( \pi^1 \mp i \pi^2 \right) \eolc \qquad \pi^0 \equiv \pi^3 \eolc
\end{equation}
and consequently have no definite third isospin component. We will choose the set of fields and states that proves more
suitable in a given situation. The physical pion states can be expressed in terms of the states $\ket{\pi^i}$ as
\begin{equation}
	\ket{\pi^\pm} = \inv{\sqrt{2}} \left( \ket{\pi^1} \pm i \ket{\pi^2} \right) \eolc \qquad
		\ket{\pi^0} = \ket{\pi^3} \eolp
\end{equation}
That the relative sign between the states is not the same as between the fields ensures that indeed the field operators
$\pi^\pm$ annihilate the states $\ket{\pi^\pm}$ as they should. For the conjugated states we find
\begin{equation}
	\bra{\pi^\pm} = \inv{\sqrt{2}} \left( \bra{\pi^1} \mp i \bra{\pi^2} \right) \eolc \qquad 
		\bra{\pi^0} = \bra{\pi^3} \eolc
\end{equation}
where the relative sign warrants the correct orthogonality relations.

\section{Two-pion states} \label{sec:2pionStates}

A generic two pion state has no definite isospin, but it can be decomposed into a sum of eigenstates of $I$ and $I_3$.
From the three pion flavours, one can construct nine different two-pion states and their decomposition in to
isospin states is given by
\begin{equation}
	\begin{array}{l@{\,=\,}ll@{\,=\,}l}
		\ket{\pi^\pm\pi^\pm} & \ket{2,\pm2} \eolc & 
		\ket{\pi^0\pi^0} & \sqrt{\dfrac{2}{3}} \ket{2,0} - \dfrac{1}{\sqrt{3}} \ket{0,0} \eolc\\[6mm]
		\ket{\pi^\pm\pi^0} & \dfrac{1}{\sqrt{2}} \big( \ket{2,\pm1} \pm \ket{1,\pm1} \big) \eolc &
		\ket{\pi^0\pi^\pm} & \dfrac{1}{\sqrt{2}} \big( \ket{2,\pm1} \mp \ket{1,\pm1} \big) \eolc\\[6mm]
		\ket{\pi^\pm\pi^\mp} & 
		\multicolumn{3}{l}
			{\dfrac{1}{\sqrt{6}} \ket{2,0} \pm \dfrac{1}{\sqrt{2}} \ket{1,0} + \dfrac{1}{\sqrt{3}} \ket{0,0} \eolp}
	\end{array}
	\label{eq:2pionStates}
\end{equation}
Of these states, only two have definite total isospin. Note, however, that two pions always form eigenstates of $I_3$,
because the third components can simply be added. The numbers appearing in the above expressions are the so-called
Clebsch-Gordan coefficients. They are tabulated, e.g., in Ref.~\cite{PDG2010}. Equation~\eqref{eq:2pionStates} can be
inverted leading to the decomposition of the isospin states into two pion states:
\begin{equation}
	\begin{array}{l@{\,=\,}ll@{\,=\,}l}
		\ket{2,\pm2} & \ket{\pi^\pm\pi^\pm}\eolc &
		\ket{2,\pm1} & \dfrac{1}{\sqrt{2}} \big( \ket{\pi^\pm \pi^0} + \ket{\pi^0 \pi^\pm} \big) \eolc\\[5mm]
		\ket{2,0} & \multicolumn{3}{l}{\dfrac{1}{\sqrt{6}} \big( \ket{\pi^+\pi^-} + 2 \ket{\pi^0\pi^0} 
							+ \ket{\pi^-\pi^+} \big) \eolc} \\[5mm]
		\ket{1,\pm1} & \dfrac{1}{\sqrt{2}} \big( \pm \ket{\pi^\pm \pi^0} \mp \ket{\pi^0 \pi^\pm} \big) \eolc &
		\ket{1,0} & \dfrac{1}{\sqrt{2}} \big( \ket{\pi^+\pi^-} - \ket{\pi^-\pi^+} \big) \eolc  \\[5mm]
		\ket{0,0} & \multicolumn{3}{l}
					{\dfrac{1}{\sqrt{3}} \big( \ket{\pi^+\pi^-} - \ket{\pi^0\pi^0} + \ket{\pi^-\pi^+} \big) \eolp}
	\end{array}
\label{eq:2pionIsospinStates}
\end{equation}
Since the pions are bosons, they obey Bose statistics, which requires that the wave function of the state must be even
(odd) under the exchange of the pions if the angular momentum quantum number $l$ is even (odd). From
Eq.~\eqref{eq:2pionIsospinStates} we can read off that the angular momentum is even for the states with $I = 0$ or $2$ 
and odd for the states with $I = 1$.

We will now treat this problem more formally in terms of the fields $\pi^i$. The single pion states form an isospin
triplet and hence transform under the representation $D^1$ of $SU(2)$:
\begin{equation}
	\pi^i \mapsto R^{ik} \pi^k \eolc \quad R \in SO(3) \eolp
	\label{eq:singlePionTransf}
\end{equation}
A two pion state is then a tensor with two isospin indices, $v^{ij}$, that transforms according to the representation
$D^1 \otimes D^1$ of $SU(2)$:
\begin{equation}
	v^{ij} \mapsto R^{ik} R^{jl} v^{kl} \eolc \quad R \in SO(3) \eolp
	\label{eq:twoPionTransf}
\end{equation}
Because an isospin rotation by $R$ only changes the third component of the isospin, but not the total isospin, the
isospin states in Eq.~\eqref{eq:2pionIsospinStates} with equal value for $I$ only transform among themselves. This
means that the space of two pion states contains three subspaces that are closed under isospin transformations. In
other words, the representation $D^1 \otimes D^1$ is reducible. The size of the subspaces can be determined by simply
counting the number of states they contain. From Eq.~\eqref{eq:2pionIsospinStates} follows that the $I = 0$ state is a
singlet, the \mbox{$I = 1$} states form a triplet, and the $I = 2$ states a quintet. Accordingly, they transform under
the $SU(2)$ representations $D^0$, $D^1$, and $D^2$, respectively.

The decomposition of the isospin states into two pion states is equivalent to the decomposition of the reducible
representation $D^1 \otimes D^1$ into the irreducible representations $D^0$, $D^1$, and $D^2$. This, in turn,
corresponds to the decomposition of the tensor $v^{ij}$ into tensors $v^{ij}_I$ that transform only among themselves
under isospin rotations. The singlet part of the tensor must be proportional to the identity matrix, since it remains
unchanged under any rotation. From Eq.~\eqref{eq:2pionIsospinStates} we know that the triplet part must be
anti-symmetric and the quintet part symmetric. Of the latter we have to remove the part that is proportional to the
identity matrix as this belongs to the singlet thus rendering the quintet traceless. The decomposition that possesses
all these properties is given by
\begin{equation}\begin{split}
	v^{ij}_0 &= \frac{\left<v\right>}{3} \delta^{ij} \eolc\\
	v^{ij}_1 &= \inv{2} \left( v^{ij} - v^{ji} \right) \eolc\\
	v^{ij}_2 &= \inv{2} \left( v^{ij} + v^{ji} \right) - \frac{\left<v\right>}{3} \delta^{ij}\eolc
\end{split}\end{equation}
where $\left<v\right>$ denotes the trace of $v^{ij}$.
One can check that the multiplicities of these subspaces are correct, that they sum up to $v^{ij}$ and also that they
are closed under isospin transformations. We can now define projection operators
\begin{equation}\begin{split}
	P_0^{kl,ij} &= \inv{3} \delta^{kl} \delta^{ij} \eolc\\[1mm]
	P_1^{kl,ij} &= \inv{2} \left( \delta^{ki} \delta^{lj} - \delta^{li} \delta^{kj} \right) \eolc\\[1mm]
	P_2^{kl,ij} &= \inv{2} \left( \delta^{ki} \delta^{lj} + \delta^{li} \delta^{kj} \right) 
								- \inv{3}\delta^{kl} \delta^{ij} \eolc
	\label{eq:2pionIsospinProjection}
\end{split}\end{equation}
with $v_I^{kl} = P_I^{kl,ij} v^{ij}$, $P_I^{mn,kl} P_J^{kl,ij} = \delta_{IJ} P_I^{mn,ij}$ and 
$P_I^{kl,ij} = P_I^{ij,kl}$, which can be checked by direct calculation. A two pion state with definite total
isospin~$I$ can now simply be written as
\begin{equation}
	\ket{\pi^k \pi^l}_I = P_I^{kl,ij} \ket{\pi^i \pi^j} \eolp
\end{equation}

\section{Kinematics and isospin structure of the amplitude} \label{sec:isoStructurePiPi}

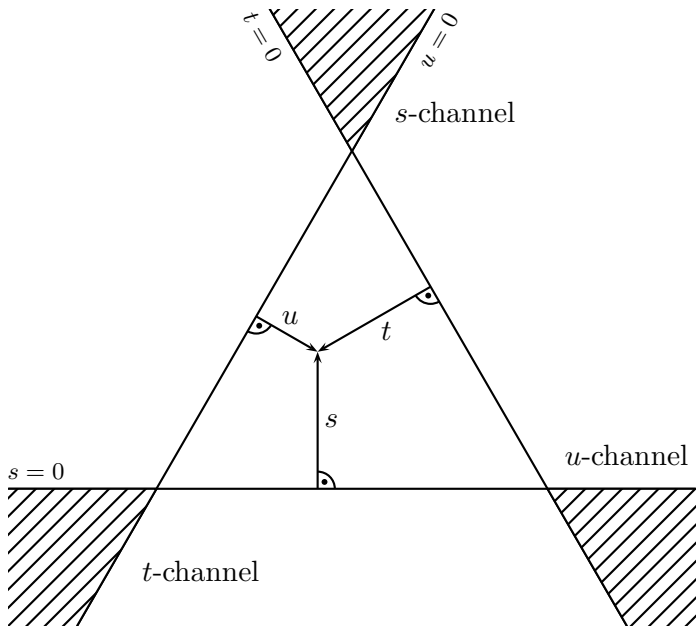
\begin{figure}[tb]
\psset{unit=.9cm}
\begin{center}
	\begin{pspicture*}(0,0)(10,9)
		\psset{linewidth=\mylw}
		\psline(0,2)(10,2)
		\psline(1,0)(6.77,10)
		\psline(9,0)(3.23,10)
		\pspolygon[fillstyle=hlines*](0,2)(2.155,2)(1,0)(1,-2)(-2,-2)(-2,2)
		\pspolygon[fillstyle=hlines*](10,2)(7.845,2)(9,0)(9,-2)(12,-2)(12,2)
		\pspolygon[fillstyle=hlines*](3.23,10)(5,6.928)(6.77,10)
		\rput(9,2.5){$u$-channel}
		\rput(6.5,7.5){$s$-channel}
		\rput(2.8,0.8){$t$-channel}
		\rput(0.4,2.25){\footnotesize{$s=0$}}
		\rput{-60}(3.7,8.6){\footnotesize{$t=0$}}
		\rput{60}(6.25,8.55){\footnotesize{$u=0$}}
		\psline{->}(4.5,2)(4.5,4)
		\psarc{-}(4.5,2){0.25}{0}{90}  
		\psdot[linewidth=0.1pt](4.6,2.1)
		\psline{->}(6.143,4.949)(4.5,4)
		\psarc{-}(6.143,4.949){0.25}{210}{300}
		\psdot[linewidth=0.1pt](6.11,4.82)
		\psline{->}(3.607,4.516)(4.5,4)
		\psarc{-}(3.607,4.516){0.25}{-120}{-30}
		\psdot[linewidth=0.1pt](3.65,4.38)
		\rput(4.7,3){$s$}
		\rput(5.5,4.3){$t$}
		\rput(4.1,4.5){$u$}
	\end{pspicture*}
\end{center}
\caption{The Mandelstam diagram for $\pi \pi \to \pi \pi$ scattering. The shaded areas represent
the physical region, the height of the triangle is $4 \mpi^2$.}
\label{fig:pipiMandelstamDiagram}
\end{figure}

We have discussed the kinematics and the crossing relations for the general $2 \to 2$ scattering process in
Sec.~\ref{sec:strongScattering}. If we neglect the pion mass difference, $\pi \pi$ scattering includes only particles
of equal mass, such that $|\vec{p}_1| = |\vec{p}_2| = |\vec{p}_3| = |\vec{p}_4| \equiv p$ and $E_1 = E_2 = E_3 = E_4 =
\sqrt{\mpi^2 + p^2} \equiv E_\text{CM}$. In the centre-of-mass frame of the $s$-channel, the Mandelstam variables then
read
\begin{equation}\begin{split}
	s &= E_\text{CM}^2 = 4 (\mpi^2 + p^2) \eolc\\[1mm]
	t &= -2 p^2 (1 - \cos \theta) \eolc\\[1mm]
	u &= -2 p^2 (1 + \cos \theta) \eolp
\end{split}\end{equation}
The Mandelstam diagram for this process with the physical region shaded is shown in
Fig.~\ref{fig:pipiMandelstamDiagram}.

The $T$-matrix element for the scattering process $\pi^i \pi^j \to \pi^k \pi^l$ is given by
\begin{multline}
	\bra{\pi^k(p_3) \pi^l(p_4)} iT \ket{\pi^i(p_1) \pi^j(p_2)}\\ = i(2\pi)^4 \delta^4(p_1+p_2-p_3-p_4) 
			\A_{\pi\pi \to \pi\pi}^{kl,ij}(p_1,p_2,p_3,p_4) \eolc
\end{multline}
where $i$, $j$, $k$, and $l$ are the isospin indices of the pion states. The matrix element $\A_{\pi\pi \to
\pi\pi}^{kl,ij}$ is invariant under isospin transformations which implies (see Eq.~\eqref{eq:singlePionTransf})
\begin{equation}
	\A_{\pi\pi \to \pi\pi}^{kl,ij} = R^{kc} R^{ld} R^{ia} R^{jb} \A_{\pi\pi \to \pi\pi}^{cd,ab} \eolp
	\label{eq:pipiAmplitudeTransformation}
\end{equation}
For the following argument, it is helpful to rewrite this equation in terms of row vectors of $R$. We define the
three-vectors $u^a = R^{ia}, v^b = R^{jb}, w^c = R^{kc}$, and $x^d = R^{ld}$ and then define
\begin{equation}
	f(\vec{u},\vec{v},\vec{w},\vec{x}) = w^c x^d u^a v^b \A_{\pi\pi \to \pi\pi}^{cd,ab} \eolc
\end{equation}
which is nothing else than the right-hand side of Eq.~\eqref{eq:pipiAmplitudeTransformation}.
Performing another isospin rotation by $R$,
\begin{equation}\begin{split}
	f(\vec{u},\vec{v},\vec{w},\vec{x}) \mapsto  &w^c x^d u^a v^b R^{cp} R^{dq} R^{am} R^{bn} 
																\A_{\pi\pi \to \pi\pi}^{pq,mn}\\[2mm]
			& = (R^T \vec{w})^p (R^T \vec{x})^q (R^T \vec{u})^m (R^T \vec{v})^n \A_{\pi\pi \to \pi\pi}^{pq,mn}\\[2mm]
			&= f(R^T\vec{u},R^T\vec{v},R^T\vec{w},R^T\vec{x}) \eolc
\end{split}\end{equation}
thus amounts to a rotation by $R^T$ of the arguments of $f$. As $f$ is required to be invariant under this
transformation, it must only depend on scalar products of the four vectors. Furthermore, each term must contain
\emph{all} of them linearly. We can thus conclude that
\begin{equation}
	f(\vec{u},\vec{v},\vec{w},\vec{x}) = A_1 \cdot (\vec{u} \cdot \vec{v})(\vec{w} \cdot \vec{x}) 
													+ A_2 \cdot (\vec{u} \cdot \vec{w})(\vec{v} \cdot \vec{x})
													+ A_3 \cdot (\vec{u} \cdot \vec{x})(\vec{v} \cdot \vec{w}) \eolc
\end{equation}
where $A_1$, $A_2$ and $A_3$ are functions of the pion momenta. The scalar products can be simplified, e.g., 
\begin{equation}
	(\vec{u} \cdot \vec{v})(\vec{w} \cdot \vec{x}) = (R^{ia} R^{ja})(R^{kc} R^{lc}) 
					= (R^{ia} (R^T)^{aj})(R^{kc} (R^T)^{cl}) = \delta^{ij} \delta^{kl} \eolc
\end{equation}
and similarly for the others. The amplitude is thus brought to the form
\begin{equation}
	\A_{\pi\pi \to \pi\pi}^{kl,ij}(s,t,u) = A_1(s,t,u)\, \delta^{ij} \delta^{kl} + A_2(s,t,u)\, \delta^{ik} \delta^{jl} 
				+ A_3(s,t,u)\, \delta^{il} \delta^{jk} \eolp
	\label{eq:pipiIsospinStructure0}
\end{equation}
$A_1$, $A_2$ and $A_3$ are the scattering amplitudes in the $s$-, $t$-, and $u$-channel, respectively. We show now that
they are related with each other due to the symmetries of the process.

From Sec.~\ref{sec:isoStructurePiPi} we know that the amplitude is invariant under $t \leftrightarrow u$ and
$i \leftrightarrow j$, which here implies that
\begin{equation}
	A_1(s,t,u) = A_1(s,u,t) \eolp
	\label{eq:pipiCross1}
\end{equation}
Furthermore, due to crossing symmetry the amplitude is invariant under simultaneous exchange of $i$ with $l$ and $p_1$
with $-p_4$ (or $s$ with $t$), as well as of $i$ with $k$ and $p_1$ with $-p_3$ (or $s$ with $u$). This requires
\begin{equation}
	A_2(s,t,u) = A_1(t,s,u) = A_1(t,u,s) \eolc \ \ A_3(s,t,u) = A_1(u,t,s) = A_1(u,s,t) \eolp
	\label{eq:pipiCross2}
\end{equation}
In both relations, we used the symmetry in the second and third argument in the last equality. This means that the
amplitude can be written in terms of a single function $A(s,t,u) \equiv A_1(s,t,u)$ as
\begin{equation}
	\A_{\pi\pi \to \pi\pi}^{kl,ij}(s,t,u) = A(s,t,u)\, \delta^{ij} \delta^{kl} + A(t,u,s)\, \delta^{ik} \delta^{jl} 
						+ A(u,s,t)\, \delta^{il} \delta^{jk} \eolp
	\label{eq:pipiIsospinStructure}
\end{equation}
As expected, crossing symmetry allows us to use the same function to describe the scattering process in
all three channels.

\section{Isospin and partial wave decomposition} \label{sec:pipiIsospinDecomp}

We have discussed in Sec.~\ref{sec:2pionStates} that two pion states can be decomposed into isospin states with total
isospin $I = 0, 1, 2$ and derived the corresponding projection operators in Eq.~\eqref{eq:2pionIsospinProjection}. We
will now make use of these to decompose the $\pi \pi$ scattering amplitude into isospin amplitudes. This is most
conveniently achieved by using the $T$-matrix element in the form of Eq.~\eqref{eq:pipiIsospinStructure0}.

We decompose the two pion states in the $T$-matrix element into isospin states, $\ket{\pi^i \pi^j}_I = P_I^{ij,kl}
\ket{\pi^k \pi^l}$, and obtain
\begin{equation}\begin{split}
	\bra{\pi^k \pi^l} T \ket{\pi^i \pi^j} &= \sum_{I,J} {}_J \bra{\pi^k \pi^l} T \ket{\pi^i \pi^j}_I 
				= \sum_{I} {}_I \bra{\pi^k \pi^l} T \ket{\pi^i \pi^j}_I\\[1mm]
				&= \sum_I  P_I^{kl,rs} P_I^{ij,mn} \bra{\pi^r \pi^s} T \ket{\pi^m \pi^n} \eolp
	\label{eq:pipiIsospinDecompTMatrix}
\end{split}\end{equation}
We define the isospin amplitudes $F_I^{kl,ij}$ by
\begin{multline}
	{}_I \bra{\pi^k(p_3) \pi^l(p_4)} iT \ket{\pi^i(p_1) \pi^j(p_2)}_I \\
						= i(2\pi)^4 \delta^4(p_1+p_2-p_3-p_4) F_I^{kl,ij}(p_1,p_2,p_3,p_4)
\end{multline}
which are, according to Eq.~\eqref{eq:pipiIsospinDecompTMatrix}, related to the amplitude by
\begin{equation}
	F_I^{kl,ij} = P_I^{kl,rs} P_I^{ij,mn} \A_{\pi\pi \to \pi\pi}^{rs,mn} = P_I^{kl,ij} F_I \eolc
	\label{eq:pipiFIDefinition}
\end{equation}
where the last equality defines the functions $F_I$ given by
\begin{equation}\begin{split}
	F_0 &= 3 A_1 + A_2 + A_3\eolc\\
	F_1 &= A_2 - A_3\eolc\\
	F_2 &= A_2 + A_3 \eolp
\label{eq:pipiIsospinDecomp}
\end{split}\end{equation}
We have now decomposed the $\pi \pi$ scattering amplitude into three amplitudes with fixed isospin as
\begin{equation}
	\A_{\pi\pi \to \pi\pi}^{kl,ij}(s,t,u) = P_0^{kl,ij} F_0(s,t,u) + P_1^{kl,ij} F_1(s,t,u) + P_2^{kl,ij} F_2(s,t,u)\eolp
	\label{eq:pipiIsospinDecompAmplitude}
\end{equation}

The next step is to make use of the fact that the $S$-matrix is unitary. This leads to the unitarity condition in
Eq.~\eqref{eq:strongUnitarityCondition2}, which we can apply to $\A_{\pi\pi \to \pi\pi}^{kl,ij}$. For energies $s < 16
\mpi^2$, only two pion states contribute to the sum over intermediate states, because no additional pion pair can be
created. Since QCD does not violate parity and the pions are pseudoscalars, it is not possible to create only a single
additional pion. In this low energy region, the intermediate states $\ket{n}$ are thus of the form $\ket{\pi^m(q_1)
\pi^n(q_2)}$ and the sum is given by
\begin{equation}
	\sum_n = \inv{2} \sum_{m,n=1}^{3} \int \frac{d^3q_1}{2 q_1^0 (2 \pi)^3} \frac{d^3q_2}{2 q_2^0 (2 \pi)^3}\eolc
\end{equation}
The states $\ket{\pi^m(q_1) \pi^n(q_2)}$ and $\ket{\pi^n(q_2) \pi^m(q_1)}$ both appear in the sum and, as they are
identical, each state is therefore counted twice. This is compensated by the factor $1/2$ in front of the sum.
Together with the explicit expression for the sum over the intermediate states, the unitarity condition takes the form
\begin{equation}\begin{split}
	\Im\, \A_{\pi\pi \to \pi\pi}^{kl,ij} 
				= \frac{(2 \pi)^4}{4} \sum_{m,n=1}^{3} &\int \frac{d^3q_1 d^3q_2}{4 q_1^0 \, q_2^0 \, (2 \pi)^6} \\[1mm]
					&\!\!\! \times \delta^4(q_1 + q_2 - p_1 - p_2) \A_{\pi\pi \to \pi\pi}^{kl,mn} 
					\A_{\pi\pi \to \pi\pi}^{\ast\,mn,ij} \eolp
\end{split}\end{equation}
On both sides of the equation we now insert the decomposition in Eq.~\eqref{eq:pipiIsospinDecompAmplitude}. For the
left-hand side we find
\begin{equation}
	P_0^{kl,ij} \,\Im\,F_0 + P_1^{kl,ij} \,\Im\,F_1 + P_2^{kl,ij} \,\Im\,F_2\eolc
\end{equation}
while the right-hand side is given by
\begin{equation}
	\inv{64 \pi^2} \int \frac{d^3q_1 d^3q_2}{q_1^0 \, q_2^0} \delta^4(q_1 + q_2 - p_1 - p_2)
		\left(P_0^{kl,ij} F_0^{} F_0^\ast + P_1^{kl,ij} F_1^{} F_1^\ast + P_2^{kl,ij} F_2^{} F_2^\ast \right) \eolp
\end{equation}
In order to perform the summation over $m$ and $n$ we used the orthogonality condition on the projection operators 
$P_I^{kl,ij}$. Equating and restoring the arguments of the functions $F_I$ leads to
\begin{multline}
	\Im\, F_I(p_1,p_2,p_3,p_4) = \\ \inv{64 \pi^2} \int \frac{d^3q_1 d^3q_2}{q_1^0 \, q_2^0}
							\delta^4(q_1 + q_2 - p_1 - p_2) F_I^{}(q_1,q_2,p_3,p_4) F_I^\ast(p_1,p_2,q_1,q_2) \eolp
	\label{eq:pipiImFI}
\end{multline}
We go on by changing to the centre of mass frame with $\vec{p}_1 = -\vec{p}_2$ and $\vec{p}_3 = -\vec{p}_4$, which
allows to rewrite the delta function as $\delta(\sqrt{s} - q_1^0 - q_2^0) \delta^3(\vec{q}_1 + \vec{q}_2)$.
Evaluation of the integration over $\vec{q}_2$ sets $\vec{q}_1 = -\vec{q}_2$ and, as the masses are equal, $q_1^0 =
q_2^0$. We define
\begin{equation}
	q \equiv |\vec{q}_1| = \sqrt{\left(q_1^0\right)^2-\mpi^2}
\end{equation}
and change the integration variable by means of $d^3q_1 = d\Omega\, dq\, q^2 = d\Omega\, dq_1^0\, q_1^0\, q$. We showed
in Sec.~\ref{sec:isoStructurePiPi} that the amplitude can be written as a function of $s$ and the scattering angle only.
Defining the angles $\theta_s = \angle(\vec{p}_1,\vec{p}_3)$, $\theta_1 = \angle(\vec{q}_1,\vec{p}_3)$ and $\theta_2 =
\angle(\vec{p}_1,\vec{q}_1)$ allows therefore to write Eq.~\eqref{eq:pipiImFI} as
\begin{equation}\begin{split}
	\Im\, F_I(s,\theta_s) &= \inv{64 \pi^2} \int d\Omega\,dq_1^0\, \frac{\sqrt{\left(q_1^0\right)^2-\mpi^2}}{q_1^0}\,
								\delta(\sqrt{s} - 2 q_1^0) F_I^{}(s,\theta_1) F_I^\ast(s,\theta_2)\\
								&= \inv{128 \pi^2} \int d\Omega\, \sqrt{\frac{s-4\mpi^2}{s}} 
											F_I^{}(s,\theta_1) F_I^\ast(s,\theta_2) \eolc
\label{eq:pipiImFI2}
\end{split}\end{equation}
where the integration over $q_1^0$ has been evaluated in the second line. The $F_I$ are expanded into partial waves as
\begin{equation}
	F_I(s,\theta) = \sum_{\ell=0}^\infty (2\ell+1) P_\ell(\cos \theta) f_\ell^I(s) \eolc
\end{equation}
where the functions $P_\ell$ are the Legendre polynomials. They are normalised as
\begin{equation}
	\int\limits_{-1}^{1} dx \; P_\ell(x) P_{\ell'}(x) = \frac{2}{2\ell+1}\, \delta_{\ell \ell'} \eolc
\end{equation}
and satisfy
\begin{equation}
	\int d\Omega \; P_\ell(\cos \theta_1) P_{\ell'}(\cos \theta_2) 
					= \frac{4 \pi}{2\ell+1}\, \delta_{\ell \ell'}\, P_\ell(\cos \theta_s) \eolp
	\label{eq:LegendreRelation}
\end{equation}
The Legendre polynomials are tabulated (e.g., in Ref.~\cite{Gradshteyn+2007}, Section 8.91) and we list here only the
first few of them:
\begin{equation}
	P_0(x) = 1, \quad P_1(x) = x, \quad P_2(x) = \inv{2} (3 x^2 - 1), \quad  P_3(x) = \inv{2} (5 x^3 - 3 x)\eolp
\end{equation}
Equation~\eqref{eq:LegendreRelation} and the fact that the Legendre polynomials are linearly independent allow
us to obtain from Eq.~\eqref{eq:pipiImFI2} that
\begin{equation}
	\Im f_\ell^I(s) = \frac{\sqrt{1-4 \mpi^2/s}}{32 \pi} f_\ell^I(s) f_\ell^I{}^\ast(s) \eolc
\end{equation}
which requires the partial waves $f_\ell^I(s)$ to be of the form
\begin{equation}
	f_\ell^I(s) = \frac{32 \pi}{\sqrt{1-4 \mpi^2/s}} \sin \delta_\ell^I(s) e^{i \delta_\ell^I(s)} \eolc
\end{equation}
where the $\delta_\ell^I(s)$ are real functions. Recall that the $T$-matrix element for $\pi \pi$ scattering remains
unchanged under simultaneous interchange of $i$ with $j$ and $p_1$ with $p_2$ (see Sec.~\ref{sec:isoStructurePiPi}).
The consequences of this exchange are
\begin{equation}\begin{split}
	\cos \theta_s &\to -\cos \theta_s \eolc \\[1mm]
	P_I^{kl,ij} &\to P_I^{kl,ji} = (-1)^I P_I^{kl,ij} \eolc \\[1mm]
		P_\ell(\cos \theta_s) &\to P_\ell(-\cos \theta_s) = (-1)^\ell P_\ell(\cos \theta_s) \eolc
\end{split}\end{equation}
and it follows by comparison with Eq.~\eqref{eq:pipiIsospinDecompAmplitude} that $\ell$ and $I$ must either both be 
odd or both be even because other terms would destroy the symmetry. The same result we have already derived from Bose
statistics for the two pion states in Eq.~\eqref{eq:2pionIsospinStates}. Restricting the sum over partial wave to $S$-
and $P$-waves, i.e., to $\ell = 0, 1$, we finally obtain for the isospin amplitudes:
\begin{equation}\begin{split}
	F_0(s,\cos \theta_s) &= \frac{32 \pi}{\sqrt{1-4 \mpi^2/s}} \sin \delta_0(s) e^{i \delta_0(s)}\eolc\\
	F_1(s,\cos \theta_s) &= \frac{96 \pi \cos \theta_s}{\sqrt{1-4 \mpi^2/s}} \sin \delta_1(s) e^{i \delta_1(s)}\eolc\\
	F_2(s,\cos \theta_s) &= \frac{32 \pi}{\sqrt{1-4 \mpi^2/s}} \sin \delta_2(s) e^{i \delta_2(s)}\eolp
	\label{eq:pipiFIResult}
\end{split}\end{equation}
$\delta_0 \equiv \delta_0^0$, $\delta_1 \equiv \delta_1^1$ and $\delta_2 \equiv \delta_0^2$ are the so-called phase  
shifts and will be discussed in detail in Sec.~\ref{sec:pipiPhaseShifts}.

We summarise the approximations that entered the decomposition. All intermediate states apart from two pion states
have been neglected. The sum over partial waves has been limited to the $S$- and $P$-waves and the pions are treated in
the isospin limit, where $\mpip = \mpiz \equiv \mpi$.

\section{Reconstruction theorem} \label{sec:pipiReconstructionTheorem}

In this section, we derive a useful decomposition of the $\pi\pi$ scattering amplitude into partial waves of fixed
isospin that will play a very important role in the construction of the dispersion relations for $\etapi$.
It was first derived for $\pi \pi$ scattering in Ref.~\cite{Stern+1993}.  Later, a generalised proof for arbitrary $2
\to 2$ scattering processes of scalar particles was given in Ref.~\cite{Zdrahal+2008}. That formula does,
however, not include the decomposition into contributions of fixed isospin. Nevertheless, the present discussion is
based on the procedure used for the general proof but extends it to also account for isospin.

The principle of maximal analyticity implies that the scattering amplitude $\A_{\pi\pi \to \pi\pi}^{kl,ij}$ is analytic
in the complex plane up to cuts starting at $4 \mpi^2$ in each channel. To simplify the notation we perform the
decomposition for a scattering process involving four definite pion species. As the proof does in no way rely on
specific properties of the pion species that take part in the scattering, it is then nevertheless valid for any possible
$\pi \pi$ scattering process. This allows us to omit all the indices on the amplitude and simply denote it by $\A
(s,t,u)$.

We state now a three-times subtracted fixed $t$ dispersion relation for the scattering amplitude in the form of
Eq.~\eqref{eq:fixedtDispRel1}. The Roy equations \cite{Roy1971} suggest that only two subtractions are enough, but using
three suppresses the unknown high-energy tails of the dispersion integrals more strongly. Later it will be shown that
as a consequence they can even be neglected. From the three subtractions arises a quadratic subtraction polynomial
in $s$ and $u$ with coefficients that may depend on $t$, denoted by $P^t_2(s,u)$. The dispersive representation at fixed
$t$ then reads
\begin{align}
	\A(s,t,u) = P^t_2(s,u)
		&+ \frac{s^3}{\pi} \int\limits_{4 \mpi^2}^\infty ds'\, \frac{\Im\, \A(s',t,u(s'))}{s'^3 (s'-s)} 
		\nonumber \displaybreak[0] \\[2mm] 
		&+ \frac{u^3}{\pi} \int\limits_{4 \mpi^2}^\infty du'\, \frac{\Im\, \A(s(u'),t,u')}{u'^3 (u'-u)} \eolp
\end{align}
As usual, one of the Mandelstam variables is understood to be defined from the other two, that is
\begin{equation}\begin{split}
	s(u') &= 4 \mpi^2 - t - u' = s + u - u' \eolc \\[3mm]
	u(s') &= 4 \mpi^2 - t - s' = s + u - s' \eolp
\end{split}\end{equation}

The polynomial $P^t_2(s,u)$ can alternatively be expressed in terms of $(s+u)$ and $(s-u)$. But due to
Eq.~\eqref{eq:strongScatMandelstamSum}, the former is equal to $4 \mpi^2 - t$ which simply contributes to the constant
term of the polynomial. $P^t_2(s,u)$ is thus a polynomial in $(s-u)$ only. Furthermore, as $t$ can be eliminated from
$P^t_2(s,u)$ by means of $t = 4 \mpi^2 - s - u$ and since it should still be a quadratic polynomial in $s$ and $u$,
$P^t_2(s,u)$ is at most quadratic in all three Mandelstam variables.

The integrals are now split into a low- and a high-energy part at some scale $\Lambda^2$ that is much larger than $s$
and $u$. Accordingly, we can use an expansion in $s/s'$ in the high energy integral over $s'$, namely
\begin{equation}
	\inv{s'-s} =\inv{s'} + \inv{s'} \frac{s}{s'} + \O \left( (s/s')^2 \right) \eolp
	\label{pipi:s'sExpansion}
\end{equation}
As the Mandelstam variables are quadratic in the momenta they are counted as $p^2$. In the high-energy region, $s' \geq
\Lambda^2$ and thus the second term in the expansion is $\O\left( (p/\Lambda)^2 \right)$ or $\O(p^2)$ for short. With
the use of this expansion, the high energy integral over $s'$ becomes
\begin{equation}
	\frac{s^3}{\pi} \int\limits_{\Lambda^2}^\infty ds'\, \frac{\Im\, \A(s',t,u(s'))}{s'^3 (s'-s)} = 
			s^3 H^s_\Lambda(t) + \O(p^8) \eolp
\end{equation}
The first term in the expansion in Eq.~\eqref{pipi:s'sExpansion} leads to $s^3 H^s_\Lambda(t)$, which is $\O(p^6)$. The
remainder of the expansion is suppressed by even higher powers of $s/\Lambda^2$ and thus at least $\O(p^8)$.
In the function $H^s_\Lambda(t)$ the amplitude in the integrand is expanded in $t/s' = \O(p^2)$ such that we find
\begin{equation}\begin{split}
	s^3 H^s_\Lambda(t) 
			&= s^3 \inv{\pi} \int\limits_{\Lambda^2}^\infty ds'\, \frac{\Im\,\A(s',t,u(s'))}{s'^4}\\[1mm]
			&= s^3 \left( \inv{\pi} \int\limits_{\Lambda^2}^\infty ds'\, \frac{\Im\, \A(s',0,u(s'))}{s'^4} + 
				\inv{\pi} \int\limits_{\Lambda^2}^\infty ds'\, \frac{\O(t/s')}{s'^4}  \right)\\[2mm]
			 &= s^3 H^s_\Lambda(0) + \O(p^8) \eolc
\end{split}\end{equation}
where $H^s_\Lambda$ is now independent from all the Mandelstam variables. An analogous expression can be found for the
integral over $u'$ and both high-energy integrals can therefore, up to terms of $\O(p^8)$, be absorbed in the
polynomial, which is now of cubic order:
\begin{equation}
	P^t_3(s,u) = P^t_2(s,u) + s^3 H^s_\Lambda(0) + u^3 H^u_\Lambda(0) \eolp
\end{equation}
If we had started out from an only twice subtracted dispersion relation, the high-energy integrals would be a polynomial
merely at $\O(p^4)$. The representation for the scattering amplitude has now become
\begin{equation}\begin{split}
	\A(s,t,u) = P^t_3(s,u)
		&+ \frac{s^3}{\pi} \int\limits_{4 \mpi^2}^{\Lambda^2} ds'\, \frac{\Im\, \A(s',t,u(s'))}{s'^3 (s'-s)} \\[2mm]
		&+ \frac{u^3}{\pi} \int\limits_{4 \mpi^2}^{\Lambda^2} du'\, \frac{\Im\, \A(s(u'),t,u')}{u'^3 (u'-u)} + \O(p^8)
	\eolp
\end{split}\end{equation}
The next step is to decompose the amplitude in the integrand into isospin amplitudes. To that end, we invert the isospin
decomposition in Eq.~\eqref{eq:pipiIsospinDecomp} and find
\begin{equation}
	A_1 = \inv{3} ( F_0 - F_2 ) \eolc \qquad A_2 = \inv{2} ( F_1 + F_2 ) \eolc \qquad A_3 = -\inv{2} ( F_1 - F_2 ) \eolp
\end{equation}
We have now expressed the scattering amplitude for each channel in terms of amplitudes with fixed isospin. Each one of
these isospin amplitudes $F_I$ is then decomposed into partial waves $f^I_\ell$ of the appropriate channel. For example,
the expansion in the $s$-channel reads
\begin{equation}
	F_I(s,t,u) = f^I_0 (s) + 3 \cos \theta_s\, f^I_1(s) + f^I_{\ell \geq 2} (s,t,u) \eolc
\end{equation}
and similarly for the other channels.
In the integrand, only the imaginary parts of the partial waves will appear, and we argue now that the imaginary parts
of the partial waves with $\ell \geq 2$ are $\O(p^8)$. The leading order contribution to the scattering amplitude is
$\O(p^2)$ and thus, according to the unitarity condition~\eqref{eq:strongUnitarityCondition2}, the imaginary part starts
at $\O(p^4)$. The contribution at $\O(p^2)$ must therefore be real and contains no poles or cuts, such that it can be
written as a polynomial. This polynomial is at most linear in $s$, $t$ and $u$ because only the linear term can be made
dimensionless by the factor $1/F_\pi^2$ appearing at this order. Any higher power of the Mandelstam variables must be
compensated by masses in the denominator. However, such terms blow up in the chiral limit and do hence not appear. But
the linear polynomial is part of the $S$- and $P$-waves. The constant term and the term linear in $s$
clearly belong to the $S$-wave. The linear term in $t$ and $u$ can be rewritten in terms of $(t+u)$ and $(t-u)$. 
The former is equal to $4 \mpi^2 - s$, i.e., $S$-wave, while the latter is $(s-4 \mpi^2) \cos \theta_s$ and thus
$P$-wave. The $D$- and higher waves start to contribute at $\O(p^4)$ only and their imaginary part is, again in
accordance with the unitarity condition, at least $\O(p^8)$.

Furthermore, we have discussed in Sec.~\ref{sec:pipiIsospinDecomp} that only partial waves contribute where $I$ and
$\ell$ are either both odd or both even and as a consequence $f^0_1 = f^1_0 = f^2_1 = 0$. We find for the
imaginary part of the $s$ and $u$ channel amplitude
\begin{equation}\begin{split}
	\Im\, \A(s',t,u) &\equiv \Im\, A_1(s',t,u) = \inv{3} \left( \Im\, F_0(s',t,u) - \Im\, F_2(s',t,u) \right) \\[1mm]
						&= \inv{3} \left( \Im\, f_0^0(s') - \Im\, f^2_0(s') \right) + \O(p^8) \eolc \\[3mm]
		\Im\, \A(s,t,u') &\equiv \Im\, A_3(s,t,u') = -\inv{2} \left( \Im\, F_1(s,t,u') - \Im\, F_2(s,t,u') \right) \\[1mm]
						&= -\inv{2} \left( 3 \cos \theta_{u'}\, \Im\, f_1^1(u') - \Im\, f^2_0(u') \right) + \O(p^8) \eolp
\end{split}\end{equation}
The $u$-channel scattering angle can be read off from Eq.~\eqref{eq:strongScatsChannelScatteringAngle} by exchanging
particles 2 and 4 as well as $s$ and $u$ and is given by
\begin{equation}
	\cos \theta_{u'} = \frac{t-s}{u' - 4 \mpi^2} \eolp
\end{equation}
After inserting the explicit expression for the scattering angle, the integral over the $P$-wave can be simplified
further using the definition of $s(u')$, namely
\begin{equation}\begin{split}
	-\frac{3 u^3}{2 \pi} \int\limits_{4 \mpi^2}^{\Lambda^2} du'\, 
				\frac{(t-s(u'))\, \Im\, f^1_1(u')}{(u'-4 \mpi^2) u'^3 (u'-u)} = 
	-\frac{3 u^3}{2 \pi} \int\limits_{4 \mpi^2}^{\Lambda^2} du'\,
				\frac{\Im\, f^1_1(u')}{(u'-4 \mpi^2) u'^3}\\[1mm]
	+(s-t)\, \frac{3 u^3}{2 \pi} \int\limits_{4 \mpi^2}^{\Lambda^2} du'\,
				\frac{\Im\, f^1_1(u')}{(u'-4 \mpi^2) u'^3 (u'-u)} \eolp
\end{split}\end{equation}
The first term contains an integral that is independent of the Mandelstam variables and can thus be absorbed
in the polynomial. Putting everything together, we arrive at
\begin{equation}\begin{split}
	\A(s,t,u) = &P_3^t(s,u)
			+ \frac{s^3}{3\pi} \int\limits_{4 \mpi^2}^{\Lambda^2} ds'\, \frac{\Im\, f^0_0(s')}{s'^3(s'-s)}
			- \frac{s^3}{3\pi} \int\limits_{4 \mpi^2}^{\Lambda^2} ds'\, \frac{\Im\, f^2_0(s')}{s'^3(s'-s)} \\[1mm]
			&+(s-t)\, \frac{3 u^3}{2 \pi} \int\limits_{4 \mpi^2}^{\Lambda^2} du'\,
					\frac{\Im\, f^1_1(u')}{(u'-4\mpi^2) u'^3 (u'-u)}\\[1mm]
			&+ \frac{u^3}{2\pi} \int\limits_{4 \mpi^2}^{\Lambda^2} du'\, \frac{\Im\, f^2_0(u')}{u'^3(u'-u)}
			+ \O(p^8) \eolp
\label{pipi:reconstrFixedt}
\end{split}\end{equation}
We argued that subtraction polynomial $P_2^t(s,u)$ is at most quadratic in all the Mandelstam variables. Since the terms
that we have in addition absorbed into the polynomial are cubic in $s$ and $u$ with coefficients that are independent of
$t$, also $P_3^t(s,u)$ is at most cubic in all the Mandelstam variables.

The amplitude $\A(s,t,u)$ is invariant under the interchange of $t$ and $u$, which must also be true for the dispersive
representation even though it is not at all manifest in Eq.~\eqref{pipi:reconstrFixedt}. This can be remedied as
follows. In exactly the same way as above, we can derive a dispersive representation for $\A(s,t,u)$ at fixed $u$,
instead of at fixed $t$. For this purpose, we need the partial wave expansion of the $t$-channel amplitude,
\begin{equation}\begin{split}
	\Im\, \A(s,t',u) &\equiv \Im\, A_2(s,t',u) = \inv{2} \left( \Im\, F_1(s,t',u) + \Im\, F_2(s,t',u) \right) \\
						&= \inv{2} \left( 3 \cos \theta_{t'}\, \Im\, f_1^1(t') + \Im\, f^2_0(t') \right) + \O(p^8) \eolp
\end{split}\end{equation}
The result is exactly the same as before but with $t$ and $u$ interchanged:
\begin{equation}\begin{split}
		\A(s,t,u) = &P_3^u(s,t)
			+ \frac{s^3}{3\pi} \int\limits_{4 \mpi^2}^{\Lambda^2} ds'\, \frac{\Im\, f^0_0(s')}{s'^3(s'-s)}
			- \frac{s^3}{3\pi} \int\limits_{4 \mpi^2}^{\Lambda^2} ds'\, \frac{\Im\, f^2_0(s')}{s'^3(s'-s)} \\[1mm]
			&+(s-u)\, \frac{3 t^3}{2 \pi} \int\limits_{4 \mpi^2}^{\Lambda^2} dt'\,
					\frac{\Im\, f^1_1(t')}{(t'-4\mpi^2) t'^3 (t'-t)}\\[1mm]
			&+ \frac{t^3}{2\pi} \int\limits_{4 \mpi^2}^{\Lambda^2} dt'\, \frac{\Im\, f^2_0(t')}{t'^3(t'-t)}
			+ \O(p^8) \eolp
\label{pipi:reconstrFixedu}
\end{split}\end{equation}
The integrals over the $s$-channel are the same as before, but the integrals over the $u$-channel have been replaced by
integrals over the $t$-channel. But at fixed $t$, these terms simply form a linear polynomial in $s$ and $u$ with
coefficients that depend on $t$ and thus they can be absorbed in the polynomial $P_3^t(s,u)$. We can therefore
symmetrise the representation as
\begin{equation}\begin{split}
		\A(s,t,u) = &P_3(s,t,u)
			+ \frac{s^3}{3\pi} \int\limits_{4 \mpi^2}^{\Lambda^2} ds'\, \frac{\Im\, f^0_0(s')}{s'^3(s'-s)}
			- \frac{s^3}{3\pi} \int\limits_{4 \mpi^2}^{\Lambda^2} ds'\, \frac{\Im\, f^2_0(s')}{s'^3(s'-s)} \\
			&+(s-u)\, \frac{3 t^3}{2 \pi} \int\limits_{4 \mpi^2}^{\Lambda^2} dt'\,
					\frac{\Im\, f^1_1(t')}{(t'-4\mpi^2) t'^3 (t'-t)}\\
			&+(s-t)\, \frac{3 u^3}{2 \pi} \int\limits_{4 \mpi^2}^{\Lambda^2} du'\,
					\frac{\Im\, f^1_1(u')}{(u'-4\mpi^2) u'^3 (u'-u)}\\
			&+ \frac{t^3}{2\pi} \int\limits_{4 \mpi^2}^{\Lambda^2} dt'\, \frac{\Im\, f^2_0(t')}{t'^3(t'-t)}
			+ \frac{u^3}{2\pi} \int\limits_{4 \mpi^2}^{\Lambda^2} du'\, \frac{\Im\, f^2_0(u')}{u'^3(u'-u)}
			+ \O(p^8) \eolc
\label{pipi:reconstrSymmetric}
\end{split}\end{equation}
where $P_3(s,t,u)$ is now a polynomial in all the Mandelstam variables, because none is kept fixed. We have already
argued before that it is of third order in $s$, $t$ and $u$. At fixed $t$, $P_3(s,t,u)$ as well as the integrals over
$t'$ can be understood as a third order polynomial in $s$ and $u$ with $t$-dependent coefficients and thus
constitute the polynomial $P^t_3(s,u)$. The symmetric representation has therewith been reduced to the representation at
fixed $t$ in Eq.~\eqref{pipi:reconstrFixedt}. In a similar way, its equivalence with the representation at fixed $u$
can be shown.

As the non-polynomial part of Eq.~\eqref{pipi:reconstrSymmetric} clearly is invariant under the interchange of $t$ and
$u$, also the polynomial must have this property in order for the entire amplitude to be symmetric as well. It is thus
of the form
\begin{equation}
	P_3(s,t,u) = a + b s + c s^2 + d s^3 + e (t-u)^2 + f s (t-u)^2
	\label{pipi:P3stu}
\end{equation}
which is simply the most general third order polynomial in $s$, $t$ and $u$ that is symmetric in $t$ and $u$.

Introducing the three functions
\begin{equation}\begin{split}
	M_0(s) &= a_0 + b_0 s + c_0 s^2 + d_0 s^3 + 
					\frac{s^3}{3\pi} \int\limits_{4 \mpi^2}^{\Lambda^2} ds'\, \frac{\Im\,f^0_0(s')}{s'^3(s'-s)} \eolc \\
	M_1(s) &= a_1 + b_1 s + c_1 s^2 +
					\frac{3 s^3}{2 \pi} \int\limits_{4 \mpi^2}^{\Lambda^2} ds'\, 
							\frac{\Im\,f^1_1(s')}{(s'-4\mpi^2) s'^3(s'-s)} \eolc \\
	M_2(s) &= a_2 + b_2 s + c_2 s^2 + d_2 s^3 + 
					\frac{s^3}{2\pi} \int\limits_{4 \mpi^2}^{\Lambda^2} ds'\, \frac{\Im\, f^2_0(s')}{s'^3(s'-s)} \eolc \\
\end{split}\end{equation}
and assuming that $P_3(s,t,u)$ can be absorbed in the polynomials contained in the $M_I$ (as will be shown below), the
$\pi \pi$ scattering amplitude has, up to corrections of $\O(p^8)$, the form
\begin{equation}
	\A(s,t,u) = M_0(s) + (s-u) M_1(t) + (s-t) M_1(u) + M_2(t) + M_2(u) - \frac{2}{3} M_2(s) \eolp
	\label{eq:pipiReconstructionTheorem}
\end{equation}

In order to show that also the polynomial can be split into the functions $M_I$, one simply has to compare the
expression
for $P_3(s,t,u)$ in Eq.~\eqref{pipi:P3stu} with the polynomial that results from the polynomial parts of the $M_I$. The
polynomials are not unique due to the relation among the Mandelstam variables, and the comparison is easiest achieved
if one of the Mandelstam variables is eliminated. As the polynomials in the $M_I$ contain totally eleven coefficients,
while there are only six in $P_3(s,t,u)$, a total of five coefficients can be set to arbitrary values. The decomposition
into the functions $M_I$ is therefore not unique. A simple but tedious calculation shows that the relations between the
coefficients of the two polynomials are
\begin{equation}\begin{split}
	a_0 &= a + 9 s_0^2 e + 3 s_0 a_1 - \tfrac{4}{3} a_2 - 3 s_0 b_2 - 9 s_0^2 c_2 - 27 s_0^3 d_2 \eolc \\[1mm]
	b_0 &= b - 12 s_0 e - 9 s_0^2 f - 3 a_1 + \tfrac{5}{3} b_2 + 9 s_0 c_2 + 27 s_0^2 d_2 \eolc \\[1mm]
	c_0 &= c + 3 e + 12 s_0 f - \tfrac{4}{3} c_2 \eolc \\[1mm]
	d_0 &= d - 3 f - \tfrac{4}{3} d_2 \eolc \\[1mm]
	b_1 &= 2 e - 6 s_0 f - c_2 - 9 s_0 d_2 \eolc \\[1mm]
	c_1 &= 4 f + 3 d_2 \eolc
\end{split}\end{equation}
with $3s_0 \equiv s+t+u = 4 \mpi^2$, and $a_1$, $a_2$, $b_2$, $c_2$, and $d_2$ are left free. From these expressions,
one can deduce that indeed $\A(s,t,u)$ is left unchanged if the $M_I$ are all shifted as
\begin{equation}\begin{split}
	M_0(s) &\mapsto \begin{aligned}[t] M_0(s) &- \tfrac{4}{3} \alpha s^3 + 27 s_0^2 \alpha (s-s_0) 
					- \tfrac{4}{3} \beta s^2 + 9 s_0 \beta (s-s_0) \\
					&+ \tfrac{5}{3} \gamma s - 3 s_0 \gamma - \tfrac{4}{3} \delta - 3 \epsilon (s-s_0) \eolc \end{aligned}\\
	M_1(s) &\mapsto M_1(s) + 3 \alpha s^2 - 9 s_0 \alpha s - \beta s + \epsilon \eolc \\
	M_2(s) &\mapsto M_2(s) + \alpha s^3 + \beta s^2 + \gamma s + \delta \eolp
	\label{eq:pipiMIShift}
\end{split}\end{equation}

\section{Phase shifts} \label{sec:pipiPhaseShifts}

Finding a reliable representation for the scattering phases shifts $\delta_\ell^I$ that we introduced in
Eq.~\eqref{eq:pipiFIResult} is a difficult task. Over the years, many results using different approaches have been
published in the literature. A rather extensive overview of these works can be found in
Ref.~\cite{Ananthanarayan+2001}. Two more recent calculations, which are not listed in the aforementioned review, have
been presented in Refs.~\cite{Descotes-Genon+2002,Garcia-Martin+2011}.

One can, for example, obtain the phase shifts from \chpt. At tree-level, the amplitude $A(s,t,u)$ is given by
\begin{equation}
	A\tree(s,t,u) = \frac{s-\mpi^2}{\Fpi^2} \eolp
\end{equation}
Because the leading contribution to the phase shifts is $\O(p^2)$, we can expand
\begin{equation}
	\sin \delta_I(s)\, e^{i \delta_I(s)} = \delta_I(s) + \O(\delta_I^2) = \delta_I + \O(p^4) \eolp
\end{equation}
From Eqs.~\eqref{eq:pipiIsospinDecomp} and \eqref{eq:pipiFIResult}, we then obtain for the phase shifts at leading order
\begin{equation}\begin{split}
	\delta_0\tree(s) &= \sqrt{\frac{s-4\mpi^2}{s}} \, \frac{2s-\mpi^2}{32 \pi \Fpi^2} \eolc \qquad
	\delta_1\tree(s) = \sqrt{\frac{s-4\mpi^2}{s}} \, \frac{s-4\mpi^2}{96 \pi \Fpi^2} \eolc \\[2mm]
	\delta_2\tree(s) &= \sqrt{\frac{s-4\mpi^2}{s}} \, \frac{2\mpi^2-s}{32 \pi \Fpi^2} \eolp
	\label{eq:pipiPhaseTree}
\end{split}\end{equation}
But as we plan to use the phases shifts in a dispersion integral that runs up to infinity, a leading-order low-energy
approximation is definitely not a suitable choice. One could, of course, obtain the phase shifts also at higher order
in the chiral expansion. But using phenomenological inputs, much more accurate results have been calculated.

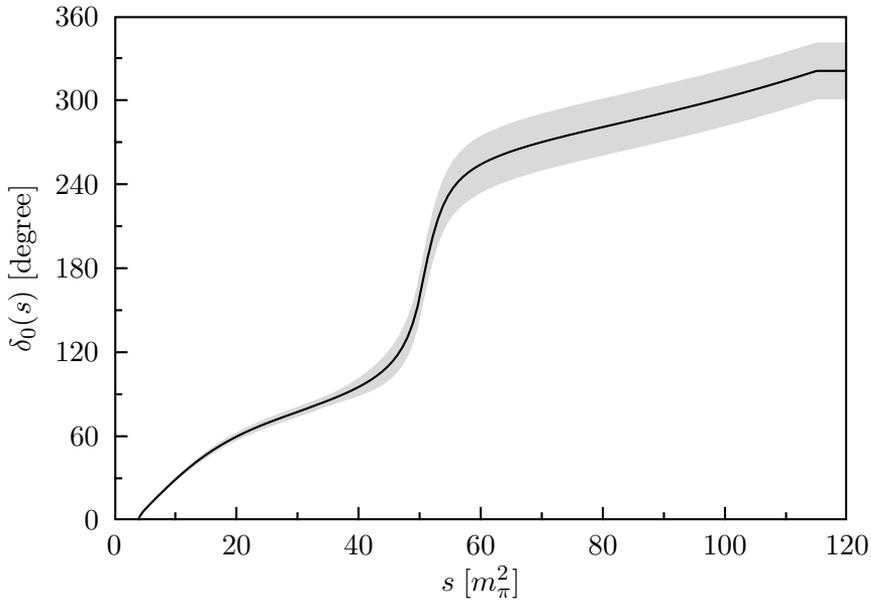
\begin{figure}[tb]
\psset{xunit=0.8mm,yunit=0.184mm}
\begin{center}\begin{pspicture*}(-18,-56)(125,367)
\psset{linewidth=\mylw}
\fileplot[plotstyle=polygon,fillstyle=solid,fillcolor=shadegray,linecolor=shadegray,linewidth=0]{data/delta0errors.dat}
\fileplot[linewidth=\mylw]{data/delta0.dat}
\psframe(0,0)(120,360)
\multips(20,0)(20,0){5}{\psline(0,5pt)}
\multips(10,0)(20,0){6}{\psline(0,3pt)}
\multido{\n=0+20}{7}{\uput{.2}[270](\n,0){\n}}
\multips(0,60)(0,60){5}{\psline(5pt,0)}
\multips(0,30)(0,60){7}{\psline(3pt,0)}
\multido{\n=0+60}{7}{\uput{.2}[180](0,\n){\n}}
\uput{.6}[270](60,0){$s \; [\mpi^2]$} 
\uput{1}[180]{90}(0,180){$\delta_0(s) \; [\text{degree}]$}
\end{pspicture*}\end{center}

\caption{The figure shows the the $\pi \pi$ scattering phase shift $\delta_0(s) \equiv \delta_0^0(s)$. The uncertainty
band is shaded in grey.}
\label{fig:pipiPhaseShifts1}
\end{figure}

\begin{figure}[p]
\psset{xunit=0.8mm,yunit=0.331mm}
\begin{center}\begin{pspicture*}(-18,-31)(125,203)
\psset{linewidth=\mylw}
\fileplot[plotstyle=polygon,fillstyle=solid,fillcolor=shadegray,linecolor=shadegray,linewidth=0]{data/delta1errors.dat}
\fileplot[linewidth=\mylw]{data/delta1.dat}
\psframe(0,0)(120,200)
\multips(20,0)(20,0){5}{\psline(0,5pt)}
\multips(10,0)(20,0){6}{\psline(0,3pt)}
\multido{\n=0+20}{7}{\uput{.2}[270](\n,0){\n}}
\multips(0,30)(0,30){6}{\psline(5pt,0)}
\multips(0,15)(0,30){8}{\psline(3pt,0)}
\multido{\n=0+30}{8}{\uput{.2}[180](0,\n){\n}}
\uput{.6}[270](60,0){$s \; [\mpi^2]$}
\uput{1}[180]{90}(0,90){$\delta_1(s) \; [\text{degree}]$}
\end{pspicture*}\end{center}

\psset{xunit=0.8mm,yunit=0.110cm}
\begin{center}\begin{pspicture*}(-18,-71)(125,2)
\psset{linewidth=\mylw}
\fileplot[plotstyle=polygon,fillstyle=solid,fillcolor=shadegray,linecolor=shadegray,linewidth=0]{data/delta2errors.dat}
\fileplot[linewidth=\mylw]{data/delta2.dat}
\psframe(0,-60)(120,0)
\multips(20,-60)(20,0){5}{\psline(0,5pt)}
\multips(10,-60)(20,0){6}{\psline(0,3pt)}
\multido{\n=0+20}{7}{\uput{.2}[270](\n,-60){\n}}
\multips(0,-50)(0,10){5}{\psline(5pt,0)}
\multips(0,-55)(0,10){7}{\psline(3pt,0)}
\multido{\n=-60+10}{7}{\uput{.2}[180](0,\n){\n}}
\uput{.6}[270](60,-60){$s \; [\mpi^2]$}
\uput{1}[180]{90}(0,-30){$\delta_2(s) \; [\text{degree}]$}
\end{pspicture*}\end{center}

\caption{The figures show the the $\pi \pi$ scattering phase shifts $\delta_1(s) \equiv \delta_1^1(s)$ and $\delta_2(s)
\equiv \delta_0^2(s)$. The uncertainty band is shaded in grey.}
\label{fig:pipiPhaseShifts2}
\end{figure}
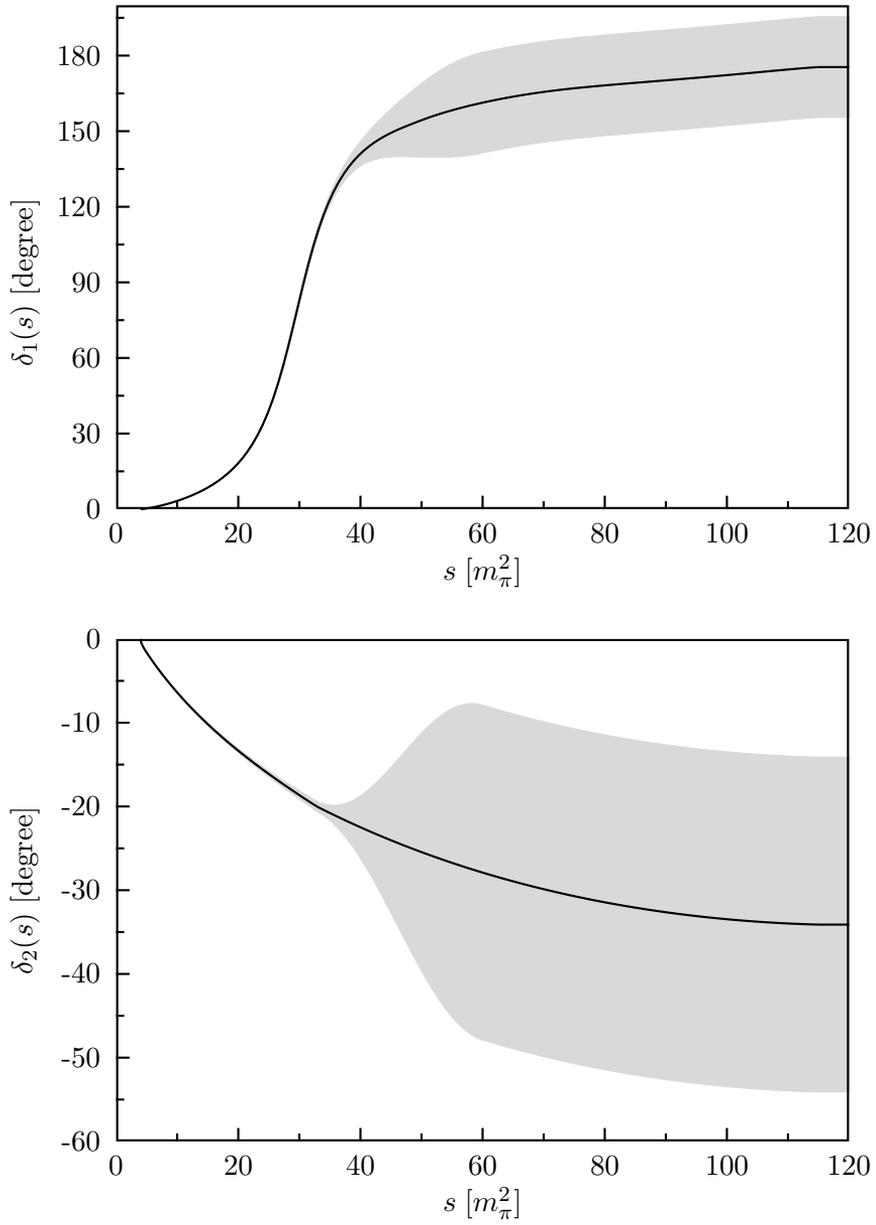

Schenk proposed the parametrisation \cite{Schenk1991}
\begin{equation}
	\tan \delta_\ell^I = \sqrt{\frac{s-4\mpi^2}{s}}\, q^{2\ell} \left( \frac{4 \mpi^2 - s_\ell^I}{s-s_\ell^I} \right) 
									\left\{ A_\ell^I + B_\ell^I q^2 + C_\ell^I q^4 + D_\ell^I q^6 \right\} \eolc
	\label{eq:pipiSchenk}
\end{equation}
with $q^2 = s/4 - \mpi^2$. Ananthanarayan et al.~\cite{Ananthanarayan+2001} have determined
the Schenk parameters from an extensive analysis based on the Roy equations~\cite{Roy1971}, using experimental inputs
for the phase shifts at high energies. The Roy equations are an infinite set of doubly subtracted dispersion relations
for the partial waves of $\pi \pi$ scattering. The subtraction constants can be identified with the $S$-wave scattering
lengths $a_0^0$ and $a_0^2$. The result from Ref~\cite{Ananthanarayan+2001} has been combined with theoretical inputs
from \chpt{} for the scattering lengths in Ref.~\cite{Colangelo2001}, where it was found that the Schenk parameters are
given by
\begin{equation}\begin{split}
	\begin{aligned}[t]
		A_0^0 &= 0.220 \eolc\\ B_0^0 &= 0.268 \eolc\\ C_0^0 &= -0.139\ee{-1} \eolc\\ D_0^0 &= -0.139\ee{-2} \eolc
	\end{aligned}
	\qquad
	\begin{aligned}[t]
		A_1^1 &= 0.379 \ee{-1} \eolc\\ B_1^1 &= 0.140\ee{-4} \eolc\\ C_1^1 &= -0.673\ee{-4} \eolc\\
		D_1^1 &= 0.163\ee{-7} \eolc
	\end{aligned}
	\qquad
	\begin{aligned}[t]
		A_0^2 &= -0.444\ee{-1} \eolc\\ B_0^2 &= -0.857\ee{-1} \eolc\\ C_0^2 &= -0.221\ee{-2} \eolc\\ 
		D_0^2 &= -0.129\ee{-3} \eolc
	\end{aligned}
\end{split}\end{equation}
in units of $\mpi$. The constants $A_\ell^I$ are the scattering lengths of their respective partial wave and the
$B_\ell^I$ are related to the effective ranges by
\begin{equation}
	b_\ell^I = B_\ell^I + \frac{4 \mpi^2}{s_\ell^I - 4 \mpi^2} a_\ell^I - (a_\ell^I)^3 \delta_{\ell 0} \eolp
\end{equation}
The parameters $s_\ell^I$ specify the values of $s$ where the phases shifts go through $90^\circ$. They are
\begin{equation}\begin{split}
	s_0^0 &= 36.77\, \mpi^2 \approx (818~\MeV)^2 \eolc \qquad 
	s_1^1 = 30.72\, \mpi^2 \approx (748~\MeV)^2 \eolc \\[1mm]
	s_0^2 &= -21.62\, \mpi^2 \eolp
\end{split}\end{equation}
The negative sign of $s_0^2$ indicates that $\delta_0^2$ stays below $90^\circ$.	The Schenk parametrisation is used up
to $s = (0.8~\GeV)^2 \approx 35\, \mpi^2$. In this region, the uncertainties on the phase shifts follow from the
uncertainties on the Schenk parameters. The lower boundary is described by
\begin{equation}\begin{split}
	\begin{aligned}[t]
		A_0^0 &= 0.215 \eolc\\ B_0^0 &= 0.262 \eolc\\ C_0^0 &= -0.171\ee{-1} \eolc\\ D_0^0 &= -0.122\ee{-2} \eolc\\
		s_0^0 &= 37.70\, \mpi^2\eolc
	\end{aligned}
	\qquad
	\begin{aligned}[t]
		A_1^1 &= 0.375 \ee{-1} \eolc\\ B_1^1 &= 0.148\ee{-3} \eolc\\ C_1^1 &= -0.864\ee{-4} \eolc\\
		D_1^1 &= 0.223\ee{-5} \eolc\\ s_1^1 &= 30.92\, \mpi^2\eolc
	\end{aligned}
	\qquad
	\begin{aligned}[t]
		A_0^2 &= -0.455\ee{-1} \eolc\\ B_0^2 &= -0.848\ee{-1} \eolc\\ C_0^2 &= -0.270\ee{-3} \eolc\\ 
		D_0^2 &= -0.228\ee{-3} \eolc\\ s_0^2 &= -28.48\, \mpi^2\eolc
	\end{aligned}
\end{split}\end{equation}
while the upper boundary is obtained from
\begin{equation}\begin{split}
	\begin{aligned}[t]
		A_0^0 &= 0.225 \eolc\\ B_0^0 &= 0.273 \eolc\\ C_0^0 &= -0.111\ee{-1} \eolc\\ D_0^0 &= -0.148\ee{-2} \eolc\\
		s_0^0 &= 35.37\, \mpi^2\eolc
	\end{aligned}
	\qquad
	\begin{aligned}[t]
		A_1^1 &= 0.385 \ee{-1} \eolc\\ B_1^1 &= -0.126\ee{-3} \eolc\\ C_1^1 &= -0.478\ee{-4} \eolc\\
		D_1^1 &= -0.223\ee{-5} \eolc\\ s_1^1 &= 30.51\, \mpi^2\eolc
	\end{aligned}
	\qquad
	\begin{aligned}[t]
		A_0^2 &= -0.434\ee{-1} \eolc\\ B_0^2 &= -0.866\ee{-1} \eolc\\ C_0^2 &= -0.443\ee{-2} \eolc\\ 
		D_0^2 &= -0.493\ee{-5} \eolp\\ s_0^2 &= -16.59\, \mpi^2\eolc
	\end{aligned}
\end{split}\end{equation}
For energies beyond $0.8~\GeV$ we rely on the same experimental input that was used in the Roy equation analysis. The
representation for the $S$ wave in the $I = 0$ channel comes from Ref.~\cite{Au+1987}, while the $P$ wave is taken from
Ref.~\cite{Hyams+1973}. For the $I = 2$ phases shift we us the representation from Ref.~\cite{Ananthanarayan+2001} that
is based on experimental data from Refs.~\cite{Losty+1974,Hoogland+1977}. For $s \geq 115\, \mpi^2 \approx 2~\GeV^2$,
these representations are no longer valid and the phase is kept at a constant value. The phase shifts are plotted in
Figs.~\ref{fig:pipiPhaseShifts1} and~\ref{fig:pipiPhaseShifts2}. The point, where the phase shifts is kept constant, is
most pronounced in $\delta_0(s)$.  The uncertainty band is shaded in grey. Up to $35\, \mpi^2$, it is calculated from
the errors on the Schenk parameters, for $s>60\, \mpi^2$ we have simply assumed a rather generous error of $\pm
20^\circ$ to account for the poor knowledge of the high-energy behaviour of the phase. In between, we have used an
interpolation that links the error on the Schenk parametrisation with the constant error at high energies.

The results from the more recent determinations of the $\pi\pi$ phase
shifts~\cite{Descotes-Genon+2002,Garcia-Martin+2011} are in agreement with the representation that we use here.
Currently, a new analysis of the Roy equations is in progress that will result in an even more accurate
representation of the phase shifts~\cite{newPipiPhase}. It has not been made use of in this thesis.

\addtocontents{toc}{\protect\pagebreak}

%% file: main/eta3pi.tex
\chapter{\texorpdfstring{$\etapi$}{Eta to 3 pions}} \label{chp:eta3pi}

The decay $\etapi$ is the main topic of this thesis and obviously deserves an extended introduction. In this chapter we
establish the basis for the later discussion of the dispersion relations. We start by proving that the
decay only occurs due to isospin violation. After the discussion of the kinematics and the structure of the decay
amplitude, we present the one-loop result from \chpt{} together with its dispersive representation. We then derive a
low-energy theorem that states that the amplitude has two zeros that are protected by chiral symmetry. Finally, we
discuss the decay rate and some experimental and theoretical results for the Dalitz plot parameters. We put special
emphasis on the rather recent measurement by the KLOE collaboration that will serve as input to the dispersive analysis.

\section{Three pion states and \texorpdfstring{$\eta$}{eta} decay} \label{sec:eta3pi3pion}

The decay of an $\eta$ into three pions is forbidden by isospin symmetry. Since the $\eta$ is a spin zero isospin
singlet state, the three pions must couple to vanishing isospin and angular momentum, in order not to violate a
symmetry. We will prove that this is impossible by showing that the $\ket{0, 0}$ state built from three pions cannot
have zero angular momentum. To that end, we decompose this state into three pion states, similarly to what
we did in Section~\ref{sec:2pionStates} for the two pion states. We refrain from giving the decomposition of all the
other isospin states, as they are not relevant for our purpose. 

We start with the third component of the isospin. The quantum number $I_3$ of a three pion state is simply the sum
of the third isospin components of the three pions. Only the states $\ket{\pi^0 \pi^0 \pi^0}$ and $\ket{\pi^+ \pi^-
\pi^0}$ can have a vanishing third component and are the only $\eta$ decay channels allowed by isospin. Of course, also
charge conservation prohibits all the other channels.

We construct a three pion state by coupling a two pion state to a single pion. As the single pion has total isospin
$I = 1$, a three pion state with vanishing total isospin can only be achieved, if the two pion state has total isospin
$I = 1$ as well. The decomposition of the $\ket{0,0}$ state must therefore be of the form
\begin{equation}
	\ket{0,0}_3 = \inv{\sqrt{3}} \big( \ket{1,+1}_1\ket{1,-1}_2 - \ket{1,0}_1\ket{1,0}_2 
			+ \ket{1,-1}_1\ket{1,+1}_2 \big) \eolc
\end{equation}
where the subscript denotes the number of pions in the state. From Eqs.~\eqref{eq:1pionStates}
and~\eqref{eq:2pionIsospinStates}, we can read off the one and two pion states and obtain
\begin{multline}
	\ket{0,0} = \inv{\sqrt{6}} \big( \ket{\pi^+ \pi^0 \pi^-} - \ket{\pi^+ \pi^- \pi^0} + \ket{\pi^0 \pi^- \pi^+} \\
							-\ket{\pi^0 \pi^+ \pi^-} + \ket{\pi^- \pi^+ \pi^0} - \ket{\pi^- \pi^0 \pi^+} \big) \eolp
\label{eq:003pions}
\end{multline}
The state containing three $\pi^0$ does not contribute at all because two $\pi^0$ cannot form a $\ket{1,0}$ state.
As the state is anti-symmetric in the pions, it has, according to Bose statistic, odd angular momentum. In particular,
the angular momentum must be nonzero and we conclude that the decay is not allowed, if isospin is conserved. Since
it is nevertheless observed in nature, isospin symmetry must be broken.

\section{Kinematics, crossing and the Dalitz plot} \label{sec:eta3piKinematics}

\begin{figure}[tb]
\SetScale{0.55}
\setlength{\unitlength}{0.55pt}
\begin{center}
\fcolorbox{white}{white}{
  \begin{picture}(468,338) (7,-29)
    \SetWidth{2.0}
    \SetColor{Black}
    \Line[arrow,arrowpos=0.5,arrowlength=8.333,arrowwidth=3.333,arrowinset=0.2](280,178)(390,288)
    \Line[arrow,arrowpos=0.5,arrowlength=8.333,arrowwidth=3.333,arrowinset=0.2](300,138)(440,138)
    \Line[arrow,arrowpos=0.5,arrowlength=8.333,arrowwidth=3.333,arrowinset=0.2](40,138)(190,138)
    \Line[arrow,arrowpos=0.5,arrowlength=8.333,arrowwidth=3.333,arrowinset=0.2](280,98)(390,-12)
    \Text(110,148)[b]{\Large{\Black{$p_1$}}}
    \Text(350,238)[lt]{\Large{\Black{$p_2$}}}
    \Text(370,148)[b]{\Large{\Black{$p_3$}}}
    \Text(350,38)[lb]{\Large{\Black{$p_4$}}}
    \Text(30,138)[r]{\Large{\Black{$A$}}}
    \Text(395,288)[lt]{\Large{\Black{$B$}}}
    \Text(450,138)[l]{\Large{\Black{$C$}}}
    \Text(395,-12)[lb]{\Large{\Black{$D$}}}
    \GOval(245,138)(60,60)(0){0.882}
  \end{picture}
}
\end{center}
\caption{The decay process $A(p_1) \to  B(p_2) C(p_3) D(p_4)$.}
\label{fig:13decay}
\end{figure}
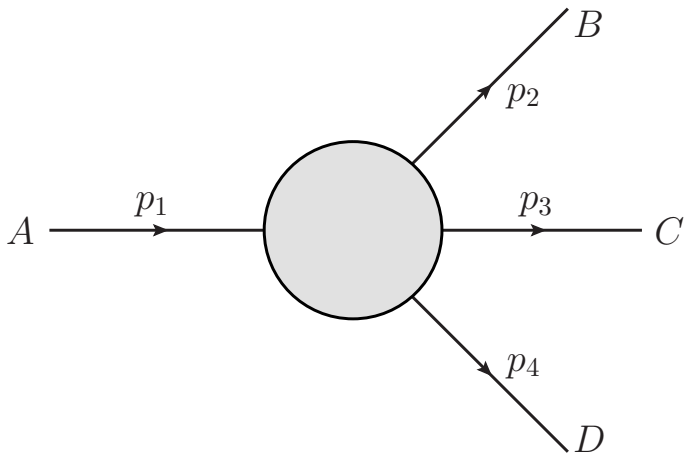

We first consider the general $1 \to 3$ decay process $A(p_1) \to  B(p_2) C(p_3) D(p_4)$ as depicted in
Fig.~\ref{fig:13decay}. We proceed in analogy to the discussion of scattering processes in
Sec.~\ref{sec:isoStructurePiPi}. The invariant amplitude $\A$ is defined from the $T$-matrix element by
\begin{equation}
	\bra{B(p_2) C(p_3) D(p_4)} T \ket{A(p_1)} = (2 \pi)^4 \, \delta^4(p_1 - p_2 - p_3 - p_4)\, \A(p_1,p_2,p_3,p_4) \eolc
\end{equation}
and the Mandelstam variables are
\begin{equation}\begin{split}
	s &= (p_2 + p_3)^2 = (p_1 - p_4)^2\eolc\\
	t &= (p_2 + p_4)^2 = (p_1 - p_3)^2\eolc\\
	u &= (p_3 + p_4)^2 = (p_1 - p_2)^2 \eolp
	\label{eq:eta3piMandelstam}
\end{split}\end{equation}
Also for the decay process, they are related according to Eq.~\eqref{eq:strongScatMandelstamSum} such that the amplitude
only depends on two independent kinematic variables.

The decay process is related by crossing to the scattering processes
\begin{alignat}{2}
	A(p_1) \bar{D}(-p_4) &\to B(p_2) C(p_3),&&\text{($s$-channel)} \label{eq:etapisChannel}\eolc\\
	A(p_1) \bar{C}(-p_3) &\to B(p_2) D(p_4),&&\text{($t$-channel)} \eolc\\
	A(p_1) \bar{B}(-p_2) &\to C(p_3) D(p_4),&\hspace{3em}&\text{($u$-channel)}\eolc
\end{alignat}
and thus it can described by the same analytic function. Indeed, every scattering process is related by crossing to
several decay processes that are, however, not necessarily physically possible. Thanks to crossing, there is no need to
discuss the kinematics for decay processes in detail, since the relevant relations can be obtained from the formul\ae{}
in Sec.~\ref{sec:strongScattering} by appropriate redefinitions of the external momenta.

\begin{figure}[tb]
\begin{center}
\psset{unit=.9}
	\begin{pspicture*}(-1,-1)(11,10)
		\psset{linewidth=\mylw}
		\psline(-1,2)(11,2)
		\psline(0.42,-1)(6.77,10)
		\psline(9.58,-1)(3.23,10)
		\rput(9.7,2.5){$u$-channel}
		\rput(7.0,8.0){$s$-channel}
		\rput(2.3,0.2){$t$-channel}
		\rput(5,2.6){decay channel}
		\rput(-.6,2.25){\footnotesize{$s=0$}}
		\rput{-60}(3.2,9.5){\footnotesize{$t=0$}}
		\rput{60}(6.7,9.45){\footnotesize{$u=0$}}
	
\psccurve[fillstyle=hlines](5.,8.49)(5.59,8.77)(5.86,9.04)(6.09,9.32)(6.3,9.6)(6.49,9.88)(6.68,10.16)(6.86,10.43)(7.04,
10.71)(7.21,10.99)(7.39,11.27)(7.56,11.55)(2.44,11.55)(2.61,11.27)(2.79,10.99)(2.96,10.71)(3.14,10.43)(3.32,10.16)(3.51,
9.88)(3.7,9.6)(3.91,9.32)(4.14,9.04)(4.41,8.77)(5.,8.49)
	
\psccurve[fillstyle=hlines](0.81,1.22)(0.27,1.6)(-0.11,1.69)(-0.46,1.75)(-0.8,1.79)(-1.14,1.82)(-1.48,1.84)(-1.81,
1.86)(-2.14,1.88)(-2.47,1.89)(-2.79,1.9)(-3.12,1.91)(-0.56,-2.52)(-0.41,-2.23)(-0.25,-1.94)(-0.1,-1.65)(0.05,-1.36)(0.2,
-1.06)(0.35,-0.76)(0.49,-0.45)(0.63,-0.14)(0.76,0.2)(0.86,0.58)(0.81,1.22)
	
\psccurve[fillstyle=hlines](9.19,1.22)(9.73,1.6)(10.11,1.69)(10.46,1.75)(10.8,1.79)(11.14,1.82)(11.48,1.84)(11.81,
1.86)(12.14,1.88)(12.47,1.89)(12.79,1.9)(13.12,1.91)(10.56,-2.52)(10.41,-2.23)(10.25,-1.94)(10.1,-1.65)(9.95,-1.36)(9.8,
-1.06)(9.65,-0.76)(9.51,-0.45)(9.37,-0.14)(9.24,0.2)(9.14,0.58)(9.19,1.22)
	
\psccurve[fillstyle=hlines](5.,3.01)(5.35,3.07)(5.47,3.12)(5.55,3.18)(5.6,3.24)(5.63,3.29)(5.66,3.35)(5.67,3.4)(5.68,
3.46)(5.69,3.51)(5.68,3.57)(5.67,3.63)(5.66,3.68)(5.65,3.74)(5.63,3.79)(5.6,3.85)(5.58,3.9)(5.55,3.96)(5.51,4.01)(5.48,
4.07)(5.43,4.13)(5.38,4.18)(5.32,4.24)(5.25,4.29)(5.13,4.35)(4.87,4.35)(4.75,4.29)(4.68,4.24)(4.62,4.18)(4.57,4.13)(4.52
,4.07)(4.49,4.01)(4.45,3.96)(4.42,3.9)(4.4,3.85)(4.37,3.79)(4.35,3.74)(4.34,3.68)(4.33,3.63)(4.32,3.57)(4.31,3.51)(4.32,
3.46)(4.33,3.4)(4.34,3.35)(4.37,3.29)(4.4,3.24)(4.45,3.18)(4.53,3.12)(4.65,3.07)(5.,3.01)
	\end{pspicture*}
\end{center}
\caption{Representation of the Mandelstam variables $s$, $t$ and $u$ in a Mandelstam diagram. The shaded areas represent
the physical regions for the decay $\eta \to \pi^+ \pi^- \pi^0$ and the related scattering processes. The height of the
triangle is $\meta^2 + 2 \mpip^2 + \mpiz^2$.}
\label{fig:eta3piMandelstamDiagram}
\end{figure}
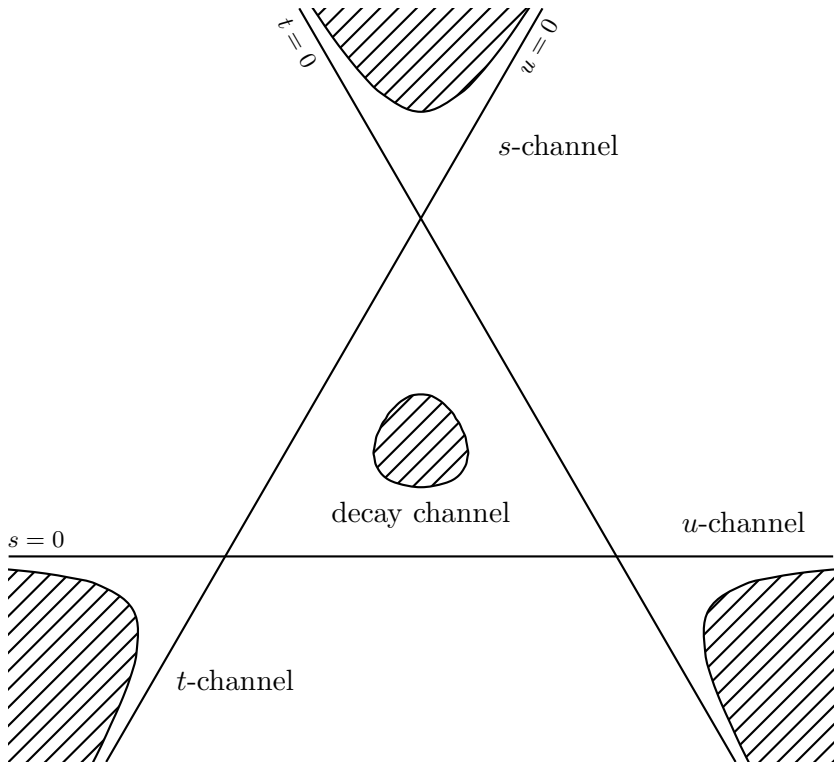

For the remainder, we treat specifically the decay $\etapi$ and the related scattering processes. We will later need
to know the phase space also for physical masses such that we discuss the charged channel here. We will stick to the
convention that $p_2$ is assigned to the $\pi^+$, $p_3$ to the $\pi^-$, and $p_4$ to the $\pi^0$ and accordingly set
$m_1 = \meta$, $m_2 = m_3 = \mpip$, and $m_4 = \mpiz$. All the results obtained from the charged channel in this way can
easily be adapted to the neutral channel or to degenerate pion masses. One simply has to use the appropriate
masses in the formul\ae.

In each channel, the physical region starts at some minimal value for the centre-of-mass momentum. For the scattering
processes, we find:
\begin{equation}\begin{split}
	s\text{-channel:} \quad &(\meta + \mpiz)^2 \leq s \eolc \\
	t\text{-channel:} \quad &(\meta + \mpip)^2 \leq t \eolc \\
	u\text{-channel:} \quad &(\meta + \mpip)^2 \leq u \eolp
\end{split}\end{equation}
The bounds for the other two Mandelstam variables can be calculated from Eqs.~\eqref{eq:strongScattu},
\eqref{eq:strongScatsu} and~\eqref{eq:strongScatst}, respectively. Figure~\ref{fig:eta3piMandelstamDiagram} shows the
physical regions for the three channels in a Mandelstam diagram. The height of the triangle is now given by $\meta^2 +
2
\mpip^2 + \mpiz^2$ and the shape of the three physical regions clearly differs from $\pi \pi$ scattering. For large
energies, the
mass differences become irrelevant and the physical regions approach the elongated sides of the triangle
asymptotically.

Let us now apply the kinematic formula that we derived for scattering processes also to the decay process. Again, we
can choose to express the amplitude in terms of one of the Mandelstam variables and an angle. Since there is only one
allowed decay channel, the three possible choices correspond to the same process and lead to equivalent results. We
decide to work in the rest frame of the two charged pions, where $\vec{p}_2 = -\vec{p}_3$, and express $t$ and
$u$ in terms of $s$ and $\theta_s = \angle(\vec{p}_1,\vec{p}_3) = \angle(\vec{p}_2,-\vec{p}_4)$. Comparison of
Eq.~\eqref{eq:etapisChannel} with Eq.~\eqref{eq:strongScatuChannel} reveals that what is called $s$-channel here
actually corresponds to the $u$-channel in the general treatment of scattering processes such that we must use
Eq.~\eqref{eq:strongScatst} with $s \leftrightarrow u$. That way, we obtain
\begin{equation}\begin{split}
	t &= \inv{2} \big( 3 s_0 - s + \kappa(s) \cos \theta_s \big) \eolc \\[1mm]
	u &= \inv{2} \big( 3 s_0 - s - \kappa(s) \cos \theta_s \big) \eolc
	\label{eq:eta3pitu}
\end{split}\end{equation}
where
\begin{equation}\begin{split}
	\kappa(s) &\equiv \frac{\lambda^{1/2}(s,\meta^2,\mpiz^2)\, \lambda^{1/2}(s,\mpip^2,\mpip^2)}{s}\\[3mm]
					&= \sqrt{ \frac{s-4 \mpip^2}{s} } \sqrt{\Big( (\meta+\mpiz)^2 - s \Big) \Big( (\meta-\mpiz)^2 - s \Big)}
	\eolc
	\label{eq:eta3piKappa}
\end{split}\end{equation}
and $3 s_0 \equiv s+t+u= \meta^2 + 2 \mpip^2 + \mpiz^2$. For the angle we obtain
\begin{equation}
	\cos \theta_s = \frac{t-u}{\kappa(s)} \eolp
	\label{eq:eta3piScatteringAngle}
\end{equation}
As usual, one can obtain the corresponding $t$- and $u$-channel expressions by appropriate variable exchanges, but
additional care is required here. The right-hand side of both formul\ae\ in Eq.~\eqref{eq:eta3pitu} contains a term 
$\pm \Delta_{\pi^+ \pi^+}\Delta_{\eta \pi^0}/2s$ that vanishes. But in the other channels, the corresponding term is
nonzero and must be included. Our choice of reference frame has thus the advantage that it leads to a particularly
simple form for the kinematic relations.

The physical region for the decay process is distinct from the physical regions for the three scattering channels as
can be seen in the Mandelstam diagram in Fig.~\ref{fig:eta3piMandelstamDiagram}. The limits of the physically allowed
region
for $s$ can be obtained from its definition in Eq.~\eqref{eq:eta3piMandelstam}. The minimal value $s_\text{min} = 4
\mpip^2$ is achieved, if the three-momenta of the charged pions vanish, while the maximum value $s_\text{max} =
(\meta-\mpiz)^2$ occurs, if the three-momentum of the neutral pion is zero. For each $s$, the physical region for $t$
and $u$ follows from Eq.~\eqref{eq:eta3pitu}.

Three-body decays are usually plotted in a Dalitz plot~\cite{Dalitz1953}, which is particularly useful also for
displaying experimental results. It is drawn in a rectangular coordinate system, where the axes are chosen to
be any two independent kinematic variables. Sometimes, it is simply used to visualise the physical region in
a similar way as we used the Mandelstam diagrams, but often the squared decay amplitude is shown as well.
Common choices for the kinematic variables are two of the Mandelstam variables, the energies of two particles or often
the so-called Dalitz-plot variables defined as
\begin{equation}
	X = \sqrt{3}\, \frac{T_3 - T_2}{Q_1} \eolc \qquad Y = 3\, \frac{T_4}{Q_1} - 1 \eolp
\end{equation}
The $T_i$ are the energies of the corresponding particles in the rest frame of the decaying particle, $T_i = E_i - m_i$,
and $Q_1 = T_2 + T_3 + T_4 = m_1 - m_2 - m_3 - m_4$. From Eq.~\eqref{eq:eta3piMandelstam} follows for the charged
channel of $\etapi$ that 
\begin{equation}
	E_2 = \frac{\meta^2 + \mpip^2 - u}{2 \meta} \eolc \quad E_3 = \frac{\meta^2 + \mpip^2 - t}{2 \meta} \eolc \quad
	E_4 = \frac{\meta^2 + \mpiz^2 - s}{2 \meta} \eolc
\end{equation}
from which one finds that the Dalitz plot variables are in terms of the Mandelstam variables given by 
\begin{equation}
	X = \frac{\sqrt{3}}{2 \meta Q_c} (u-t)\eolc \quad
	Y = \frac{3}{2 \meta Q_c} \left( (\meta - \mpiz)^2 - s \right) - 1 \eolc
	\label{eq:etapiDalitzVars}
\end{equation}
with $Q_c = \meta - 2 \mpip - \mpiz$. The expressions for the neutral channel are obtained from these by replacing
$\mpip$ by $\mpiz$. The Dalitz plot variables are dimensionless, Lorentz invariant and have the nice property that 
$X^2 + Y^2 \leq 1$, that is all physical values lie within the unit circle (but not all values inside the circle are
physically allowed).

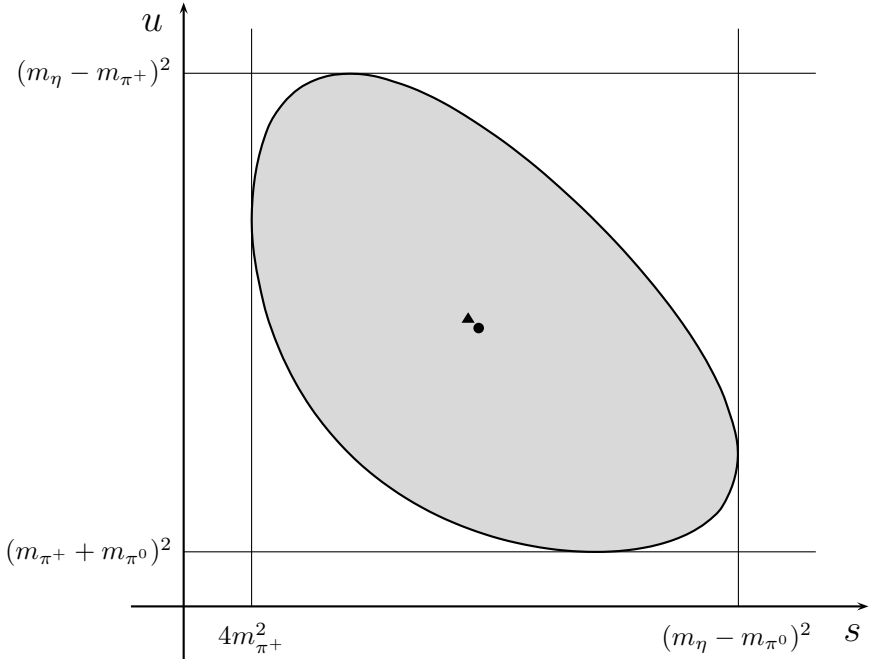
\begin{figure}[tb]
\begin{center}
\psset{unit=69}
\begin{pspicture*}(0.030,0.055)(0.195,0.18)
\psset{linewidth=\mylw}

\psline{->}(0.055,0.065)(0.195,0.065)
\rput(0.192,0.06){\Large $s$}
\psline{->}(0.065,0.055)(0.065,0.18)
\rput(0.059,0.176){\Large $u$}

\psccurve[fillstyle=solid,fillcolor=shadegray](0.07792,0.1395)(0.08092,0.1559)(0.08392,0.1608)(0.08692,0.1636)(0.08992,
0.1652)(0.09292,0.1661)(0.09592,
0.1664)(0.09892,0.1663)(0.1019,0.1658)(0.1049,0.1649)(0.1079,0.1639)(0.1109,0.1626)(0.1139,0.1611)(0.1169,0.1594)(0.1199
,0.1575)(0.1229,0.1555)(0.1259,0.1534)(0.1289,0.1511)(0.1319,0.1487)(0.1349,0.1461)(0.1379,0.1434)(0.1409,0.1406)(0.1439
,0.1376)(0.1469,0.1345)(0.1499,0.1312)(0.1529,0.1277)(0.1559,0.124)(0.1589,0.12)(0.1619,0.1156)(0.1649,0.1105)(0.1679,
0.104)(0.1702,0.09338)(0.1672,0.08395)(0.1642,0.08096)(0.1612,0.07906)(0.1582,0.07774)(0.1552,0.0768)(0.1522,
0.07614)(0.1492,0.0757)(0.1462,0.07546)(0.1432,0.07538)(0.1402,0.07545)(0.1372,0.07566)(0.1342,0.07601)(0.1312,
0.07649)(0.1282,0.0771)(0.1252,0.07784)(0.1222,0.07873)(0.1192,0.07977)(0.1162,0.08096)(0.1132,0.08231)(0.1102,
0.08386)(0.1072,0.08561)(0.1042,0.0876)(0.1012,0.08985)(0.09818,0.09243)(0.09518,0.09538)(0.09218,0.09881)(0.08918,
0.1028)(0.08618,0.1077)(0.08318,0.1139)(0.08018,0.1227)

\psset{linewidth=.4pt}
\psline(0.0779,0.065)(0.0779,0.175)
\rput(0.0779,0.059){\small $4 \mpip^2$}
\psline(0.1703,0.065)(0.1703,0.175)
\rput(0.170,0.059){\small $(\meta - \mpiz)^2$}
\psline(0.065,0.1665)(0.185,0.1665)
\rput(0.048,0.1665){\small $(\meta - \mpip)^2$}
\psline(0.065,0.0754)(0.185,0.0754)
\rput(0.047,0.0754){\small $(\mpip + \mpiz)^2$}

\pscircle*(0.121,0.118){0.001}
\pstriangle*(0.119,0.119)(0.0025,0.002)

\end{pspicture*}
\end{center}
\caption{Dalitz plot of the physical region for $\eta \to \pi^+ \pi^- \pi^0$ in terms of $s$ and $u$ with the typical
shape of a rounded triangle. The centre of the Dalitz plot is marked with a dot and the point $s=t=u=s_0$ with a small
triangle.}
\label{fig:eta3piDalitzPlotsu}
\end{figure}

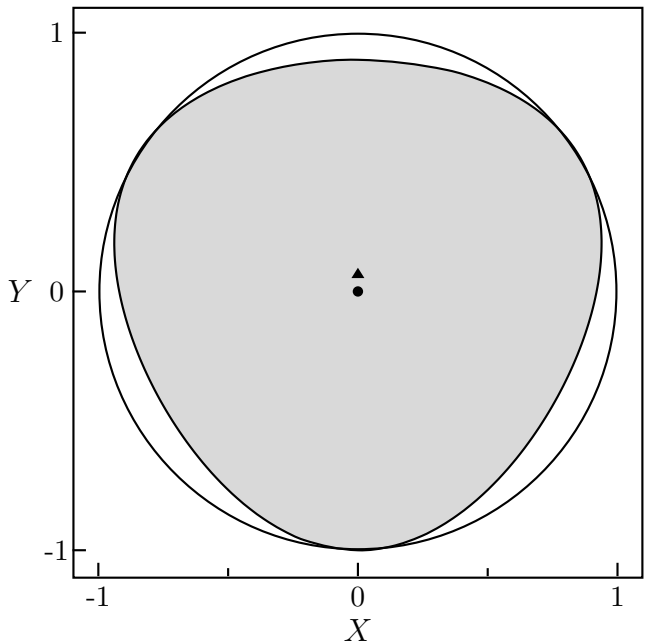
\begin{figure}[tb]
\begin{center}
\psset{unit=3.4}
\begin{pspicture*}(-1.35,-1.35)(1.11,1.11)

\psset{linewidth=\mylw}

\psframe(-1.1,-1.11)(1.1,1.1)
\rput(0,-1.31){\large $X$}
\multips(-1,-1.1)(1,0){3}{\psline(0,5pt)}
\multips(-0.5,-1.1)(1,0){2}{\psline(0,3pt)}
\multido{\n=-1+1}{3}{\rput(\n,-1.18){\n}}
\rput(-1.3,0){\large $Y$}
\multiput(-1.1,-1)(0,1){3}{\psline(5pt,0)}
\multiput(-1.1,-5)(0,1){2}{\psline(3pt,0)}
\multido{\n=-1+1}{3}{\rput(-1.16,\n){\n}}

\psccurve[fillstyle=solid,fillcolor=shadegray](0.,-1.)(-0.2092,-0.96)(-0.2953,-0.92)(-0.361,-0.88)(-0.4159,
-0.84)(-0.4639 , -0.8)(-0.5068 , -0.76)(-0.5458,
-0.72)(-0.5817,-0.68)(-0.6148,-0.64)(-0.6457,-0.6)(-0.6745,-0.56)(-0.7015,-0.52)(-0.7268,-0.48)(-0.7505,-0.44)(-0.7727,
-0.4)(-0.7934,-0.36)(-0.8128,-0.32)(-0.8308,-0.28)(-0.8475,-0.24)(-0.8629,-0.2)(-0.877,-0.16)(-0.8897,-0.12)(-0.901,
-0.08)(-0.911,-0.04)(-0.9195,0.0)(-0.9266,0.04)(-0.9321,0.08)(-0.936,0.12)(-0.9381,0.16)(-0.9385,0.2)(-0.9368
,0.24)(-0.9331,0.28)(-0.9272,0.32)(-0.9187,0.36)(-0.9075,0.4)(-0.8934,0.44)(-0.8758,0.48)(-0.8545,0.52)(-0.8288,
0.56)(-0.798,0.6)(-0.7611,0.64)(-0.717,0.68)(-0.6636,0.72)(-0.5979,0.76)(-0.5146,0.8)(-0.4018,0.84)(-0.2158,
0.88)(0.0,0.8951)(0.3456,0.8551)(0.4765,0.8151)(0.5689,0.7751)(0.6404,0.7351)(0.698,0.6951)(0.7454,
0.6551)(0.7848,0.6151)(0.8178,0.5751)(0.8453,0.5351)(0.8682,0.4951)(0.8872,0.4551)(0.9026,0.4151)(0.9148,0.3751)(0.9243,
0.3351)(0.9312,0.2951)(0.9357,0.2551)(0.9381,0.2151)(0.9385,0.1751)(0.937,0.1351)(0.9338,0.09511)(0.9289,0.05511)(0.9224
,0.01511)(0.9144,-0.02489)(0.905,-0.06489)(0.8941,-0.1049)(0.8819,-0.1449)(0.8684,-0.1849)(0.8535,-0.2249)(0.8373,
-0.2649)(0.8198,-0.3049)(0.8009,-0.3449)(0.7807,-0.3849)(0.759,-0.4249)(0.7359,-0.4649)(0.7112,-0.5049)(0.6849,
-0.5449)(0.6568,-0.5849)(0.6268,-0.6249)(0.5945,-0.6649)(0.5597,-0.7049)(0.5219,-0.7449)(0.4806,-0.7849)(0.4347,
-0.8249)(0.3827,-0.8649)(0.3218,-0.9049)(0.2454,-0.9449)(0.1287,-0.9849)

\pscircle[linewidth=\mylw](0,0){1}

\pscircle*(0,0){0.02}
\pstriangle*(0,0.0517)(0.05,0.04)

\end{pspicture*}
\end{center}
\caption{Dalitz plot of the physical region for $\eta \to \pi^+ \pi^- \pi^0$ in terms of $X$ and $Y$ with the typical
shape of a rounded triangle. The figure also shows the unit circle around $X=Y=0$. The centre of the Dalitz plot is
marked with a dot and the point $s=u=t=s_0$ with a small triangle.}
\label{fig:eta3piDalitzPlotXY}
\end{figure}

Figure~\ref{fig:eta3piDalitzPlotsu} shows a Dalitz plot for the charged channel of $\etapi$. The axes have been chosen
to be the Mandelstam variables $s$ and $u$. The physical region is shaded in grey and also the minimally and maximally
allowed physical values for $s$ and $u$ are indicated. Because of the symmetry under interchange of $t$ and $u$, the
picture looks the same if the $u$-axis is replaced by the $t$-axis.

As a second example, Fig.~\ref{fig:eta3piDalitzPlotXY} shows a Dalitz plot for the same processes, but with the Dalitz
plot variables $X$ and $Y$ as the axes. One can immediately see that the shape is symmetric under $X \to -X$, which is
a consequence of the symmetry under $t \leftrightarrow u$. The figure also shows the unit circle around $X = Y = 0$
and it is clearly visible that the physical region is completely contained within the circle, but does not fully cover
it. For an example of the graphical representation of a measured momentum distribution in a Dalitz plot, see
Fig.~\ref{fig:eta3piDalitzKLOE}.

The kinematic point, where the kinetic energies of all the decay products are the same, is called the centre of the
Dalitz plot. For the decay $\eta \to \pi^+ \pi^- \pi^0$, in the rest frame of the $\eta$, this corresponds to
\begin{equation}
	E_2 - \mpip = E_3 - \mpip = E_4 - \mpiz = \frac{Q_c}{3} \eolp
\end{equation}
In terms of the Mandelstam variables, the centre of the Dalitz plot is thus at
\begin{equation}
	s = (\meta - \mpiz)^2 - \frac{2 \meta Q_c}{3} \eolc \qquad 
	t = u = (\meta - \mpip)^2 - \frac{2 \meta Q_c}{3} \eolc
\end{equation}
while it is at $X = Y = 0$ in terms of the Dalitz plot variables. In both figures, the centre is marked with a dot. If
one considers the neutral channel or works with degenerate pion masses in the charged channel, all the pion masses are
equal, such that the centre of the Dalitz plot in terms of the Mandelstam variables is simply at $s=t=u=s_0$. But if the
masses are distinct, the point of equal Mandelstam variables differs from the centre of the Dalitz plot, as can be seen
from the figures, where it has been marked with a small triangle. Since we almost exclusively work with degenerate
pion masses, we will often refer to $s=t=u=s_0$ as the centre of the Dalitz plot, but one should always keep in mind
that this is only true if all the decay products have the same~mass.

\section{Isospin structure and the relation to \texorpdfstring{$\pi \pi$}{pion pion} scattering}
\label{sec:eta3piIsoStructure}

According to a low-energy theorem by Suther\-land~\cite{Bell+1968,Sutherland1966}, the electromagnetic contributions to
the $\etapi$ decay amplitude are strongly suppressed. This has later been confirmed by explicit one-loop calculations in
\chpt~\cite{Baur+1996,Ditsche+2009}. Neglecting these small effects, the decay is entirely due to the isospin breaking
part of the QCD Lagrangian. By means of Eq.~\eqref{prel:Ma}, the mass matrix $\M$ can be decomposed as
\begin{equation}
	\M = \frac{m_u+m_d+m_s}{\sqrt{6}} \lambda^0 + \frac{m_u-m_d}{2} \lambda^3
		 + \frac{\hat{m} - m_s}{\sqrt{3}} \lambda^8 \eolp
\end{equation}
From the term proportional to $\lambda_3$ comes a contribution to the Lagrangian given by
\begin{equation}
	\lib = -\frac{m_u - m_d}{2}\, \bar{q} \lambda^3 q  = -\frac{m_u - m_d}{2}\, ( \bar{u} u - \bar{d}d ) \eolc
	\label{eq:LIB}
\end{equation}
which is a $\Delta I = 1$ operator, as we will show now. Under a rotation of the light quark fields of the form 
$q \mapsto U q,\, U \in SU(3)$, it transforms as
\begin{equation}
	\bar{q} \lambda^i q \mapsto \bar{q} U\dega \lambda^i U q \eolp
\end{equation}
From $\langle U\dega \lambda^i U \rangle = \langle \lambda^i \rangle = 0$ follows that
\begin{equation}
	U\dega \lambda^i U = R^{ij}\lambda^j \eolc
	\label{eq:eta3piUlambdaU}
\end{equation}
for some $8\times8$ matrix $R$, because all traceless matrices can be written as a linear combination of the
Gell-Mann matrices.
Isospin transformations form a subgroup of $SU(3)$ with
\begin{equation}
	U = \left( \begin{array}{cc}V&0\\0&1\end{array} \right), \quad V \in SU(2) \eolp
\end{equation}
In this case, only the Gell-Mann matrices $\lambda^1$, $\lambda^2$, and $\lambda^3$ appear on the right-hand side of
Eq.~\eqref{eq:eta3piUlambdaU} and the matrix $R$ is reduced to a $3 \times 3$ matrix. Furthermore, we have
\begin{equation}
	\langle U\dega \lambda^i U U\dega \lambda^j U \rangle = \langle \lambda^i \lambda^j \rangle = 2 \delta^{ij} \quad
\end{equation}
and
\begin{equation}
	\langle U\dega \lambda^i U U\dega \lambda^j U \rangle = \langle R^{ik} \lambda^k R^{jl} \lambda^l \rangle 
			= R^{ik} R^{jl} 2 \delta^{kl} = 2 (R R^T)^{ij}
\end{equation}
and from equating the two expressions follows $R \in SO(3)$. $\bar{q} \lambda^m q$ transforms under isospin
rotation in exactly the same way as a pion, and $\lib$ is thus an operator that carries isospin quantum numbers $I = 1$
and $I_3 = 0$. It generates transitions between states with $\Delta I = 1$ and $\Delta I_3 = 0$. The strong
interaction thus violates the total isospin, but not its third component. In Sec.~\ref{sec:eta3pi3pion}, we have shown
that three pions can only couple to odd total isospin. Now that we know that the decay is due to a $\Delta I = 1$
operator, it is clear that they must form a $\ket{1,0}$ state.

The $S$-matrix element for the decay $\eta \to \pi^i \pi^j \pi^k$ is
\begin{multline}
	\bra{\pi^i(p_2) \pi^j(p_3) \pi^k(p_4)} S \ket{\eta(p_1)} \\
			= \bra{\pi^i(p_2) \pi^j(p_3) \pi^k(p_4)} \exp\left( i \intlib \right) \ket{\eta(p_1)} \eolc
\end{multline}
where we have used the fact that $\lib$ is the only part of the QCD Lagrangian that can generate this transition.
Expanding the exponential and taking only first order isospin breaking into account yields
\begin{equation}
	\bra{\pi^i \pi^j \pi^k} S \ket{\eta} 
			= \braket{\pi^i \pi^j \pi^k}{\eta} +i \bra{\pi^i \pi^j \pi^k} \intlib \ket{\eta} \eolc
	\label{eq:eta3piTmatrix}
\end{equation}
where we have omitted the momenta for simplicity. Comparison with Eq.~\eqref{eq:strongTmatrixElement} reveals that the
second term is nothing else than the $T$-matrix element. The amplitude is defined in the usual way
\begin{equation}
	\bra{\pi^i(p_2) \pi^j(p_3) \pi^k(p_4)} iT \ket{\eta(p_1)} = i(2\pi)^4 \delta^4(p_1-p_2-p_3-p_4)
					\A_\etapi^{ijk}(p_1,p_2,p_3,p_4) \eolc
	\label{eq:eta3piAmplitude}
\end{equation}
and in addition, we also define the function $\A_\etapi^{ijk,l}$ by
\begin{equation}
	-i \frac{m_u - m_d}{2}\, \bra{\pi^i \pi^j \pi^k} \int \! d^4 x\; \bar{q} \lambda^l q \ket{\eta} = i(2\pi)^4 
				\delta^4(p_1-p_2-p_3-p_4) \A^{ijk,l}_\etapi\eolp
	\label{eq:eta3piMatrixElement}
\end{equation}
From the prior observation about the $T$-matrix element follows that this matrix element reduces to the $T$-matrix
element of $\etapi$ for $l = 3$. Furthermore, the space integral in Eq.~\eqref{eq:eta3piTmatrix} simply results in the
momentum conserving delta function, such that we have
\begin{equation}
	\A_\etapi^{ijk} = \A^{ijk,3}_\etapi = \bra{\pi^i \pi^j \pi^k} \lib \ket{\eta} \eolp
	\label{eq:eta3piAmplitudeMatrixElement}
\end{equation}
Because the operator $\bar{q} \lambda^l q$ transforms in the same way as a pion under isospin, the function
$\A^{ijk,l}_\etapi$ transforms exactly as the $\pi \pi$ scattering amplitude in
Eq.~\eqref{eq:pipiAmplitudeTransformation} and they must share the same isospin structure. For the decay amplitude we
obtain
\begin{equation}
	\A_{\eta \to 3 \pi}^{ijk} = \A^{ijk,3}_\etapi = A_1(s,t,u) \delta^{ij} \delta^{k3} 
				+ A_2(s,t,u) \delta^{ik} \delta^{j3} + A_3(s,t,u) \delta^{i3} \delta^{jk} \eolp
\end{equation}
Similarly as for $\pi \pi$ scattering, the amplitude must be invariant under permutations of the particles in the final
state, leading to relations among the amplitudes $A_i$. These are
\begin{equation}\begin{split}
	p_2 \leftrightarrow p_3\; (t \leftrightarrow u)\ \text{and}\ i \leftrightarrow j 
					\quad &\Rightarrow \quad A_1(s,u,t) = A_1(s,t,u) \eolc \\
	p_3 \leftrightarrow p_4\; (s \leftrightarrow t)\ \text{and}\ j \leftrightarrow k 
					\quad &\Rightarrow \quad A_2(s,t,u) = A_1(t,s,u) \eolc \\
	p_2 \leftrightarrow p_4\; (s \leftrightarrow u)\ \text{and}\ i \leftrightarrow k 
					\quad &\Rightarrow \quad A_3(s,t,u) = A_1(u,t,s) \eolp
\end{split}\end{equation}
These results are identical to Eqs.~\eqref{eq:pipiCross1} and~\eqref{eq:pipiCross2}, which is a consequence of crossing.
The decay amplitude is then expressed in terms of the single function $A(s,t,u) \equiv A_1(s,t,u)$ as
\begin{equation}
	\A_\etapi^{ijk}(s,t,u) = A(s,t,u)\, \delta^{ij} \delta^{k3} + A(t,u,s)\, \delta^{ik} \delta^{j3} 
			+ A(u,s,t)\, \delta^{i3} \delta^{jk} \eolp
	\label{eq:eta3piIsospinStructure}
\end{equation}
In terms of the physical pions, we find the two allowed decay channels
\begin{equation}\begin{split}
	\A_\etapi^{+-0}(s,t,u) = \A_\etapi^{113}(s,t,u) &= A(s,t,u)\eolc\\[1mm]
	\A_\etapi^{000}(s,t,u) = \A_\etapi^{333}(s,t,u) &= A(s,t,u) + A(t,u,s) + A(u,s,t)\\[1mm]
		 &\equiv \bar{A}(s,t,u) \eolp
	\label{eq:eta3piChargedNeutralChannel}
\end{split}\end{equation}
The second line shows that the amplitude for the neutral channel can very easily be obtained from the amplitude of the
charged channel, such that there is no need to calculate it separately. Note, however, that this result assumes that the
decay is exclusively due to first order isospin breaking in the QCD Lagrangian. Contributions from matrix element that
contain $\lib$ to some power and electromagnetic effects violate this simple relation.

Since the $\etapi$ decay amplitude and the $\pi \pi$ scattering amplitude have identical isospin structure, and because
of $\Delta_{23} = m_2^2 - m_3^2 = 0$, the proof of the reconstruction theorem follows the same lines for both
processes. Hence, also the eta decay amplitude can be written in the form of Eq.~\eqref{eq:pipiReconstructionTheorem}.

\section{One-loop result from \chpt} \label{sec:eta3piOneLoop}

\begin{figure}[tb]
\psset{xunit=0.9cm,yunit=1.6cm}
\begin{center}\begin{pspicture*}(-2,-1)(12.25,4.1)
\psset{linewidth=\mylw}
\psframe[fillstyle=solid,fillcolor=shadegray,linecolor=shadegray,linewidth=0](5.06,-0.35)(7.74,4.0)
\psframe(0,-.35)(12,4)
\multips(2,-0.35)(2,0){5}{\psline(0,5pt)}
\multips(1,-0.35)(2,0){6}{\psline(0,3pt)}
\multido{\n=0+2}{7}{\uput{.2}[270](\n,-0.35){\n}}
\multips(0,0)(0,1){4}{\psline(5pt,0)}
\multips(0,0.5)(0,1){5}{\psline(3pt,0)}
\multido{\n=0+1}{5}{\uput{0.2}[180](0,\n){\n}}
\uput{0.3}[270](6,-0.5){$s \; [\mpi^2]$}
\uput{0.7}[180]{90}(0,1.8){$\Mbar(s,3s_0-2s,s)$}
\psset{linewidth=\mylw}
\fileplot[linestyle=dotted,dotsep=.05]{data/ChPT/Mtree.dat}
\fileplot{data/ChPT/ReM1loop.dat}
\fileplot[linestyle=dashed]{data/ChPT/ImM1loop.dat}
\end{pspicture*}\end{center}
\caption{The figure shows the tree-level result (dotted line) and the real (solid line) and imaginary (dashed line)
	part of the one-loop result for $\etapi$ along the line $s=u$. The physical region is shaded in grey.}
\label{fig:eta3pi1loop}
\end{figure}
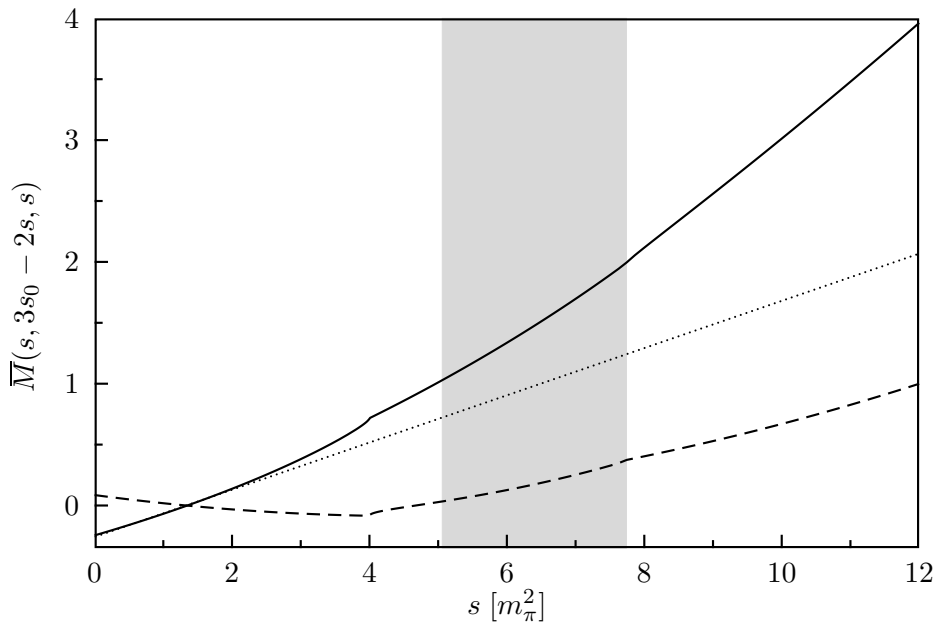

The one-loop result from \chpt\ is used as theoretical input to the dispersive analysis and we thus quote it
here~\cite{Gasser+1985a,Gasser+1985}. It has been obtained using the degenerate pion and kaon masses and does
not include electromagnetic contributions. The decay has been calculated within \chpt\ up to two-loop
order~\cite{Bijnens+2007} and the electromagnetic contributions up to $\O(e^2 \M)$~\cite{Ditsche+2009}, but these
results are
not made use of in this work. The tree-level and the one-loop result are shown in Fig.~\ref{fig:eta3pi1loop}.

From Eq.~\eqref{eq:eta3piAmplitudeMatrixElement} follows that the amplitude is proportional to 
$B_0 (m_u-m_d)$, which can be expressed in terms of $Q$ by means of Eq.~\eqref{eq:B0mumdQ}. We can therefore
extract a normalisation factor from the amplitude:
\begin{equation}
	A(s,t,u) = - \inv{Q^2} \frac{\mK^2 (\mK^2-\mpi^2)}{3 \sqrt{3} \mpi^2 \Fpi^2} M(s,t,u) \eolp
	\label{eq:eta3piNorm}
\end{equation}
It is chosen such that the tree-level result is normalised to one at the centre of the Dalitz plot:
\begin{equation}
	M\tree \equiv T(s) = \frac{3s - 4\mpi^2}{\meta^2 - \mpi^2} = 1 + 3 \frac{s-s_0}{\meta^2 - \mpi^2} \eolp
	\label{eq:eta3piMtree}
\end{equation}
The structure of this result is determined by chiral symmetry and has been known long before the formulation of
\chpt~\cite{Cronin1967, Osborn+1970}. In agreement with the reconstruction theorem in
Eq.~\eqref{eq:pipiReconstructionTheorem}, the one-loop amplitude is found to be of the form
\begin{equation}
	\Mbar(s,t,u) = \Mbar_0(s) + (s-u) \Mbar_1(t) + (s-t) \Mbar_1(u) + \Mbar_2(t) + \Mbar_2(u) - \tfrac{2}{3} \Mbar_2(s)
	\eolp
\end{equation}
The isospin amplitudes $\Mbar_0(s)$, $\Mbar_1(s)$ and $\Mbar_2(s)$ are given by
\begin{equation}\begin{split}
	\Mbar_0(s) &= \begin{aligned}[t] &T(s) + \Delta_0(s) \left( 1 + \frac{2}{3} T(s) \right) 
							+ \Delta_2(s) \left( 1 - \frac{1}{3} T(s) \right) + \Delta_3(s) \\[2mm]
						  &+ V(s) + \frac{2}{3} \Delta_\text{GMO} T(s) 
							+ \frac{8}{3} \mpi^2 \frac{\Delta_F T(s) -\Delta_\text{GMO}}{\meta^2-\mpi^2} \eolc \end{aligned}
\\[2mm]
	\Mbar_1(s) &= \frac{3}{2 (\meta^2 - \mpi^2)} \left( \Delta_1(s) - \frac{8}{3 \Fpi^2} s L_3 \right) \eolc \\[2mm]
	\Mbar_2(s) &= \frac{1}{2} \Delta_2(s) \big( 3 - T(s) \big) \eolp
	\label{eq:eta3piMbar}
\end{split}\end{equation}
The quantities
\begin{equation}
	\Delta_F \equiv \frac{\FK}{\Fpi} - 1 = 0.193(6) \eolc \hspace{3em}
		\Delta_\text{GMO} \equiv \frac{4 \mK^2 - 3 \meta^2 - \mpi^2}{\meta^2 - \mpi^2} = 0.218 \eolc
\end{equation}
have been introduced in order to remove the dependence of the amplitude on the low-energy constants $L_5$, $L_7$ and
$L_8$. The value for $\Delta_F$ is an average of lattice results for $N_f = 2+1$ taken from Ref.~\cite{Colangelo+2010a}
and $\Delta_\text{GMO}$ has been calculated from the degenerate meson masses in
Eq.~\eqref{eq:strongIsospinMesonMasses}. The amplitude only depends on a single low-energy constant from $\L_4$,
\begin{equation}
	L_3 = (-2.35 \pm 0.37)\ee{-3} \eolp
\end{equation}
The value is taken from ``fit 10'' in Ref.~\cite{Amoros+2001}. We introduce the abbreviations $\Delta_{PQ}
= m_P^2 - m_Q^2$ and $\Sigma_{PQ} = m_P^2 + m_Q^2$. The functions $\Delta_i(s)$ are expressed in terms of the
renormalised loop integrals as
\begin{align}
	\Delta_0(s) &= \inv{2 \Fpi^2} (2s - \mpi^2)\, J_{\pi\pi}^r(s)\eolc \nonumber \displaybreak[0] \\[1mm]
	\Delta_1(s) &= \frac{2s}{\Fpi^2} \left( M_{\pi \pi}^r(s) + \inv{2} M_{KK}^r(s) \right)\eolc \nonumber \\[1mm]
	\Delta_2(s) &= -\inv{2 \Fpi^2} (s - 2\mpi^2)\, J_{\pi\pi}^r(s) + \inv{4 \Fpi^2} (3s-4\mK^2) J_{KK}^r(s) 
						+ \frac{\mpi^2}{3 \Fpi^2} J_{\pi\eta}^r(s)\eolc \nonumber \\[1mm]
	\Delta_3(s) &= \begin{aligned}[t]
	               &-\inv{6 \Fpi^2 \Delta_{\eta \pi}} (s-2\mpi^2)(3s-4\mK^2) J_{\pi\pi}^r(s) 
							- \frac{s(3s-4\mpi^2)}{4 \Fpi^2 \Delta_{\eta \pi}} J_{KK}^r(s) \\[1mm]
						&+ \frac{\mpi^2}{3 \Fpi^2 \Delta_{\eta \pi}}(3s-4\mpi^2)J_{\pi\eta}^r(s) 
						- \frac{\mpi^2}{2 \Fpi^2} J_{\eta\eta}^r(s) \\[1mm]
						&- \frac{3s}{8 \Fpi^2}\frac{(3s-4\mK^2)}{(s-4\mK^2)} 
								\left(J_{KK}^r(s)-J_{KK}^r(0)-\inv{8 \pi^2} \right) \eolp
	               \end{aligned}
	\label{eq:eta3piDeltai}
\end{align}
The renormalised loop integrals are given by
\begin{equation}\begin{split}
	J^r_{PQ}(s) &= \bar{J}_{PQ}(s) - 2 k_{PQ} \eolc\\
	M^r_{PP} &= \inv{12s}(s-4 m_P^2) \bar{J}_{PP}(s) - \inv{6} k_{PP} + \inv{288\pi^2} \eolc
\end{split}\end{equation}
with
\begin{equation}\begin{split}
	k_{PQ} &= \inv{32\pi^2} \frac{m_P^2 \log (m_P^2/\mu^2) - m_Q^2 \log (m_Q^2/\mu^2)}{\Delta_{PQ}} \eolc\\[1mm]
	\bar{J}_{PQ}(s) &= \inv{32\pi^2} \left\{ 2 + \left(\frac{\Delta_{PQ}}{s} 
						- \frac{\Sigma_{PQ}}{\Delta_{PQ}} \right) \log \frac{m_Q^2}{m_P^2}
						- \frac{\nu}{s} \log \frac{(s+\nu)^2 - \Delta_{PQ}^2}{(s-\nu)^2 - \Delta_{PQ}^2} \right\} \eolc\\[2mm]
	\nu &= \lambda^{1/2}(s,m_P^2,m_Q^2) = \sqrt{\big( s - (m_P+m_Q)^2 \big) \big( s - (m_P-m_Q)^2 \big)} \eolp
\end{split}\end{equation}
In the case of $m_P = m_Q$, these expressions can be simplified considerably:
\begin{equation}\begin{split}
	k_{PP} &= \inv{32\pi^2} \left( \log \frac{m_P^2}{\mu^2} + 1 \right) \eolc\\[2mm]
	\bar{J}_{PP}(s) &= \inv{16 \pi^2} \left( \sigma \log \frac{\sigma-1}{\sigma+1} + 2 \right) \eolc\\[2mm]
	\sigma &= \sqrt{ \frac{s-4m_P^2}{s} } \eolp
\end{split}\end{equation}
Furthermore, $\bar{J}_{PQ}(0) = 0$, such that $J^r_{PQ}(0) = - 2 k_{PQ}$. The function $V(s)$ is given by
\begin{equation}\begin{split}
	V(s) = \begin{aligned}[t] &T(s) \, \big( a_1 + 3 a_2 \Delta_{\eta \pi} + a_3 (9 \meta^2-\mpi^2) \big) + a_4 \eolc 
\\[2mm]
	&a_1 = \frac{2}{3 \Delta_{\eta \pi}} \mpi^2 ( -\mu_\pi - 2 \mu_K + 3 \mu_\eta) \eolc \\[1mm]
	&a_2 = \frac{2}{3 \Delta_{\eta \pi}^2} ( -\mpi^2 \mu_\pi + 4 \mK^2 \mu_K - 3 \meta^2 \mu_\eta) \eolc \\[1mm]
	&a_3 = \frac{1}{128 \pi^2 \Fpi^2} \left( \log \frac{\mK^2}{\mu^2} + 1 \right) \eolc \\[1mm]
	&a_4 = -8 \mpi^2 a_2 - 12 \mK^2 a_3 
		+ \frac{3 \mpi^2}{32 \pi^2 \Fpi^2} \left( 1 - \frac{\mpi^2}{\Delta_{K\pi}} \log \frac{\mK^2}{\mpi^2} \right) \eolp
	\end{aligned}
\end{split}\end{equation}
The chiral logs $\mu_P$ are given by
\begin{equation}
	\mu_P = \inv{32 \pi^2 \Fpi^2}\, m_P^2 \log \frac{m_P^2}{\mu^2} \eolp
\end{equation}

This representation of the $\etapi$ amplitude is valid only at leading order in the quark mass expansion. Because the
kaon mass does not appear in the tree-level approximation, it can be replaced by means of the Gell-Mann--Okubo relation
\eqref{eq:strongGellMannOkubo}, which is valid to leading order in the quark mass expansion. The amplitude is then
independent of the renormalisation scale $\mu$. Note, however, that the isospin amplitudes $\Mbar_0(s)$, $\Mbar_1(s)$
and $\Mbar_2(s)$ are scale dependent nevertheless.

\section{Dispersive representation for the \texorpdfstring{$\etapi$}{eta to 3 pions} amplitude at one loop}

For later use, we derive here a dispersive representation for the $\etapi$ amplitude at one loop in \chpt. $\Mbar_0(s)$,
$\Mbar_1(s)$ and $\Mbar_2(s)$ are functions of one variable and they are real below the threshold at $s = 4 \mpi^2$,
thus satisfying the Schwarz reflection principle. This is exactly the situation that was discussed in section
\ref{sec:dispRelOneVar} and the dispersion relations will accordingly be of the form of
Eq.~\eqref{eq:disprelUnsubtrIm} with an appropriate number of subtractions applied.

In order to decide upon the required number of subtractions, we have to consider the asymptotic behaviour of the
one-loop results for the $\Mbar_I(s)$. From the asymptotic behaviour of the loop functions,
\begin{equation}
	J^r_{PQ} \asymp \frac{\sigma - \log s}{16 \pi^2}\eolc \qquad
	M^r_{PP} \asymp \frac{\frac{2}{3}+\sigma-\log s}{192 \pi^2} \eolc
\end{equation}
where $\sigma = \log \mu^2 + 1 + i\pi$, follows
\begin{equation}\begin{split}
	\Mbar_0(s) &\asymp -\inv{2} \frac{s^2 \log s-\sigma s^2}{16 \pi^2 \Fpi^2 (\meta^2 - \mpi^2)} \eolc\\[2mm]
	\Mbar_1(s) &\asymp -\frac{3}{8}  \frac{s \log s-\sigma s}{16 \pi^2 \Fpi^2 (\meta^2 - \mpi^2)}
										-\frac{(4 L_3-\inv{64 \pi^2})s}{\Fpi^2 (\meta^2 - \mpi^2)} \eolc\\[2mm]
	\Mbar_2(s) &\asymp \frac{3}{8}  \frac{s^2 \log s-\sigma s^2}{16 \pi^2 \Fpi^2 (\meta^2 - \mpi^2)}\eolp
\label{eq:OneLoopAsymp}
\end{split}\end{equation}
Only $\sigma$ contributes to the imaginary part, which thus grows quadratically for  $\Mbar_0(s)$ and $\Mbar_2(s)$
and linearly for $\Mbar_1(s)$. We obtain the dispersion relations
\begin{equation}\begin{split}
	\Mbar_0(s) &= a_0 + b_0 s + c_0 s^2 + \frac{s^3}{\pi} \int\limits_{4 \mpi^2}^\infty ds'
\frac{\Im\Mbar_0(s')}{s'^3(s'-s)} \eolc \\[1mm]
	\Mbar_1(s) &= a_1 + b_1 s + \frac{s^2}{\pi} \int\limits_{4 \mpi^2}^\infty ds'
\frac{\Im\Mbar_1(s')}{s'^2(s'-s)} \eolc \\[1mm]
	\Mbar_2(s) &= a_2 + b_2 s + c_2 s^2 + \frac{s^3}{\pi} \int\limits_{4 \mpi^2}^\infty ds'
\frac{\Im\Mbar_2(s')}{s'^3(s'-s)} \eolp
\label{eq:OneLoopDispRel}
\end{split}\end{equation}
From Eq.~\eqref{eq:OneLoopAsymp}, one verifies that
\begin{equation}
	\Mbar_0(s) + \frac{4}{3} \Mbar_2(s) \asymp \O(s \log s)\eolc \quad
	s \Mbar_1(s) + \Mbar_2(s) \asymp -\frac{4 L_3-\inv{64 \pi^2}}{\Fpi^2 (\meta^2 - \mpi^2)}\; s^2  \eolc
\end{equation}
which allows to fix two combinations of the subtraction constants as follows. We add up the dispersion relations to the
same combinations and require the terms that grow too fast to cancel. For the first combination we get
\begin{equation}
	\Mbar_0(s) + \frac{4}{3} \Mbar_2(s) \asymp c_0 s^2 + \frac{4}{3} c_2 s^2
				+ \frac{s^3}{\pi} \int\limits_{4 \mpi^2}^\infty ds'\; \frac{\Im\Mbar_0(s') + \frac{4}{3}
\Im\Mbar_2(s')}{s'^3(s'-s)} \eolp
	\label{eq:eta3piM0M2}
\end{equation}
The integral is a three-times-subtracted representation of $\Mbar_0(s) + \frac{4}{3} \Mbar_2(s)$. This function grows
like $s \log s$ and so should the dispersive representation. But a three-times subtracted dispersion relation requires
a subtraction polynomial of quadratic order, such that the integral in Eq.~\eqref{eq:eta3piM0M2} contains a
contribution of $\O(s^2)$. We can remove this term by noting that the integral is oversubtracted. We remove
one subtraction by means of Eq.~\eqref{eq:subtraction} with $s_1 = 0$ and find
\begin{equation}
	\frac{s^2}{\pi} \int\limits_{4 \mpi^2}^\infty ds'\; \frac{\Im\Mbar_0(s') + \frac{4}{3} \Im\Mbar_2(s')}{s'^2(s'-s)}
	- \frac{s^2}{\pi} \int\limits_{4 \mpi^2}^\infty ds'\; \frac{\Im\Mbar_0(s') + \frac{4}{3} \Im\Mbar_2(s')}{s'^3} \eolp
\end{equation}
The first term is now a twice-subtracted dispersion integral and thus grows as $s \log  s$ up to a linear subtraction
polynomial. It therefore contains no contribution that grows faster than $s \log s$. The second term is simply
proportional to $s^2$ because the integral converges and is independent of $s$. This implies that
\begin{equation}\begin{split}
	\Mbar_0(s) + \frac{4}{3} \Mbar_2(s) \asymp \; &\O(s \log s) + c_0 s^2 + \frac{4}{3} c_2 s^2 \\[1mm]
				&- \frac{s^2}{\pi} \int\limits_{4 \mpi^2}^\infty ds'\; \frac{\Im\Mbar_0(s') 
					+ \frac{4}{3} \Im\Mbar_2(s')}{s'^3} \eolp
\end{split}\end{equation}
The requirement that the terms growing faster than $s \log s$ must cancel then finally leads to
\begin{equation}
	c_0 + \frac{4}{3} c_2 = \inv{\pi} \int\limits_{4 \mpi^2}^\infty ds'\; \frac{\Im\Mbar_0(s') + \frac{4}{3}
\Im\Mbar_2(s')}{s'^3} \eolp
	\label{eq:oneLoopDispRelComb1}
\end{equation}
For the second combination a similar argument holds. Even though $s \Mbar_1(s) + \Mbar_2(s)$ is quadratic in $s$, the
dispersion integral is still oversubtracted, because $s^2$ is multiplied by a real number and the imaginary part is
thus $\O(s \log s)$. Proceeding as before, we obtain
\begin{equation}
	b_1 + c_2 = -\frac{4 L_3-\inv{64 \pi^2}}{\Fpi^2 (\meta^2 - \mpi^2)}
				+ \inv{\pi} \int\limits_{4 \mpi^2}^\infty ds'\; \frac{s'\, \Im\Mbar_1(s') + \Im\Mbar_2(s')}{s'^3} \eolp
	\label{eq:oneLoopDispRelComb2}
\end{equation}
Putting the dispersion relations together to $\Mbar(s,t,u)$ we find for the polynomial contribution,
\begin{equation}\begin{split}
P(s,t,u) &= a + b s + c s^2 - d (s^2 + 2 t u) \eolc\\
a &= a_0 + \frac{4}{3} a_2 - 3 s_0 (a_1 - b_2 - 3 s_0 c_2) \eolc\\
b &= b_0 + 3 a_1 - \frac{5}{3} b_2 + 3 s_0(b_1-2c_2)\eolc\\
c &= c_0 + \frac{4}{3} c_2\eolc\\
d &= b_1 + c_2\eolc
\end{split}\end{equation}
where we made use of $s+t+u=3 s_0$. Only four combinations of the 8 subtraction constants are of physical relevance.
This is in agreement with the invariance of the reconstruction theorem under the shift in
Eq.~\eqref{eq:pipiMIShift}. Because the polynomial in $\Mbar_1(s)$ is linear, $\alpha$ must vanish, leaving only four
subtraction constants that can be arbitrarily chosen without affecting the total amplitude. $c$ and $d$ are fixed
by Eqs.~\eqref{eq:oneLoopDispRelComb1} and~\eqref{eq:oneLoopDispRelComb2}, respectively, and the dispersive
representation of the one-loop result is hence fully determined by $a$, $b$, and $L_3$.

\section{A soft-pion theorem: the Adler zero} \label{sec:eta3piAdler}

A soft-pion theorem is a statement on the scattering or decay amplitude for a process involving a pion in the limit
where the four-momentum of the pion vanishes. It is based exclusively on symmetry principles and does not require an
explicit Lagrangian. Indeed, these theorems were first developed in the days of current algebra, when neither the
Lagrangian for \chpt{} nor for QCD was known. We first derive the general form of a soft-pion theorem and then apply the
result to $\etapi$.

For the derivation of the general soft-pion theorem we follow the approach of Ref.~\cite{Alfaro+1973} and start with
the amplitude
\begin{equation}
	T_\mu^a = i \int d^4x\, e^{iqx} \bra{B} T \big\{ A_\mu^a(x) \Lambda(0) \big\} \ket{A} \eolc
	\label{eq:eta3piTmuaDef}
\end{equation}
where $\ket{A}$ and $\bra{B}$ are arbitrary hadronic states and $\Lambda(0)$ is an external field. From the LSZ
reduction formula%
\footnote{A readable proof of the LSZ formula in a form that is suitable for our purposes can be found in
Ref.~\cite{Weinberg1995}, sections 10.2 and 10.3.}~\cite{Lehmann+1955} follows
\begin{equation}\begin{split}
	T_\mu^a =\; &i \frac{i}{q^2 - \mpi^2 + i\epsilon} \sum_b \bra{0} A_\mu^a(0) \ket{\pi^b(q)}
					\bra{B \, \pi^b(q)} \Lambda(0) \ket{A} \\
					&+ \text{terms regular at } q^2 \to \mpi^2 \eolp
	\label{eq:eta3piTmuaLSZ}
\end{split}\end{equation}
The matrix element of the axial current is given in Eq.~\eqref{eq:strongF0} and that it is non-vanishing means that the
octet axial vector current is a so-called \emph{interpolating field} for the pion: it can play the role of the pion
field in matrix elements even if no canonical pion field exists. Inserting the explicit expression for this matrix
element we find
\begin{equation}
	T_\mu^a = i F_0 \frac{q_\mu}{q^2 - \mpi^2 + i\epsilon} T^a(q) + \overline{T}^a_\mu \eolc
	\label{eq:eta3piTmuaLSZFinal}
\end{equation}
where we have introduced the off-shell amplitude $T^a(q)$ that corresponds to the hadronic matrix element in
Eq.~\eqref{eq:eta3piTmuaLSZ} and denoted the remaining terms by $\overline{T}^a_\mu$. The latter contains all the
singularities of $T_\mu^a$ with the exception of the pion pole. If the pion momentum $q$ is put on-shell, $T^a$ is the
amplitude for the pion production process $A + (\Lambda) \to B + \pi^a$, i.e.,
\begin{equation}
	\text{Amp}( A + (\Lambda) \to B + \pi^a ) = \lim_{q^2 \to \mpi^2} T^a(q) \eolp
\end{equation}
We switch now to the chiral limit, where the pion is massless. Contracting Eq.~\eqref{eq:eta3piTmuaLSZFinal} with
$q^\mu$ and solving for $T^a$ yields
\begin{equation}
	T^a(q) = \frac{i}{F_0} q^\mu T^a_\mu - \frac{i}{F_0} q^\mu \overline{T}^a_\mu \eolp
	\label{eq:eta3piTa}
\end{equation}
In the chiral limit, putting the pion momentum on-shell corresponds to sending it to zero, and it is in this limit that
the soft-pion theorem is valid.

The next goal is now to bring $q^\mu T^a_\mu$ to a suitable form. To that end, we contract
Eq.~\eqref{eq:eta3piTmuaDef} with $q^\mu$ and, after partial integration, find
\begin{equation}
	q^\mu T^a_\mu = - \int d^4x\, e^{iqx} \partial^\mu \Big\{ 
								\bra{B} A_\mu^a(x) \Lambda(0) \ket{A} \theta(x_0) + 
								\bra{B} \Lambda(0) A_\mu^a(x) \ket{A} \theta(-x_0) \Big\} \eolp
\end{equation}
The current $A_\mu^a(x)$ is conserved in the chiral limit, as becomes immediately clear from
Eq.~\eqref{eq:strongCurrentDivergences} and therefore, only the terms where the derivative acts on the step functions
survive. By means of $\partial^0 \theta(x_0) = \delta(x_0)$, we obtain
\begin{equation}
	q^\mu T^a_\mu = - \int d^4x\, e^{iqx} \bra{B} [ A_0^a(x), \Lambda(0) ] \ket{A}\, \delta(x_0) \eolc
\end{equation}
and, taking the limit $q^\mu \to 0$, finally arrive at
\begin{equation}
	\lim_{q^\mu \to 0} q^\mu T^a_\mu = - \bra{B} [ Q_A^a(0), \Lambda(0) ] \ket{A} \eolc
\end{equation}
where $Q_A^a(0)$ is the axial charge operator defined as
\begin{equation}
	Q_A^a(t) = \int d^3x\, A_0^a(x) \eolp
\end{equation}
The soft-pion theorem in its general form now follows from Eq.~\eqref{eq:eta3piTa}. In the chiral limit and for
vanishing pion momentum the amplitude satisfies
\begin{equation}
	\text{Amp}( A + (\Lambda) \to B + \pi^b ) = - \frac{i}{F_0} \bra{B} [ Q_A^a(0), \Lambda(0) ] \ket{A} 
															- \lim_{q^\mu \to 0} \frac{i}{F_0} q^\mu \overline{T}^a_\mu \eolp
	\label{eq:eta3piSoftPionGeneral}
\end{equation}

Let us now apply the soft-pion theorem to the $\etapi$ amplitude. In the chiral limit, the amplitude $A(s,t,u)$
vanishes, because it is proportional to $B_0 (m_u-m_d)$, but from $M(s,t,u)$ this factor has been extracted. It is
proportional to the matrix element
\begin{equation}
	\bra{\pi \pi \pi^a(q)} S_3(0) \ket{\eta} \eolc
\end{equation}
where we have introduced the scalar density $S_3(x) = \bar{q}(x) \lambda^3 q(x)$ that is contained in the isospin
breaking operator $\lib$. From Eq.~\eqref{eq:eta3piSoftPionGeneral} follows in this case
\begin{equation}
	\text{Amp}( \eta \to \pi \pi + \pi^a ) \propto 
												\bra{\pi \pi \pi^a(q)} [ Q_A^a(0), S_3(0) ] \ket{\eta} 
												+ \lim_{q^\mu \to 0} q^\mu \overline{T}^a_\mu \eolp
\end{equation}
Because the pions are the only massless particles in the hadron spectrum, the amplitude $\overline{T}^a_\mu$ must be
finite at $q^\mu = 0$ and its contribution to the soft-pion theorem vanishes. In order to calculate the commutator, we
examine the behaviour of the scalar density $S_3(0)$ under infinitesimal axial $SU(2) \times SU(2)$ transformations.
On the one hand, we have
\begin{equation}
	S_3 \mapsto e^{-i \epsilon_a Q^a_A}\, S_3\, e^{i \epsilon_a Q^a_A} = S_3 - i \epsilon_a [Q^a_A,S_3] \eolp
	\label{eq:strongS3AxialTransf1}
\end{equation}
On the other hand, we can also let the transformation act on the quark fields. An axial transformation rotates the
left- and the right-handed quark fields in opposite direction:
\begin{equation}
	q_L \mapsto e^{i \epsilon_a \frac{\lambda^a}{2}}\, q_L \eolc \qquad 
	q_R \mapsto e^{-i \epsilon_a \frac{\lambda^a}{2}}\, q_R \eolp
\end{equation}
Using the projection operators~\eqref{eq:strongPLR}, one can show that this is equivalent to 
\begin{equation}
	q \mapsto e^{-i \epsilon_a \frac{\lambda^a}{2} \gamma_5}\, q \eolc \qquad
	\bar{q} \mapsto \bar{q}\, e^{-i \epsilon_a \frac{\lambda^a}{2} \gamma_5} \eolc
\end{equation}
under which $S_3$ transforms as
\begin{equation}
	S_3 \mapsto \bar{q}\, e^{-i \epsilon_a \frac{\lambda^a}{2} \gamma_5} \lambda^3 
						e^{-i \epsilon_a \frac{\lambda^a}{2} \gamma_5}\, q
			= S_3 - \frac{i}{2} \epsilon_a \bar{q} \{ \lambda^a,\lambda^3 \} \gamma_5 q \eolp
	\label{eq:strongS3AxialTransf2}
\end{equation}
From comparison of Eqs.~\eqref{eq:strongS3AxialTransf1} and~\eqref{eq:strongS3AxialTransf2}, one reads off the
commutator
\begin{equation}
	[Q^a_A,S_3] = \frac{1}{2} \bar{q} \{ \lambda^a,\lambda^3 \} \gamma_5 q \eolp
\end{equation}
The anti-commutator on the right-hand side vanishes except if $a = 3$. Consequently, the $\etapi$ amplitude has a zero,
if the four momentum of one of the charged pions vanishes, but not for vanishing $\pi^0$ momentum. These zeros that are
protected by chiral $SU(2) \times SU(2)$ symmetry are called Adler zeros~\cite{Adler1965,Adler1965a}. The amplitude is
not subject to large corrections of order $m_s$ at these points. Note that the protection only applies to the position
of the Adler zero, but not to the slope of the decay amplitude.

In terms of the Mandelstam variables, the limit of vanishing $\pi^+$ momentum, $p_2 \to 0$, corresponds to $s = t = 0$,
while $p_3 \to 0$ corresponds to $s = u = 0$. The Adler zeros are thus related by crossing symmetry.
The tree-level result for $M(s,t,u)$ in Eq.~\eqref{eq:eta3piMtree} has, for vanishing pion mass, even a zero along the
entire line with $s = 0$. If the pion mass is turned on, the line of Adler zeros is moved to $s = \frac{4}{3} \mpi^2$.

The higher order corrections distort the line of zeros. These contributions are, however, of order $\mpi^2$ such that
the line still passes through the vicinity of $s = t = 0$ and $s = u = 0$. Along the line $s = u$, the amplitude has no
zero, but its real part vanishes at $s_A = 1.35\, \mpi^2$ and we take this point as an
approximation of the Adler zero. The distance to the actual Adler zero, as well as the imaginary part of the amplitude
at $s = u = s_A$ are both of order~$\mpi^2$.

\section{Decay rate}

The transition matrix element $\bra{\pi^i(p_2) \pi^j(p_3) \pi^k(p_4)} T \ket{\eta(p_1)}$ or the corresponding amplitude
are by themselves not very useful quantities because they cannot be directly measured. Experiments on decay processes
rather deliver decay rates. The total decay rate (or decay width) is defined as
\begin{equation}
	\Gamma \equiv \frac{\text{number of decays per unit time}}{\text{number of particles present}} \eolc
\end{equation}
and has accordingly unit $\eV$. Its inverse is the lifetime of the particle. One can also measure the decay rate into a
given channel, or the fraction of decays that go into that channel, which is called a branching ratio. The differential
decay rate $d\Gamma$ counts the number of decays per unit time into some small momentum region $d^3p_1 \cdots d^3p_n$.
For the charged $\etapi$ channel, for example, it is defined as
\begin{equation}
	d\Gamma = \frac{P(\eta(p_1) \to \pi^+(p_2) \pi^-(p_3) \pi^0(p_4))}{dt} \eolc
\end{equation}
where
\begin{multline}
	P(\eta(p_1) \to \pi^+(p_2) \pi^-(p_3) \pi^0(p_4)) = \\[1mm]
					|\bra{\pi^i(p_2) \pi^j(p_3) \pi^k(p_4)} T \ket{\eta(p_1)}|^2 \;\,
					\widetilde{dp}_2\, \widetilde{dp}_3\, \widetilde{dp}_4
\end{multline}
is the transition probability, which is, by definition, to be evaluated in the rest frame of the decaying particle. For
brevity, we have introduced the Lorentz invariant phase-space differential
\begin{equation}
	\widetilde{dp} = \frac{d^3p}{(2 \pi)^3 2 p^0} \eolp
\end{equation}
In this normalisation, we have for single-particle states
\begin{equation}
	\braket{\vec{p}\,}{\vec{q}\,} = (2 \pi)^3 2 p^0 \delta^3(\vec{p} - \vec{q}) \eolp
	\label{eq:eta3piSingleParticleNorm}
\end{equation}

We must now find an expression for the squared matrix element in terms of the decay amplitude. Naively inserting the
definition of the amplitude in Eq.~\eqref{eq:eta3piAmplitude} leads to a indefinite result proportional to the square
of a delta function. In order to obtain a valid expression, the incoming momentum eigenstate must be replaced by a
wave package
\begin{equation}
	\ket{\eta} = \int \widetilde{dk}\,  \phi(k) \ket{\eta(\vec{k})} \eolp
\end{equation}
By $\ket{\eta(\vec{k})}$ we still denote one-particle states of momentum $\vec{k}$. The function $\phi(k)$ is sharply
peaked around the $\eta$ momentum $\vec{p}_1$ and satisfies
\begin{equation}
	1 = \braket{\eta}{\eta} = \int \widetilde{dk}\, \widetilde{dq}\, \phi^*(q) \phi(k)
			\braket{\eta(\vec{q})}{\eta(\vec{k})} =  \int \widetilde{dk}\,  |\phi(k)|^2 = 1 \eolc
	\label{eq:eta3piPhiNorm}
\end{equation}
where we made use of the normalisation~\eqref{eq:eta3piSingleParticleNorm} for the single-particle states. With the
incoming wave packet, the squared $T$-matrix element becomes
\begin{equation}\begin{split}
	|\bra{3\pi}T\ket{\eta}|^2 &= \int \widetilde{dk}\, \widetilde{dq}\, \phi^*(q) \phi(k) 
						\bra{3\pi}T\ket{\eta(\vec{q})}^* \bra{3\pi}T\ket{\eta(\vec{k})}\\[1mm]
					&= (2\pi)^8\int \widetilde{dk}\, \widetilde{dq}\, \phi^*(q) \phi(k) \delta^4(p_f - q) \delta^4(p_f - k)
						A^*(q,p_f) A(k,p_f) \eolc
	\label{eq:eta3piSqrMat1}
\end{split}\end{equation}
where $p_f$ is an abbreviation for the momenta of the final-state pions and $A(k,p_f)$ stands for the decay amplitude
of an $\eta$ with momentum $k$ into three pions with momenta $p_f$. In the second delta function, we replace $p_f$ by
$q$ and rewrite it as
\begin{equation}\begin{split}
	\delta^4(q-k) &= \delta(q^0-k^0) \delta^3(\vec{q} - \vec{k}) 
						= \inv{2\pi} \int dt\, e^{i(q^0-k^0)t}\, \delta^3(\vec{q} - \vec{k})\\[2mm]
						&= \inv{2\pi} \int dt\, \delta^3(\vec{q} - \vec{k}) \eolp
\end{split}\end{equation}
In the last equality we have used that the three-dimensional $\delta$-function not only implies $\vec{q} = \vec{k}$,
but also $q^0 = k^0$. Inserting this result into Eq.~\eqref{eq:eta3piSqrMat1} and integrating over $q$ yields
\begin{equation}
	|\bra{3\pi}T\ket{\eta}|^2 = (2\pi)^4 \int \frac{\widetilde{dk}}{2 k^0}\, dt\, |\phi(k)|^2\, 
			\delta^4(p_f - k) |A(k,p_f)|^2 \eolp
\end{equation}
Because the wave packet $\phi(k)$ is sharply peaked at $k = p_1$, we can replace $k$ by $p_1$ everywhere except in the
argument of $\phi$. By means of the normalisation condition~\eqref{eq:eta3piPhiNorm}, the $k$-integration can then be
evaluated. Going over to the rest frame of the $\eta$, where $p_1^0 = \meta$, and  inserting the resulting
expression into the definition of the differential decay rate gives
\begin{equation}
	d\Gamma = \frac{(2\pi)^4}{2 \meta} \delta^4(p_1-p_2-p_3-p_4)\, |A(s,t,u)|^2 \,
					\widetilde{dp}_2\, \widetilde{dp}_3\, \widetilde{dp}_4 \eolp
	\label{eq:eta3piDiffDecayRate}
\end{equation}
Clearly, the amplitude is all we need to know in order to calculate the decay rate. In order to find the total decay
rate $\Gamma$, we must now integrate over the entire phase space. Owing to the delta function, the integral over
$\vec{p}_2$ can be calculated easily, leading to $\vec{p}_2 = -\vec{p}_3 - \vec{p}_4$ because of $\vec{p}_1 = 0$. For
the remaining integrals, we introduce spherical coordinates by means of $d^3q = d\Omega\, dq^0q^0\, |\vec{q}\,|$. The
integration over $\vec{p}_4$ is evaluated in a frame, where $\vec{p}_3$ points along the $z$-axis. The corresponding
angles are denoted by $\varphi_{34}$ and $\theta_{34} = \angle(\vec{p}_3,\vec{p}_4)$. In this coordinate system we find
for the momentum volume elements
\begin{equation}
	d^3p_3\,d^3p_4 = dp^0_3\, dp^0_4\, p^0_3\, |\vec{p}_3|\, p^0_4\, |\vec{p}_4|\, d\Omega_3\, d\varphi_{34}\,
							\dcos \theta_{34} \eolp
\end{equation}
The argument of the remaining delta function, $\delta(\meta - p^0_2 - p^0_3 - p^0_4)$, depends on the angle
$\theta_{34}$ through 
\begin{equation}
	p^0_2 = \sqrt{ \mpip^2 + |\vec{p}_3|^2 + |\vec{p}_4|^2 + |\vec{p}_3|\,|\vec{p}_4|\, \cos \theta_{34}} \eolp
\end{equation}
By means of
\begin{equation}
	\frac{\partial}{\partial\,\cos \theta_{34}} (\meta - p^0_2 - p^0_3 - p^0_4) = \frac{|\vec{p}_3|\,|\vec{p}_4|}{p^0_2}
	\eolc
\end{equation}
it can be rewritten to
\begin{equation}
	\delta(\meta - p^0_2 - p^0_3 - p^0_4) = \frac{p^0_2}{|\vec{p}_3|\,|\vec{p}_4|}\,
				\delta(\cos \theta_{34} - \cos \tilde \theta_{34}) \eolc
\end{equation}
where $\cos \tilde \theta_{34}$ denotes the zero of the argument of the delta function. After the insertion of this
expression, we have arrived at
\begin{equation}
	\Gamma = \inv{16 (2\pi)^5 \meta} \int dp^0_3\, dp^0_4\, d\Omega_3\, d\varphi_{34}\, \dcos \theta_{34}\, 
					\delta(\cos \theta_{34} - \cos \tilde \theta_{34})\, |A(s,t,u)|^2 \eolp
\end{equation}
The only remaining angular dependence is in the argument of the delta function and, possibly, in the Mandelstam
variables. In the rest frame of the $\eta$, the latter are given by
\begin{equation}\begin{split}
	s &= \meta^2 + \mpiz^2 - 2 \meta p^0_4 \eolc \qquad
	t = \meta^2 + \mpip^2 - 2 \meta p^0_3 \eolc \\[1mm]
	u &= \meta^2 + \mpip^2 - 2 \meta p^0_2 \eolc
	\label{eq:eta3pistuRestFrame}
\end{split}\end{equation}
such that only $u$ depends on the angles, namely through $p_2^0$.
Due to the standard relation among the Mandelstam variables, $u$ can be replaced by $s$ and $t$ and the amplitude thus
becomes independent of the angles. The angular integration can then be evaluated trivially and, changing the
integration variables to $s$ and $t$ according to Eq.~\eqref{eq:eta3pistuRestFrame}, we finally find
\begin{equation}
	\Gamma = \inv{256 \pi^3 \meta^3} \int ds\,dt\, |A(s,t,u)|^2 \eolp
\end{equation}
The integral runs over the entire physical region for the decay. In Sec.~\ref{sec:eta3piKinematics} we have found that
the boundaries of the physical region in $s$ are given by
\begin{equation}
	s_\text{min} = 4 \mpip^2 \eolc \quad \text{and} \quad s_\text{max} = (\meta - \mpiz)^2 \eolc
\end{equation}
and from Eq.~\eqref{eq:eta3pitu} we can read-off that for a given $s$, the physically allowed values for $t$ lie in the
interval $[ t_\text{min}(s), t_\text{max}(s)]$ with

\begin{equation}
	t_\text{min/max}(s) = \inv{2} \big( 3 s_0 - s \mp \kappa(s) \big) \eolp
\end{equation}

The integral can also be expressed with the Dalitz plot variables $X$ and $Y$ as integration variables. It then reads
\begin{equation}
	\Gamma = \frac{Q_c^2}{384 \sqrt{3}\, \pi^3 \meta} \int dY\,dX\, |A(X,Y)|^2 \eolc
\end{equation}
and the integration boundaries are given by
\begin{align}
	Y_\text{min} =  -1 \eolc \qquad Y_\text{max} =  \frac{\meta - 3 \mpiz + 6 \mpip}{2 \meta} \eolc
\end{align}
and
\begin{align}\begin{split}
	X_\text{min}(s) &=  -\frac{\sqrt{3}}{2 \meta Q_c}\, \kappa\left( (\meta-\mpiz)^2 - \frac{2 \meta Q_c}{3}(Y+1)
\right)
 			\eolc \\[2mm]
	X_\text{max}(s) &= -X_\text{min}(s) \eolp
\end{split}\end{align}

\begin{table}[tb]
	\renewcommand{\arraystretch}{1.6}
	\begin{center}
	\begin{tabular}{lr@{\,$\pm$\,}lr@{\,$\pm$\,}l}
		Channel 					& \multicolumn{2}{c}{$\Gamma_i/\Gamma_{tot}$ in \%}
																		& \multicolumn{2}{c}{$\Gamma_i$ in eV} \\ \hline
		total						& \multicolumn{2}{c}{---}	& 1300&70 \\
		neutral modes			& 71.90&0.34					&  935&51 \\
		$2 \gamma$				& 39.31&0.20					&  511&28 \\
		$3 \pi^0$				& 32.57&0.23					&  423&23 \\
		charged modes			& 28.10&0.34					&  365&20 \\
		$\pi^+ \pi^- \pi^0$	& 22.74&0.28					&  296&16 \\
		$\pi^+ \pi^- \gamma$	&  4.60&0.16					&   60&4
	\end{tabular}
	\end{center}
	\caption{PDG averages for the total and some partial decay rates of the $\eta$~\cite{PDG2010}.}
	\label{tab:eta3piDecayRate}
\end{table}

Expressing the decay rate in terms of the normalised amplitude $M(s,t,u)$ reveals its dependence on the quark mass
double ratio $Q$:
\begin{equation}
	\Gamma = \inv{Q^4} \frac{\mK^4(\mK^2-\mpi^2)^2}{6912 \pi^3 \meta^3 \mpi^4 \Fpi^4} \int ds\, dt\, |M(s,t,u)|^2 \eolp
	\label{eq:eta3piDecayRateQ}
\end{equation}
Since the decay rate can be measured, a theoretical representation of the decay amplitude $M(s,t,u)$ allows to
determine $Q$ from this equation.

So far, we have discussed the situation in the charged channel, but the formul\ae{} can easily be translated to be used
in the neutral channel. Of course, we must then use the neutral channel decay amplitude and integrate over the correct
phase space, which is easily obtained from the phase space of the charged channel by replacing $\mpip$ by $\mpiz$. In
addition, we have to multiply by a factor $1/3!$ to account for the indistinguishability of the neutral pions. 

In Table~\ref{tab:eta3piDecayRate} we list the current PDG averages for the total decay rate of the $\eta$ as well as
decay rates of the most important channels. The branching ratio of the two $\etapi$ channels is
\begin{equation}
	r = \frac{\Gamma(3\pi^0)}{\Gamma(\pi^+ \pi^- \pi^0)} = 1.432 \pm 0.026 \eolp
\end{equation}

\section{Dalitz plot parametrisation}

It is rather common to parametrise the squared decay amplitude of a three-particle decay as a power series in the Dalitz
plot variables $X$ and $Y$. The coefficients of the series are called the Dalitz plot parameters. In particular, also
measurements are often represented in this form.

Including terms of at most cubic order, the Dalitz plot parametrisation for the charged eta decay channel reads
\begin{align}\begin{split}
	\Gamma(X,Y) = |A(s,t,u)|^2 =
		\begin{aligned}[t]
		 	N ( 1 &+ a Y + b Y^2 + c X + d X^2 + e XY\\ &+ f Y^3 + g X^3 + h X^2 Y + l X Y^2) \eolc
		\end{aligned}
	\label{eq:eta3piDalitzParam}
\end{split}\end{align}
where $N$ is the overall normalisation and $a, b, \ldots, l$ are the Dalitz plot parameters. Several nomenclatures are
in use and we follow here to convention from Ref.~\cite{Ambrosino+2008}.
Since the amplitude is symmetric under the exchange $t \leftrightarrow u$, or correspondingly $X \leftrightarrow -X$, 
all the terms odd in $X$ must vanish. This implies $c = e = g = l = 0$.

The neutral channel is even totally symmetric in the Mandelstam variables, leading to additional simplifications. To
find a suitable form for the Dalitz plot parametrisation in this case, we count the number of independent,
fully-symmetric polynomials at a given order in the Mandelstam variables. At linear order, there is only one, $s+t+u$,
but it is equal to a constant. Accordingly, no linear term appears in the Dalitz plot parametrisation. At quadratic
order, there are two symmetric terms,
\begin{equation}
	s^2 + t^2 + u^2 \qquad \text{and} \qquad st+tu+us \eolc
\end{equation}
but they are not independent, since
\begin{equation}
	(s+t+u)^2 = s^2 + t^2 + u^2 + 2 (st+tu+us) = (\meta^2+3 \mpiz^2)^2 \eolc
\end{equation}
such that only one term remains. At cubic order, one finds three terms:
\begin{equation}
	s^3 + t^3 + u^3 \eolc \quad s t^2 + s u^2 + t s^2 + t u^2 + u s^2 + u t^2 \eolc \quad stu \eolp
\end{equation}
They are related with each other through $(s+t+u)^3$, but this still leaves two independent polynomials.

At quadratic order, the amplitude can therefore be parametrised in terms of a single variable. It is convenient to use a
definition, where the symmetry is explicit, namely
\begin{equation}
	Z = \frac{2}{3} \sum_{i=2}^4 \left( \frac{3 T_i}{Q_n} - 1 \right)^2 = X^2 + Y^2 
		= \frac{3}{2 \meta^2 Q_n^2} (s^2+t^2+u^2-3 s_0^2)	\eolc
\end{equation}
where $Q_n = T_2+T_3+T_4 = \meta - 3 \mpiz$. The Dalitz plot parametrisation up to quadratic order is then
\begin{equation}
	\Gamma(X,Y) = |\bar{A}(s,t,u)|^2 = N (1 + 2 \alpha Z ) \eolp
\end{equation}
At cubic order, two independent terms are needed, such that it cannot be expressed in $Z$ only. The slope parameter
$\alpha$ has been determined by many experiments to excellent accuracy, but none of the higher order coefficients have
been measured so far.

Due to the relation~\eqref{eq:eta3piChargedNeutralChannel} between the amplitudes for the charged and the neutral
channel, also the two Dalitz plot parametrisations are related. The amplitudes for both channels are expanded in the
Dalitz plot variables as
\begin{align}
	A(s,t,u) &= N_c (1 + \bar a Y + \bar b Y^2 + \bar d X^2 + \ldots ) \eolc \\
	\bar A (s,t,u) &= N_n(1+ \bar \alpha Z + \ldots) \eolp 
\end{align}
Comparing the absolute square of these expansions with the respective Dalitz plot parametrisations including terms up
to quadratic order, we obtain
\begin{equation}
	a = 2\, \Re \bar a \eolc \quad b = 2\, \Re \bar b + |\bar a|^2 \eolc \quad d = 2\, \Re \bar d \eolc \quad
	\alpha = \Re \bar \alpha \eolp
	\label{eq:eta3piDalitParamsRel}
\end{equation}
We express the amplitude for the neutral channel in terms of the charged channel amplitude by means of
Eq.~\eqref{eq:eta3piChargedNeutralChannel}. For the prefactors in the definition of the Dalitz plot variables
we introduce the abbreviations
\begin{equation}
	R_{c/n} = \frac{2}{3} \meta Q_{c/n} \eolc 
\end{equation}
with $Q_c = \meta - 2 \mpip - \mpiz$ and $Q_n = \meta - 3 \mpiz$. The quantity $R_n-R_c = 3.35\ee{-3}~\GeV^2$ is small
and we only keep terms at most linear therein. This leads to
\begin{equation}
	\bar A(s,t,u) = N_c \left\{ 3\left( 1-\frac{R_n-R_c}{R_c}\, \bar a \right) 
						+\frac{3}{2} \frac{Q_n^2}{Q_c^2} (\bar b + \bar d \,) Z \right\} + \O\left( (R_n-R_c)^2 \right) \eolp
\end{equation}
Neglecting the term proportional to $\bar a$, we obtain
\begin{equation}
	N_n = 3 N_c \eolc \qquad \text{and} \qquad \bar \alpha = \frac{Q_n^2}{2 Q_c^2} (\bar b + \bar d \,) \eolp
\end{equation}
Together with the expressions from Eq.~\eqref{eq:eta3piDalitParamsRel} this finally yields~\cite{Schneider+2011}
\begin{equation}
	\alpha = \frac{Q_n^2}{2 Q_c^2} ( \Re \bar b + \Re \bar d \,) 
				= \frac{Q_n^2}{4 Q_c^2} (b + d - \inv{4} a^2 - (\Im \bar a)^2 ) \eolp
	\label{eq:eta3piAlphaFromCharged}
\end{equation}
The slope parameter $\alpha$ can thus not be obtained from the Dalitz plot parameters for the charged channel, since 
this requires knowledge of $(\Im \bar a)$. Still, we have found an upper limit for $\alpha$:
\begin{equation}
	\alpha \leq \frac{Q_n^2}{4 Q_c^2} (b + d - \inv{4} a^2) \eolc
	\label{eq:eta3piAlphaLimit}
\end{equation}
with equality only if $\Im \bar a = 0$. In the isospin limit, where $Q_n = Q_c$, the equation is reduced to the formula
given in Ref.~\cite{Bijnens+2007}.

\section[Overview of theoretical and experimental results]{Overview of theoretical and experimental results for the
Dalitz plot parameters} \label{sec:eta3piResults}

We present here a number of experimental and theoretical results for the Dalitz plot parameters in both channels.

\begin{table}[tb]
	\renewcommand{\arraystretch}{1.6}
	\newcommand{\mc}[1]{\multicolumn{2}{c}{#1}}
	 { \footnotesize
	\begin{center}\begin{tabular}{l@{\hspace{.6em}}r@{\;}lr@{\;}lr@{\;}lr@{\;}lc}
		
				& \mc{$a$}		& \mc{$b$}		& \mc{$d$}		& \mc{$f$}		& \\ \hline
		\chpt\ $\O(p^4)$				& $-$1.33&			& 0.42&				& 0.08&				& \mc{---}				&
				\cite{Gasser+1985a} \\
		\chpt\ $\O(p^6)$				& $-$1.271&$\pm$ 0.075	& 0.394&$\pm$ 0.102	& 0.055&$\pm$ 0.057	& 0.025&$\pm$ 0.160	
					& \cite{Bijnens+2007} \\
		Dispersive						& $-$1.16&	& \mc{0.24 $\ldots$ 0.26}	& \mc{0.09 $\ldots$ 0.10}	& \mc{---}
			& \cite{Kambor+1996} \\
		NREFT								& $-$1.213&$\pm$ 0.014	& 0.308 &$\pm$ 0.023	& 0.050 &$\pm$ 0.003	& 	0.083 &$\pm$ 0.019
					& \cite{Schneider+2011}\\ \hline
		Gormley et al.		& $-$1.17&$\pm$ 0.02	& 0.21&$\pm$ 0.03	& 0.06&$\pm$ 0.04	& \mc{---}
					& \cite{Gormley+1970} \\
		Layter et al. 		& $-$1.08&$\pm$ 0.014	& 0.034&$\pm$ 0.027	& 0.046&$\pm$ 0.031	& \mc{---}	
					& \cite{Layter+1973} \\
		Crystal Barrel		& $-$1.21&$\pm$ 0.07 & 0.21&$\pm$ 0.11 & 0.046&	& \mc{---}	& \cite{Abele+1998a}\\
		KLOE					& $-$1.090&${}^{+0.009}_{-0.020}$	& 0.124&$\pm$ 0.012	& 0.057&${}^{+0.009}_{-0.017}$	
					& 0.14&$\pm$ 0.02 & \cite{Ambrosino+2008} \\ \hline
	\end{tabular} \end{center} }
	\caption{Compilation of theoretical and experimental results for the Dalitz plot parameters of the charged channel.
In the result from Crystal Barrel, $c$ has been kept fixed to the value from Layter et al. The KLOE collaboration have
been the first to measure $f$.}
	\label{tab:eta3piDalitzCharged}
\end{table}

Table~\ref{tab:eta3piDalitzCharged} lists theoretical and experimental results for the charged decay
channel. Presently, the only existing modern measurement of the Dalitz plot for the charged channel comes from the KLOE
collaboration~\cite{Ambrosino+2007b,Ambrosino+2008}. All the other experiments have been performed rather long ago and
had much
lower statistics. Having only one experimental result at hand is not very satisfactory as we have no
possibility for comparison. This would be particularly welcome for the parameter $b$, which turns out
surprisingly small in this measurement. Luckily, other experiments will present measurements of this decay in the near
future (WASA-at-COSY~\cite{Adam+2004,Adlarson+2011}, BES-III~\cite{Li2009}). Also the KLOE collaboration is
soon to deliver a new result~\cite{Amelino-Camelia+2010}. Compared to their recent measurement, statistics will be
increased by an order of magnitude and systematic uncertainties reduced due to an improved event selection process.

The result from the KLOE collaboration enters our calculation as input and we thus discuss it in some more detail here.
The KLOE detector is situated at the DA$\Phi$NE $e^+ e^-$ collider~\cite{Zobov2008} in Frascati. It operates at a
center-of-mass energy of $1020~\GeV$, which corresponds to the mass of the $\phi$ meson. The $\eta$ mesons are then
produced through the decay process $\phi \to \eta \gamma$.

\begin{figure}[t]
	\begin{center}
		\includegraphics[width=\textwidth]{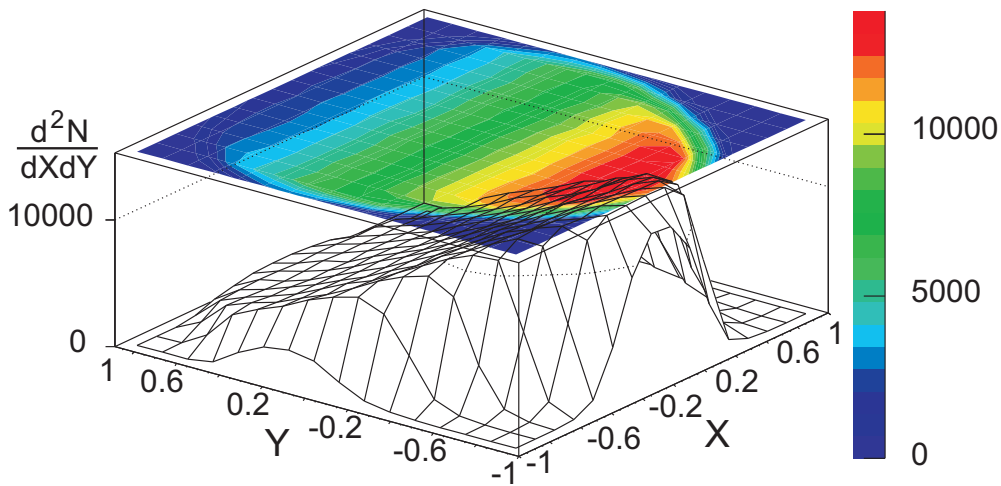}
	\end{center}
	\caption{Measured Dalitz plot distribution for the charged channel from the KLOE collaboration
\cite{Ambrosino+2008}. The plot contains $1.34\ee{6}$ events in 246 bins. The Dalitz plot variables $X$ and $Y$ have
been defined in Eq.~\eqref{eq:etapiDalitzVars}.}
	\label{fig:eta3piDalitzKLOE}
\end{figure}

The analysis we make use of is based on about $450~\text{pb}^{-1}$ collected in the years 2001 and 2002. About $1.4
\ee{9}$ $\phi$ mesons were produced from which some four million $\eta \to \pi^+ \pi^- \pi^0$ events are expected. This
follows from the branching ratios of the two processes involved:
\begin{equation}
	BR(\phi \to \eta \gamma) BR(\eta \to \pi^+ \pi^- \pi^0) =  1.3\% \cdot 22.74\% = 3\ee{-3} \eolp
\end{equation}
Since this is a two particle decay, the recoil photons from $\phi$ decays are almost monochromatic with $E_\gamma
\approx 363~\MeV$. They are thus well separated from the softer photons coming from $\pi^0$ decays and can be used to
identify the production of an $\eta$. Using Monte Carlo simulations, the selection efficiency was found to be $\epsilon
= (33.4 \pm 0.2)\%$ and the background contamination is expected to be 0.3 per cent. After background subtraction,
$1.34\ee{6}$ events are left that are collected in 264 bins with $\Delta X = \Delta Y = 0.125$. The resulting Dalitz
distribution is shown in Fig.~\ref{fig:eta3piDalitzKLOE}.

The measured distribution is then fitted to the Dalitz plot parametrisation. Only those 154 bins that lie entirely
within the physical region were used and various forms of the Dalitz plot parametrisation were fitted. It was found
that the coefficients of terms that are odd in $X$, i.e., $c$, $e$, $g$ and $l$, are consistent with zero in agreement
with charge conjugation symmetry. Leaving them out of the fit does not affect the result for the other parameters.
Including only terms up to quadratic order in the parametrisation leads to results with very low $p$-value. As it
turns out, the coefficient $h$ is consistent with zero, but $f$ must be included as it differs substantially from zero.
The $p$-value of the fit is then 74 per cent and the Dalitz plot parameters are
\begin{align}\begin{split}
	a &= -1.090 \pm 0.005\; \text{(stat)} {}^{+0.008}_{-0.019}\; \text{(syst)} \eolc \\[1mm]
	b &= 0.124 \pm 0.006\; \text{(stat)} \pm 0.010 \; \text{(syst)} \eolc \\[1mm]
	d &= 0.057 \pm 0.006\; \text{(stat)} {}^{+0.007}_{-0.016}\; \text{(syst)} \eolc \\[1mm]
	f &= 0.14 \pm 0.01\; \text{(stat)} \pm 0.02 \; \text{(syst)} \eolc 
\end{split}\end{align}
with the correlation matrix
\begin{equation}
	\begin{array}{ccccc}
		  & a & b      & d      & f 		\\[1mm]
		a & 1 & -0.226 & -0.405 & -0.795 \\[1mm]
		b &   & 1		& 0.358  & 0.261  \\[1mm]
		d &   & 			& 1		& 0.113  \\[1mm]
		f &   &			&			& 1	
	\end{array}
\end{equation}
All the parameters that are not listed were not included in the fit but kept fixed at zero. The systematic error
contains an estimate of the uncertainties coming from event selection, binning, and background contamination. Other
systematic effects were found to be negligible.

\begin{table}[t!]
	\renewcommand{\arraystretch}{1.55}
	\newcommand{\mc}[1]{\multicolumn{2}{c}{#1}}
	 { \small
	\begin{center}\begin{tabular}{lr@{\;}lc}
		
												& \mc{$\alpha$}							& \\ \hline
		\chpt\ $\O(p^4)$ 					&0.015&										&\cite{Bijnens+2002} \\
 		\chpt\ $\O(p^6)$ 					&0.013&$\pm$ 0.032						&\cite{Bijnens+2007} \\
 		Kambor et al. 						&$-$0.014&$\ldots$\;$-$0.007			&\cite{Kambor+1996} \\
		Bijnens \& Gasser					&$-$0.007&									&\cite{Bijnens+2002} \\
		Kampf et al.						&$-$0.044&$\pm$ 0.004					&\cite{Kampf+2011} \\
 		NREFT									&$-$0.025&$\pm$ 0.005					&\cite{Schneider+2011} \\ \hline
 		GAMS-2000 (1984)					&$-$0.022&$\pm$ 0.023					&\cite{Alde+1984} \\
 		Crystal Barrel@LEAR (1998)		&$-$0.052&$\pm$ 0.020					&\cite{Abele+1998} \\
 		Crystal Ball@BNL (2001)			&$-$0.031&$\pm$ 0.004					&\cite{Tippens+2001} \\
 		SND (2001)							&$-$0.010&$\pm$ 0.023					&\cite{Achasov+2001} \\
 		WASA@CELSIUS (2007)				&$-$0.026&$\pm$ 0.014					&\cite{Bashkanov+2007} \\
 		WASA@COSY (2008)					&$-$0.027&$\pm$ 0.009					&\cite{Adolph+2009} \\
 		Crystal Ball@MAMI-B (2009)		&$-$0.032&$\pm$ 0.003					&\cite{Unverzagt+2009} \\
 		Crystal Ball@MAMI-C (2009)		&$-$0.032&$\pm$ 0.003					&\cite{Prakhov+2009} \\
		KLOE (2010)							&$-$0.0301&${}_{-0.0049}^{+0.0041}$	&\cite{Ambrosino+2010} \\ \hline
		PDG average							&$-$0.0317&$\pm$ 0.0016					&\cite{PDG2010}
	\end{tabular} \end{center} }
	\caption{Various theoretical and experimental results for the slope parameter $\alpha$. We have added systematic
and statistical uncertainties in quadrature. The PDG average contains all the experimental results given here with the
exception of the value by the KLOE collaboration. The same results are also visualised in Fig.~\ref{fig:eta3piAlpha}.}
	\label{tab:eta3piAlpha}
\end{table}

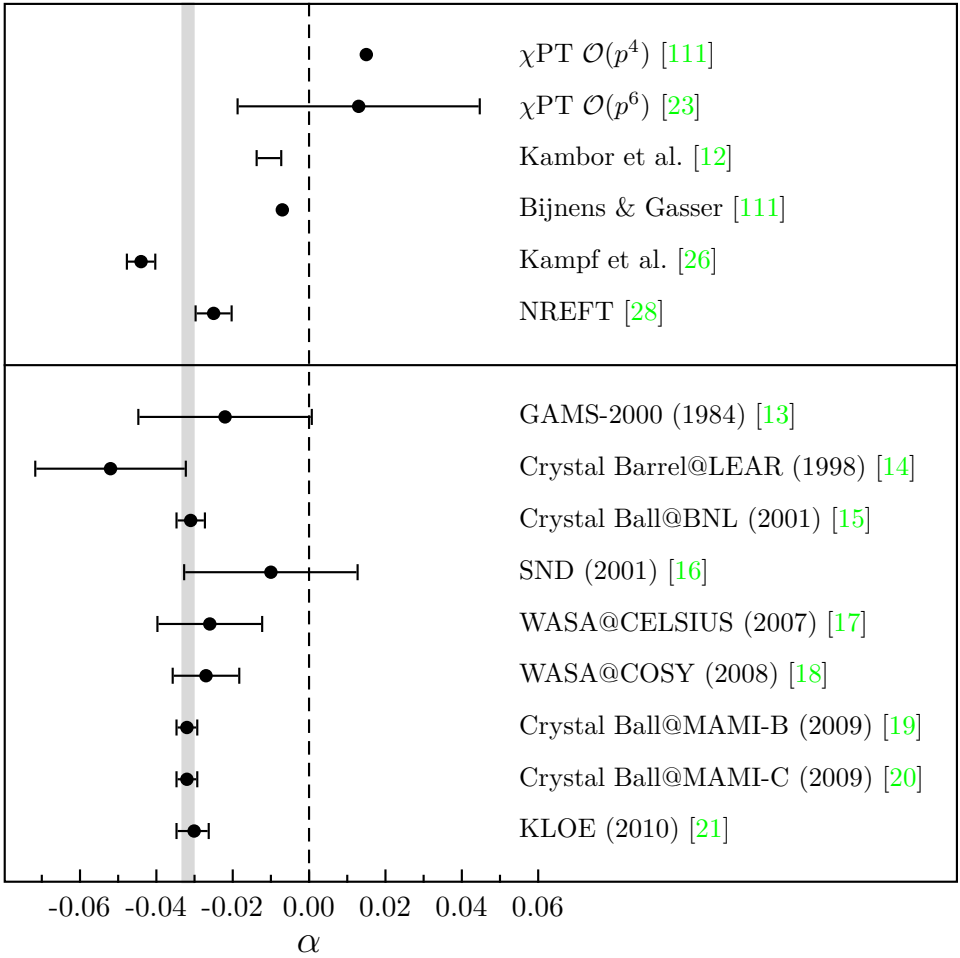
\begin{figure}[p]
\begin{center}
	\psset{xunit=50cm, yunit=.68cm}
	\begin{pspicture}(-0.08,0)(0.17,-18)

		\psset{linewidth=\mylw}
		\setcounter{entrynum}{0}
		\renewcommand{\entryNum}{17}
		\renewcommand{\entryTextPos}{0.055}
		\setlength{\dotsize}{2.5pt}

		\psline[linestyle=dashed](0,0)(0,-\entryNum)

		\psframe[fillstyle=solid,fillcolor=shadegray,linecolor=shadegray,linewidth=0](-0.0333,-\entryNum)(-0.0301,0)

		\psframe(-0.08,0)(0.17,-\entryNum)
		\multips(-0.06,-\entryNum)(0.02,0){7}{\psline(0,5pt)}
		\multips(-0.07,-\entryNum)(0.02,0){7}{\psline(0,3pt)}
		\multido{\n=-0.06+0.02}{7}{\uput{.2}[270](\n,-\entryNum){\n}}
		\uput{.7}[270](0,-\entryNum){\Large $\alpha$}

		\entry{\chpt\ $\O(p^4)$~\cite{Bijnens+2002}}{0.015}{}{}
		\entry{\chpt\ $\O(p^6)$~\cite{Bijnens+2007}}{0.013}{0.032}{0.032}
		\entry{Kambor et al.~\cite{Kambor+1996}}{}{-0.014}{-0.007}
		\entry{Bijnens \& Gasser~\cite{Bijnens+2002}}{-0.007}{}{}
		\entry{Kampf et al.~\cite{Kampf+2011}}{-0.044}{0.004}{0.004}
		\entry{NREFT~\cite{Schneider+2011}}{-0.025}{0.005}{0.005}
		
		\addtocounter{entrynum}{-1}
		\psline(-0.08,\theentrynum)(0.17,\theentrynum)

		\entry{GAMS-2000 (1984)~\cite{Alde+1984}}{-0.022}{0.023}{0.023}
		\entry{Crystal Barrel@LEAR (1998)~\cite{Abele+1998}}{-0.052}{0.020}{0.020}
		\entry{Crystal Ball@BNL (2001)~\cite{Tippens+2001}}{-0.031}{0.004}{0.004}
		\entry{SND (2001)~\cite{Achasov+2001}}{-0.010}{0.023}{0.023}
		\entry{WASA@CELSIUS (2007)~\cite{Bashkanov+2007}}{-0.026}{0.014}{0.014}
		\entry{WASA@COSY (2008)~\cite{Adolph+2009}}{-0.027}{0.009}{0.009}
		\entry{Crystal Ball@MAMI-B (2009)~\cite{Unverzagt+2009}}{-0.032}{0.003}{0.003}
		\entry{Crystal Ball@MAMI-C (2009)~\cite{Prakhov+2009}}{-0.032}{0.003}{0.003}
		\entry{KLOE (2010)~\cite{Ambrosino+2010}}{-0.0301}{0.0049}{0.0041}
	\end{pspicture}
\end{center}

\caption{Visualisation of the theoretical and experimental results listed in Table~\ref{tab:eta3piAlpha}. The shaded
interval corresponds to the PDG average.}
\label{fig:eta3piAlpha}

\end{figure}

For the neutral channel, the experimental situation is much clearer, as the Dalitz plot has been measured in recent
years by several experiments in excellent agreement. The present PDG average is $\alpha = -0.0317 \pm 0.0016$.
Some  experimental results, in particular also those contained in the average, are compiled in
Table~\ref{tab:eta3piAlpha} and visualised in Fig.~\ref{fig:eta3piAlpha}. We do, however, not rely on any of these
results as input. While the experimental value for $\alpha$ is well established, its prediction from theory
has proved rather difficult as can be seen from the results that are also given in the table. Chiral perturbation
theory favours a positive value, even though the uncertainty on the two-loop result is large enough to also encompass
negative values. The dispersive analysis of Kambor et al.~\cite{Kambor+1996} leads to a negative result, but with
too small magnitude. A similar result comes from Bijnens and Gasser, who simplified the method by Anisovich and
Leutwyler by only considering rescattering effects in the direct channel (i.e., $\hat{M}_I=0$, see
Chapter~\ref{chp:etaDisp}).
Quite recently, Schneider et al.~\cite{Schneider+2011} have obtained a value
that is
compatible with experiment in a non-relativistic effective field theory (NREFT) framework. Kampf et
al.~\cite{Kampf+2011} found a negative value from an analytical dispersive analysis using experimental data on the
charged channel plus two-loop \chpt\ as inputs.

We check now whether the results compiled in Tables~\ref{tab:eta3piDalitzCharged} and~\ref{tab:eta3piAlpha} satisfy the
inequality in Eq.~\eqref{eq:eta3piAlphaLimit}. With the exception of the measurement by Layter et al., which leads to
$\alpha \leq -0.060$, from all the results for the Dalitz plot parameters of the charged channel follows an
upper limit for $\alpha$ consistent with the PDG average. The KLOE collaboration are the only one to deliver Dalitz
plot parameters for both channels and their findings are indeed in agreement with the inequality. Also the
theoretical results for the charged channel lead to an upper limit that is obeyed by the corresponding values for
$\alpha$%
\footnote{In case of the $\O(p^6)$ result, the inequality seems to be violated, which is, however, a consequence of
round-off errors.}.
With the exception of the value from NREFT, all the upper limits from theory are positive. The reason for the
failure of \chpt{} seems to be the size of $b$ which turns out quite large compared to the experiments.

The authors of Ref.~\cite{Schneider+2011} have found a tension between their analysis, the measurement of the charged
channel by the KLOE collaboration, and the well-established experimental value for $\alpha$. In the NREFT framework, it
is possible to explicitly calculate $\Im \bar a$. This value is inserted into Eq.~\eqref{eq:eta3piAlphaFromCharged}
together with the Dalitz plot parameters from KLOE, leading to $\alpha = -0.062^{+0.006}_{-0.007}$, in rather strong
disagreement with the PDG average and KLOE's own result. This confirms the strong need for more experimental information
on the charged channel.

%% file: main/eta3piDispersionRelations.tex
\chapter{Dispersion relations for \texorpdfstring{$\etapi$}{eta to 3 pions}} \label{chp:etaDisp}

We have now set up the mathematical and physical framework required to construct the dispersion relations for $\etapi$.
We follow the procedure that was outlined by Anisovich and Leutwyler~\cite{Anisovich+1996,Anisovich+1996b} and
implemented by Walker~\cite{Walker1998}. At the time of the publication of the former article, Kambor, Wiesendanger,
and Wyler~\cite{Kambor+1996} also carried out a dispersive analysis of this process, using a somewhat differing
approach.

We start with the derivation of the unitarity condition for the imaginary part of the decay amplitude that serves
as input for the dispersion integrals. It turns out that an approximation used in this procedure is actually
inconsistent with unitarity and we must discuss, how unitarity can be restored. The chapter closes with the construction
of the dispersion relations with special emphasis on the required number of subtraction constants.

\section{Unitarity Condition}

We denote the transition amplitude of an $\eta$ into some three pion state $n$ by $\A_\etapi^n$. The decay occurs
entirely through the isospin breaking operator $\L_\text{IB}$ and, according to
Eq.~\eqref{eq:eta3piAmplitudeMatrixElement}, the amplitude is given by
\begin{equation}
	\A_\etapi^n = \bra{n \outs} \lib \ket{\eta} \eolp
\end{equation}
In the absence of isospin breaking, the $\eta$ is a stable particle and the operator $\lib$ can be understood as a small
perturbation that generates transitions between eigenstates of the unperturbed system. Time reversal symmetry then
relates the transition matrix elements of $\lib$ between a stable $\eta$ state and an arbitrary outgoing or incoming
state as
\begin{equation}
 \bra{n\outs} \lib \ket{\eta} = \bra{\eta} \lib \ket{n\ins} = \bra{n\ins} \lib \ket{\eta}^* \eolc
\end{equation}
and from this follows for the imaginary part of the transition matrix element
\begin{equation}\begin{split}
 \Im \A_\etapi^n &= \inv{2 i} \big( \bra{n\outs} \lib \ket{\eta} - \bra{n\ins} \lib \ket{\eta} \big)\\[2mm]
			&= \inv{2 i} \sum_{n'} \big( \braket{n\outs}{n'\outs} \bra{n'\outs} \lib \ket{\eta}\\
				& \qquad \qquad \qquad - \braket{n\ins}{n'\outs} \bra{n'\outs} \lib \ket{\eta} \big)\\[1mm]
			&= \inv{2 i} \sum_{n'} \big( \delta_{nn'} - \braket{n'\outs}{n\ins}^\ast \big) \A_\etapi^{n'} \eolc
\end{split}\end{equation}
where a complete set of ``out'' states has been inserted in the second line. We introduce the scattering amplitude
$T_{n' n}$ and obtain

\begin{equation}\begin{split}
	\Im \A_\etapi^n &= \inv{2 i} \sum_{n'} \Big( \delta_{nn'} 
				- \big( \delta_{nn'} + i (2 \pi)^4 \delta(p_n-p_{n'})\, \T_{n'n} \big)^\ast  \Big) \A_\etapi^{n'}\\
				&= \frac{1}{2} \sum_{n'} (2 \pi)^4  \delta(p_n-p_n') \T^*_{n' n} \A_\etapi^{n'} \eolp
\end{split}\end{equation}
In the physical region, the sum over $n'$ is restricted to three pion states. Furthermore, the $3 \pi \to 3 \pi$
transition matrix $\T_{n' n}$ contains disconnected contributions, where two pions scatter, while the third one is only
a spectator, as well as connected contributions, where all pions participate in the scattering process. Chiral
perturbation theory shows that the latter are suppressed, as they only enter at the two-loop level. We therefore omit
the connected part of the $\pi \pi$ scattering amplitude and indicate this with the superscript $d$. The unitarity
condition then reads
\begin{equation}
	\Im \A_\etapi^n \big|_d = \frac{1}{2} \sum_{n'} (2 \pi)^4  \delta(p_n-p_n') (\T^d_{n' n})^* \A_\etapi^{n'} \eolp
  \label{eq:etaDispAnUnitarity}
\end{equation}
We found in Eq.~\eqref{eq:eta3piChargedNeutralChannel} for the decay amplitude in the charged channel that
\begin{equation}
	\A_\etapi^{+-0}(s,t,u) = \A_\etapi^{113}(s,t,u) = A(s,t,u) \eolp
\end{equation}
In the following, we derive a dispersive representation for $A(s,t,u)$, from which the amplitudes for the charged and
the neutral channel can easily be obtained by means of Eq.~\eqref{eq:eta3piChargedNeutralChannel}.
The unitarity condition~\eqref{eq:etaDispAnUnitarity} imposes a linear constraint on $A(s,t,u)$ such that we may
extract the usual normalisation factor,
\begin{equation}
	A(s,t,u) = - \frac{1}{Q^2} \frac{m_K^2 (m_K^2 - m_\pi^2)}{3 \sqrt{3} m_\pi^2 F_\pi^2}\; M(s,t,u) \;,
\end{equation}
where $Q$ is the quark mass double ratio defined in Eq.~\eqref{eq:strongQDefinition}. In the following, we will work
with the normalised amplitude $M(s,t,u)$, which at $\O(p^4)$ coincides with the one-loop amplitude $\Mbar(s,t,u)$.
It can be decomposed into isospin amplitudes $M_I(s)$ according to the reconstruction theorem in
Eq.~\eqref{eq:pipiReconstructionTheorem},
\begin{equation}
	M(s,t,u) = M_0(s) + (s-u) M_1(t) + (s-t) M_1(u) + M_2(t) + M_2(u) - \frac{2}{3} M_2(s) \eolp
	\label{eq:etaDispReconstructionTheorem}
\end{equation}
Recall that this decomposition was obtained neglecting the high energy tails in the dispersion integrals as well as the
imaginary parts of $D$- and higher waves. In the chiral counting, the neglected contributions are of $\O(p^8)$.
As we do not intend to perform a loop-calculation within \chpt, we do however not need to respect chiral counting. We
simply make use of the fact that the terms we omit are numerically small. Instead of formulating a dispersion relation
for the amplitude $M(s,t,u)$, we will derive one for each of the three isospin amplitudes. This procedure constitutes a
considerable simplification since the $M_I(s)$ are functions of one variable that only possess the right-hand cut.

We return, for now, to the decay of an $\eta$ into three arbitrary pions $\pi^i(p_2)$,  $\pi^j(p_3)$, and  $\pi^k(p_4)$.
The properties of the decay amplitude have been discussed in great detail in Chapter~\ref{chp:eta3pi} and we will make
extensive use of these results in the following. In compliance with Eq.~\eqref{eq:eta3piIsospinStructure}, the decay
amplitude $M^{ijk}$ for this process can be decomposed as
\begin{equation}
	M^{ijk}(s,t,u) = M(s,t,u)\, \delta^{ij} \delta^{k3} + M(t,u,s)\, \delta^{ik} \delta^{j3} 
			+ M(u,s,t)\, \delta^{i3} \delta^{jk} \eolp
\end{equation}
Inserting the normalised amplitude $M^{ijk}$ into the unitarity condition yields
\begin{equation}
	\Im  M^{ijk}(s,t,u) \big|_d = \inv{2} \sum_{n'} (2 \pi)^4 \, \delta^4(p_2 + p_3 + p_4 - p_{n'}) (\T^d_{n' (ijk)})^*
	M_{n'} \eolc
\end{equation}
where $M_{n'}$ is the amplitude for the decay $\eta \to n'$ and $T^d_{n' (ijk)}$ is the disconnected
part of the scattering amplitude for $\pi^i \pi^j \pi^k \to n'$. The latter only contains scattering processes
that involve two of the pions. Any pion can be the spectator that does not take part in the final-state rescattering,
such that there are a total of three contributions that must be taken into account. For example, if $\pi^k(p_4)$ is not
scattered, then the $\eta$ decays into $\pi^a(p_a)$,  $\pi^b(p_b)$, and $\pi^k(p_4)$ and the former two pions then
scatter into $\pi^i(p_2)$ and  $\pi^j(p_3)$. The sum over the intermediate states is only taken over the two pion
states $\ket{\pi^a(p_a) \pi^b(p_b)}$. These processes contribute the term
\begin{equation}
	\inv{4 (2 \pi)^2} \sum_{a,b} \int \frac{d^3 p_a \, d^3 p_b}{2 p_a^0 \, 2 p_b^0} \, 	
			\A_{\pi\pi \to \pi\pi}^{ab,ij \,*}(s, \theta_s) \delta^4(p_a+p_b-p_2-p_3) M^{abk}(s,t'_s,u'_s)
	\label{eq:etaDispPikSpectatorInt}
\end{equation}
to the unitarity condition, where the general $n' \to n$ scattering amplitude has now been replaced by the according
$\pi \pi$ scattering amplitude. The kinematic variables must be chosen very carefully in order to prevent sign
mistakes coming from incompatible definitions of the scattering angles. We follow the conventions as outlined in
Chapters~\ref{chp:pipi} and \ref{chp:eta3pi}, such that we can make use of the formul\ae\ presented there. For example,
we define the Mandelstam variables in the $s$-channel by
\begin{equation}
	s = (p_a+p_b)^2 = (p_2+p_3)^2 \eolc \quad t'_s = (p_b + p_4)^2 \eolc \quad u'_s = (p_a + p_4)^2 \eolp
\end{equation}
Furthermore, we introduce three angles:
\begin{equation}
	\theta = \angle(\vec{p}_4,\vec{p}_a) \eolc \qquad
	\theta_0 = \angle(\vec{p}_4,\vec{p}_3) \eolc \qquad
	\theta_s = \angle(\vec{p}_a,\vec{p}_3) \eolc \qquad
\end{equation}
With this choice, $t$ and $u$ ($t'_s$ and $u'_s$) can be expressed in $s$ and $\theta_0$ ($s$ and $\theta$) by means of
Eq.~\eqref{eq:eta3pitu}. We will evaluate the integral in Eq.~\eqref{eq:etaDispPikSpectatorInt} in the frame, where
$\vec{p}_4$ is oriented along the $z$-axis, while $\vec{p}_3$ lies in the $x$-$z$-plane with positive $x$- and
$z$-component. The orientation of $\vec{p}_a$ is then described by $\theta$ and the azimuth $\phi$. The angle $\theta_s$
is related to the other angles by
\begin{equation}
	\cos \theta_s = \cos \theta_0\, \cos \theta + \sin \theta_0\, \sin \theta\, \cos \phi 
						= \frac{t-u}{\kappa(s)}\, \cos \theta + \ldots \eolp
	\label{eq:etaDispThetaS}
\end{equation}
If one of the other pions is the spectator, we proceed analogously and summing up the three contributions yields
\begin{equation}\begin{split}
	\Im M^{ijk}&(s,t,u) \big|_d = \inv{4 (2 \pi)^2} \sum_{a,b} \int \frac{d^3 p_a \, d^3 p_b}{2 p_a^0 \, 2 p_b^0}
	\\[1mm]
	&\times \Big\{
	\begin{aligned}[t]
		&\A_{\pi\pi \to \pi\pi}^{ab,ij \,*}(s, \theta_s) \delta^4(p_a+p_b-p_2-p_3) M^{abk}(s,t'_s,u'_s)\\[2mm]
		&+ \A_{\pi\pi \to \pi\pi}^{ab,ik \,*}(t, \theta_t) \delta^4(p_a+p_b-p_2-p_4) M^{ajb}(s'_t,t,u'_t)\\[2mm]
		&+ \A_{\pi\pi \to \pi\pi}^{ab,jk \,*}(u, \theta_u) \delta^4(p_a+p_b-p_3-p_4) M^{iab}(s'_u,t'_u,u) \Big\} \eolp
	\end{aligned}
	\label{eq:etaDispAllSpectatorsSum}
\end{split}\end{equation}
We return now to the decay amplitude of the charged channel with
\begin{equation}
M^{+-0}(s,t,u) = M^{113}(s,t,u) = M(s,t,u) \eolc
\end{equation}
and fix the indices $i$, $j$, and $k$ accordingly in Eq.~\eqref{eq:etaDispAllSpectatorsSum}. Also the decay amplitude in
the integrand can be written as function of the corresponding centre-of-mass momentum and an angle and the integration
over the momenta can then be evaluated in the same way as in Eq.~\eqref{eq:pipiImFI}. Only the integration over the
orientation of $\vec{p}_a$ remains:
\begin{equation}\begin{split}
	\Im M(s,t,u) \big|_d = \; &\inv{128 \pi^2} \sum_{a,b} \int d\Omega \\[1mm]
	&\times \Bigg\{
	\begin{aligned}[t]
		&\sqrt{\frac{s-4\mpi^2}{s}} \, \A_{\pi\pi \to \pi\pi}^{ab,11 \,*}(s, \theta_s) M^{ab3}(s,t'_s,u'_s) \\[2mm]
		&+ \sqrt{\frac{t-4\mpi^2}{t}} \, \A_{\pi\pi \to \pi\pi}^{ab,13 \,*}(t, \theta_t) M^{a1b}(s'_t,t,u'_t) \\[2mm]
		&+ \sqrt{\frac{u-4\mpi^2}{u}} \, \A_{\pi\pi \to \pi\pi}^{ab,13 \,*}(u, \theta_u) M^{1ab}(s'_u,t'_u,u) \,
	\Bigg\} \eolc
	\end{aligned}
\end{split}\end{equation}
where $d\Omega = d\phi\, \dcos \theta$.
We decompose the $\pi\pi$ scattering amplitude as in Eq.~\eqref{eq:pipiIsospinDecompAmplitude} with
the projection operators from Eq.~\eqref{eq:2pionIsospinProjection} and the functions $F_I$ from
Eq.~\eqref{eq:pipiFIResult}. For the $\eta$ decay amplitude, we insert the decomposition from
Eq.~\eqref{eq:eta3piIsospinStructure}. The sum over $a$ and $b$ can then be evaluated explicitly and we obtain
\begin{align}
	&\Im M(s,t,u) \big|_d = \inv{4 \pi} \int d\Omega \nonumber \\[1mm]
	&\ \ \  
	\begin{aligned}[b]
		\times \bigg\{ \,
		&\inv{3} \sin \delta_0(s)\, e^{-i \delta_0(s)} \Big( 3 M(s,t'_s,u'_s) + M(t'_s,u'_s,s) + M(u'_s,s,t'_s) \Big)\\
		&-\inv{3} \sin \delta_2(s)\, e^{-i \delta_2(s)} \Big( M(t'_s,u'_s,s) + M(u'_s,s,t'_s) \Big)  \\[3mm]
		&+\frac{3}{2} \sin \delta_1(t)\, e^{-i \delta_1(t)} \cos \theta_t \Big( M(s'_t,t,u'_t) - M(u'_t,s'_t,t)
				\Big)\\[1mm]
		&+\inv{2} \sin \delta_2(t)\, e^{-i \delta_2(t)} \Big( M(s'_t,t,u'_t) + M(u'_t,s'_t,t) \Big)	
	\end{aligned}
		\displaybreak[0] \nonumber \\[3mm]
	&\ \ \  
	\begin{aligned}[b]
		&+\frac{3}{2} \sin \delta_1(u)\, e^{-i \delta_1(u)} \cos \theta_u \Big( M(s'_u,t'_u,u) - M(t'_u,u,s'_u)
				\Big)\\[1mm]
		&+\inv{2} \sin \delta_2(u)\, e^{-i \delta_2(u)} \Big( M(s'_u,t'_u,u) + M(t'_u,u,s'_u) \Big) \; \bigg\} \eolp
	\end{aligned}
	\label{eq:etaDispImM}
\end{align}
If we now insert the reconstruction theorem \eqref{eq:etaDispReconstructionTheorem}, the right-hand side must take
the form of this decomposition as well:
\begin{equation}
	G_0(s) + (s-u) G_1(t) + (s-t) G_1(u) + G_2(t) + G_2(u) - \frac{2}{3} G_2(s) \eolp
\end{equation}
The functions $G_I(s)$ are compiled from the terms that are proportional to $\delta_I(s) \exp(-i \delta_I(s))$. By
means of Eq.~\eqref{eq:etaDispThetaS}, the scattering angles can be rewritten in terms of the integration variables. The
term proportional to $\cos \phi$ vanishes, since there is no other $\phi$-dependence in the integrand. After the
$\phi$ integration has been evaluated, we are left with the integral over $z\equiv\cos \theta$. In order to
achieve that each integrand only depends on one primed Mandelstam variable, we make use of the relation
\begin{equation}
	\int \! d\Omega\, z^n M_I(t'_s) = (-1)^n \! \int \! d\Omega\, z^n M_I(u'_s) \eolc
\end{equation}
and similarly for the other channels. After all these simplifications, the functions $G_I(s)$ read
\begin{equation}\begin{split}
	G_0(s) = &\sin \delta_0(s)\, e^{-i \delta_0(s)} \inv{2} \int\limits_{-1}^1 dz \, 
				\begin{aligned}[t]
				\Big\{	&M_0(s) + \frac{2}{3} M_0(t'_s) + 2 (s-s_0) M_1(t'_s)\\[1mm]
							&+ \frac{2}{3} z\, \kappa(s) M_1(t'_s) + \frac{20}{9} M_2(t'_s) \Big\} \eolc
				\end{aligned} \\[3mm]
	G_1(s) = &\sin \delta_1(s)\, e^{-i \delta_1(s)} \inv{\kappa(s)} \inv{2} \int\limits_{-1}^1 dz \, z
				\begin{aligned}[t]
				\Big\{	&3 M_0(t'_s) + 3 z\, \kappa(s) M_1(s)\\[1mm] &+ \frac{9}{2} (s-s_0) M_1(t'_s)\\[1mm]
							&+\frac{3}{2} z \, \kappa(s) M_1(t'_s) - 5 M_2(t'_s) \Big\} \eolc
				\end{aligned} \\[3mm]
	G_2(s) = &\sin \delta_2(s)\, e^{-i \delta_2(s)} \inv{2} \int\limits_{-1}^1 dz \, 
				\begin{aligned}[t]
				\Big\{	&M_0(t'_s) - \frac{3}{2} (s-s_0) M_1(t'_s) - \inv{2} z \, \kappa(s) M_1(t'_s)\\[1mm]
							&+ M_2(s) + \inv{3} M_2(t'_s) \Big\} \eolp
				\end{aligned}
\end{split}\end{equation}
Each $G_I(s)$ contains a term proportional to $M_I(s)$, for which the integral can be evaluated explicitly. For the
remainder, we introduce angular averages of the form
\begin{equation}
  \aav{z^n M_I} = \frac{1}{2} \int_{-1}^1 dz \; z^n M_I\left( \frac{3 s_0 - s + z \kappa(s)}{2} \right) \eolp
  \label{eq:etaDispAngularAverage}
\end{equation}
Equating corresponding expressions in Eq.~\eqref{eq:etaDispImM}, that is $\Im M_I(s) = G_I(s)$, we obtain the unitarity
condition
\begin{equation}
  \Im M_I(s) \Big|_d  = \theta(s-4\mpi^2) \left\{ M_I(s) + \hat{M}_I(s) \right\} \sin \delta_I(s) e^{-i \delta_I(s)}
\eolc
  \label{eq:etaDispImMI}
\end{equation}
where the inhomogeneities are built from angular averages as
\begin{equation}\begin{split}
  \hat M_0 &= \frac{2}{3} \aav{M_0} + 2 (s-s_0) \aav{M_1} + \frac{2}{3} \kappa \aav{z M_1}  + \frac{20}{9} \aav{M_2}
			\eolc \\[1mm]
  \hat M_1 &= \frac{1}{\kappa} \left\{ 3 \aav{z M_0} + \frac{9}{2} (s-s_0) \aav{z M_1} + \frac{3}{2}
			 \kappa \aav{z^2 M_1}  - 5 \aav{z M_2} \right\} \eolc \\[1mm]
  \hat M_2 &= \aav{M_0} - \frac{3}{2} (s-s_0) \aav{M_1} - \frac{1}{2} \kappa \aav{z M_1} + \frac{1}{3} \aav{M_2} \eolp
\label{eq:etaDispMhat}\end{split}\end{equation}

\section{The unitarity condition for the one-loop result} \label{sec:etaDispUnitarityOneLoop}

We check explicitly that the one-loop result from \chpt\ verifies the unitarity condition.
The tree-level expressions for the phase shifts have been given in Eq.~\eqref{eq:pipiPhaseTree}.
From Eq.~\eqref{eq:eta3piMtree} follows for the leading order $\etapi$ amplitude
\begin{equation}
	M_0\tree(s) \equiv T(s) = \frac{3s-4\mpi^2}{\meta^2-\mpi^2} \eolc \quad 
	M_1\tree = M_2\tree = 0 \eolp
	\label{eq:etaDispMtree}
\end{equation}
Only two of the angular averages that build up the $\hat M_I$ are non vanishing:
\begin{equation}
	\aav{M_0\tree}(s) = \frac{3 - T(s)}{2} \eolc \qquad
	\aav{z\,M_0\tree}(s) = \frac{\kappa(s)}{2(\meta^2-\mpi^2)} \eolp
	\label{eq:etaDispMhattree}
\end{equation}
From these one immediately obtains
\begin{equation}
	\hat M_0\tree(s) = \frac{3 - T(s)}{3} \eolc \quad
	\hat M_1\tree(s) = \frac{3}{2(\meta^2-\mpi^2)}  \eolc \quad
	\hat M_2\tree(s) = \frac{3 - T(s)}{2} \eolp
	\label{eq:etaDispMhatTree}
\end{equation}
The tree-level expressions compose the right-hand side of the unitarity constraint~\eqref{eq:etaDispImMI}, while on the
left-hand side we must insert the imaginary part of the one-loop result:
\begin{equation}
	\Im \Mbar_I(s) = \Big\{ M_I\tree + \hat M_I\tree \Big\} \delta_I\tree + \O(p^6) \eolp
	\label{eq:etaDispUnitarityOneLoop}
\end{equation}
With the explicit tree-level expressions, this relation becomes
\begin{equation}\begin{split}
	\Im \Mbar_0(s) &= \frac{3+2T(s)}{3} \delta_0\tree(s) \eolc \quad
	\Im \Mbar_1(s) = \frac{3}{2(\meta^2-\mpi^2)} \delta_1\tree(s) \eolc \\[1mm]
	\Im \Mbar_2(s) &= \frac{3 - T(s)}{2}  \delta_2\tree(s) \eolp
\end{split}\end{equation}
The imaginary part of the one-loop result comes exclusively from the loop-integrals. The unitarity condition only
comprises pion loops and we must accordingly also omit the contributions of all the other loops in the perturbative
result. The imaginary part of the pion loop integral is given by
\begin{equation}
	\Im J_{\pi \pi}^r(s) = \theta(s-4\mpi^2)\, \inv{16 \pi}\, \sqrt{\frac{s-4\mpi^2}{s}} \eolp
\end{equation}
It enters the $\Mbar_I(s)$ through the functions $\Delta_i(s)$. From Eq.~\eqref{eq:eta3piDeltai} results
\begin{equation}
	\Im \Delta_I(s) = \delta_I\tree(s) \eolc \quad I = 0,1,2 \eolc \qquad
	\Im \Delta_3(s) = \inv{3} \delta_2\tree(s) \frac{3s - 4\mK^2}{\meta^2-\mpi^2} \eolc
\end{equation}
and from Eq.~\eqref{eq:eta3piMbar} we find
\begin{equation}\begin{split}
	\Im \Mbar_0(s) &= \frac{3+2T(s)}{3}\, \Im \Delta_0(s) + \frac{3-T(s)}{3}\, \Im \Delta_2(s) 
							+ \Im \Delta_3(s) \eolc \\[2mm]
	\Im \Mbar_1(s) &= \frac{3}{2(\meta^2-\mpi^2)}\, \Im \Delta_1(s) \eolc \qquad
	\Im \Mbar_2(s) = \frac{3-T(s)}{2}\, \Im \Delta_2(s) \eolp
\end{split}\end{equation}
If in $\Im \Mbar_0(s)$ the kaon mass is replaced by means of the Gell-Mann--Okubo mass
formula~\eqref{eq:strongGellMannOkubo}, the last two terms cancel. It is then immediately clear that the unitarity
condition is satisfied.

\section{Elastic unitarity} \label{sec:etaDispElasticUnitarity}

When going beyond leading order in the low-energy expansion, the decay amplitude picks up an imaginary part such that
the functions $\hat M_I$ must necessarily be complex, too. But this contradicts the unitarity condition in
Eq.~\eqref{eq:etaDispImMI}, because the latter implies that $\Im \hat M_I = 0$:
\begin{equation}
  \sin \delta \; \Im M = \Im \left( e^{i \delta} \; \Im M \right) = \sin \delta \; \Im \left( M+\hat M \right) \eolp
\end{equation}
In other words, the higher order corrections bring about a complex value for $\Im M_I(s)$. The problem originates in
the fact that we neglected the connected contributions to the imaginary part. While this is not in principle a bad
approximation---it works well for $\pi \pi$ scattering, where it is referred to as \emph{elastic
unitarity}---it is in the present situation not compatible with unitarity.

In the case of three-body decays the consistent formulation of the analogue of elastic unitarity is more complicated
\cite{Anisovich+1966}.
The solution is to start with the elastic $\pi \pi$ scattering amplitude, where the problem with the imaginary part does
not occur. Given that it has the same structure as the $\etapi$ amplitude, it can be decomposed according to the
reconstruction theorem as well and the $M_I(s)$ satisfy the same unitarity condition. The crucial difference is that the
eta mass $\meta$ has been replaced by the pion mass $\mpi$ and thus the argument of $M_I$ in the integrand of the
angular average in Eq.~\eqref{eq:etaDispAngularAverage} is reduced to
\begin{equation}
 \frac{1}{2} (4 \mpi^2 - s)(1-z) \eolp
\end{equation}
The integration thus runs only over values of $s$ that lie on the negative real axis, where $M_I$ is real to
all orders and consequently also $\hat{M}_I$ remains real in accordance with the unitarity condition. In the case of 
$\pi \pi$ scattering it is thus compliant with unitarity to drop the disconnected part of the amplitude. Because the
amplitude is real for $s \leq 4 \mpi^2$, the Schwarz reflection principle~\eqref{eq:Schwarz} applies and the imaginary
part can be written as
\begin{equation}
  \Im M_I(s) = \frac{M_I(s + i \epsilon) - M_I(s - i \epsilon)}{2 i} \equiv \disc M_I(s) \eolp
\end{equation}
The imaginary part on the left hand side of Eq.~\eqref{eq:etaDispImMI} can thus be replaced by the discontinuity.
If one of the incoming pions is replaced by an $\eta$ with a mass $\tilde{m}_\eta = \mpi$, these arguments still hold
and, due to crossing, they remain valid also for the decay of the $\eta$. Of course, with this mass the $\eta$ is too
light for the decay to actually happen, but we can analytically continue the amplitude in the mass $\tilde{m}_\eta$,
until it reaches the physical eta mass $\meta$. Because the argument of $M_I$ in the angular averages depends on
$\meta$, doing this results in a deformation of the integration path. All we have to do is to make sure that this path
never crosses the cut and then, the unitarity condition retains the same form for $\eta \to 3 \pi$ as it had for elastic
$\pi \pi$ scattering, namely
\begin{equation}
\disc M_I(s) = \theta(s-4\mpi^2) \left\{ M_I(s) + \hat{M}_I(s) \right\} \sin \delta_I(s) e^{-i \delta_I(s)} \eolp
\label{eq:discMI}
\end{equation}
The discontinuity does not need to be real, such that it is not in contradiction with unitarity any longer that the
$M_I(s)$ pick up an imaginary part at higher orders. Because neglecting the connected contributions does no longer lead
to complications, we have omitted the label $d$ in the unitarity condition. Note, however, that this procedure has only
saved unitarity, but not restored the connected part of the amplitude.

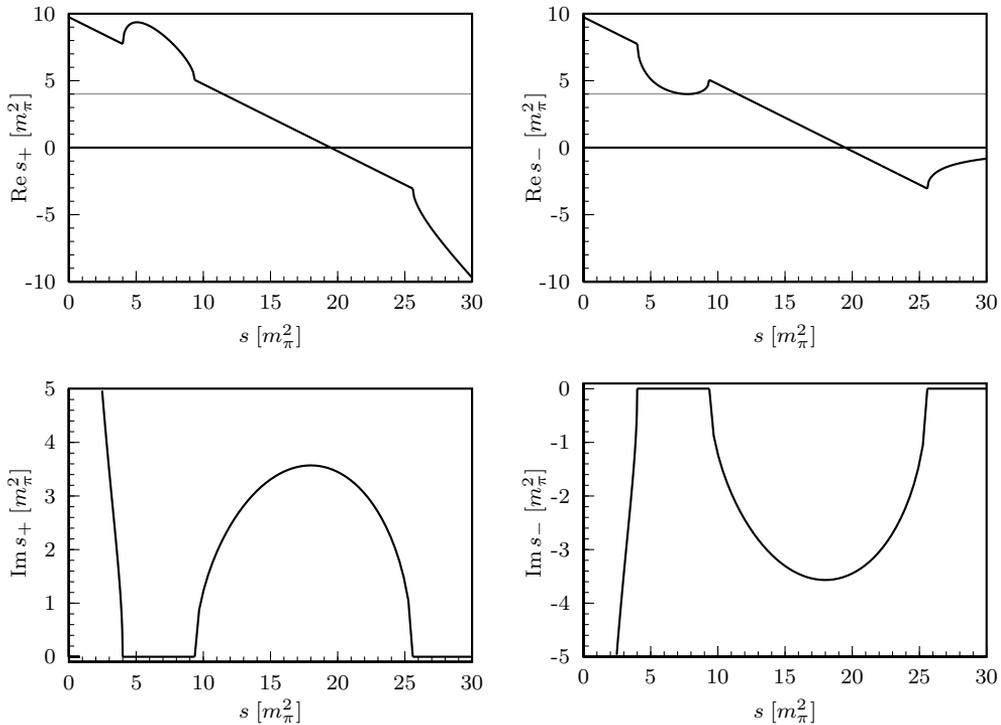
\begin{figure}[tb]


\def\pshlabel#1{\scriptsize #1}
\def\psvlabel#1{\scriptsize #1}

\psset{unit=.176cm, linewidth=.4pt}
\begin{center}
\begin{pspicture*}(-4.55,-17)(33,11)
\psset{linewidth=\mylw}
\psline(0,0)(30,0)
\psline[linecolor=gray,linewidth=.5\mylw](0,4)(30,4)
\psaxes[Dx=5,dx=5,Dy=5,dy=5,Oy=-10, axesstyle=frame,labels=all,showorigin=true,tickstyle=top,
ticksize=4pt](0,-10)(30,10)
\psaxes[dx=1,dy=1, labels=none, tickstyle=top,ticksize=2pt](0,-10)(30,10)
\uput{16pt}[270](15,-10){$\scriptstyle s \; [ \mpi^2 ]$}
\uput{14pt}[180]{90}(0,0){$\scriptstyle \Re s_+\; [ \mpi^2 ]$}

\psplot{0}{4}{19.47 x neg add .5 mul}
\psplot{4}{9.36}{19.47 x neg add
		  1 -4 x div add sqrt
		  9.36 x neg add
		  25.59 x neg add mul sqrt mul
		  add .5 mul}
\psplot{9.36}{25.59}{19.47 x neg add .5 mul}
\psplot{25.59}{30}{19.47 x neg add
		  1 -4 x div add sqrt
		  9.36 x neg add
		  25.59 x neg add mul sqrt mul
		  neg add .5 mul}
\end{pspicture*}
%
\begin{pspicture*}(-4.55,-17)(31,11)
\psset{linewidth=\mylw}
\psline(0,0)(30,0)
\psline[linecolor=gray,linewidth=.5\mylw](0,4)(30,4)
\psaxes[Dx=5,dx=5,Dy=5,dy=5,Oy=-10, axesstyle=frame,labels=all,showorigin=true,tickstyle=top,
ticksize=4pt](0,-10)(30,10)
\psaxes[dx=1,dy=1, labels=none, tickstyle=top,ticksize=2pt](0,-10)(30,10)
\uput{16pt}[270](15,-10){$\scriptstyle s \; [ \mpi^2 ]$}
\uput{14pt}[180]{90}(0,0){$\scriptstyle \Re s_-\; [ \mpi^2 ]$}

\psplot{0}{4}{19.47 x neg add .5 mul}
\psplot{4}{9.36}{19.47 x neg add
		  1 -4 x div add sqrt
		  9.36 x neg add
		  25.59 x neg add mul sqrt mul
		  neg add .5 mul}
\psplot{9.36}{25.59}{19.47 x neg add .5 mul}
\psplot{25.59}{30}{19.47 x neg add
		  1 -4 x div add sqrt
		  9.36 x neg add
		  25.59 x neg add mul sqrt mul
		  add .5 mul}
\end{pspicture*}

\psset{yunit=4 \psunit}
\begin{pspicture*}(-4.55,-1.4)(33,5.2)
\psset{linewidth=\mylw}
\psaxes[Dx=5,dx=5, axesstyle=frame,labels=x,ticks=x,showorigin=true,tickstyle=top, ticksize=4pt](0,-.1)(30,5)
\psaxes[dx=1,tickstyle=top,ticksize=2pt,labels=none](0,-.1)(30,-.1)
\psaxes[Dy=1,dy=1,Oy=0,tickstyle=top,labels=y,ticksize=4pt,showorigin=true](0,0)(.8,5)
\psaxes[dy=.2,Oy=0,tickstyle=top,labels=none,ticksize=2pt](0,0)(.1,5)
\uput{16pt}[270](15,0){$\scriptstyle s \; [ \mpi^2 ]$}
\uput{14pt}[180]{90}(0,2.5){$\scriptstyle \Im s_+\; [ \mpi^2 ]$}

\psplot{2.47}{4}{1 -4 x div add neg sqrt
		  9.36 x neg add
		  25.59 x neg add mul sqrt mul
		  .5 mul}
\psplot{4}{9.36}{0}
\psplot{9.36}{25.59}{1 -4 x div add sqrt
		  9.36 x neg add
		  25.59 x neg add mul neg sqrt mul
		  .5 mul}
\psplot{25.59}{30}{0}
\end{pspicture*}
%
\begin{pspicture*}(-4.55,-6.4)(31,.2)
\psset{linewidth=\mylw}
\psaxes[Dx=5,dx=5,Dy=1,dy=1, Oy=-5, axesstyle=frame,labels=all,showorigin=true,tickstyle=top, ticksize=4pt](0,-5)(30,.1)
\psaxes[Dx=1,dx=1,Dy=.2,dy=.2, labels=none, tickstyle=top,ticksize=2pt](0,-5)(30,.1)
\uput{16pt}[270](15,-5){$\scriptstyle s \; [ \mpi^2 ]$}
\uput{14pt}[180]{90}(0,-2.5){$\scriptstyle \Im s_-\; [ \mpi^2 ]$}

\psplot{2.47}{4}{1 -4 x div add neg sqrt
		  9.36 x neg add
		  25.59 x neg add mul sqrt mul
		  .5 mul neg}
\psplot{4}{9.36}{0}
\psplot{9.36}{25.59}{1 -4 x div add sqrt
		  9.36 x neg add
		  25.59 x neg add mul neg sqrt mul
		  .5 mul neg}
\psplot{25.59}{30}{0}
\end{pspicture*}
\end{center}

\caption{Real and imaginary part of the start point $s_-$ and the end point $s_+$ of the integration path.
The thin grey lines in the figures for the real part mark the position of the branch point at
$4 \mpi^2$. The black curves cross it at $s = \meta^2 - 5 \mpi^2$. At $s = \frac{1}{2}(\meta^2 - \mpi^2)$, $\Re
s_-$ touches the grey line: it is there that $s_-$ moves from the lower to the upper rim of the cut. }
\label{fig:etaDispSpm}
\end{figure}

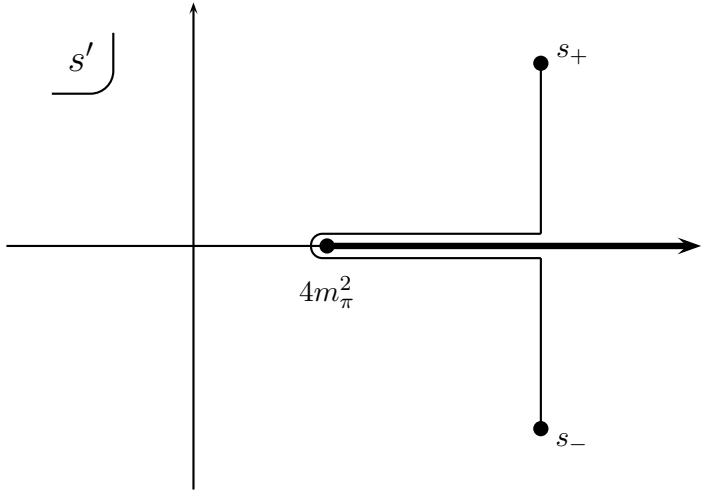
\begin{figure}[t]
	\psset{xunit=0.7cm,yunit=0.4cm}
	\begin{center}\begin{pspicture*}(-3,-8)(10,8)

		\psset{linewidth=\mylw}

		\psline(-3,0)(5,0)
		\psline{->}(0.5,-8)(0.5,8)
		\psline[linewidth=3\mylw]{->}(3,0)(10,0)
		\pscircle*(3,0){0.1}
		\uput{0.4}[-90](3,0){$4 \mpi^2$}

		\psline[linewidth=\mylw,linearc=0.3](-1,7)(-1,5)(-2.15,5)
		\rput(-1.6,6.2){\Large $s'$}

		\pscircle*(7,-6){0.1}
		\uput{0.2}[-30](7,-6){$s_-$}
		\pscircle*(7,6){0.1}
		\uput{0.2}[20](7,6){$s_+$}
		\psline(7,-6)(7,-.4)
		\psline(7,6)(7,.4)
		\psline[linearc=0.15](7,.4)(2.7,.4)(2.7,-0.4)(7,-0.4)
		
	\end{pspicture*}\end{center}

	\caption{The horseshoe shaped integration path that is used in order to calculate $\hat{M}_I(s)$ for 
				$\meta^2 - 5 \mpi^2 < s < (\meta + \mpi)^2$.}
	\label{fig:etaDisHorseShoe}
\end{figure}

The angular averages $\aav{z^n M_I}(s)$ deserve a detailed discussion. They are integrals in the complex plane
along a path that starts at $s_-$ and ends at $s_+$ with
\begin{equation}
  s_\pm = \frac{1}{2}(3 s_0 - s \pm \kappa(s)) \eolp
	\label{eq:detaDispSpm}
\end{equation}
The points move around in the complex plane depending on the value of $s$. The angular averages build up the functions
$\hat{M}_I(s)$ that are part of the integrand in the dispersion integrals and accordingly, we need them for 
$s \in [4 \mpi^2, \infty )$. In Fig.~\ref{fig:etaDispSpm} the position of $s_\pm$ is plotted in the
interesting part of this region. For $s \geq (\meta + \mpi)^2$, $s_\pm$ are real and negative, and thus away
from the cut. This is also true for $s$ outside of the range shown in the figure, because for $s \to \infty$,
the end points behave as $s_+ \to -\infty$ and $s_- \to 0$. The integration can be performed along the real axis without
problems. For $s = (\meta + \mpi)^2$, the endpoints even coincide because $\kappa(s)$ vanishes there,
leading to a zero in the angular averages. For $(\meta + \mpi)^2 > s > (\meta - \mpi)^2$, $s_\pm$ move away from the
real axis and, because $\kappa(s)$ is purely imaginary in this region, they are related by complex conjugation. In this
situation, we have to be more careful and at all times make sure that the integration path avoids the cut.
As long as $s$ stays larger than $\meta^2 - 5 \mpi^2$, the real parts of $s_\pm$ are still smaller than
$4 \mpi^2$ and we can integrate along a straight line connecting them. But as soon as $s$ drops below this value,
this is no longer possible, since the straight integration path crosses the cut: it is here that the consequences of
the omission of the connected contributions show up. The integration path has to be deformed in some way, e.g., by
following a line parallel to the imaginary axis until the real axis is reached, then going along the real axis on the
lower rim of the cut, turning around the branch point, then following the real axis again, but on the upper rim of the
cut, and finally going along a parallel to the imaginary axis again until $s_+$ is reached. An example of such an
integration path is shown in Fig.~\ref{fig:etaDisHorseShoe}. At $s = (\meta - \mpi)^2$, $\kappa(s)$ again vanishes and
the endpoints thus coincide, but this time they lie on opposite sides of the cut and the angular average has no zero.
Indeed, for $(\meta - \mpi)^2 \geq s > \frac{1}{2}(\meta^2 - \mpi^2)$ we still have to integrate along the horseshoe
shaped path described above, even though the parts parallel to the imaginary axis are now missing. With decreasing $s$,
$s_-$ is moving towards the branch point at $4 \mpi^2$. For $s = \frac{1}{2}(\meta^2 - \mpi^2)$, $s_-$ is equal to $4
\mpi^2$ and moves on the upper rim of the cut, such that the integration can now again be performed along a straight
line.

If $s$ lies between $(\meta - \mpi)^2$ and $4 \mpi^2$, where $s_\pm$ are real,  it is not immediately clear on which
side of the cut the end points actually lie. This can be decided by equipping the eta mass with a small positive
imaginary part, which then shows that the above description is indeed correct. In particular, one can clearly see that
the imaginary part of $s_-$ changes its sign at $s = \frac{1}{2}(\meta^2 - \mpi^2)$.

\section{Integral equations and ambiguity}

Once the unitary condition in Eq.~\eqref{eq:discMI} is known, we can immediately write down an $n$ times
subtracted dispersion relation for the functions $M_I(s)$:
\begin{equation}
 M_I(s) = P_I(s) + \frac{s^n}{\pi} \int_{4 \mpi^2}^\infty \frac{ds'}{s'^n} 
	    \frac{ \sin \delta_I(s') e^{-i \delta_I(s')} }{ (s' - s - i \epsilon ) } 
	    \left\{ M_I(s') + \hat{M}_I(s') \right\} \eolc
  \label{eq:dispRelAmb}
\end{equation}
where $P_I(s)$ is a polynomial of order $n-1$ in $s$. Its coefficients are not constrained by this equation,
which thus contains $n-1$ free parameters.

This dispersion relation, although correct, is not a good choice as it does not determine the amplitude uniquely.
This can be seen easiest by dropping $\hat M_I$ in the unitarity condition, which is then
reduced to 
\begin{equation}
	\disc M_I(s) =  \theta(s-4 \mpi^2) \sin \delta_I(s) e^{-i \delta_I(s)} M_I(s) \eolp
\end{equation}
This equation is of the form of Eq.~\eqref{eq:discPhi} and in Sec.~\ref{sec:dispOmnes} we have derived its solution to
be
\begin{equation}
	M_I(s) = \Omega_I(s) m_I(s) \eolp
\end{equation}
where $\Omega(s)$ is the Omn\`es function that was introduced in Eq.~\eqref{eq:dispOmnes} and $m_I(s)$ is a polynomial.
The unitarity condition allows for an arbitrary polynomial that can be further constrained, if we have some knowledge
on the asymptotic behaviour of $M_I(s)$. Here, the Omn\`es function is given by
\begin{equation}
  \Omega_I(s) = \exp \left\{ \frac{s}{\pi} \int_{4 \mpi^2}^{\infty} \frac{\delta_I(s')}{s' (s' - s)} d s' \right\} \eolp
	\label{eq:etaDispOmnes}
\end{equation}
Assuming that the phase shifts reach their asymptotic values of $k \pi$ at some large scale $\Lambda^2$, with $k=1$
for $I = 0,1$ and $k = 0$ for $I=2$, the asymptotic behaviour of the Omn\`es functions is according to
Eq.~\eqref{eq:dispOmnesAsymptotic} given by
\begin{equation}
  \Omega_0(s) \asymp \frac{1}{s} \;, \qquad \Omega_1(s) \asymp \frac{1}{s} \;, \qquad \Omega_2(s) \asymp 1 \eolp
  \label{eq:etaDispOmnesAsymptotic}
\end{equation}
Assuming $M_I(s)$ to grow asymptotically at most with some power
of $s$ limits the order of the polynomial $m_I(s)$. The origin of the aforementioned ambiguity is the asymptotic
behaviour of the Omn\`es functions for $I = 0,1$ (which itself is due to the asymptotic behaviour of the scattering
phase shifts). It implies that, if $M_I(s)$ is asymptotically bounded by $s^{n-1}$, $m_I(s)$ must be
of order $n$. The general solution of the unitarity condition thus contains a total of $n$ free parameters. The integral
representation of $M_I(s)$ given in Eq.~\eqref{eq:dispRelAmb} indeed grows like $s^{n-1}$, but it contains only $n-1$
free parameters in the polynomial, i.e., one parameter less than the general solution. The dispersion relations for
$M_0(s)$ and $M_1(s)$ therefore allow for a one parameter family of solutions. The dispersion relation for $M_2(s)$ can,
on the other hand, be solved uniquely, because $m_2(s)$ and the subtraction polynomial are both of order $n-1$.

We return now to the unitarity condition including the functions $\hat{M}_I(s)$ and search for a set of dispersion
relations that can be solved uniquely. Since we have found before that $m_I(s) \equiv M_I(s)/\Omega_I(s)$ contains the
correct number of degrees of freedom, we derive dispersion relations for these functions. Their discontinuity can be
calculated to be
\begin{equation}\begin{split}
  \disc m_I(s) &= \frac{M_I(s + i\epsilon)}{2i\, \Omega_I(s + i\epsilon)} 
						- \frac{M_I(s - i\epsilon)}{2i\, \Omega_I(s - i\epsilon)}\\[3.3mm]
					&= \frac{M_I(s + i\epsilon) e^{- i \delta_I(s)} - M_I(s - i\epsilon) e^{i \delta_I(s)}}
							{2i | \Omega_I(s) |}\\[3.7mm]
					&= \frac{-M_I(s) \sin \delta_I(s) + e^{i \delta_I(s)} \disc M_I(s)}{ | \Omega_I(s) |}
					= \frac{ \sin \delta_I(s) \hat{M}_I(s) }{ |\Omega_I(s)|} \eolc
\end{split}\end{equation}
where we have used Eq.~\eqref{eq:dispOmnesCut} in the second equality. This leads immediately to a set of dispersion
relations for $m_I(s)$ that can be solved for $M_I(s)$:
\begin{equation}
 M_I(s) = \Omega_I(s) \left\{ P_I(s) + \frac{s^n}{\pi} \int_{4 \mpi^2}^\infty \frac{ds'}{s'^n} 
	    \frac{ \sin \delta_I(s') \hat{M}_I(s') }{ |\Omega_I(s')| (s' - s - i \epsilon ) } \right\} \eolp
  \label{eq:dispRel}
\end{equation}
The polynomials $P_0(s)$ and $P_1(s)$ are now of order $n$, one order larger than in the standard dispersion relation, 
while $P_2(s)$ is still of order $n-1$. The solution of these equations is indeed unique, as $M_0(s)$ and $M_1(s)$
now each contain one free parameter more than in Eq.~\eqref{eq:dispRelAmb}.

\section{Number of subtraction constants}

The number of subtractions needed for the dispersion integral in Eq.~\eqref{eq:dispRel}, or in other words the value of
$n$, is a rather delicate matter. From a purely mathematical point of view, subtracting the dispersion integral is
simply a rearrangement of the equation. Once we have performed enough subtractions to ensure that the integral along
the infinite circle vanishes, any further subtraction does not effectively alter the equation any more, because the
change in the dispersion integral is absorbed in an additional parameter in the subtraction polynomial.

But from a physical point of view, there is more to it than this. If few subtractions are used, the high energy
behaviour of the integrand is less suppressed. In the present case, this would mean that the contributions above the $K
\Kbar$-threshold are more important. Using more subtraction constants, on the other hand, reduces the predictive power
of the solution, because more \mbox{parameters} have to be determined using other methods. For example, if we use
one-loop chiral perturbation theory to fix the subtraction polynomial, having too many subtractions means that the
solution will not substantially differ from the one-loop result itself.

The Roy equations \cite{Roy1971} for $\pi \pi$ scattering indicate that the decay amplitude $M(s,t,u)$ obeys a fixed $t$
dispersion relation with two subtractions. Chiral perturbation theory thus leads to an oversubtracted dispersion
relation: at one-loop level it requires 4 subtractions and at higher orders even more. This is due to the fact that the
integration extends until far outside of the region of validity of \chpt. The large number of subtractions is
required by the unphysical singularity of the low-energy result at infinity.

We fix the number of subtraction constants by requiring the amplitude to satisfy the Froissart bound
\cite{Froissart1961}, which here amounts to the restriction that the polynomial part of the amplitude $M(s,t,u)$
asymptotically grows at most linearly in $s$, $t$ and $u$. For the individual isospin components this implies $M_0(s) =
\O (s)$, $M_1(s) = \O (1)$ and $M_2(s) = \O (s)$. Each subtraction polynomial is multiplied by the Omn\`es function and,
recalling the asymptotic behaviour of the latter from Eq.~\eqref{eq:etaDispOmnesAsymptotic}, they are
\begin{equation}
  P_0(s) = \alpha_0 + \beta_0 s + \gamma_0 s^2 \; , \quad P_1(s) = \alpha_1 + \beta_1 s \; , \quad P_2(s) = \alpha_2 +
\beta_2 s \; ,
\end{equation}
containing a total number of 7 subtraction constants. This number can be further reduced because the isospin
decomposition \eqref{eq:etaDispReconstructionTheorem} is not unique. Shifting the $M_I(s)$ simultaneously according to
Eq.~\eqref{eq:pipiMIShift} leaves the total amplitude $M(s,t,u)$ unchanged. We can now use such a transformation to
shift some of the subtraction constants to zero. In order to leave the asymptotic behaviour of the $M_I(s)$
intact, we must set $\alpha = \beta = 0$ and the shift is then determined by the remaining three parameters:
\begin{equation}\begin{split}
	M_0(s) &\mapsto M_0(s) + \frac{5}{3} \gamma s - 3 s_0 \gamma - \frac{4}{3} \delta - 3 \epsilon (s-s_0) \eolc\\[1mm]
	M_1(s) &\mapsto M_1(s) + \epsilon \eolc \\[2mm]
	M_2(s) &\mapsto M_2(s) + \gamma s + \delta \eolp
	\label{eq:etaDispMIShift}
\end{split}\end{equation}
We can choose $\gamma$, $\delta$ and $\epsilon$ such that $\alpha_1 = \alpha_2 = \beta_2 = 0$. So finally, we are left
with four undetermined subtraction constants and we can now state the dispersion relations in the form
\begin{align}
  M_0(s) = &\Omega_0(s) \left\{ \alpha_0 + \beta_0 s + \gamma_0 s^2 + \frac{s^2}{\pi} \int_{4 \mpi^2}^\infty
				\frac{ds'}{s'^2} 	    \frac{ \sin \delta_0(s') \hat{M}_0(s') }{ |\Omega_0(s')| (s' - s - i \epsilon ) }
				\right\} \eolc \nonumber \\[3mm]
  M_1(s) = &\Omega_1(s) \left\{ \beta_1 s + \frac{s}{\pi} \int_{4 \mpi^2}^\infty \frac{ds'}{s'} 
	   \frac{ \sin \delta_1(s') \hat{M}_1(s') }{ |\Omega_1(s')| (s' - s - i \epsilon ) } \right\} \eolc
		\label{eq:dispRelFull} \\[3mm]
  M_2(s) = &\Omega_2(s) \, \frac{s^2}{\pi} \int_{4 \mpi^2}^\infty \frac{ds'}{s'^2} 
	    \frac{ \sin \delta_2(s') \hat{M}_2(s') }{ |\Omega_2(s')| (s' - s - i \epsilon ) } \eolp \nonumber
\end{align}
The dispersion integrals have been subtracted at $s = 0$, but also some other point would do, provided we change the
subtraction constants accordingly. As already indicated in Eq.~\eqref{eq:dispRel}, the subtraction polynomial in
$M_0(s)$ and $M_1(s)$ is one order higher than would be expected from the number of subtractions actually performed in
the dispersion integral. This is related to the fact that we have written the dispersion relation for $M_I(s) /
\Omega_I(s)$. In principle, we could even subtract these integrals once more and absorb the difference into the
subtraction constants $\gamma_0$ and $\beta_1$ respectively. But keeping the relations in their present form leads to
simpler expressions for these two subtraction constants, as will be discussed below.

%% file: main/eta3piSolution.tex
\chapter{Solution of the dispersion relations} \label{chp:Solution}

In the last chapter, we have derived a set of coupled dispersion relations for the isospin amplitudes $M_I(s)$. In the
present chapter, we discuss our iterative solution algorithm for this problem. We present two different ways to
determine the subtraction constants: from a matching to \chpt{} at one loop~\cite{Anisovich+1996} and from a combined
fit to experimental data and one-loop \chpt. This latter procedure had not been possible at the time of the original
publication by Anisovich and Leutwyler due to the lack of suitable experimental data and is thus an entirely new
approach in this context. A detailed account of the implementation into a numerical computer program and the results
thereof will be given in Chapters~\ref{chp:Numerics} and~\ref{chp:Results}, respectively.

\section{Iterative solution algorithm}

\begin{figure}[tb]
\begin{center}

\psset{linewidth=1.2pt, framearc=.3, unit=.8cm}

\begin{pspicture*}(0,-4.5)(15.5,3)
  \rput[lb]{0}(0,0){%
    \psframe(0,0)(3.5,2)
    \rput{0}(1.8,1){ \parbox{1.7cm}{\raggedright set $M_I$ to tree-level} }
    \psline{->}(3.5,1)(6,1)
  }
  \rput[lb]{0}(6,0){%
    \psframe(0,0)(3.5,2)
    \rput{0}(1.85,1){ \parbox{2.3cm}{\raggedright calculate $\hat{M}_I$ from $M_I$} }
    \psline{->}(3.5,1)(6,1)
  }
  \rput[lb]{0}(12,0){%
    \psframe(0,0)(3.5,2)
    \rput{0}(1.85,1){ \parbox{2.3cm}{\raggedright calculate $M_I$ from $\hat{M}_I$} }
    \psline{->}(1.75,0)(1.75,-2)
  }
  \rput[lb]{0}(12,-4){%
    \psframe(0,0)(3.5,2)
    \rput{0}(1.85,1){ \parbox{2.1cm}{\raggedright determine subtraction constants} }
    \psline{->}(0,1)(-2.25,1)
  }
  \rput[lb]{0}(6,-4){%
    \psdiamond(1.75,1)(2,1.5)
    \rput{0}(1.75,1){ \parbox{1.5cm}{\raggedright accuracy reached?} }
    \psline{->}(-.25,1)(-2.5,1)
    \uput[90](-1.13,1){ Yes }
    \psline{->}(1.75,2.5)(1.75,4)
    \uput[0](1.75,3.1){ No }
  }
  \rput[lb]{0}(1,-3.5){%
    \psframe(0,0)(2.5,1)
    \rput{0}(1.25,.5){ \psframebox*{done} }
  }

\end{pspicture*}
\end{center}

\caption{Iteration algorithm used to solve the dispersion relations in Eq.~\eqref{eq:dispRelFull}.}
\label{fig:iteration}

\end{figure}
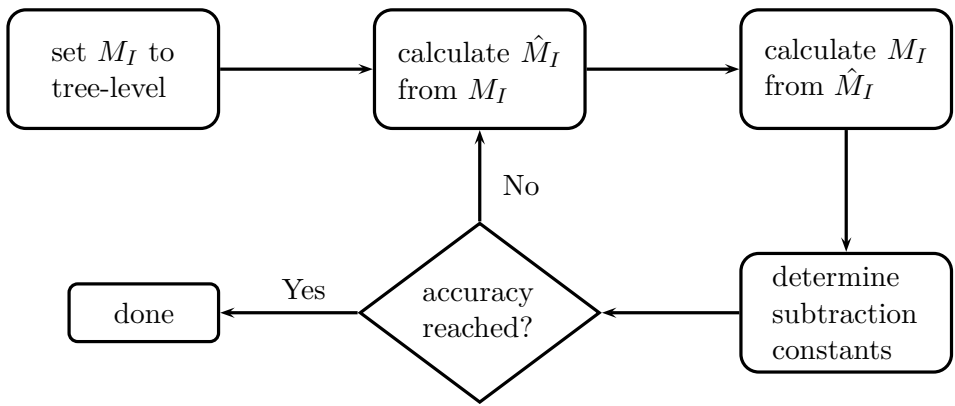

In order to solve the integral equations~\eqref{eq:dispRelFull}, we apply the iterative procedure illustrated
in Fig.~\ref{fig:iteration}. We start the iterations by setting the functions $M_I(s)$ to their tree-level expressions
from \chpt. The configuration we start with does not influence the result of the
calculation (but probably its efficiency), and thus other choices are possible. We have checked that starting with
$M_I(s) = 0$ does indeed lead to the same result, but takes an iteration step more to achieve the same
accuracy. From the initial $M_I(s)$ we then calculate the $\hat{M}_I(s)$ according to Eq.~\eqref{eq:etaDispMhat}. When
starting with $M_I(s) = 0$ or with the tree-level approximation, this can even be done analytically the first time (see
Eq.~\eqref{eq:etaDispMhatTree}). We insert the resulting $\hat{M}_I$ into the dispersion integrals in
Eq.~\eqref{eq:dispRelFull} and evaluate them numerically. The only remaining unknowns in the isospin amplitudes are then
the four subtraction constants, whose determination will be discussed in the next section. With the new representation
of the $M_I(s)$, the entire procedure is repeated until, after a sufficient number of iterations, the algorithm has
converged to the desired accuracy and the process is stopped.

\section{Determination of the subtraction constants}

We have solved the dispersion relation with two different methods. Both apply the iteration procedure as described
above, but they differ in the way the subtraction constants are determined. The first method follows Anisovich and
Leutwyler~\cite{Anisovich+1996}, who have matched the subtraction constants to the one-loop result from \chpt{} in every
iteration step. In their article, the possibility to determine the subtraction constants with the help of experimental
data was already mentioned. At the time, suitable data was simply not available, but the recent results from the KLOE
collaboration~\cite{Ambrosino+2008} can indeed be used for this purpose. However, the overall normalisation of the
amplitude cannot be obtained from the data and is in either case obtained from \chpt.

Even though we cannot do entirely without input from perturbation theory, the dispersive approach has the advantage that
we can rely on the perturbative result in a region, where it converges rather quickly. From this low-energy input, the
dispersion relations then extrapolate the amplitude to the physical region, where \chpt, at least up to $\O(p^4)$, fails
to give an accurate prediction. Roiesnel and Truong~\cite{Roiesnel+1981} have shown that the discrepancy between the
decay width as obtained from one-loop \chpt{} and the measured value is due to large final-state rescattering effects.
The advantage of the dispersive approach over perturbation theory thus comes from the fact that it can complement the
one-loop approximation by an arbitrary number of final-state rescattering processes.

Of course, it is still desirable to reduce the influence of \chpt{} on our result to a minimum. If the subtraction
constants are fixed from a pure matching to the one-loop result, even its high energy contributions enter to some
extent. On the other hand, if we use experimental data in addition, we must rely on the perturbative expression only at
very low energy, where it is expected to converge much quicker than in the physical region. Both approaches are
discussed in detail in the following sections.

\subsection{Matching to 1-loop \chpt} \label{sec:solMatch}

Two of the subtraction constants, $\gamma_0$ and $\beta_1$, can be fixed by sum rules. We have found in
Eqs.~\eqref{eq:oneLoopDispRelComb1} and \eqref{eq:oneLoopDispRelComb2} that two combinations of subtraction constants
are related to integrals over the imaginary parts of the functions $\Mbar_I(s)$ at one loop. Taking derivatives of the
dispersion relations for the one-loop approximation in Eq.~\eqref{eq:OneLoopDispRel}, we find for the same combinations
\begin{align}
	c_0 + \frac{4}{3} c_2 &= \inv{2} \Mbar_0''(0) + \frac{2}{3} \Mbar_2''(0) 
			= \inv{\pi} \int\limits_{4 \mpi^2}^\infty ds'\; \frac{\Im\Mbar_0(s') + \frac{4}{3}
\Im\Mbar_2(s')}{s'^3}\eolc \label{eq:solMatchCondChPT1} \\
	b_1 + c_2 &= \Mbar_1'(0) + \inv{2} \Mbar_2''(0) \nonumber\\[1mm]
			&= -\frac{4 L_3-\inv{64 \pi^2}}{\Fpi^2 (\meta^2 - \mpi^2)}
				+ \inv{\pi} \int\limits_{4 \mpi^2}^\infty ds'\; \frac{s'\, \Im\Mbar_1(s') + \Im\Mbar_2(s')}{s'^3}\eolp
	\label{eq:solMatchCondChPT2}
\end{align}
The same derivatives can also be calculated from the dispersive representation~\eqref{eq:dispRelFull}, and we find
\begin{align}
	\inv{2} M_0''(0) + \frac{2}{3} M_2''(0) =\; &\gamma_0
			+ \inv{\pi} \int\limits_{4 \mpi^2}^\infty \frac{ds'}{s'^3} 
				\left\{ \frac{\sin \delta_0 (s') \hat M_0(s')}{|\Omega_0(s')|} 
			+ \frac{4}{3} \frac{\sin \delta_2 (s') \hat M_2(s')}{|\Omega_2(s')|} \right\} \nonumber\\[1mm]
			&+ \inv{2} \alpha_0 \, \Omega_0''(0) + \beta_0 \, \Omega_0'(0) \eolc
			\label{eq:solMatchCondDisp1} \displaybreak[1] \\[2mm]
	M_1'(0) + \inv{2} M_2''(0) =\; &\beta_1
			+ \inv{\pi} \int\limits_{4 \mpi^2}^\infty \frac{ds'}{s'^3} 
				\left\{ \frac{s' \sin \delta_1 (s') \hat M_1(s')}{|\Omega_1(s')|} 
			+ \frac{\sin \delta_2 (s') \hat M_2(s')}{|\Omega_2(s')|} \right\}
			\label{eq:solMatchCondDisp2}
\end{align}
These relations are independent of the transformation in Eq.~\eqref{eq:etaDispMIShift}, since the
shifts in $M_0(s)$ and $M_2(s)$ are linear in $s$, while the one in $M_1(s)$ is a constant, such that they vanish in
the derivatives. Matching the corresponding expressions,
\begin{equation}\begin{split}
	\inv{2} M_0''(0) + \frac{2}{3} M_2''(0) &\simeq \inv{2} \Mbar_0''(0) + \frac{2}{3} \Mbar_2''(0) \eolc \\[3mm]
	M_1'(0) + \inv{2} M_2''(0) &\simeq \Mbar_1'(0) + \inv{2} \Mbar_2''(0) \eolc
\end{split}\end{equation}
thus yields conditions for the subtraction constants $\gamma_0$ and $\beta_1$ that are independent of this convention.
Each side of the matching conditions consists of a subtraction term and an integral over the discontinuities of the
amplitude. The integrals account for the contributions from low lying states, while all other singularities, including
those at infinity, are contained in the subtraction terms. The difference between the two representations comes from
the integrals. While the integrals in Eqs.~\eqref{eq:solMatchCondDisp1} and \eqref{eq:solMatchCondDisp2} only comprise
the $\pi \pi$ discontinuities, the one-loop approximations also takes $K \Kbar$, $\eta \pi$ and $\eta \eta$
singularities into account. On the other hand, the same integrals are not restricted to $\O(p^4)$ but rather include
terms of arbitrary chiral order. 

The integrals must be compared at $\O(p^4)$, since other terms are beyond the accuracy of the one-loop approximation.
To do this, we expand the integrals on both sides of the matching conditions to this order. Taking only the $\pi \pi$
discontinuities into account---the others will be discussed below---the imaginary part of the $\Mbar_I(s)$ is
given by the unitarity condition in Eq.~\eqref{eq:etaDispUnitarityOneLoop}, where only tree-level expressions enter on
the right-hand side.

In order to go on, we must discuss in detail the chiral order of all the parts of the dispersion relations. We already
know from Sec.~\ref{sec:etaDispUnitarityOneLoop} that the phase shifts, the $\hat{M}_I(s)$, and $M_0(s)$ are 
$\O(p^2)$, while $M_1(s)$ and $M_2(s)$ only start at $\O(p^4)$. Clearly, due to the exponential, the leading
contribution to the Omn\`es function is $\O(1)$. From comparison of the dispersive representation for $M_0(s)$ with
the tree-level expression in Eq.~\eqref{eq:etaDispMtree} follows that $M_I\tree(s)$ is entirely determined by the two
subtraction constants $\alpha_0$ and $\beta_0$, which are hence $\O(p^2)$. Accordingly, the remaining subtraction
constants, $\gamma_0$ and $\beta_1$, obtain their leading contribution at $\O(p^4)$.

We use the unitarity condition in the form of Eq.~\eqref{eq:etaDispUnitarityOneLoop} together with $M_0\tree(s) =
\alpha_0 + \beta_0 s$ to replace the imaginary parts of the amplitudes in Eq.~\eqref{eq:solMatchCondChPT1}. The
right-hand side of this equation thus becomes
\begin{multline}
	\alpha_0 \inv{\pi} \int\limits_{4 \mpi^2}^\infty \frac{ds'}{s'^3} \, \delta_0\tree(s')
	+ \beta_0 \inv{\pi} \int\limits_{4 \mpi^2}^\infty \frac{ds'}{s'^2} \, \delta_0\tree(s') \\[1mm]
	+ \inv{\pi} \int\limits_{4 \mpi^2}^\infty \frac{ds'}{s'^3}
		\left( \delta_0\tree(s') \hat{M}_0\tree(s') + \frac{4}{3} \delta_2\tree(s') \hat{M}_2\tree(s') \right) \eolp
	\label{eq:solMatchChPTIntermediate1}
\end{multline}
The terms multiplying the subtraction constants can be identified with the derivatives of the Omn\`es function at
$\O(p^4)$:
\begin{align}
	\Omega_I'(0)  &= \inv{\pi} \int\limits_{4 \mpi^2}^\infty \frac{ds'}{s'^2} \, \delta_I\tree(s') + \O(p^4) \\[1mm]
	\Omega_I''(0) &= \frac{2}{\pi} \int\limits_{4 \mpi^2}^\infty \frac{ds'}{s'^3} \, \delta_I\tree(s') + 
						\left( \inv{\pi} \int\limits_{4 \mpi^2}^\infty \frac{ds'}{s'^2} \, \delta_I\tree(s') \right)^{\!\!2}
						 + \O(p^4)
\end{align}
The second term in $\Omega_I''(0)$ is also $\O(p^4)$ and drops out together with all the other terms of that order
because the subtraction constants are $\O(p^2)$. Equation~\eqref{eq:solMatchChPTIntermediate1} then becomes
\begin{equation}
	\inv{2} \alpha_0 \, \Omega_0\tree{}''(0) + \beta_0 \, \Omega_0\tree{}'(0) + 
		\inv{\pi} \int\limits_{4 \mpi^2}^\infty \Big( \cdots \Big) \eolc
\end{equation}
where the integrand has remained the same. The corresponding integral in Eq.~\eqref{eq:solMatchCondDisp1} is easily
expanded by means of
\begin{equation}
	\frac{\sin \delta_I (s) \hat M_I(s)}{|\Omega_I(s)|} = \delta_I\tree(s) \hat M_I\tree(s) + \O(p^6) \eolc
\end{equation}
and it is then clear that the two integrals are the same up to corrections of $\O(p^6)$. Since the one-loop
approximation does not contain a subtraction term, the matching condition implies that $\gamma_0 = 0$.

Similarly, the integral in Eq.~\eqref{eq:solMatchCondChPT2} can be rewritten as
\begin{equation}
	\inv{\pi} \int\limits_{4 \mpi^2}^\infty \frac{ds'}{s'^3} \, 
			\left( s' \delta_1\tree(s') \hat M_1\tree(s') + \delta_2\tree(s') \hat M_2\tree(s') \right) \eolp
\end{equation}
The only subtraction constant contained in $M_1(s)$ and $M_2(s)$, $\beta_1$, is $\O(p^4)$ and does therefore not enter.
The integral in Eq.~\eqref{eq:solMatchCondDisp2} is again easily expanded and found to be identical. This time, also
the one-loop expression contains a subtraction term, such that the matching condition yields
\begin{equation}
	\beta_1 = -\frac{4 L_3 - \inv{64\pi^2}}{\Fpi^2 (\meta^2-\mpi^2)} \approx 4.6~\GeV^{-4} \eolp
\end{equation}

These results have been obtained neglecting all the discontinuities but those from two-pion intermediate states. 
In the following we estimate the contributions from the other singularities to the integrals in
Eqs.~\eqref{eq:solMatchCondChPT1} and~\eqref{eq:solMatchCondChPT2}. The first one to set in is the cut due to $\eta
\pi$ states for $s \geq (\meta+\mpi)^2$. However, from Eq.~\eqref{eq:eta3piDeltai} we see that the corresponding loop
integrals are multiplied by $\mpi^2$ and hence lead to a tiny contribution. Indeed, taking only these intermediate
states into account, the integral over the interval \mbox{$(\meta+\mpi)^2 \leq$} $s \leq 4 \mK^2$ amounts to less
than $0.01~\GeV^{-4}$. The only relevant effects from inelastic channels come from $s > 4 \mK^2$. The entire
contribution of this interval to the integrals amounts to a shift of $+5.3~\GeV^{-4}$ in $\gamma_0$ and 
$+1.7~\GeV^{-4}$ in $\beta_1$. We do not add these corrections to the subtraction constants, but rather use them as
an estimate for the uncertainty due to the inelastic channels:
\begin{equation}
	\gamma_0 \approx (0.0\pm5.3)~\GeV^{-4} \eolc \qquad
	\beta_1 \approx (4.6\pm1.7)~\GeV^{-4} \eolp
	\label{eq:solGamma0Beta1}
\end{equation}
In every iteration step, $\gamma_0$ and $\beta_1$ are fixed to these values.

In order to determine the remaining two subtraction constants, $\alpha_0$ and $\beta_0$, we make use of the soft-pion
theorem for the $\etapi$ amplitude. We have found in Sec.~\ref{sec:eta3piAdler} that the decay amplitude has two Adler
zeros that are protected by chiral $SU(2) \times SU(2)$ symmetry, such that the amplitude is not subject to large
corrections of $\O(m_s)$ at these points. Along the line $s = u$, at leading order in the low-energy expansion, the
Adler zero appears at $s = 4/3\, \mpi^2$. One-loop corrections shift it away from the real $s$-axis, but
they are of order $\mpi^2$ and thus strongly suppressed. Since we still want to have the Adler zero on the line $s=u$,
we choose to use the zero of the real part on this line, the deviation from the actual Adler zero being $\O(\mpi^2)$. 
At one-loop order, this point lies at $s = 1.35\, \mpi^2$, very close to the tree-level Adler zero.

Because we need to fix two complex subtraction constants, the position of the Adler zero alone is not sufficient and
also the slope of the amplitude must be used. The slope is, however, not protected by a symmetry.

In every iteration step, $\alpha_0$ and $\beta_0$ are fixed such that the amplitude agrees with the one-loop
approximation at the Adler zero up to linear order in an expansion in $s$ around $s_A$ along the line $s = u$. For the
one-loop result we find
\begin{equation}
	\Mbar(s,3 s_0-2s,s) = M_A + (s-s_A) S_A + \O((s-s_A)^2) \eolc
\end{equation}
with
\begin{equation}
	s_A = 1.35\, \mpi^2 \eolc \hspace{0.7em} M_A = -i\,0.850\ee{-3} \eolc \hspace{0.7em} S_A = (0.200 - i\,0.0532)\,
\mpi^{-2} \eolp
\end{equation}
For comparison, we also give the corresponding tree-level results:
\begin{equation}
	s_A\tree = \frac{4}{3}\, \mpi^2 \eolc \quad M_A\tree = 0 \eolc \quad 
	S_A\tree = \frac{3}{\meta^2-\mpi^2} = 0.194\, \mpi^{-2} \eolp
\end{equation}
Clearly, all the values obtain only a small contribution from one-loop corrections in agreement with the low-energy
theorem. 

Denoting by $\tilde M(s,t,u)$ the amplitude with vanishing subtraction constants, that is
\begin{equation}\begin{split}
	M(s,t,u) =\;\, &\Omega_0(s) (\alpha_0 + \beta_0 s + \gamma_0 s^2) + (s-u)\,t\, \Omega_1(t) \beta_1\\[1mm]
					&+ (s-t)\,u\, \Omega_1(u) \beta_1 + \tilde M(s,t,u) \eolc
\end{split}\end{equation}
we find two equations for the subtraction constants $\alpha_0$ and $\beta_0$:
\begin{equation}\begin{split}
	M_A = \;\, &\Omega_0(s_A) \alpha_0 + \Omega_0(s_A) s_A\, \beta_0 + \Omega_0(s_A) s_A^2\, \gamma_0\\[1mm]
					&+ 3(s_A-s_0)\,s_A\,\Omega_1(s_A) \beta_1 + \tilde M(s_A,3 s_0 - 2s_A,s_A) \eolc\\[3mm]
	S_A = \;\, &\Omega_0'(s_A) \alpha_0 + (\Omega_0(s_A) + \Omega_0'(s_A) s_A) \beta_0\\[1mm]
					&+  (2 \Omega_0(s_A) s_A + \Omega_0'(s_A) s_A^2) \gamma_0\\[1mm]
					&+ (3(2 s_A - s_0) \Omega_1(s_A) + 3(s_A - s_0) s_A \Omega_1'(s_A)) \beta_1\\
					&+ \frac{\partial}{\partial s} \tilde M(s,3 s_0 - 2s,s) \Big|_{s=s_A} \eolp
	\label{eq:solAlpha0Beta0}
\end{split}\end{equation}
Once we have evaluated the dispersion integrals, we know $\tilde M(s,t,u)$. The subtraction constants $\gamma_0$ and
$\beta_1$ are then fixed to their values given in Eq.~\eqref{eq:solGamma0Beta1} and, to conclude the iteration step,
$\alpha_0$ and $\beta_0$ computed from Eq.~\eqref{eq:solAlpha0Beta0}. Since they change in every step, these two
subtraction constants can also serve as indicator for the accuracy of the result. We end the iteration procedure when
the relative change of both, their real and imaginary part, is less than~$10^{-3}$.

\subsection{Combined fit to KLOE data and one-loop \chpt} \label{sec:solFit}

Instead of using chiral perturbation theory to fix the subtraction constants, we can also rely on experimental data in
order to do the same. We stress, however, that the experimental information is not sufficient. The data describes the
function $|A(s,t,u)|^2$, which differs from the dispersive  expression for $|M(s,t,u)|^2$ by the normalisation factor
introduced in Eq.~\eqref{eq:eta3piNorm}. Since this factor depends on the unknown quantity $Q$, which we want to
extract in the end, the overall normalisation of the dispersive representation cannot be fixed from experimental data.
Nevertheless, owing to the data, the influence of \chpt{} can be reduced to a minimum.

The experimental input comes from the KLOE collaboration. Their measurement of the Dalitz distribution has been
discussed in Sec.~\ref{sec:eta3piResults}. The result is given in terms of a Dalitz plot parametrisation, which
describes the squared amplitude everywhere in the physical region. Unfortunately, the binned data set, from which the
Dalitz plot parametrisation has been computed and which would form the ideal basis for our fit, is not available.
However, Andrzej Kup\'s\'c has provided us with a binned data set that he simulated based on the Dalitz plot
parametrisation by KLOE.

The subtraction constants are obtained by fitting the dispersive parametrisation at the same time to the simulated data
set and the one-loop result from \chpt. In order to reduce the uncertainties coming from the low-energy expansion of the
amplitude, we again make use of the one-loop result in an interval around the Adler zero.

Surface fitting is a statistical standard method%
\footnote{As an introduction to various statistical techniques, the book by Lions has proved very useful
\cite{Lyons1989}.}.
The general idea of a fit is the following. Consider a function of a
set of variables $\vec g$ that in addition also depends on a set of $n$ parameters $\vec \theta$ that determine the
shape of the function $F(\vec g;\vec \theta\,)$. The parameter values $\vec \theta$ shall now be determined such, that
the resulting function gives the best possible approximation to some given set of $N$ data points $f(\vec g_i)$. Roughly
speaking, we require
\begin{equation}
	F(\vec g = \vec g_i;\vec \theta = \hat{\vec \theta}\,) \approx f(\vec g_i) \eolp
\end{equation}
Of course, this equation is not a usable formulation of the problem. What we need is a clear definition for the meaning
of the approximate equal sign or, in other words, we must say what we mean by the best approximation. This amounts to
the choice of a certain fit method. 

We will use the method of least squares, where one finds the parameter values that minimise
the function
\begin{equation}
	S(\vec \theta\,) = \sum_{i=1}^{N} \frac{(F(\vec g_i;\vec \theta\,) - f(\vec g_i))^2}{\sigma_i^2} \eolc
	\label{eq:solChi2Def}
\end{equation}
where $\sigma_i$ is the statistical uncertainty on the data point $f(\vec g_i)$. The parameters $\hat{\vec \theta}$ that
minimise this quantity are found from the set of equations
\begin{equation}
	\frac{\partial S}{\partial \theta_i} = 0 \eolp
\end{equation}
The function $S(\vec \theta\,)$ is $\chi^2$-distributed with $N-n$ degrees of freedom%
\footnote{The $\chi^2$-distribution is briefly discussed in Appendix~\ref{chp:appendixChi2}.} and accordingly often
simply called the chi square. The minimal value $S_\text{min} = S(\hat{\vec \theta}\,)$ is a measure for the quality
of the approximation. Often it is normalised by the number of degrees of freedom and denoted as
\begin{equation}
	\frac{\chi^2}{\textit{d.o.f.}} = \frac{S_\text{min}}{N-n} \eolp
\end{equation}
A value smaller than 1 for this quantity hints that the model is overfitting the data, while a value larger than 1 means
that the parametrisation is not flawless in describing the data. The ideal result is $\chi^2/\textit{d.o.f.} = 1$, where
the model curve lies within the error bars on the data. Another way to measure the accuracy of a fit is the $p$-value,
which gives the probability that, given the model function, a value for $S_\text{min}$ of that size is obtained in an
experiment. It is accordingly calculated as
\begin{equation}
	p = \int\limits_{S_\text{min}}^\infty f_{\chi^2}(N-n,t)\, dt = 1 - F_{\chi^2}(N-n,S_\text{min}) \eolc
\end{equation}
where $f_{\chi^2}(N-n,x)$ and $F_{\chi^2}(N-n,x)$ are the density and the distribution function, respectively, for the
$\chi^2$-distribution with $N-n$ degrees of freedom. The covariance matrix is estimated by the inverse of the Hessian
matrix:
\begin{equation}
	\text{Cov}(\hat \theta_i,\hat \theta_j) = (H^{-1})_{ij} \eolc \qquad
	\text{with} \quad H_{ij} = \inv{2} \frac{\partial^2 S}{\partial \theta_i \, \partial \theta_j} \eolp
\end{equation}

Before we turn our attention to the fit of the subtraction constants, let us first briefly discuss the fit of the
Dalitz plot parametrisation to the simulated data set. This will be important below, because we need to normalise the
data set to one at the centre of the Dalitz plot. Since the bin centred at $X = Y = 0$ is not part of the data set, we
determine the normalisation factor from this fit. The data set contains the number of events in each bin, and we must
therefore also integrate the Dalitz plot parametrisation over the entire bin in the definition of the $\chi^2$. The
integral over the bin centred at $(X,Y)$ with dimensions $\Delta X$ and $\Delta Y$ is given by
\begin{multline}
	\int\limits_\text{bin} dX' dY'\, \Gamma(X',Y') = \N \Delta X \Delta Y \Bigg( 
			1 +  a Y + b \bigg( Y^2 + \frac{\Delta Y^2}{12} \bigg) \\[2mm]
			 + d \bigg( X^2 + \frac{\Delta X^2}{12} \bigg) + f \bigg( Y^3 + \frac{\Delta Y^2 Y}{4} \bigg) \Bigg) \eolc
	\label{eq:solDalitzIntBin}
\end{multline}
where we have only included the parameters $a$, $b$, $d$, and $f$, which are nonzero in the KLOE result. The data set
contains 154 square bins with $\Delta X = \Delta Y = 0.125$. We find for the normalisation and the Dalitz plot
parameters
\begin{align}
	\N \Delta X \Delta Y &= 7404 \pm 13 \eolc &a &= -1.086 \pm 0.005 \eolc &b &= 0.127 \pm 0.005 \eolc
\nonumber \\[2mm]
	d &= 0.056 \pm 0.005 \eolc  &f &= 0.117 \pm 0.01 \eolp
	\label{eq:solDalitzSimData}
\end{align}
The fit routine has problems with varying very large numbers and the fit works hence much better, if we use 
$\N \Delta X \Delta Y$ as free parameter rather than the normalisation $\N$ itself.

We are now ready to discuss the fit of the subtraction constants. We have a parametrisation of the $\etapi$ decay
amplitude in terms of the complex numbers $\alpha_0$, $\beta_0$, $\gamma_0$, $\beta_1$, resulting in a
total number of eight parameters that we want to fit:
\begin{equation}
	\vec \theta = ( \Re \alpha_0,\, \Im \alpha_0,\, \Re \beta_0,\, \Im \beta_0,\,
						\Re \gamma_0,\, \Im \gamma_0,\, \Re \beta_1,\, \Im \beta_1 ) \eolp
	\label{eq:solThetaSubConst}
\end{equation}
The amplitude depends on two independent kinematic variables and we can write it for example as
\begin{equation}
	M(s,t,u;\vec \theta\,) \eolc \quad \text{or} \quad
	 M(X,Y;\vec \theta\,) \eolc
\end{equation}
where, as usual, one Mandelstam variable is understood to depend on the other two. Since the experimental data on the
Dalitz plot distribution cannot fix the overall normalisation of the amplitude, we must also fit to the one-loop
amplitude from \chpt{} and the definition of $S(\vec \theta\,)$ therefore contains several sums of the type in
Eq.~\eqref{eq:solChi2Def}. We require the model function to approximate the one-loop amplitude in the
vicinity of the Adler zero along the line $s=u$. This is achieved by choosing $\bar N = 5$ points
with $0 \leq s_i \leq 2 \mpi^2$, $t_i = 3 s_0 - 2 s_i$, and $u_i = s_i$. The square differences of the real and
imaginary part are summed separately and each term is weighed with $(\Re \bar \sigma_i)^{-2}$ or $(\Im \bar
\sigma_i)^{-2}$, respectively, with
\begin{equation}
	\bar \sigma_i = \Mbar(s_i,3 s_0 - 2 s_i,s_i) - M\tree(s_i,3 s_0 - 2 s_i,s_i) \eolp
\end{equation}
In the physical region, we fit to the simulated data set based on the KLOE experiment using $N = 154$ points at
coordinates $(X_i,Y_i)$. The number of events in the bin centred at $(X,Y)$ we denote by $N_\text{bin}(X,Y)$. The data
must be properly normalised by the number of events that are contained in the bin centred at $X=Y=0$ which, according to
Eq.~\eqref{eq:solDalitzIntBin}, is given by
\begin{equation}\begin{split}
	N_\text{bin}(0,0) &= \int\limits_\text{bin} dX' dY'\, \Gamma(X',Y') \bigg|_{X=Y=0} \\[3mm]
											&= \N \Delta X \Delta Y \left( 1 + \frac{b \Delta Y^2 + d \Delta X^2}{12} \right)\eolp
\end{split}\end{equation}
Using the numerical values given in Eq.~\eqref{eq:solDalitzSimData}, this evaluates to
$N_\text{bin}(0,0) = 7406 \pm 13$. The contribution of $b$ and $d$ to the uncertainty is negligible.
The statistical uncertainties on the number of events in the bin at $(X_i,Y_i)$, denoted by $\sigma_i$, are part of the
simulated data set as well. They must be normalised by the same factor.

Finally, in order to compare with the binned data set, we must also integrate the square of the dispersive amplitude
over the bin. Accordingly, we define the function
\begin{equation}
	M_\text{bin}(X,Y;\vec \theta\,) \equiv \int\limits_\text{bin} dX' dY'\, |M(X',Y';\vec \theta\,)|^2
\end{equation}
where the integral is taken over the bin centred at $(X,Y)$.

Taking all of this into account, we define the $\chi^2$ as
\begin{equation}\begin{split}
	S(\vec \theta\,) = \; 
		&W_A \sum_{i=1}^{\bar N} \frac{ \big( \Re M(s_i,t_i,u_i;\vec \theta\,)
				- \Re \Mbar(s_i,t_i,u_i) \big)^2}{\big(\Re \bar \sigma_i \big)^2}
	\\[2mm] &+W_A
		\sum_{i=1}^{\bar N} \frac{ \big( \Im M(s_i,t_i,u_i;\vec \theta\,)
				- \Im \Mbar(s_i,t_i,u_i) \big)^2}{\big(\Im \bar \sigma_i \big)^2}
	\\[2mm] &+
		\sum_{i=1}^{N} \frac{ \big( M_\text{bin} (X_i,Y_i;\vec \theta\,)/M_\text{bin}(0,0;\vec \theta\,)
				- N_\text{bin}(X_i,Y_i)/N_\text{bin}(0,0) \big)^2}
				{\big( \sigma_i/N_\text{bin}(0,0) \big)^2} \eolp
\end{split}\end{equation}
This quantity is $\chi^2$ distributed with $2 \bar N + N - 8 = 156$ degrees of freedom.
Note that the first two sums are multiplied with a factor $W_A$. By choosing its value properly, we can ensure that the
constraints from the
low-energy region and from the physical region have both enough weight in the fit. A too small value for $W_A$ indeed
leads to an amplitude that does not approximate the one-loop amplitude at low energies very well. The reason is that
the value of $S(\vec \theta\,)$ can then be reduced by a tiny improvement in the physical region at the cost of a large
deviation around the Adler zero. Since we use $N = 154$ points in the physical region and $2 \bar N = 10$ around the
Adler zero, $W_A = 15$ seems a natural choice. We have tried other values for $W_A$ and the results support this claim:
for $W_A \lesssim 5$ the one-loop amplitude is not satisfactorily approximated. On the other hand, we do not want to
overemphasize the one-loop result and thus prefer a small value. The uncertainty coming from the ambiguity in the
choice of $W_A$ will, of course, be accounted for in the error analysis.

For the minimisation of $S(\vec \theta\,)$ we use numerical standard routines that return central values
and covariances for the subtraction constants. 

%% file: main/numerics.tex
\chapter{Implementation of the numerical solution} \label{chp:Numerics}

This chapter contains some remarks on the implementation of the numerical solution of the dispersion relations. The
procedure has already been laid out in great detail in Ref.~\cite{Walker1998}. We follow the same approach, but give
some additional details where they are required.

Since we deal with functions that are not known as mathematical expressions, but only numerically, we start the
chapter by discussing how these are represented. This is followed by the description of the evaluation of the
Omn\`es function. The main computational problem is, of course, the computation of the integrals contained in the
definitions of $\hat M_I$ and $M_I$, which is discussed extensively.

Quite often, we are faced with numerical standard problems, such as the evaluation of a numerical integral or the
inversion of a matrix. In these situations, we make use of the numerical libraries from the Numerical Algorithms Group
(NAG) that offer many routines to solve these problems reliably and efficiently. Appendix \ref{chp:appendixFortran}
contains a list of all the routines that have been used and is also intended to show, which tasks have been delegated
to external programs.

\section{Representation of functions}

When we evaluate an integral over $\hat M_I(s)$ or $M_I(s)$, we cannot obtain an exact value for the integrand at
each point on the integration path, since these functions are recursively defined as integrals over each other. In
order to calculate an exact value, we would have to go back through all the iteration steps performed so far which,
even though not impossible in principle, would require huge amounts of computer time and is not practicable. Instead,
we only calculate the functions at a number of points and then use some interpolation procedure to obtain values also
in between. We choose to represent the functions in terms of cubic splines. The method of spline interpolation yields
curves that link the interpolation points smoothly by piecewise polynomials of third order, such that the function and
its first two derivatives are continuous at the interpolation points. Once the spline interpolant is found, the
evaluation of function values is very efficient. Even more importantly, the method leads to an accurate approximation
also close to the edge of the interval covered by the interpolation points. This is an important requirement for our
purposes, as we will see in a moment.

First of all, we have to work out for what values of their argument $\hat M_I(s)$ and $M_I(s')$ must be available. The
functions $\hat M_I(s)$ are required, in principle, from $4 \mpi^2$ up to infinity. Of course, we have to cut off the
integration at some scale $\Lambda^2$ and thus need interpolation points only in the interval $[4 \mpi^2, \Lambda^2]$.
For the $M_I(s')$ the situation is more involved, since they are contained in the integrands of the angular
averages. As we have described in detail in Sec.~\ref{sec:etaDispElasticUnitarity}, the integration follows a rather
complicated path that also depends on the point $s$, where $\hat M_I(s)$ is to be evaluated. For every $s$ that
should serve as an interpolation point for $\hat M_I(s)$, we must ensure that $M_I(s')$ can be evaluated along the
entire integration path. Let us first identify the part of the real axis, where $M_I(s')$ is needed. For 
$s \geq (\meta+ \mpi)^2$, the end points $s_\pm$ of the integration path are real and negative with $s_+ \leq s_-$. The
start point $s_+$ moves in negative direction with growing $s$ and thus reaches its smallest value at $s = \Lambda^2$,
namely
\begin{equation}
	s_+(s=\Lambda^2) = \inv{2} \left(3 s_0 - \Lambda^2 - \kappa(\Lambda^2) \right) \gtrsim -\Lambda^2 \eolp
\end{equation}
For $s \leq (\meta - \mpi)^2$, $s_\pm$ pick up an imaginary part. Their real part stays below $(\meta-\mpi)^2$, 
which implies that is enough to know $M_I(s')$ on the real axis for $-\Lambda^2 \leq s \leq (\meta-\mpi)^2$. It is,
however, useful to calculate it up to $32 \mpi^2$, since this allows for certain checks, as we will see later. Note
that we must calculate $M_I(s')$ on the upper and the lower rim of the cut separately.

Away from the real axis, we need information on $M_I(s')$ in the rectangular region $[-4\mpi^2,5\mpi^2] \times
[-4 \mpi^2,4 \mpi^2]$. We could use a grid of points to interpolate the two dimensional function, but as we need it
along the same paths in every iteration steps, we simply calculate a number of points on each path.  

By definition, a spline-curve describes a smooth function. Since $M_I(s')$ and $\hat M_I(s)$ are expected to have cusps,
a single spline-curve cannot adequately approximate them. Instead, me must use independent spline representations
on each side of such a point. The resulting spline has then continuous first and second derivative everywhere but at
these break points. Since they also appear on the integration path, we must be able to evaluate the
interpolant arbitrarily close to the break points, for which the splines are ideally suited. The $M_I(s')$ are
smooth up to a cusp at $4 \mpi^2$ such that we split the interpolation points into two regions. Instead of the $\hat
M_I(s)$, we will interpolate the functions $\tilde M_I(s)$ that are defined below. They have cusps at $(\meta\pm\mpi)^2$
and accordingly require three regions of interpolation points.

The distribution of the interpolation points must be tuned rather carefully. A large number of points leads to a more
accurate approximation, but the calculation consumes more computer time and we must hence achieve a compromise between
precision and efficiency. In regions where the functions change slowly, only few points are needed. But otherwise, for
example in the vicinity of cusps, the density must be increased considerably to ensure an accurate interpolation. While
in some regions, the interpolation points are $2 \mpi^2$ apart, we use as much as 400 points per $\mpi^2$ in the
vicinity of the cusps.

\section{Computation of the phase shifts and the Omn\`es functions}

The $\pi \pi$ phase shifts enter the equations through the Omn\`es functions but also directly in the dispersion
integrals. From the representation that we have discussed in Sec.~\ref{sec:pipiPhaseShifts}, they can be calculated
efficiently for arbitrary values of $s$, such that there is no need to use interpolation.

Once the phase shifts are available for any $s$, the Omn\`es functions can be obtained by calculating the corresponding
integrals. The numerical computation of an integral takes much more time than the evaluation of a spline interpolant
and thus one would expect that the use of interpolation could indeed save some computer time. However, as we argue now,
this is not the case. The Omn\`es functions appear in two places: as prefactors multiplying the subtraction polynomial
and the dispersion integral in the definition of the $M_I(s)$, but also as a part of the integrand inside the dispersion
integrals. Since we only calculate $M_I(s)$ at a limited number of interpolation points anyway, no significant
reduction of computing time comes from interpolating the Omn\`es functions in the prefactors. The integrand of the
dispersion integrals---and thus the Omn\`es function appearing there---is, however, evaluated much more often. But
instead of approximating the Omn\`es function by itself, the integrand is interpolated as a whole. This means that there
is no need to implement a dedicated interpolation routine for the Omn\`es functions.

The integral in Eq.~\eqref{eq:etaDispOmnes} can be solved using standard routines for numerical integration. If
$s \notin [ 4 \mpi^2, \infty )$, the integral is easy to evaluate. Otherwise, the integrand has a pole that must be
treated explicitly. By use of the Sokhotski-Plemelij formula \eqref{eq:dispS-P}, the integral can be split up as
\begin{equation}
	\Omega_I(s) = \exp \left\{ \frac{s}{\pi} \hspace{0.6em} \mathcal{P}\hspace{-1.35em} \int\limits_{4 \mpi^2}^{\infty}
ds'						\frac{\delta_I(s')}{s' (s' - s)} + i \delta_I(s) \right\} \eolc
\end{equation}
which is then solved with a numerical routine that calculates the Cauchy principal value. Since also the integrand in
the dispersion relation is inversely proportional to $(s' - s \mp i \epsilon)$, we will have to solve many integrals of
this type and the procedure will always be the one outlined here.

\section{Computation of the \texorpdfstring{$\hat M_I(s)$}{Mhat\_I(s)}}

From a given representation for the $M_I(s)$, we can calculate the $\hat M_I(s)$ by means of Eq.~\eqref{eq:etaDispMhat}.
In order to simplify the numerical treatment of the angular averages, it is convenient to introduce the variable
transformation
\begin{equation}
	z = \frac{2}{\kappa(s)} \left( s' - \sigma \right) \eolc \quad \text{with} \quad 
	\sigma = \frac{3}{2} s_0 - \inv{2} s
\end{equation}
in the angular averages. This leads to
\begin{equation}
	\aav{z^n M_I} = \frac{2^n}{\kappa(s)^{n+1}} \int_\gamma ds'\, (s'-\sigma)^n M_I(s') \eolp
\end{equation}
The integration path $\gamma$ has been discussed in detail in Sec.~\ref{sec:etaDispElasticUnitarity}. In terms of the
new integration variable, the $\hat M_I(s)$ are then found to be
\begin{align}
	\hat M_0(s) &= \inv{\kappa(s)} \int_\gamma ds'\, \bigg\{
				\begin{aligned}[t]
					 & \frac{2}{3} M_0(s') + 2 (s-s_0)M_1(s') \\
					&+ \frac{4}{3} (s'-\sigma) M_1(s') + \frac{20}{9} M_2(s') \bigg\} \eolc
				\end{aligned} \nonumber \displaybreak[0] \\[2mm]
	\hat M_1(s) &= \inv{\kappa(s)^3} \int_\gamma ds'\, \Big\{
				\begin{aligned}[t]
					 & 6(s'-\sigma) M_0(s') + 9(s-s_0)(s'-\sigma) M_1(s') \\[1mm]
					& + 6 (s'-\sigma)^2 M_1(s') - 10 (s'-\sigma) M_2(s') \Big\} \eolc
				\end{aligned} \nonumber \\[2mm]
	\hat M_2(s) &= \inv{\kappa(s)} \int_\gamma ds'\, \bigg\{
				\begin{aligned}[t]
					 & M_0(s') - \frac{3}{2} (s-s_0)M_1(s') \\
					&-(s'-\sigma) M_1(s') + \inv{3} M_2(s') \bigg\} \eolp
				\end{aligned}
\end{align}
Since the function $\kappa(s)$ has zeros that could lead to problems in the integrals over the $\hat M_I(s)$, we
define the functions $\tilde M_I(s)$ by
\begin{equation}
	\hat M_0(s) \equiv \frac{\tilde M_0(s)}{\kappa(s)} \eolc \quad 
	\hat M_1(s) \equiv \frac{\tilde M_1(s)}{\kappa(s)^3} \eolc \quad 
	\hat M_2(s) \equiv \frac{\tilde M_2(s)}{\kappa(s)} \eolp
\end{equation}
They are finite everywhere on the real axis, but we expect them to have cusps at the zeros of $\kappa(s)$. The only
difficulty involved in the calculation of the $\tilde M_I(s)$ is the complicated shape of the integration path. The
numerical integration itself poses no problems and can be solved by standard integration routines.

\section{Computation of the \texorpdfstring{$M_I(s)$}{M\_I(s)}}

The functions $M_I(s)$ are fully determined by a subtraction polynomial and a dispersion integral according to
Eq.~\eqref{eq:dispRelFull}. The dispersion integrals are of the form
\begin{equation}
	\int\limits_{4 \mpi^2}^\infty ds' \frac{F_I(s')}{\kappa(s')^{n_I} (s'-s \mp i \epsilon)} \eolc
\end{equation}
with
\begin{equation}
	F_I(s) = \frac{\sin \delta_I(s) \tilde M_I(s)}{|\Omega_I(s)| s^{m_I}} \eolc
\end{equation}
and $n_0 = n_2 = 1$, $n_1 = 3$,  $m_0 = m_2 = 2$, and $m_1 = 1$. We have used the definition of $\tilde M_I(s)$
from the last section, which enabled us to write $\kappa(s)$ explicitly in the integrand. The functions $F_I(s)$ pose
no substantial problems. They have cusps at $(\meta \pm \mpi)^2$ because they contain the $\tilde M_I(s)$. In order to
make sure that this behaviour is taken care of properly, $F_I(s)$ is interpolated separately in the three regions
defined by these break points.

The difficulty in the evaluation of the integral comes from the zeros of $\kappa(s)$, which may lead to poles in the
integrand. They are at
\begin{equation}
	s = 4 \mpi^2 \eolc \qquad s = (\meta - \mpi)^2 \equiv a \eolc \qquad s = (\meta + \mpi)^2 \equiv b \eolp
\end{equation}
The integrand has no pole at $s = 4 \mpi^2$. For $s$ close to $4 \mpi^2$, the end points of the integration path in the
angular averages lie on the upper rim of the cut close to each other. The integrand is then approximately constant over
the path, such that the result of the integration is proportional to the path length:
\begin{equation}
	\tilde M_I(s) = \O \left( | \kappa(s) | \right) \eolp
\end{equation}
From the Schenk parametrisation \eqref{eq:pipiSchenk} follows the behaviour of $\sin \delta_I(s)$ close to threshold:
\begin{equation}
	\sin \delta_I(s) = \O \left( (1 - 4 \mpi^2/s)^{n_I/2} \right) = \O\left( \kappa(s)^{n_I} \right) \eolc
\end{equation}
which cancels the factor $\kappa(s)^{-n_I}$. The Omn\`es function is equal to one at $s = 4 \mpi^2$ and the integrand is
thus $\O( | \kappa(s) | )$ and tends to zero. The two other zeros of $\kappa(s)$, on the other hand, generate
singularities that must be handled with care. To that end, the integration path is divided into four parts as
\begin{equation}
	\int\limits_{4 \mpi^2}^{\infty} = \int\limits_{4 \mpi^2}^{a - g} + \int\limits_{a - g}^{a + g} + 
			\int\limits_{b - h}^{b + h} + \int\limits_{b+h}^\infty \eolp
\end{equation}
The numbers $g$ and $h$ must be chosen such that $a+g = b-h$ and $s$ never lies on the boundary of any of these
intervals, with the exception of $s = 4 \mpi^2$. The first and the last integral are easily evaluated
with standard integration routines that also take care of the pole at $s$. But the other two go over the poles at $a$
and $b$ and need special treatment. The fist step for both of them is to put the part of the integrand that has no
singularities in a function $G_I(s)$. The function $\kappa(s)$ is the product of three terms whereof only one
effectuates a pole. It turns out that the splitting
\begin{equation}
	\kappa(s) = \sqrt{1-4\mpi^2/s}\, \sqrt{a-s}\, \sqrt{b-s}
\end{equation}
reproduces the correct prescription that was discussed in Sec.~\ref{sec:etaDispElasticUnitarity}, if the square
roots are evaluated on the upper rim of the cut along the negative real axis. This form is thus well suited for the
numerical treatment. For convenience, we define the functions
\begin{equation}\begin{split}
	G_I^a(s) = \frac{F_I(s)}{\left(\sqrt{1-4\mpi^2/s}\, \sqrt{b-s}\right)^{n_I}} \eolc \\[1mm]
	G_I^b(s) = \frac{F_I(s)}{\left(\sqrt{1-4\mpi^2/s}\, \sqrt{a-s}\right)^{n_I}} \eolp
\end{split}\end{equation}
The first one is finite in the interval $[a-g,a+g]$, the other one in $[b-h,b+h]$. In terms of these, the two integrals
can be written as
\begin{equation}\
	\int\limits_{a - g}^{a + g} ds' \frac{G^a_I(s')}{\sqrt{a-s'}^{\,n_I}(s'-s \mp i\epsilon)} \eolc \qquad
			\int\limits_{b - h}^{b + h} ds' \frac{G^b_I(s')}{\sqrt{b-s'}^{\,n_I}(s'-s \mp i\epsilon)} \eolp
\end{equation}
They can both be solved in exactly the same way and only the integral over the pole at $a$ is discussed in detail here.
We first consider $I = 0$ and $I = 2$, where $n_I = 1$. The integral is split up in two parts as
\begin{multline}
	\int\limits_{a - g}^{a + g} ds' \frac{G^a_I(s')}{\sqrt{a-s'}\,(s'-s \mp i\epsilon)} = \\[1mm]
	\int\limits_{a - g}^{a + g} ds' \frac{G^a_I(s') - G^a_I(a)}{\sqrt{a-s'}\,(s'-s \mp i\epsilon)} 
	+ G^a_I(a) \int\limits_{a - g}^{a + g} \frac{ds'}{\sqrt{a-s'}\,(s'-s \mp i\epsilon)} \eolc
	\label{eq:numIntSplit0}
\end{multline}
The first integral is finite at $a$ and can be evaluated numerically as it stands. It is, however, difficult to
evaluate the integrand numerically close to the singularity at $s' = a$, and we will discuss later, how this problem is
solved. The second integral has an analytical solution. We change the integration variable to $x = s'-a$ and then
distinguish two cases:
\begin{equation}
	\int\limits_{-g}^g \frac{dx}{\sqrt{-x}\, (x + a - s \mp i\epsilon)} \equiv \left\{
	\begin{array}{lr}
		Q_0^a(s) \eolc \quad & (a-s) \notin [-g,g] \eolc \\[2mm]
		I_0^a(s) \eolc \quad & (a-s) \in [-g,g] \eolp
	\end{array} \right.
\end{equation}
The integral is solved using Eq.~\eqref{eq:appendixIntegral1} in both cases. The pole at $x = 0$ is integrable
and needs no special attention. Due to the relation
\begin{equation}
	\arctan(y) = \frac{i}{2} \log \left( \frac{1 - i y}{1 + i y} \right) \eolc
\end{equation}
the results can be rewritten in terms of real quantities. For $(a-s) \notin [-g,g]$, we find
\begin{equation}
	Q_0^a(s) = \left\{
	\begin{array}{lr}
		-\inv{\sqrt{a-s}} \left\{ 2 i \arctan \sqrt{\frac{g}{a-s}} 
				+ \log \frac{\sqrt{a-s} - \sqrt{g}}{\sqrt{a-s} + \sqrt{g}} \right\} \eolc \quad
		& s < a \eolc \\[5mm]
		-\inv{\sqrt{s-a}} \left\{ 2 \arctan \sqrt{\frac{g}{s-a}} 
				+ i \log \frac{\sqrt{s-a} - \sqrt{g}}{\sqrt{s-a} + \sqrt{g}} \right\} \eolc \quad
		& s > a \eolp
	\end{array} \right.
\end{equation}
In the case of $(a-s) \in [-g,g]$, the argument of the logarithm is a negative number. By means of
\begin{equation}
	\log \left( -y^2 \mp i\epsilon \right) = \log \left( y^2 \right) \mp i \pi \eolc
\end{equation}
we obtain
\begin{equation}
	I_0^a(s) = \left\{
	\begin{array}{lr}
		-\inv{\sqrt{a-s}} \left\{ 2 i \arctan \sqrt{\frac{g}{a-s}} 
				+ \log \frac{\sqrt{g} - \sqrt{a-s}}{\sqrt{g} + \sqrt{a-s}} \mp i \pi \right\} \eolc \
		& s < a \eolc \\[5mm]
		-\inv{\sqrt{s-a}} \left\{ 2 \arctan \sqrt{\frac{g}{s-a}} 
				+ i \log \frac{\sqrt{g} - \sqrt{s-a}}{\sqrt{g} + \sqrt{s-a}} \mp \pi \right\} \eolc \
		& s > a \eolc
	\end{array} \right.
\end{equation}
where the minus (plus) sign must be applied for $s$ on the upper (lower) rim of the cut. Also $s = a$ is contained in
the domain of $I_0^a(s)$ and the function is indeed finite on the upper rim of the cut. On the lower rim, however, it
is singular, but since this point never lies on the integration path, the pole causes no problems. 

If $I = 1$, we have $n_1 = 3$ and must also subtract a linear term in the denominator to ensure that the integrand
is finite at $s' = a$:
\begin{multline}
	\int\limits_{a - g}^{a + g} ds' \frac{G^a_1(s')}{(a-s')^{3/2}\,(s'-s \mp i\epsilon)} =
	\int\limits_{a - g}^{a + g} ds' \frac{G^a_1(s') - G^a_1(a) - (s'-a) G_1^a{}'(s')}{(a-s')^{3/2}\,(s'-s \mp i\epsilon)}
\\[1mm] 
	- G_1^a{}'(a) \int\limits_{a - g}^{a + g} \frac{ds'}{\sqrt{a-s'}\,(s'-s \mp i\epsilon)} \
	+ G^a_1(a) \int\limits_{a - g}^{a + g} \frac{ds'}{(a-s')^{3/2}\,(s'-s \mp i\epsilon)} \eolp
	\label{eq:numIntSplit1}
\end{multline}
The first integral can be solved numerically, but the problem with the evaluation of the integrand close to $s'=a$ is
even more severe here. The second integral we have solved above but the last one is new. With the variable
transformation $x = s'-a$ it reads
\begin{equation}
	\int\limits_{-g}^g \frac{dx}{(-x)^{3/2}\, (x + a - s \mp i\epsilon)} \equiv \left\{
	\begin{array}{lr}
		Q_1^a(s) \eolc \quad & (a-s) \notin [-g,g] \eolc \\[2mm]
		I_1^a(s) \eolc \quad & (a-s) \in [-g,g] \eolp
	\end{array} \right.
\end{equation}
This time, the pole at $x = 0$ is not integrable and in its vicinity, the integration path must be deformed into a
little half-circle. The shape of the half-circle depends on the position of the singularity at $x = s-a$: if this pole
lies in the upper half-plane, i.e., if we chose $-i\epsilon$ in the denominator, the half-circle must lie in the lower
half-plane and vice versa. It turns out that the integral along the half-circle gives no contribution. Using the
indefinite integral from Eq.~\eqref{eq:appendixIntegral2}, we find in a similar way as before:
\begin{equation}\begin{split}
	Q_1^a(s) = - \frac{2 + 2i}{(a-s) \sqrt{g}} + \frac{Q_0^a(s)}{a-s} \eolc \\[3mm]
	I_1^a(s) = - \frac{2 + 2i}{(a-s) \sqrt{g}} + \frac{I_0^a(s)}{a-s} \eolp
\end{split}\end{equation}
The evaluation of the same integral for $s = a$ would require an additional subtraction in the numerator of the
integrand in Eq.~\eqref{eq:numIntSplit1}. Instead of evaluating complicated integrals, we make use of the fact that the
integral is finite also in this point and simply avoid it. The value of $M_I(a)$ is then obtained from the
interpolation.

Along the same lines, one can evaluate the integration over the interval $(b-h,b+h)$. In particular, the same analytically
solvable integrals will appear in the process.

We have hinted above that the evaluation of the numerically solvable integrals in Eqs.~\eqref{eq:numIntSplit0} and
\eqref{eq:numIntSplit1}, as well as their counterparts over the interval $(b-h,b+h)$,  poses a problem because of the
singularities at $s' = a$ and $s' = b$, respectively. At these points, the denominator of the integrand has a zero. The
integrand, however, is finite, because we have made sure that the numerator vanishes fast enough. But if the
computer is asked to evaluate the integrands in the vicinity of these points, it is bound to fail. The problem is caused
by the numerator, which contains the difference of two numbers that almost cancel. A small relative error in each of
these numbers can lead to a large relative error in their difference. Since the result is divided by a very small
number, the relative error is blown up to be a large absolute error as well.

This problem is solved by using an approximative representation of the inte\-grand---excluding the factor $1/(s'-s\mp
i\epsilon)$, which must be kept explicitly---in the vicinity of the singularities. For $s' \lesssim a$, real and
imaginary part are approximated by functions of the form
\begin{equation}
	 A + B (a-s') \quad \text{and} \quad C + D \sqrt{a-s'} +E (a-s') \eolc
\end{equation}
respectively, with real coefficients, while for $s' \gtrsim a$, the shape of the approximation is
\begin{equation}
	  F + G \sqrt{s'-a} +H (s'-a) \quad \text{and} \quad J + K (s'-a) \eolp
\end{equation}
The coefficients of these functions are fixed by requiring the approximation to reproduce the integrand at an
appropriate number of points. In the vicinity of $b$, the integrand can be well approximated by a linear function.

Once the dispersion integrals have been evaluated, the only unknowns in the $M_I(s)$ are the subtraction constants. We
have applied two different methods in order to determine them, which have been discussed in the previous chapter.

%% file: main/results.tex
\chapter{Results} \label{chp:Results}

In this chapter, we present the numerical results of the dispersive analysis. Let us briefly summarise, what
quantities we calculate. From the iteration algorithm comes a dispersive representation for the isospin amplitudes
$M_I(s)$, and from these one can build the decay amplitudes for the charged as well as for the neutral decay channel.
In the physical region, the squares of both amplitudes can be represented in terms of a few Dalitz plot parameters. From
the integral of the squared amplitude over the entire phase space then follow the quark mass ratio $Q$ and the branching
ratio $r$.

Our results for all these quantities coming from the matching to the one-loop result and from the fit to the Dalitz
distribution from the KLOE collaboration are discussed in two separate sections. Afterwards, we compare our results to
other authors' findings.

\section{Matching to the one-loop result}

We start with a presentation of the results obtained, when the subtraction constants are determined from a matching to
the one-loop result from \chpt. The procedure has been described in detail in Sec.~\ref{sec:solMatch}. This basically
constitutes an update of the calculation in Ref.~\cite{Anisovich+1996} using new inputs. Note that the update
that was recently presented by Leutwyler~\cite{Leutwyler2009} only accounts for the new value of the decay width, but
does not rely on a re-evaluation of the dispersion relations.

%
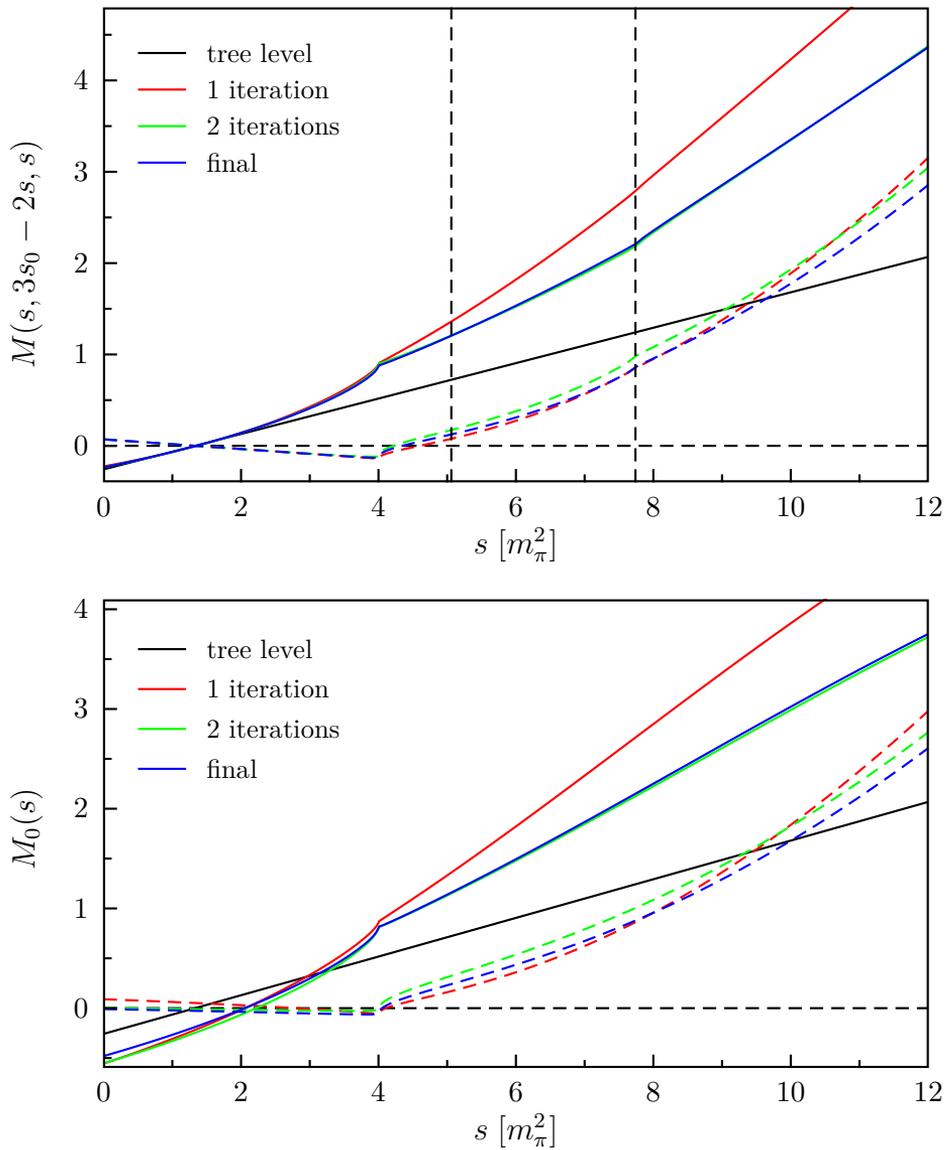
\begin{figure}[p]
\psset{xunit=.9cm,yunit=1.2cm}
\begin{center}\begin{pspicture*}(-1.6,-1.3)(12.2,4.8005)

\psset{linewidth=\mylw,plotstyle=line}

\psline(0,-0.258)(12,2.068)
\psline[linestyle=dashed](0,0)(12,0)
\readdata{\data}{data/match/ReMStep1.dat}
\dataplot[linecolor=red]{\data}
\readdata{\data}{data/match/ImMStep1.dat}
\dataplot[linestyle=dashed,linecolor=red]{\data}
\readdata{\data}{data/match/ReMStep2.dat}
\dataplot[linecolor=green]{\data}
\readdata{\data}{data/match/ImMStep2.dat}
\dataplot[linestyle=dashed,linecolor=green]{\data}
\readdata{\data}{data/match/ReMCentral.dat}
\dataplot[linecolor=blue]{\data}
\readdata{\data}{data/match/ImMCentral.dat}
\dataplot[linestyle=dashed,linecolor=blue]{\data}

\psframe(-0.01,-0.4)(12.01,4.8)
\multips(2,-0.4)(2,0){5}{\psline(0,5pt)}
\multips(1,-0.4)(2,0){6}{\psline(0,3pt)}
\multido{\n=0+2}{7}{\uput{.2}[270](\n,-0.4){\n}}
\multips(0,0)(0,1){5}{\psline(5pt,0)}
\multips(0,0.5)(0,1){5}{\psline(3pt,0)}
\multido{\n=0+1}{5}{\uput{0.2}[180](0,\n){\n}}
\uput{0.6}[270](6,-0.4){\large $s\; [\mpi^2]$}
\uput{0.8}[180]{90}(0,2.1){\large $M(s,3 s_0 - 2s,s)$}

\psline[linestyle=dashed](5.06,-0.4)(5.06,4.8)
\psline[linestyle=dashed](7.74,-0.4)(7.74,4.8)

\psline[linecolor=black](0.5,4.3)(1.2,4.3) \rput[l](1.5,4.3){\small tree level}
\psline[linecolor=red](0.5,3.9)(1.2,3.9) \rput[l](1.5,3.9){\small 1 iteration}
\psline[linecolor=green](0.5,3.5)(1.2,3.5) \rput[l](1.5,3.5){\small 2 iterations}
\psline[linecolor=blue](0.5,3.1)(1.2,3.1) \rput[l](1.5,3.1){\small final}
 
\end{pspicture*}\end{center}
\psset{xunit=.9cm,yunit=1.31cm}
\begin{center}\begin{pspicture*}(-1.6,-1.5)(12.2,4.1)

\psset{linewidth=\mylw,plotstyle=line}

\psline(0,-0.258)(12,2.068)
\psline[linestyle=dashed](0,0)(12,0)
\readdata{\data}{data/match/ReM0Step1.dat}
\dataplot[linecolor=red]{\data}
\readdata{\data}{data/match/ImM0Step1.dat}
\dataplot[linestyle=dashed,linecolor=red]{\data}
\readdata{\data}{data/match/ReM0Step2.dat}
\dataplot[linecolor=green]{\data}
\readdata{\data}{data/match/ImM0Step2.dat}
\dataplot[linestyle=dashed,linecolor=green]{\data}
\readdata{\data}{data/match/ReM0Central.dat}
\dataplot[linecolor=blue]{\data}
\readdata{\data}{data/match/ImM0Central.dat}
\dataplot[linestyle=dashed,linecolor=blue]{\data}

\psframe(-0.01,-0.6)(12.01,4.1)
\multips(2,-0.6)(2,0){5}{\psline(0,5pt)}
\multips(1,-0.6)(2,0){6}{\psline(0,3pt)}
\multido{\n=0+2}{7}{\uput{.2}[270](\n,-0.6){\n}}
\multips(0,0)(0,1){5}{\psline(5pt,0)}
\multips(0,-0.5)(0,1){5}{\psline(3pt,0)}
\multido{\n=0+1}{5}{\uput{0.2}[180](0,\n){\n}}
\uput{0.6}[270](6,-0.6){\large $s\; [\mpi^2]$}
\uput{0.8}[180]{90}(0,1.8){\large $M_0(s)$}

\psline[linecolor=black](0.5,3.6)(1.2,3.6) \rput[l](1.5,3.6){\small tree level}
\psline[linecolor=red](0.5,3.2)(1.2,3.2) \rput[l](1.5,3.2){\small 1 iteration}
\psline[linecolor=green](0.5,2.8)(1.2,2.8) \rput[l](1.5,2.8){\small 2 iterations}
\psline[linecolor=blue](0.5,2.4)(1.2,2.4) \rput[l](1.5,2.4){\small final}

\end{pspicture*}\end{center}

\caption{The upper panel shows the development of the decay amplitude $M(s,t,u)$ along the
line $s=u$ during the iteration procedure, the lower panel the same for the isospin amplitude $M_0(s)$.
In both cases, the real part is plotted as a solid line, while the imaginary part is dashed. The vertical dashed lines
mark the boundaries of the physical region.}
\label{fig:resultsMatchMIterate1}
\end{figure}

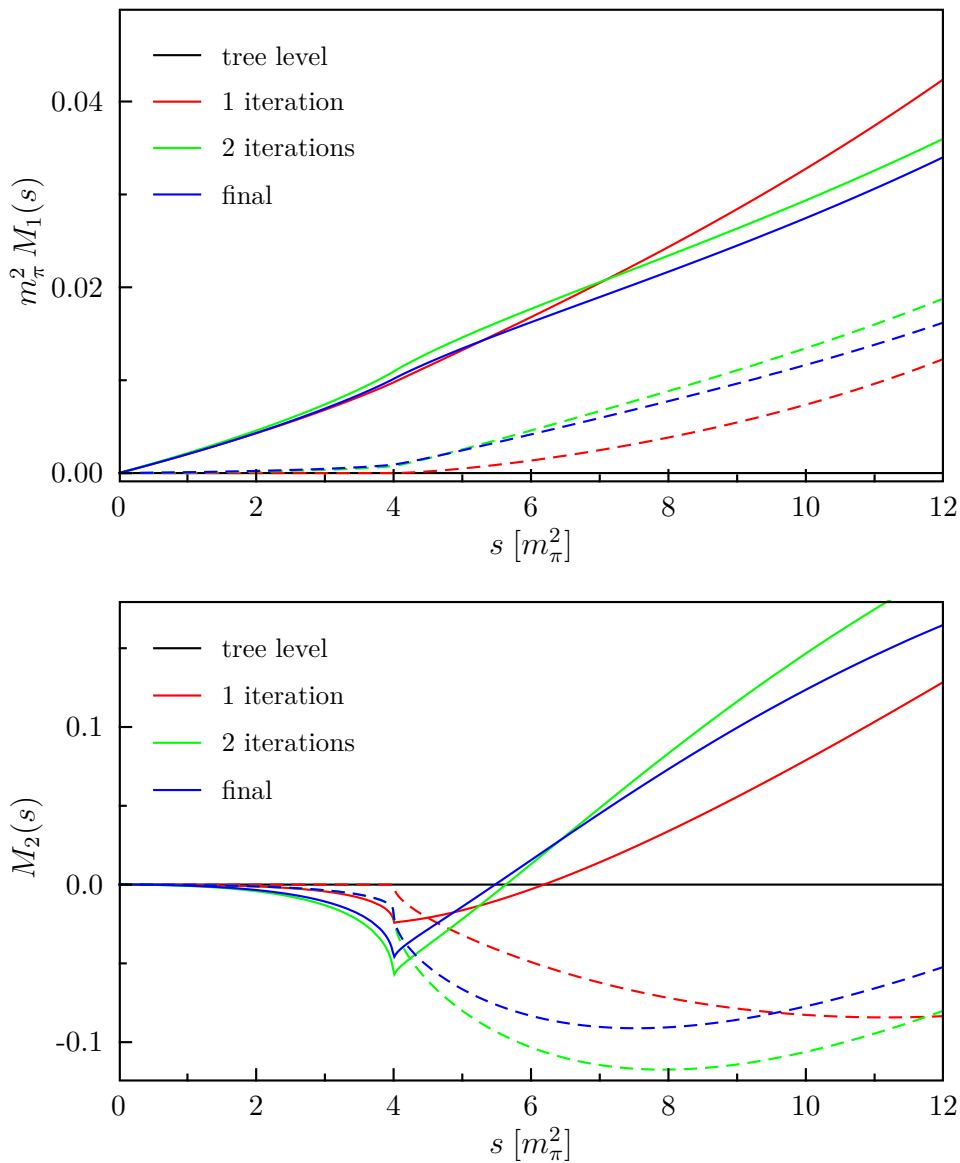
\begin{figure}[p]
\psset{xunit=.9cm,yunit=122cm}
\begin{center}\begin{pspicture*}(-1.6,-0.01)(12.2,0.05)

\psset{linewidth=\mylw,plotstyle=line}

\psline(0,0)(12,0)
\readdata{\data}{data/match/ReM1Step1.dat}
\dataplot[linecolor=red]{\data}
\readdata{\data}{data/match/ImM1Step1.dat}
\dataplot[linestyle=dashed,linecolor=red]{\data}
\readdata{\data}{data/match/ReM1Step2.dat}
\dataplot[linecolor=green]{\data}
\readdata{\data}{data/match/ImM1Step2.dat}
\dataplot[linestyle=dashed,linecolor=green]{\data}
\readdata{\data}{data/match/ReM1Central.dat}
\dataplot[linecolor=blue]{\data}
\readdata{\data}{data/match/ImM1Central.dat}
\dataplot[linestyle=dashed,linecolor=blue]{\data}

\psframe(0,-0.001)(12.01,0.05)
\multips(2,-0.001)(2,0){5}{\psline(0,5pt)}
\multips(1,-0.001)(2,0){6}{\psline(0,3pt)}
\multido{\n=0+2}{7}{\uput{.2}[270](\n,-0.001){\n}}
\multips(0,0.00)(0,0.02){3}{\psline(5pt,0)}
\multips(0,0.01)(0,0.02){2}{\psline(3pt,0)}
\multido{\n=0.00+0.02}{3}{\uput{0.2}[180](0,\n){\n}}
\uput{0.6}[270](6,-0.001){\large $s\; [\mpi^2]$}
\uput{0.98}[180]{90}(0,0.025){\large $\mpi^2\, M_1(s)$}

\psline[linecolor=black](0.5,0.045)(1.2,0.045) \rput[l](1.5,0.045){\small tree level}
\psline[linecolor=red](0.5,0.040)(1.2,0.040) \rput[l](1.5,0.040){\small 1 iteration}
\psline[linecolor=green](0.5,0.035)(1.2,0.035) \rput[l](1.5,0.035){\small 2 iterations}
\psline[linecolor=blue](0.5,0.030)(1.2,0.030) \rput[l](1.5,0.030){\small final}

\end{pspicture*}\end{center}
\psset{xunit=.9cm,yunit=20.7cm}
\begin{center}\begin{pspicture*}(-1.6,-.175)(12.2,.1801)

\psset{linewidth=\mylw,plotstyle=line}

\psline(0,0)(12,0)
\readdata{\data}{data/match/ReM2Step1.dat}
\dataplot[plotstyle=line,linecolor=red]{\data}
\readdata{\data}{data/match/ImM2Step1.dat}
\dataplot[linestyle=dashed,linecolor=red]{\data}
\readdata{\data}{data/match/ReM2Step2.dat}
\dataplot[plotstyle=line,linecolor=green]{\data}
\readdata{\data}{data/match/ImM2Step2.dat}
\dataplot[linestyle=dashed,linecolor=green]{\data}
\readdata{\data}{data/match/ReM2Central.dat}
\dataplot[plotstyle=line,linecolor=blue]{\data}
\readdata{\data}{data/match/ImM2Central.dat}
\dataplot[linestyle=dashed,linecolor=blue]{\data}

\psframe(0,-0.125)(12.01,0.18)
\multips(2,-0.125)(2,0){5}{\psline(0,5pt)}
\multips(1,-0.125)(2,0){6}{\psline(0,3pt)}
\multido{\n=0+2}{7}{\uput{.2}[270](\n,-0.125){\n}}
\multips(0,-0.1)(0,.1){3}{\psline(5pt,0)}
\multips(0,-0.05)(0,0.1){3}{\psline(3pt,0)}
\multido{\n=-0.1+0.1}{5}{\uput{0.2}[180](0,\n){\n}}
\uput{0.6}[270](6,-0.125){\large $s\; [\mpi^2]$}
\uput{1.0}[180]{90}(0,0.03){\large $M_2(s)$}

\psline[linecolor=black](0.5,0.15)(1.2,0.15) \rput[l](1.5,0.15){\small tree level}
\psline[linecolor=red](0.5,0.12)(1.2,0.12) \rput[l](1.5,0.12){\small 1 iteration}
\psline[linecolor=green](0.5,0.09)(1.2,0.09) \rput[l](1.5,0.09){\small 2 iterations}
\psline[linecolor=blue](0.5,0.06)(1.2,0.06) \rput[l](1.5,0.06){\small final}

\end{pspicture*}\end{center}

\caption{The upper panel shows the development of the isospin amplitude $M_1(s)$ during the iteration procedure, 
the lower panel the same for the isospin amplitude $M_2(s)$. $M_1(s)$ is multiplied by $\mpi^2$ in order to make
it dimensionless. In both cases, the real part is plotted as a solid line, while the imaginary part is dashed.}
\label{fig:resultsMatchMIterate2}
\end{figure}

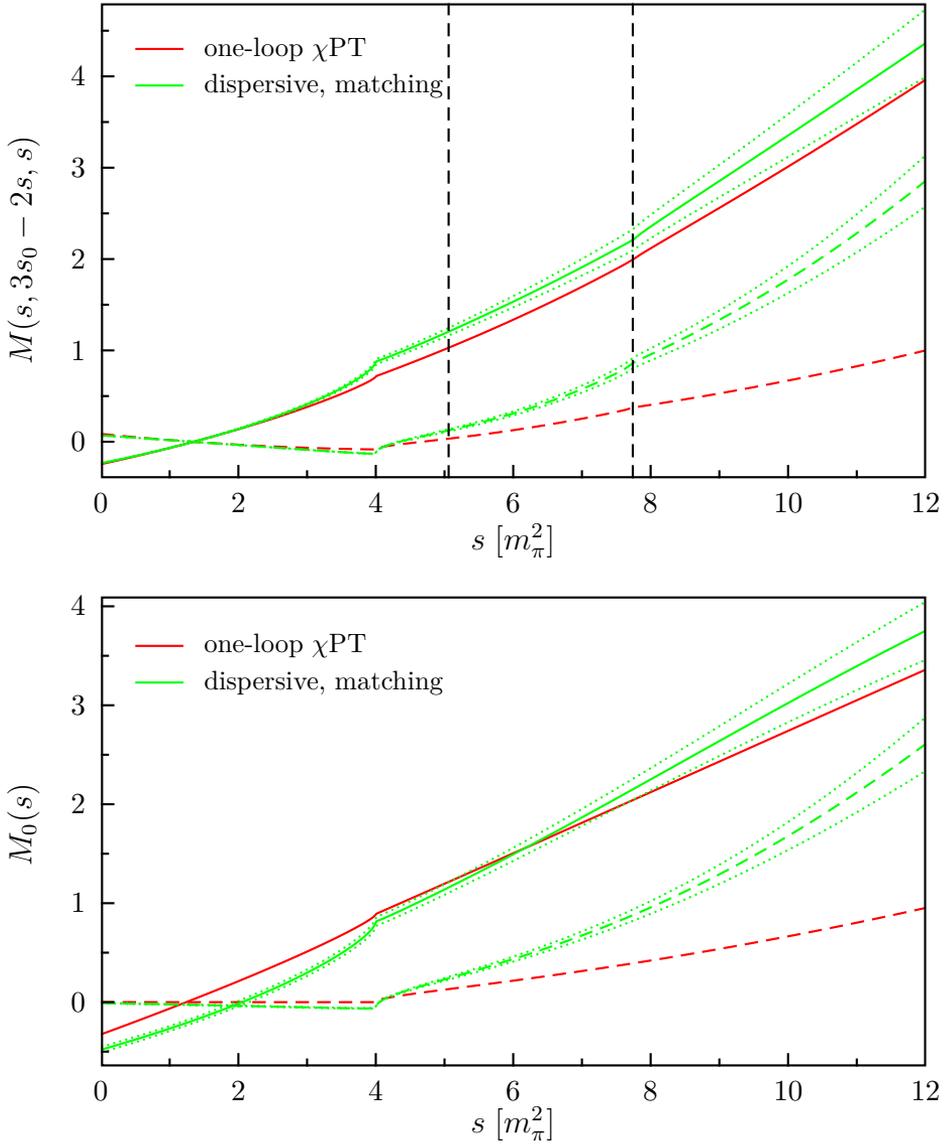
\begin{figure}[p]
\psset{xunit=.9cm,yunit=1.2cm}
\begin{center}\begin{pspicture*}(-1.4,-1.3)(12.2,4.81)

\psset{linewidth=\mylw,plotstyle=line}

\psset{linecolor=red}
\readdata{\data}{data/ChPT/ReM1loop.dat}
\dataplot{\data}
\readdata{\data}{data/ChPT/ImM1loop.dat}
\dataplot[linestyle=dashed]{\data}

\psset{linecolor=green}
\readdata{\data}{data/match/ReMCentral.dat}
\dataplot{\data}
\readdata{\data}{data/match/ReMLower.dat}
\dataplot[linestyle=dotted,dotsep=.05,linewidth=\mydotlw]{\data}
\readdata{\data}{data/match/ReMUpper.dat}
\dataplot[linestyle=dotted,dotsep=.05,linewidth=\mydotlw]{\data}
\readdata{\data}{data/match/ImMCentral.dat}
\dataplot[linestyle=dashed]{\data}
\readdata{\data}{data/match/ImMLower.dat}
\dataplot[linestyle=dotted,dotsep=.05,linewidth=\mydotlw]{\data}
\readdata{\data}{data/match/ImMUpper.dat}
\dataplot[linestyle=dotted,dotsep=.05,linewidth=\mydotlw]{\data}

\psset{linecolor=black}
\psframe(0,-0.4)(12.01,4.8)
\multips(2,-0.4)(2,0){5}{\psline(0,5pt)}
\multips(1,-0.4)(2,0){6}{\psline(0,3pt)}
\multido{\n=0+2}{7}{\uput{.2}[270](\n,-0.4){\n}}
\multips(0,0)(0,1){5}{\psline(5pt,0)}
\multips(0,0.5)(0,1){5}{\psline(3pt,0)}
\multido{\n=0+1}{5}{\uput{0.2}[180](0,\n){\n}}
\uput{0.6}[270](6,-0.4){\large $s\; [\mpi^2]$}
\uput{0.8}[180]{90}(0,2.1){\large $M(s,3 s_0 - 2s,s)$}

\psline[linestyle=dashed](5.06,-0.4)(5.06,4.8)
\psline[linestyle=dashed](7.74,-0.4)(7.74,4.8)

\psline[linecolor=red](0.5,4.3)(1.2,4.3) \rput[l](1.5,4.3){\small one-loop \chpt}
\psline[linecolor=green](0.5,3.9)(1.2,3.9) \rput[l](1.5,3.9){\small dispersive, matching}

\end{pspicture*}\end{center}
\psset{xunit=.9cm,yunit=1.3cm}
\begin{center}\begin{pspicture*}(-1.4,-1.5)(12.2,4.11)

\psset{linewidth=\mylw,plotstyle=line}

\psset{linecolor=red}
\readdata{\data}{data/ChPT/ReM01loop.dat}
\dataplot{\data}
\readdata{\data}{data/ChPT/ImM01loop.dat}
\dataplot[linestyle=dashed]{\data}

\psset{linecolor=green}
\readdata{\data}{data/match/ReM0Central.dat}
\dataplot{\data}
\readdata{\data}{data/match/ReM0Lower.dat}
\dataplot[linestyle=dotted,dotsep=.05,linewidth=\mydotlw]{\data}
\readdata{\data}{data/match/ReM0Upper.dat}
\dataplot[linestyle=dotted,dotsep=.05,linewidth=\mydotlw]{\data}
\readdata{\data}{data/match/ImM0Central.dat}
\dataplot[linestyle=dashed]{\data}
\readdata{\data}{data/match/ImM0Lower.dat}
\dataplot[linestyle=dotted,dotsep=.05,linewidth=\mydotlw]{\data}
\readdata{\data}{data/match/ImM0Upper.dat}
\dataplot[linestyle=dotted,dotsep=.05,linewidth=\mydotlw]{\data}

\psset{linecolor=black}
\psframe(0,-0.65)(12.01,4.1)
\multips(2,-0.65)(2,0){5}{\psline(0,5pt)}
\multips(1,-0.65)(2,0){6}{\psline(0,3pt)}
\multido{\n=0+2}{7}{\uput{.2}[270](\n,-0.65){\n}}
\multips(0,0)(0,1){5}{\psline(5pt,0)}
\multips(0,-0.5)(0,1){5}{\psline(3pt,0)}
\multido{\n=0+1}{5}{\uput{0.2}[180](0,\n){\n}}
\uput{0.6}[270](6,-0.6){\large $s\; [\mpi^2]$}
\uput{0.8}[180]{90}(0,1.8){\large $M_0(s)$}

\psline[linecolor=red](0.5,3.6)(1.2,3.6) \rput[l](1.5,3.6){\small one-loop \chpt}
\psline[linecolor=green](0.5,3.23)(1.2,3.23) \rput[l](1.5,3.23){\small dispersive, matching}

\end{pspicture*}\end{center}

\caption{The figure shows the decay amplitude $M(s,t,u)$ along the line $s=u$ in the upper panel and the isospin
amplitude $M_0(s)$ in the lower panel. In both cases, solid and dashed lines stand for the central values of the real
and imaginary part, respectively, dotted lines for the error band. The vertical dashed lines mark the boundaries of the
physical region.}
\label{fig:resultsMatchMFinal1}
\end{figure}

The immediate outcome of the iteration procedure are the isospin amplitudes $M_I(s)$ as well as the decay amplitude
$M(s,t,u)$. In Figs.~\ref{fig:resultsMatchMIterate1} and \ref{fig:resultsMatchMIterate2} we show how these functions
develop from step to step. The tree-level result is plotted in black and non-vanishing only for
$M_0(s)$ and $M(s,t,u)$. The figures also show the situation after one and two iterations steps as well as the final
result. The total number of iterations is determined by the requirement that the subtraction constants have converged at
the level of one per mille. This happens after 12 iterations, which is, however, far more than is necessary in order
to obtain precise results. Since efficiency is no relevant issue in this calculation, there is nevertheless no need to
reduce the number of iterations. Figures~\ref{fig:resultsMatchMFinal1} and \ref{fig:resultsMatchMFinal2} show only the
final results for the amplitudes together with the error band. A detailed discussion of the error analysis will follow
later. All the other results are based on the decay amplitude $M(s,t,u)$ and its error band.

%

\begin{figure}[p]
\psset{xunit=.9cm,yunit=122cm}
\begin{center}\begin{pspicture*}(-1.6,-0.01)(12.2,0.05)

\psset{linewidth=\mylw,plotstyle=line}

\psset{linecolor=red}
\readdata{\data}{data/ChPT/ReM11loop.dat}
\dataplot{\data}
\readdata{\data}{data/ChPT/ImM11loop.dat}
\dataplot[linestyle=dashed]{\data}

\psset{linecolor=green}
\readdata{\data}{data/match/ReM1Central.dat}
\dataplot{\data}
\readdata{\data}{data/match/ReM1Lower.dat}
\dataplot[linestyle=dotted,dotsep=.05,linewidth=\mydotlw]{\data}
\readdata{\data}{data/match/ReM1Upper.dat}
\dataplot[linestyle=dotted,dotsep=.05,linewidth=\mydotlw]{\data}
\readdata{\data}{data/match/ImM1Central.dat}
\dataplot[linestyle=dashed]{\data}
\readdata{\data}{data/match/ImM1Lower.dat}
\dataplot[linestyle=dotted,dotsep=.05,linewidth=\mydotlw]{\data}
\readdata{\data}{data/match/ImM1Upper.dat}
\dataplot[linestyle=dotted,dotsep=.05,linewidth=\mydotlw]{\data}

\psset{linecolor=black}
\psframe(0,-0.001)(12.01,0.05)
\multips(2,-0.001)(2,0){5}{\psline(0,5pt)}
\multips(1,-0.001)(2,0){6}{\psline(0,3pt)}
\multido{\n=0+2}{7}{\uput{.2}[270](\n,-0.001){\n}}
\multips(0,0.00)(0,0.02){3}{\psline(5pt,0)}
\multips(0,0.01)(0,0.02){2}{\psline(3pt,0)}
\multido{\n=0.00+0.02}{3}{\uput{0.2}[180](0,\n){\n}}
\uput{0.6}[270](6,-0.001){\large $s\; [\mpi^2]$}
\uput{0.98}[180]{90}(0,0.025){\large $\mpi^2\, M_1(s)$}

\psline[linecolor=red](0.5,0.045)(1.2,0.045) \rput[l](1.5,0.045){\small one-loop \chpt}
\psline[linecolor=green](0.5,0.041)(1.2,0.041) \rput[l](1.5,0.041){\small dispersive, matching}

\end{pspicture*}\end{center}
\psset{xunit=.9cm,yunit=20.7cm}
\begin{center}\begin{pspicture*}(-1.6,-.175)(12.2,.181)

\psset{linewidth=\mylw,plotstyle=line}

\psset{linecolor=red}
\readdata{\data}{data/ChPT/ReM21loop.dat}
\dataplot{\data}
\readdata{\data}{data/ChPT/ImM21loop.dat}
\dataplot[linestyle=dashed]{\data}

\psset{linecolor=green}
\readdata{\data}{data/match/ReM2Central.dat}
\dataplot{\data}
\readdata{\data}{data/match/ReM2Lower.dat}
\dataplot[linestyle=dotted,dotsep=.05,linewidth=\mydotlw]{\data}
\readdata{\data}{data/match/ReM2Upper.dat}
\dataplot[linestyle=dotted,dotsep=.05,linewidth=\mydotlw]{\data}
\readdata{\data}{data/match/ImM2Central.dat}
\dataplot[linestyle=dashed]{\data}
\readdata{\data}{data/match/ImM2Lower.dat}
\dataplot[linestyle=dotted,dotsep=.05,linewidth=\mydotlw]{\data}
\readdata{\data}{data/match/ImM2Upper.dat}
\dataplot[linestyle=dotted,dotsep=.05,linewidth=\mydotlw]{\data}

\psset{linecolor=black}
\psframe(0,-0.125)(12.01,0.18)
\multips(2,-0.125)(2,0){5}{\psline(0,5pt)}
\multips(1,-0.125)(2,0){6}{\psline(0,3pt)}
\multido{\n=0+2}{7}{\uput{.2}[270](\n,-0.125){\n}}
\multips(0,-0.1)(0,.1){3}{\psline(5pt,0)}
\multips(0,-0.05)(0,0.1){3}{\psline(3pt,0)}
\multido{\n=-0.1+0.1}{5}{\uput{0.2}[180](0,\n){\n}}
\uput{0.6}[270](6,-0.125){\large $s\; [\mpi^2]$}
\uput{1.0}[180]{90}(0,0.03){\large $M_2(s)$}

\psline[linecolor=red](0.5,0.15)(1.2,0.15) \rput[l](1.5,0.15){\small one-loop \chpt}
\psline[linecolor=green](0.5,0.125)(1.2,0.125) \rput[l](1.5,0.125){\small dispersive, matching}

\end{pspicture*}\end{center}

\caption{The figure shows the isospin amplitude $M_1(s)$ in the upper panel and the isospin amplitude $M_2(s)$ in
the lower panel. $M_1(s)$ is multiplied by $\mpi^2$ in order to make it dimensionless. In both cases, solid and dashed
lines stand for the central values of the real and imaginary part, respectively, dotted lines for the error band.}
\label{fig:resultsMatchMFinal2}
\end{figure}

%
\begin{figure}[tb]

\begin{center}
	\scalebox{.95}{\input{figs/dalitz1.tex}}
\end{center}
	\caption{The Dalitz plot distribution obtained from the dispersive analysis.}
\label{fig:resultsMatchDalitz}
\end{figure}

In order to compare the decay amplitude with experimental results, we calculate its square as a function of the
Dalitz plot variables $X$ and $Y$ inside the physical region. The result is visualised in
Fig.~\ref{fig:resultsMatchDalitz}. The rough shape of the Dalitz distribution does
indeed agree with the experimental result in Fig.~\ref{fig:eta3piDalitzKLOE}, but for a more precise comparison, we
have also plotted two cuts through the Dalitz plot in Fig.~\ref{fig:resultsMatchDalitzXYc}. The plots show the
results from \chpt, from our dispersive analysis, and from the KLOE experiment, and they reveal an obvious
discrepancy between the latter two. We find an amplitude that is much closer to the one-loop result than to
experiment. The dispersive result for the Dalitz plot distribution for the neutral channel, plotted in
Fig.~\ref{fig:resultsMatchDalitzXY0}, shows the same well-known problem that also the one-loop result has:
it is curved in the wrong direction compared to experiment. We must conclude that, if the subtraction constants are
obtained from a matching to one-loop \chpt{} in this way, the dispersive method fails to overcome the discrepancy
between experiment and perturbation theory.

%
\begin{figure}[p]
\psset{xunit=5cm,yunit=61.8cm}

\begin{center}\begin{pspicture*}(-1.36,0.825)(1.05,0.938)

\psset{linewidth=\mylw,plotstyle=curve}

\readdata{\data}{data/ChPT/dalitzXc1loop.dat}
\dataplot[linecolor=red]{\data}

\readdata{\data}{data/exp/dalitzXcKLOECentral.dat}
\dataplot[linecolor=black]{\data}
\readdata{\data}{data/exp/dalitzXcKLOEUpper.dat}
\dataplot[linestyle=dotted,dotsep=.05,linecolor=black,linewidth=\mydotlw]{\data}
\readdata{\data}{data/exp/dalitzXcKLOELower.dat}
\dataplot[linestyle=dotted,dotsep=.05,linecolor=black,linewidth=\mydotlw]{\data}

\readdata{\data}{data/match/dalitzXcCentral.dat}
\dataplot[linecolor=green]{\data}
\readdata{\data}{data/match/dalitzXcLower.dat}
\dataplot[linestyle=dotted,dotsep=.05,linecolor=green,linewidth=\mydotlw]{\data}
\readdata{\data}{data/match/dalitzXcUpper.dat}
\dataplot[linestyle=dotted,dotsep=.05,linecolor=green,linewidth=\mydotlw]{\data}

\psframe(-1.002,0.84)(1.002,0.937)
\multips(-.5,0.84)(0.5,0){3}{\psline(0,5pt)}
\multips(-0.75,0.84)(0.5,0){4}{\psline(0,3pt)}
\multido{\n=-1.0+0.5}{5}{\uput{.2}[270](\n,0.84){\n}}
\multips(-1,0.86)(0,0.02){4}{\psline(5pt,0)}
\multips(-1,0.85)(0,0.02){5}{\psline(3pt,0)}
\multido{\n=0.84+0.02}{5}{\uput{0.2}[180](-1,\n){\n}}
\uput{0.6}[270](0,0.84){\large $X$}
\uput{1.2}[180]{90}(-1,0.89){\large $\Gamma(X,0.125)$}

\psline[linestyle=dashed](-0.936,0.84)(-0.936,0.937)
\psline[linestyle=dashed](0.936,0.84)(0.936,0.937)

\psline[linecolor=red](-0.6,0.93)(-0.45,0.93) \rput[l](-0.4,0.93){\small one-loop \chpt}
\psline[linecolor=black](-0.6,0.922)(-0.45,0.922) \rput[l](-0.4,0.922){\small KLOE}
\psline[linecolor=green](-0.6,0.914)(-0.45,0.914) \rput[l](-0.4,0.914){\small dispersive, matching}

\end{pspicture*}\end{center}

\vfill

\psset{xunit=5cm,yunit=2.08cm}
\begin{center}\begin{pspicture*}(-1.36,-0.45)(1.05,2.99)

\psset{linewidth=\mylw,plotstyle=curve}

\readdata{\data}{data/ChPT/dalitzYc1loop.dat}
\dataplot[linecolor=red]{\data}

\readdata{\data}{data/exp/dalitzYcKLOECentral.dat}
\dataplot[linecolor=black]{\data}
\readdata{\data}{data/exp/dalitzYcKLOEUpper.dat}
\dataplot[linestyle=dotted,dotsep=.05,linecolor=black,linewidth=\mydotlw]{\data}
\readdata{\data}{data/exp/dalitzYcKLOELower.dat}
\dataplot[linestyle=dotted,dotsep=.05,linecolor=black,linewidth=\mydotlw]{\data}

\readdata{\data}{data/match/dalitzYcCentral.dat}
\dataplot[linecolor=green]{\data}
\readdata{\data}{data/match/dalitzYcLower.dat}
\dataplot[linestyle=dotted,dotsep=.05,linecolor=green,linewidth=\mydotlw]{\data}
\readdata{\data}{data/match/dalitzYcUpper.dat}
\dataplot[linestyle=dotted,dotsep=.05,linecolor=green,linewidth=\mydotlw]{\data}

\psframe(-1.002,0.1)(1.002,2.98)
\multips(-.5,0.1)(0.5,0){3}{\psline(0,5pt)}
\multips(-0.75,0.1)(0.5,0){4}{\psline(0,3pt)}
\multido{\n=-1.0+0.5}{5}{\uput{.2}[270](\n,0.1){\n}}
\multips(-1,0.5)(0,0.5){5}{\psline(5pt,0)}
\multips(-1,0.75)(0,0.5){5}{\psline(3pt,0)}
\multido{\n=0.5+0.5}{5}{\uput{0.2}[180](-1,\n){\n}}
\uput{0.7}[270](0,0.1){\large $Y$}
\uput{1.2}[180]{90}(-1,1.45){\large $\Gamma(0.125,Y)$}

\psline[linestyle=dashed](-0.986,0.1)(-0.986,2.98)
\psline[linestyle=dashed](0.890,0.1)(0.890,2.98)

\psline[linecolor=red](0,2.7)(0.15,2.7) \rput[l](0.2,2.7){\small one-loop \chpt}
\psline[linecolor=black](0,2.47)(0.15,2.47) \rput[l](0.2,2.47){\small KLOE}
\psline[linecolor=green](0,2.24)(0.15,2.24) \rput[l](0.2,2.24){\small dispersive, matching}

\end{pspicture*}\end{center}
\caption{The figure shows two cuts through the Dalitz plot for $\eta \to \pi^+ \pi^- \pi^0$. The upper panel depicts
the distribution along the line with $Y = 0.125$, the lower panel along the line with $X = 0.125$. The solid lines
represent the central values and the dotted lines the error band. The dashed lines mark the boundaries of the physical
region.}
\label{fig:resultsMatchDalitzXYc}
\end{figure}

\begin{figure}[tb]
\psset{xunit=5cm,yunit=34.3cm}
\begin{center}\begin{pspicture*}(-1.36,0.875)(1.05,1.081)

\psset{linewidth=\mylw,plotstyle=curve}

\readdata{\data}{data/ChPT/dalitzX01loop.dat}
\dataplot[linecolor=red]{\data}

\readdata{\data}{data/exp/dalitzX0PDGCentral.dat}
\dataplot[linecolor=black]{\data}
\readdata{\data}{data/exp/dalitzX0PDGUpper.dat}
\dataplot[linestyle=dotted,dotsep=.05,linecolor=black,linewidth=\mydotlw]{\data}
\readdata{\data}{data/exp/dalitzX0PDGLower.dat}
\dataplot[linestyle=dotted,dotsep=.05,linecolor=black,linewidth=\mydotlw]{\data}

\readdata{\data}{data/match/dalitzX0Central.dat}
\dataplot[linecolor=green]{\data}
\readdata{\data}{data/match/dalitzX0Lower.dat}
\dataplot[linestyle=dotted,dotsep=.05,linecolor=green,linewidth=\mydotlw]{\data}
\readdata{\data}{data/match/dalitzX0Upper.dat}
\dataplot[linestyle=dotted,dotsep=.05,linecolor=green,linewidth=\mydotlw]{\data}

\psframe(-1.002,0.905)(1.002,1.08)
\multips(-.5,0.905)(0.5,0){3}{\psline(0,5pt)}
\multips(-0.75,0.905)(0.5,0){4}{\psline(0,3pt)}
\multido{\n=-1.+0.5}{5}{\uput{.2}[270](\n,0.905){\n}}
\multips(-1,0.94)(0,0.04){4}{\psline(5pt,0)}
\multips(-1,0.92)(0,0.04){4}{\psline(3pt,0)}
\multido{\n=0.94+0.04}{4}{\uput{0.2}[180](-1,\n){\n}}
\uput{0.65}[270](0,0.905){\large $X$}
\uput{1.2}[180]{90}(-1,1.005){\large $\Gamma(X,0)$}

\psline[linestyle=dashed](-0.941,0.905)(-0.941,1.08)
\psline[linestyle=dashed](0.941,0.905)(0.941,1.08)

\psline[linecolor=red](-0.45,1.067)(-0.3,1.067) \rput[l](-0.25,1.067){\small one-loop \chpt}
\psline[linecolor=black](-0.45,1.053)(-0.3,1.053) \rput[l](-0.25,1.053){\small PDG}
\psline[linecolor=green](-0.45,1.039)(-0.3,1.039) \rput[l](-0.25,1.039){\small dispersive, matching}

\end{pspicture*}\end{center}
\caption{The figure shows a cut along the line $Y = 0$ through the Dalitz plot for $\eta \to 3 \pi^0$. The solid lines
represent the central values and the dotted lines the error band. The vertical dashed lines mark the boundaries of the
physical region.}
\label{fig:resultsMatchDalitzXY0}
\end{figure}

The Dalitz distribution can be formulated in terms of the Dalitz plot parameters. From a fit of the Dalitz plot
parametrisation to our result in the charged channel, we find
\begin{align}\begin{aligned}
	a &= -1.266 \pm 0.042 \eolc \qquad 	\\[1mm]	b &= 0.516 \pm 0.065 \eolc \\[1mm]
	d &= 0.047 \pm 0.011 \eolc \qquad 	\\[1mm]	f &= -0.052 \pm 0.031 \eolc \\[1mm]
	h &= -0.050 \pm 0.007 \eolp
\end{aligned}\end{align}
The other Dalitz plot parameters must vanish due to charge conjugation symmetry. If they are also included in the fit,
they turn out to be very small and consistent with zero, while all the other parameters remain unchanged. The fit also
delivers the correlation matrix:
\begin{equation}
	\begin{array}{cccccc}
		  & a & b      & d      & f 		& h		\\[2mm]
		a & 1 & -0.794 & -0.137	& 0.617	& 0.254	\\[2mm]
		b &   & 1		& -0.493	& -0.968 & 0.386	\\[2mm]
		d &   & 			& 1		& 0.692  & -0.993	\\[2mm]
		f &   &			&			& 1		& -0.602	\\[2mm]
		h &	&			&			&			& 1
	\end{array}
\end{equation}
Our result for the Dalitz plot parameters, of course, also reflects the discrepancy with the experimental result.

In the neutral channel, we find for the single slope parameter
\begin{equation}
	\alpha = 0.030 \pm 0.011 \eolc
\end{equation}
which agrees with experiment in its absolute value, but not in its sign. Note that according to
Eq.~\eqref{eq:eta3piAlphaFromCharged}, $\alpha$ grows with $b$. We have found a large value for $b$, even larger than 
at one and two loop, and it is thus not surprising to find $\alpha$ positive.

\begin{figure}[tb]

\psset{xunit=1.5cm,yunit=1.8cm}

\begin{center}\begin{pspicture*}(-1,21.1)(7.1,25.51)

\psset{linewidth=\mylw}

\psframe[fillstyle=solid,fillcolor=shadegray,linewidth=1pt,linecolor=shadegray](0,22.07)(7,23.42)
\psline(0,22.74)(7,22.74)

\pscircle*(1,25.04){0.1}
\pscircle*(2,22.79){0.1}
\pscircle*(3,22.66){0.1}
\pscircle*(4,22.68){0.1}
\pscircle*(5,22.73){0.1}
\pscircle*(6,22.74){0.1}

\psframe(0,21.8)(7,25.5)
\multips(1,21.8)(1,0){6}{\psline(0,5pt)}
\multido{\n=1+1}{6}{\uput{.2}[270](\n,21.8){\n}}
\multips(0,22)(0,1){4}{\psline(5pt,0)}
\multips(0,22.5)(0,1){3}{\psline(3pt,0)}
\multido{\n=22+1}{4}{\uput{0.2}[180](0,\n){\n}}
\uput{0.8}[270](3.5,21.8){\large iteration steps}
\uput{1.0}[180]{90}(0,23.9){\large $Q(\pi^+ \pi^- \pi^0)$}

\end{pspicture*}\end{center}

\caption{The figure shows the development of $Q(\pi^+ \pi^- \pi^0)$ over the iteration steps. The
solid black line and the grey band indicate the final result and the corresponding error band, respectively.}
\label{fig:resultsMatchQIterations}
\end{figure}
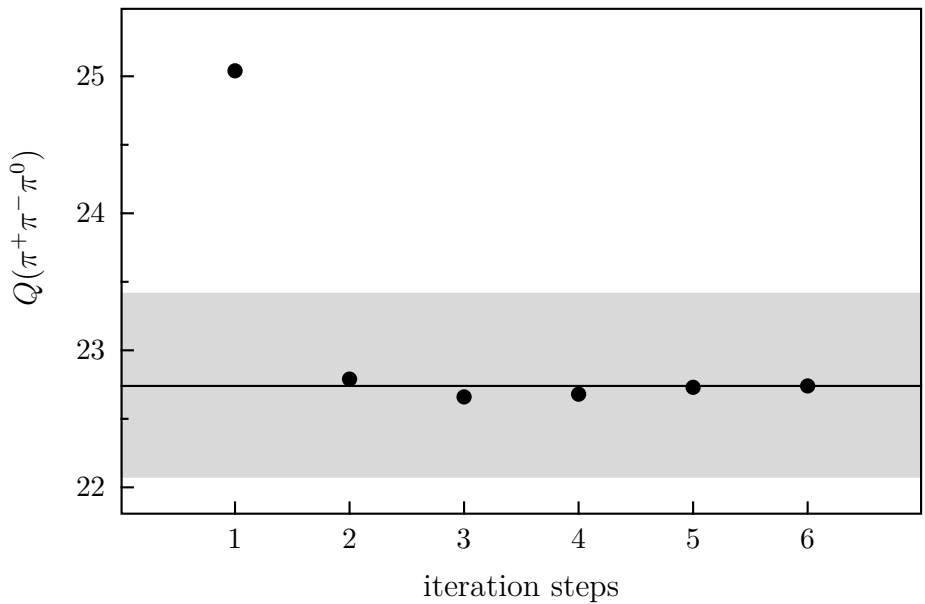

We integrate the square of the dispersive amplitude over the entire phase space and then, by means of
Eq.~\eqref{eq:eta3piDecayRateQ}, calculate $Q$ from the charged and from the neutral channel separately. We
obtain
\begin{equation}
	Q(\pi^+ \pi^- \pi^0) = 22.74{}^{+0.68}_{-0.67} \eolc \qquad Q(3 \pi^0) = 22.72{}^{+0.65}_{-0.64} \eolp
\end{equation}
As an illustration of the effects of the iterations, Fig.~\ref{fig:resultsMatchQIterations} shows how the value for $Q$
obtained from the charged channel varies over the iteration steps. Already after two steps, $Q$ lies well within the
error band of the final result, however, it takes a few more to make the uncertainty due to the finite number of
iteration steps entirely negligible.

From the ratio of the phase space integrals we obtain the branching ratio
\begin{equation}
	r = 1.428 \pm 0.01 \eolp
\end{equation}
It is in excellent agreement with the the experimental number, $r = 1.432 \pm 0.026$. This was to be expected, since
we have found almost identical values for $Q$ from the two decay channels.

\subsection{Error analysis}

Several error sources contribute to the uncertainties on the amplitudes and the physical quantities discussed in the
previous section. In the following, we list all of them and describe how their influence on the final results has been
estimated. Most of these sources are uncertainties on the input parameters, namely on $\gamma_0$ and $\beta_1$ (see
Eq.~\eqref{eq:solGamma0Beta1}), $L_3$, and the $\pi \pi$ phase shifts (see Sec.~\ref{sec:pipiPhaseShifts}). There are
also theoretical uncertainties coming from the interpolation of functions, the finite number of iterations, and the
cut-off in the dispersion integrals. Only the latter leads to a non-negligible effect and is included in the
error analysis. The error bands on the decay amplitude, the isospin amplitudes $M_I(s)$ and the Dalitz plot distribution
as well as the errors on the Dalitz plot parameters and on the branching ratio $r$ consist of these contributions. There
is an additional error on the quark mass ratio $Q$ due to the uncertainty on the experimental decay width. Furthermore,
inelastic processes have been completely neglected in the dispersive analysis. In Ref.~\cite{Walker1998} it was
found that this leads to an uncertainty of $\pm 0.2$ on $Q$. Since we have not explicitly calculated the effects of
inelasticity, these are not included in the error bands on the amplitudes and in the uncertainties on $r$ and the
Dalitz plot parameters.

\begin{table}[tb]
\renewcommand{\arraystretch}{1.6}
\begin{center}\begin{tabular}{lcccc}
								& $Q (\pi^+ \pi^- \pi^0)$	& $Q (3 \pi^0)$				& $r$				
								& $\alpha$						\\ \hline
$\Gamma$						& $\pm$ 0.31					& $\pm$ 0.31					& ---				
								& ---								\\
$\gamma_0$					& $\pm$ 0.38					& $\pm$ 0.36					& $\pm$ 0.0069	
								& $\pm$ 0.0096					\\
$\beta_1$					& $\pm$ 0.36					& $\pm$ 0.35					& $\pm$ 0.0039	
								& $\pm$ 0.0026					\\
$L_3$							& ${}^{+0.025}_{-0.023}$	& ${}^{+0.036}_{-0.033}$	& $\pm$ 0.0026 	
								& $\pm$ 0.0009					\\
$\delta_I(s)$				& ${}^{+0.18}_{-0.15}$		& ${}^{+0.17}_{-0.13}$		& ${}^{+0.0027}_{-0.0032}$	
								& $\pm$ 0.0040					\\
inelasticity				& $\pm$ 0.2 					& $\pm$ 0.2						& ---	
								& ---								\\
cut-off						& $\pm$ 0.09					& $\pm$ 0.09					& $\pm$ 0.002		
								& $\pm$ 0.0026					\\ \hline
total uncertainty			& ${}^{+0.68}_{-0.67}$		& ${}^{+0.65}_{-0.64}$		& ${}^{+0.0090}_{-0.0092}$
								& $\pm$ 0.011
\end{tabular}\end{center}
\caption{All the contributions to the uncertainties on $Q$, $r$, and $\alpha$.}
\label{tab:resultsMatchErrorsQr}
\end{table}

For each of these sources we show now, how its error contribution is estimated. The dependence of $Q$ on the decay
width $\Gamma$ is explicitly known, and the error can be calculated analytically to be
\begin{equation}
	dQ_\Gamma = \left| \frac{dQ}{d\Gamma} \right| \, d\Gamma = \frac{Q}{4 \Gamma}\, d\Gamma \eolp
\end{equation}
In order to estimate the uncertainty coming from a certain input, the latter is set to one of
the extremal values that are allowed by its uncertainty and the dispersive analysis is repeated with this modified
input. All the desired numerical results are evaluated and the deviation from the central values is taken to be the
uncertainty due to this source. The subtraction constants $\gamma_0$ and $\beta_1$ are only
set to their maximum values and the error is then taken to be symmetric. The low-energy constant $L_3$ and the phase
shifts, on the other hand, are set to both extremal values and the resulting error is thus asymmetric. The uncertainty
due to the cut-off is estimated by shifting it from the usual $\Lambda^2 = 200\, \mpi^2$ up to $\Lambda^2 = 300\,
\mpi^2$.

There is an additional complication involved in the evaluation of the uncertainty on the Dalitz plot parameters: they
are correlated. For each set of inputs, we obtain the Dalitz plot parameters from a fit to the square of the resulting
amplitude in the physical region. Let $a_i(j)$ be the result for the $i^\text{th}$ Dalitz plot parameter calculated from
the $j^\text{th}$ set of inputs and denote the central values by $a_j(0)$. We calculate the uncertainty on each Dalitz
plot parameter by
\begin{equation}
	\sigma(a_i) = \sqrt{\sum_k \big(a_i(k) - a_i(0)\big)^2} \eolc
	\label{eq:resultsDalitzParamsCov}
\end{equation}
and then estimate the correlation matrix by
\begin{equation}
	\text{Corr}(a_i,a_j) = \frac{\sum_k \big(a_i(k) - a_i(0)\big)\big(a_j(k) - a_j(0)\big)}%
								{\sqrt{\sum_k \big(a_i(k) - a_i(0)\big)^2}\, \sqrt{\sum_k \big(a_j(k) - a_j(0)\big)^2} } \eolp
	\label{eq:resultsDalitzParamsCorr}
\end{equation}
We have checked that the error band on the Dalitz plot parametrisation obtained in this manner agrees well with the
actual error band on the Dalitz distribution.

Table~\ref{tab:resultsMatchErrorsQr} lists all the separate contributions to the uncertainties on $Q$, $r$, and
$\alpha$ together with the total uncertainty which is the sum of their squares.

\subsection{Comparison with the result of the original calculation}

In this section, we compare our results to those of the original calculation by Walker~\cite{Walker1998}.
While the procedure is still the same, several ingredients differ between his and our computation.
We have a new representation of the $\pi \pi$ phase shifts, and  the values of the low-energy constant $L_3$ and other
physical constants have changed. The total decay width of the $\eta$ has been modified considerably because recently,
measurements that make use of the Primakoff effect have been removed from the PDG average \cite{PDG2004}. We have tried
to reproduce Walker's inputs as accurately as possible and found the results in the second row of
Table~\ref{tab:resultsWalkerComparison} in almost perfect agreement with Ref.~\cite{Walker1998}. The remainder of
the table lists the effect that the change in each input to the calculation has on the final results, which are given
in the last row.

Leutwyler has recently presented an update of the analysis by Walker, where he took only the new value for the decay
width into account but did not recalculate the solution of the dispersion relations with new
input~\cite{Leutwyler2009}. The table explains, why his value, $Q = 22.3 \pm 0.8$, is different from the outcome of our
analysis: While the new decay width lowers the value of $Q$ considerably, all the other new inputs work in the opposite
direction, such that in the end, $Q$ is almost not changed at all.

\begin{table}[tb]
\renewcommand{\arraystretch}{1.6}
\begin{center}\begin{tabular}{lccc}
								& $Q (\pi^+ \pi^- \pi^0)$	& $Q (3 \pi^0)$	& $r$			\\ \hline
Results from Ref.~\cite{Walker1998}
								& 22.8							& 22.9				& 1.43		\\
Our reproduction			& 22.74							& 22.87				& 1.425		\\ \hline
$\delta_I(s)$				& +0.14							& +0.13				& $-$0.004	\\
$L_3$							& +0.07							& +0.11				& +0.008		\\
$m_K$							& +0.22							& +0.21				& +0.000		\\
$\mpi$, $\meta$, $\Fpi$, $\Delta_F$
								& +0.02							& +0.02				& $-$0.001	\\
$\Gamma$ 					& $-$0.45						& $-$0.62			& ---		\\ \hline
Our result					& 22.74							& 22.72				& 1.428
\end{tabular}\end{center}

\caption{Comparison of the result from Ref.~\cite{Walker1998} with our updated result. We have tried to reproduce the
inputs used by Walker as accurately as possible and obtained the results in row two. The next rows list the change in
the result coming from updating the inputs, leading to our final result in the last row.}
\label{tab:resultsWalkerComparison}
\end{table}

That the two values for $Q$ have moved closer together is due to the fact that not only the total decay width of the
$\eta$ has changed but also the branching ratios of the charged and the neutral channel. While the branching ratio of
the neutral channel has increased, the one for the charged channel has decreased such that the shift in $Q (3 \pi^0)$
is larger than in $Q (\pi^+ \pi^- \pi^0)$. The branching ratio $r$ is calculated from the ratio of the phase space
integrals and thus not affected by the experimental decay width.

\section{Fit to experimental data}

%
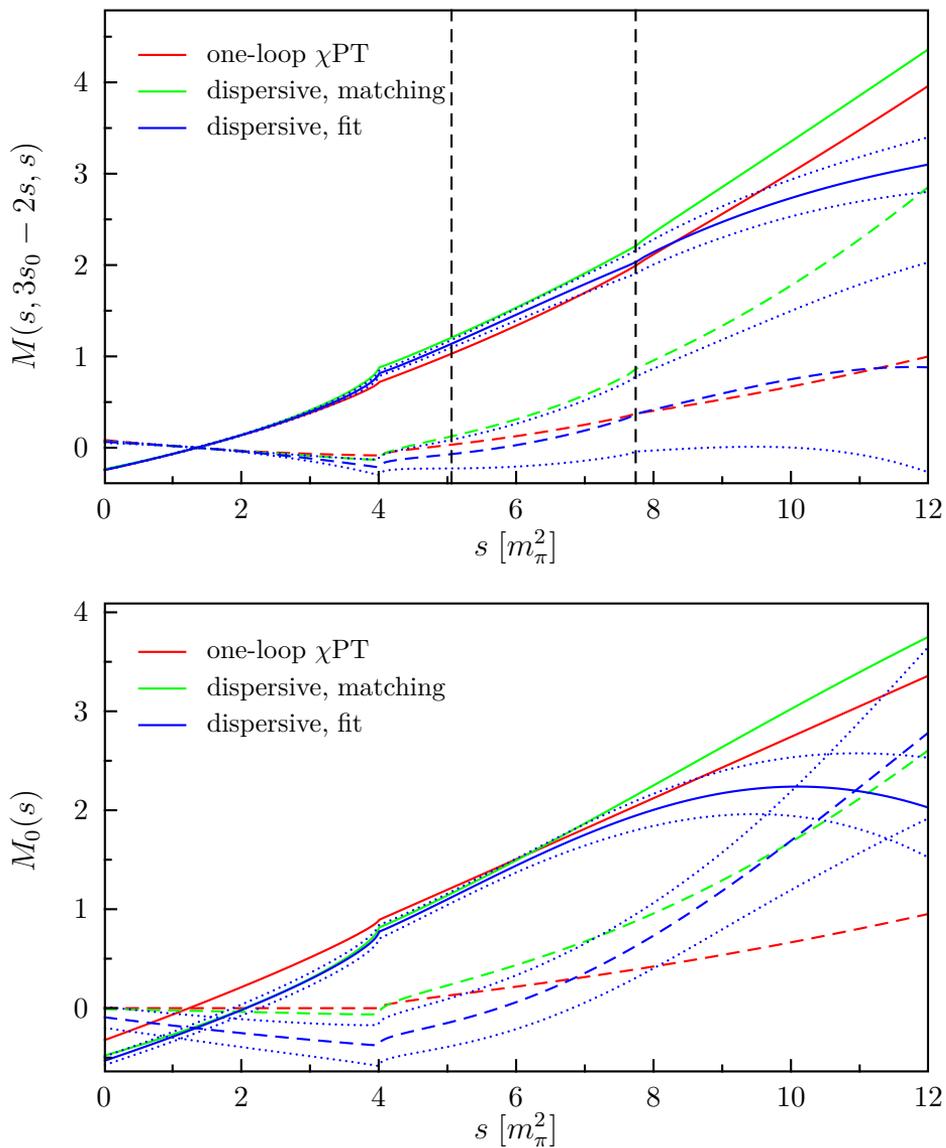
\begin{figure}[p]
\psset{xunit=.9cm,yunit=1.2cm}
\begin{center}\begin{pspicture*}(-1.4,-1.3)(12.2,4.81)

\psset{linewidth=\mylw,plotstyle=line}

\psset{linecolor=red}
\readdata{\data}{data/ChPT/ReM1loop.dat}
\dataplot{\data}
\readdata{\data}{data/ChPT/ImM1loop.dat}
\dataplot[linestyle=dashed]{\data}

\psset{linecolor=green}
\readdata{\data}{data/match/ReMCentral.dat}
\dataplot{\data}
\readdata{\data}{data/match/ImMCentral.dat}
\dataplot[linestyle=dashed]{\data}

\psset{linecolor=blue}
\readdata{\data}{data/fit/ReMCentral.dat}
\dataplot{\data}
\readdata{\data}{data/fit/ReMLower.dat}
\dataplot[linestyle=dotted,dotsep=.05,linewidth=\mydotlw]{\data}
\readdata{\data}{data/fit/ReMUpper.dat}
\dataplot[linestyle=dotted,dotsep=.05,linewidth=\mydotlw]{\data}
\readdata{\data}{data/fit/ImMCentral.dat}
\dataplot[linestyle=dashed]{\data}
\readdata{\data}{data/fit/ImMLower.dat}
\dataplot[linestyle=dotted,dotsep=.05,linewidth=\mydotlw]{\data}
\readdata{\data}{data/fit/ImMUpper.dat}
\dataplot[linestyle=dotted,dotsep=.05,linewidth=\mydotlw]{\data}

\psset{linecolor=black}
\psframe(0,-0.4)(12.01,4.8)
\multips(2,-0.4)(2,0){5}{\psline(0,5pt)}
\multips(1,-0.4)(2,0){6}{\psline(0,3pt)}
\multido{\n=0+2}{7}{\uput{.2}[270](\n,-0.4){\n}}
\multips(0,0)(0,1){5}{\psline(5pt,0)}
\multips(0,0.5)(0,1){5}{\psline(3pt,0)}
\multido{\n=0+1}{5}{\uput{0.2}[180](0,\n){\n}}
\uput{0.6}[270](6,-0.4){\large $s\; [\mpi^2]$}
\uput{0.8}[180]{90}(0,2.1){\large $M(s,3 s_0 - 2s,s)$}

\psline[linestyle=dashed](5.06,-0.4)(5.06,4.8)
\psline[linestyle=dashed](7.74,-0.4)(7.74,4.8)

\psline[linecolor=red](0.5,4.3)(1.2,4.3) \rput[l](1.5,4.3){\small one-loop \chpt}
\psline[linecolor=green](0.5,3.9)(1.2,3.9) \rput[l](1.5,3.9){\small dispersive, matching}
\psline[linecolor=blue](0.5,3.5)(1.2,3.5) \rput[l](1.5,3.5){\small dispersive, fit}

\end{pspicture*}\end{center}
\psset{xunit=.9cm,yunit=1.3cm}
\begin{center}\begin{pspicture*}(-1.4,-1.5)(12.2,4.11)

\psset{linewidth=\mylw,plotstyle=line}

\psset{linecolor=red}
\readdata{\data}{data/ChPT/ReM01loop.dat}
\dataplot{\data}
\readdata{\data}{data/ChPT/ImM01loop.dat}
\dataplot[linestyle=dashed]{\data}

\psset{linecolor=green}
\readdata{\data}{data/match/ReM0Central.dat}
\dataplot{\data}
\readdata{\data}{data/match/ImM0Central.dat}
\dataplot[linestyle=dashed]{\data}

\psset{linecolor=blue}
\readdata{\data}{data/fit/ReM0Central.dat}
\dataplot{\data}
\readdata{\data}{data/fit/ReM0Lower.dat}
\dataplot[linestyle=dotted,dotsep=.05,linewidth=\mydotlw]{\data}
\readdata{\data}{data/fit/ReM0Upper.dat}
\dataplot[linestyle=dotted,dotsep=.05,linewidth=\mydotlw]{\data}
\readdata{\data}{data/fit/ImM0Central.dat}
\dataplot[linestyle=dashed]{\data}
\readdata{\data}{data/fit/ImM0Lower.dat}
\dataplot[linestyle=dotted,dotsep=.05,linewidth=\mydotlw]{\data}
\readdata{\data}{data/fit/ImM0Upper.dat}
\dataplot[linestyle=dotted,dotsep=.05,linewidth=\mydotlw]{\data}

\psset{linecolor=black}
\psframe(0,-0.65)(12.01,4.1)
\multips(2,-0.65)(2,0){5}{\psline(0,5pt)}
\multips(1,-0.65)(2,0){6}{\psline(0,3pt)}
\multido{\n=0+2}{7}{\uput{.2}[270](\n,-0.65){\n}}
\multips(0,0)(0,1){5}{\psline(5pt,0)}
\multips(0,-0.5)(0,1){5}{\psline(3pt,0)}
\multido{\n=0+1}{5}{\uput{0.2}[180](0,\n){\n}}
\uput{0.6}[270](6,-0.6){\large $s\; [\mpi^2]$}
\uput{0.8}[180]{90}(0,1.8){\large $M_0(s)$}

\psline[linecolor=red](0.5,3.6)(1.2,3.6) \rput[l](1.5,3.6){\small one-loop \chpt}
\psline[linecolor=green](0.5,3.23)(1.2,3.23) \rput[l](1.5,3.23){\small dispersive, matching}
\psline[linecolor=blue](0.5,2.86)(1.2,2.86) \rput[l](1.5,2.86){\small dispersive, fit}

\end{pspicture*}\end{center}

\caption{The figure shows the decay amplitude $M(s,t,u)$ along the line $s=u$ in the upper panel and the isospin
amplitude $M_0(s)$ in the lower panel. In both cases, solid and dashed lines stand for the central values of the real
and imaginary part, respectively, dotted lines for the error band. The vertical dashed lines mark the boundaries of the
physical region.}
\label{fig:resultsFitMFinal1}
\end{figure}

\begin{figure}[p]
\psset{xunit=.9cm,yunit=23.1cm}
\begin{center}\begin{pspicture*}(-1.85,-0.235)(12.2,0.0865)

\psset{linewidth=\mylw,plotstyle=line}

\psset{linecolor=red}
\readdata{\data}{data/ChPT/ReM11loop.dat}
\dataplot{\data}
\readdata{\data}{data/ChPT/ImM11loop.dat}
\dataplot[linestyle=dashed]{\data}

\psset{linecolor=green}
\readdata{\data}{data/match/ReM1Central.dat}
\dataplot{\data}
\readdata{\data}{data/match/ImM1Central.dat}
\dataplot[linestyle=dashed]{\data}

\psset{linecolor=blue}
\readdata{\data}{data/fit/ReM1Central.dat}
\dataplot{\data}
\readdata{\data}{data/fit/ReM1Lower.dat}
\dataplot[linestyle=dotted,dotsep=.05,linewidth=\mydotlw]{\data}
\readdata{\data}{data/fit/ReM1Upper.dat}
\dataplot[linestyle=dotted,dotsep=.05,linewidth=\mydotlw]{\data}
\readdata{\data}{data/fit/ImM1Central.dat}
\dataplot[linestyle=dashed]{\data}
\readdata{\data}{data/fit/ImM1Lower.dat}
\dataplot[linestyle=dotted,dotsep=.05,linewidth=\mydotlw]{\data}
\readdata{\data}{data/fit/ImM1Upper.dat}
\dataplot[linestyle=dotted,dotsep=.05,linewidth=\mydotlw]{\data}

\psset{linecolor=black}
\psframe(0,-0.185)(12.01,0.086)
\multips(2,-0.185)(2,0){5}{\psline(0,5pt)}
\multips(1,-0.185)(2,0){6}{\psline(0,3pt)}
\multido{\n=0+2}{7}{\uput{.2}[270](\n,-0.185){\n}}
\multips(0,-0.15)(0,0.05){6}{\psline(5pt,0)}
\multips(0,-0.175)(0,0.05){6}{\psline(3pt,0)}
\multido{\n=-0.15+0.05}{6}{\uput{0.2}[180](0,\n){\n}}
\uput{0.6}[270](6,-0.185){\large $s\; [\mpi^2]$}
\uput{1.2}[180]{90}(0,-0.04){\large $\mpi^2\, M_1(s)$}

\psline[linecolor=red](0.5,-0.10)(1.2,-0.10) \rput[l](1.5,-0.10){\small one-loop \chpt}
\psline[linecolor=green](0.5,-0.12)(1.2,-0.12) \rput[l](1.5,-0.12){\small dispersive, matching}
\psline[linecolor=blue](0.5,-0.14)(1.2,-0.14) \rput[l](1.5,-0.14){\small dispersive, fit}

\end{pspicture*}\end{center}
\psset{xunit=.9cm,yunit=20.7cm}
\begin{center}\begin{pspicture*}(-1.85,-.175)(12.2,.181)

\psset{linewidth=\mylw,plotstyle=line}

\psset{linecolor=red}
\readdata{\data}{data/ChPT/ReM21loop.dat}
\dataplot{\data}
\readdata{\data}{data/ChPT/ImM21loop.dat}
\dataplot[linestyle=dashed]{\data}

\psset{linecolor=green}
\readdata{\data}{data/match/ReM2Central.dat}
\dataplot{\data}
\readdata{\data}{data/match/ImM2Central.dat}
\dataplot[linestyle=dashed]{\data}

\psset{linecolor=blue}
\readdata{\data}{data/fit/ReM2Central.dat}
\dataplot{\data}
\readdata{\data}{data/fit/ReM2Lower.dat}
\dataplot[linestyle=dotted,dotsep=.05,linewidth=\mydotlw]{\data}
\readdata{\data}{data/fit/ReM2Upper.dat}
\dataplot[linestyle=dotted,dotsep=.05,linewidth=\mydotlw]{\data}
\readdata{\data}{data/fit/ImM2Central.dat}
\dataplot[linestyle=dashed]{\data}
\readdata{\data}{data/fit/ImM2Lower.dat}
\dataplot[linestyle=dotted,dotsep=.05,linewidth=\mydotlw]{\data}
\readdata{\data}{data/fit/ImM2Upper.dat}
\dataplot[linestyle=dotted,dotsep=.05,linewidth=\mydotlw]{\data}

\psset{linecolor=black}
\psframe(0,-0.125)(12.01,0.18)
\multips(2,-0.125)(2,0){5}{\psline(0,5pt)}
\multips(1,-0.125)(2,0){6}{\psline(0,3pt)}
\multido{\n=0+2}{7}{\uput{.2}[270](\n,-0.125){\n}}
\multips(0,-0.1)(0,.1){3}{\psline(5pt,0)}
\multips(0,-0.05)(0,0.1){3}{\psline(3pt,0)}
\multido{\n=-0.1+0.1}{5}{\uput{0.2}[180](0,\n){\n}}
\uput{0.6}[270](6,-0.125){\large $s\; [\mpi^2]$}
\uput{1.0}[180]{90}(0,0.03){\large $M_2(s)$}

\psline[linecolor=red](0.5,0.15)(1.2,0.15) \rput[l](1.5,0.15){\small one-loop \chpt}
\psline[linecolor=green](0.5,0.125)(1.2,0.125) \rput[l](1.5,0.125){\small dispersive, matching}
\psline[linecolor=blue](0.5,0.1)(1.2,0.1) \rput[l](1.5,0.1){\small dispersive, fit}

\end{pspicture*}\end{center}

\caption{The figure shows the isospin amplitude $M_1(s)$ in the upper panel and the isospin amplitude $M_2(s)$ in
the lower panel. $M_1(s)$ is multiplied by $\mpi^2$ in order to make it dimensionless. In both cases, solid and dashed
lines stand for the central values of the real and imaginary part, respectively, dotted lines for the error band.}
\label{fig:resultsFitMFinal2}
\end{figure}

The Dalitz plot distribution that we obtained from matching to the one-loop result is in clear disagreement with
the measurement by the KLOE collaboration. Also, our result for the sign of the slope parameter $\alpha$ conflicts with
the very accurate PDG average. Even though we have included many final-state rescattering processes, we fail to remove
the discrepancy between experiment and \chpt, if the subtraction constants are fixed by a pure matching to the one-loop
result. In order to reduce the impact of \chpt\ to a minimum, we fit our representation of the amplitude in terms
of the subtraction constants to the measured Dalitz distribution from KLOE. Chiral perturbation theory is then only
required in order to fix the normalisation. How we proceed has been described in Chapter~\ref{chp:Solution} and we
present now our results from this approach. 

The fit successfully approximates the one-loop result around the Adler zero and the experimental Dalitz plot
distribution as can be seen from
\begin{equation}
	\frac{\chi^2}{\textit{d.o.f.}} = \frac{141.5}{156} \approx 0.91 \eolc
\end{equation}
which corresponds to a $p$-value of 79 per cent. The success of the fit is also confirmed by the the results that
we present below.

%
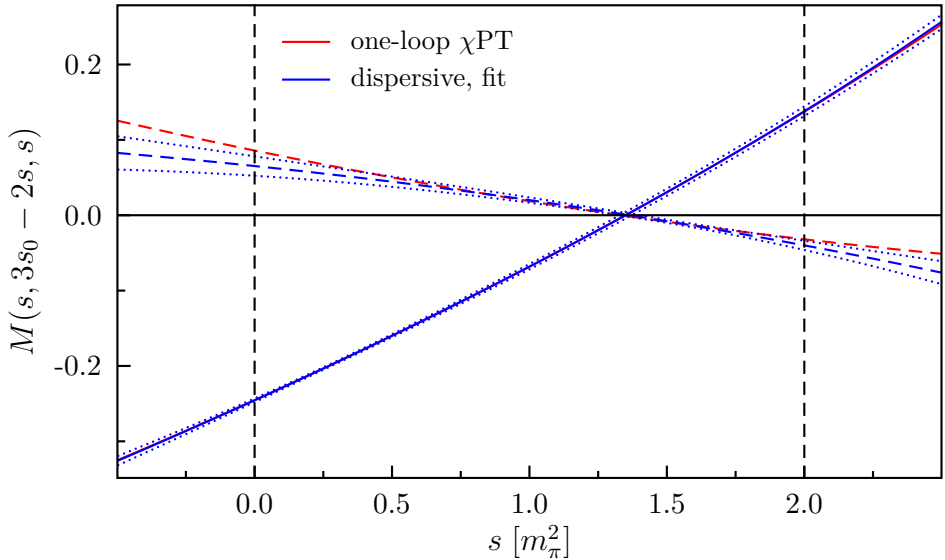
\begin{figure}[t]
\psset{xunit=3.6cm,yunit=9.9cm}
\begin{center}\begin{pspicture*}(-0.9,-0.47)(2.51,0.3)

\psset{linewidth=\mylw,plotstyle=line}

\psset{linecolor=red}
\readdata{\data}{data/ChPT/ReM1loopAdler.dat}
\dataplot{\data}
\readdata{\data}{data/ChPT/ImM1loopAdler.dat}
\dataplot[linestyle=dashed]{\data}

\psset{linecolor=blue}
\readdata{\data}{data/fit/ReMAdlerCentral.dat}
\dataplot{\data}
\readdata{\data}{data/fit/ReMAdlerLower.dat}
\dataplot[linestyle=dotted,dotsep=.05,linewidth=\mydotlw]{\data}
\readdata{\data}{data/fit/ReMAdlerUpper.dat}
\dataplot[linestyle=dotted,dotsep=.05,linewidth=\mydotlw]{\data}
\readdata{\data}{data/fit/ImMAdlerCentral.dat}
\dataplot[linestyle=dashed]{\data}
\readdata{\data}{data/fit/ImMAdlerLower.dat}
\dataplot[linestyle=dotted,dotsep=.05,linewidth=\mydotlw]{\data}
\readdata{\data}{data/fit/ImMAdlerUpper.dat}
\dataplot[linestyle=dotted,dotsep=.05,linewidth=\mydotlw]{\data}

\psset{linecolor=black}
\psframe(-0.503,-0.35)(2.503,0.28)
\psline(-0.5,0)(2.5,0)
\multips(0,-0.35)(0.5,0){5}{\psline(0,5pt)}
\multips(-0.25,-0.35)(0.5,0){6}{\psline(0,3pt)}
\multido{\n=0.0+.5}{5}{\uput{.2}[270](\n,-0.35){\n}}
\multips(-0.5,-0.2)(0,.2){3}{\psline(5pt,0)}
\multips(-0.5,-0.3)(0,.2){3}{\psline(3pt,0)}
\multido{\n=-0.2+0.2}{3}{\uput{0.2}[180](-0.5,\n){\n}}
\uput{0.6}[270](1,-0.35){\large $s\; [\mpi^2]$}
\uput{1.0}[180]{90}(-0.5,-0.035){\large $M(s,3 s_0 - 2s,s)$}

\psline[linestyle=dashed](0,-0.35)(0,0.28)
\psline[linestyle=dashed](2,-0.35)(2,0.28)

\psline[linecolor=red](0.1,0.23)(0.275,0.23) \rput[l](0.35,0.23){\small one-loop \chpt}
\psline[linecolor=blue](0.1,0.18)(0.275,0.18) \rput[l](0.35,0.18){\small dispersive, fit}

\end{pspicture*}\end{center}

\caption{The figure shows the decay amplitude $M(s,t,u)$ along the line $s=u$ in the vicinity of the Adler zero. For
the fit, we use five equally distributed points between 0 and $2 \mpi^2$. This interval is indicated by the vertical
dashed lines. One can see that the dispersive amplitude approximates the one-loop amplitude rather well and has small
uncertainties. Solid lines stand for the real part, dashed lines for the imaginary part and dotted lines for the error
band.}
\label{fig:resultsFitMAdler}
\end{figure}

In Figs.~\ref{fig:resultsFitMFinal1} and \ref{fig:resultsFitMFinal2} we compare the decay amplitude and the isospin
amplitudes we obtain from the fit with the results from the matching and the one-loop amplitude from \chpt. It is
striking that the uncertainties on the amplitude, in particular on its imaginary part, have become much larger. The
reason for this is simply that the fit to the data in the physical region fixes the absolute value of the amplitude, but
does not constrain its phase at all. Information on the shape of the amplitude itself comes exclusively from the fit
around the Adler zero and, as the plot in Fig.~\ref{fig:resultsFitMAdler} shows, in this region the uncertainties are
indeed small. From the same figure, one can also see that the one-loop amplitude is approximated very well.

%
\begin{figure}[tb]

\begin{center}
	\scalebox{.95}{\input{figs/dalitz2.tex}}
\end{center}
	\caption{The Dalitz plot distribution obtained from the dispersive analysis.}
\label{fig:resultsFitDalitz}
\end{figure}

%
\begin{figure}[p]
\psset{xunit=5cm,yunit=61.8cm}

\begin{center}\begin{pspicture*}(-1.36,0.825)(1.05,0.938)

\psset{linewidth=\mylw,plotstyle=curve}

\readdata{\data}{data/ChPT/dalitzXc1loop.dat}
\dataplot[linecolor=red]{\data}

\readdata{\data}{data/exp/dalitzXcKLOECentral.dat}
\dataplot[linecolor=black]{\data}
\readdata{\data}{data/exp/dalitzXcKLOEUpper.dat}
\dataplot[linestyle=dotted,dotsep=.05,linecolor=black,linewidth=\mydotlw]{\data}
\readdata{\data}{data/exp/dalitzXcKLOELower.dat}
\dataplot[linestyle=dotted,dotsep=.05,linecolor=black,linewidth=\mydotlw]{\data}

\readdata{\data}{data/match/dalitzXcCentral.dat}
\dataplot[linecolor=green]{\data}
\readdata{\data}{data/match/dalitzXcLower.dat}
\dataplot[linestyle=dotted,dotsep=.05,linecolor=green,linewidth=\mydotlw]{\data}
\readdata{\data}{data/match/dalitzXcUpper.dat}
\dataplot[linestyle=dotted,dotsep=.05,linecolor=green,linewidth=\mydotlw]{\data}

\readdata{\data}{data/fit/dalitzXcCentral.dat}
\dataplot[linecolor=blue]{\data}
\readdata{\data}{data/fit/dalitzXcLower.dat}
\dataplot[linestyle=dotted,dotsep=.05,linecolor=blue,linewidth=\mydotlw]{\data}
\readdata{\data}{data/fit/dalitzXcUpper.dat}
\dataplot[linestyle=dotted,dotsep=.05,linecolor=blue,linewidth=\mydotlw]{\data}

\psframe(-1.002,0.84)(1.002,0.937)
\multips(-.5,0.84)(0.5,0){3}{\psline(0,5pt)}
\multips(-0.75,0.84)(0.5,0){4}{\psline(0,3pt)}
\multido{\n=-1.0+0.5}{5}{\uput{.2}[270](\n,0.84){\n}}
\multips(-1,0.86)(0,0.02){4}{\psline(5pt,0)}
\multips(-1,0.85)(0,0.02){5}{\psline(3pt,0)}
\multido{\n=0.84+0.02}{5}{\uput{0.2}[180](-1,\n){\n}}
\uput{0.6}[270](0,0.84){\large $X$}
\uput{1.2}[180]{90}(-1,0.89){\large $\Gamma(X,0.125)$}

\psline[linestyle=dashed](-0.936,0.84)(-0.936,0.937)
\psline[linestyle=dashed](0.936,0.84)(0.936,0.937)

\psline[linecolor=red](-0.6,0.93)(-0.45,0.93) \rput[l](-0.4,0.93){\small one-loop \chpt}
\psline[linecolor=black](-0.6,0.922)(-0.45,0.922) \rput[l](-0.4,0.922){\small KLOE}
\psline[linecolor=green](-0.6,0.914)(-0.45,0.914) \rput[l](-0.4,0.914){\small dispersive, matching}
\psline[linecolor=blue](-0.6,0.906)(-0.45,0.906) \rput[l](-0.4,0.906){\small dispersive, fit}

\end{pspicture*}\end{center}

\vfill

\psset{xunit=5cm,yunit=2.08cm}
\begin{center}\begin{pspicture*}(-1.36,-0.45)(1.05,2.99)

\psset{linewidth=\mylw,plotstyle=curve}

\readdata{\data}{data/ChPT/dalitzYc1loop.dat}
\dataplot[linecolor=red]{\data}

\readdata{\data}{data/exp/dalitzYcKLOECentral.dat}
\dataplot[linecolor=black]{\data}
\readdata{\data}{data/exp/dalitzYcKLOEUpper.dat}
\dataplot[linestyle=dotted,dotsep=.05,linecolor=black,linewidth=\mydotlw]{\data}
\readdata{\data}{data/exp/dalitzYcKLOELower.dat}
\dataplot[linestyle=dotted,dotsep=.05,linecolor=black,linewidth=\mydotlw]{\data}

\readdata{\data}{data/match/dalitzYcCentral.dat}
\dataplot[linecolor=green]{\data}
\readdata{\data}{data/match/dalitzYcLower.dat}
\dataplot[linestyle=dotted,dotsep=.05,linecolor=green,linewidth=\mydotlw]{\data}
\readdata{\data}{data/match/dalitzYcUpper.dat}
\dataplot[linestyle=dotted,dotsep=.05,linecolor=green,linewidth=\mydotlw]{\data}

\readdata{\data}{data/fit/dalitzYcCentral.dat}
\dataplot[linecolor=blue]{\data}
\readdata{\data}{data/fit/dalitzYcLower.dat}
\dataplot[linestyle=dotted,dotsep=.05,linecolor=blue,linewidth=\mydotlw]{\data}
\readdata{\data}{data/fit/dalitzYcUpper.dat}
\dataplot[linestyle=dotted,dotsep=.05,linecolor=blue,linewidth=\mydotlw]{\data}

\psframe(-1.002,0.1)(1.002,2.98)
\multips(-.5,0.1)(0.5,0){3}{\psline(0,5pt)}
\multips(-0.75,0.1)(0.5,0){4}{\psline(0,3pt)}
\multido{\n=-1.0+0.5}{5}{\uput{.2}[270](\n,0.1){\n}}
\multips(-1,0.5)(0,0.5){5}{\psline(5pt,0)}
\multips(-1,0.75)(0,0.5){5}{\psline(3pt,0)}
\multido{\n=0.5+0.5}{5}{\uput{0.2}[180](-1,\n){\n}}
\uput{0.7}[270](0,0.1){\large $Y$}
\uput{1.2}[180]{90}(-1,1.45){\large $\Gamma(0.125,Y)$}

\psline[linestyle=dashed](-0.986,0.1)(-0.986,2.98)
\psline[linestyle=dashed](0.890,0.1)(0.890,2.98)

\psline[linecolor=red](0,2.7)(0.15,2.7) \rput[l](0.2,2.7){\small one-loop \chpt}
\psline[linecolor=black](0,2.47)(0.15,2.47) \rput[l](0.2,2.47){\small KLOE}
\psline[linecolor=green](0,2.24)(0.15,2.24) \rput[l](0.2,2.24){\small dispersive, matching}
\psline[linecolor=blue](0,2.01)(0.15,2.01) \rput[l](0.2,2.01){\small dispersive, fit}

\end{pspicture*}\end{center}
\caption{The figure shows two cuts through the Dalitz plot for $\eta \to \pi^+ \pi^- \pi^0$. The upper panel depicts
the distribution along the line with $Y = 0.125$, the lower panel along the line with $X = 0.125$. The solid lines
represent the central values and the dotted lines the error band. The dashed lines mark the boundaries of the physical
region.}
\label{fig:resultsFitDalitzXYc}
\end{figure}

%
\begin{figure}[p]

\psset{xunit=10cm,yunit=5.6cm}
\begin{center}\begin{pspicture*}(-0.18,-0.075)(1.04,1.06)

\psset{linewidth=\mylw,plotstyle=curve}

\readdata{\data}{data/ChPT/dalitzYc1loop.dat}
\dataplot[linecolor=red]{\data}

\readdata{\data}{data/exp/dalitzYcKLOECentral.dat}
\dataplot[linecolor=black]{\data}
\readdata{\data}{data/exp/dalitzYcKLOEUpper.dat}
\dataplot[linestyle=dotted,dotsep=.05,linecolor=black,linewidth=\mydotlw]{\data}
\readdata{\data}{data/exp/dalitzYcKLOELower.dat}
\dataplot[linestyle=dotted,dotsep=.05,linecolor=black,linewidth=\mydotlw]{\data}

\readdata{\data}{data/match/dalitzYcCentral.dat}
\dataplot[linecolor=green]{\data}
\readdata{\data}{data/match/dalitzYcLower.dat}
\dataplot[linestyle=dotted,dotsep=.05,linecolor=green,linewidth=\mydotlw]{\data}
\readdata{\data}{data/match/dalitzYcUpper.dat}
\dataplot[linestyle=dotted,dotsep=.05,linecolor=green,linewidth=\mydotlw]{\data}

\readdata{\data}{data/fit/dalitzYcCentral.dat}
\dataplot[linecolor=blue]{\data}
\readdata{\data}{data/fit/dalitzYcLower.dat}
\dataplot[linestyle=dotted,dotsep=.05,linecolor=blue,linewidth=\mydotlw]{\data}
\readdata{\data}{data/fit/dalitzYcUpper.dat}
\dataplot[linestyle=dotted,dotsep=.05,linecolor=blue,linewidth=\mydotlw]{\data}

\psframe[linecolor=white,fillstyle=solid,fillcolor=white](-0.5,0.9)(0,1.5)

\psframe(-0.001,0.1)(1.001,1.05)
\multips(0.25,0.1)(0.25,0){3}{\psline(0,5pt)}
\multips(0.125,0.1)(0.25,0){5}{\psline(0,3pt)}
\multido{\n=0.00+0.25}{5}{\uput{.2}[270](\n,0.1){\n}}
\multips(0,0.25)(0,0.25){4}{\psline(5pt,0)}
\multips(0,0.125)(0,0.25){4}{\psline(3pt,0)}
\multido{\n=0.25+0.25}{4}{\uput{0.2}[180](0,\n){\n}}
\uput{0.7}[270](0.5,0.1){\large $Y$}
\uput{1.2}[180]{90}(0,0.55){\large $\Gamma(0.125,Y)$}

\psline[linestyle=dashed](0.890,0.1)(0.890,1.05)

\psline[linecolor=red](0.45,0.94)(0.52,0.94) \rput[l](0.55,0.94){\small one-loop \chpt}
\psline[linecolor=black](0.45,0.85)(0.52,0.85) \rput[l](0.55,0.85){\small KLOE}
\psline[linecolor=green](0.45,0.76)(0.52,0.76) \rput[l](0.55,0.76){\small dispersive, matching}
\psline[linecolor=blue](0.45,0.67)(0.52,0.67) \rput[l](0.55,0.67){\small dispersive, fit}

\end{pspicture*}\end{center}
\caption{The figure shows a detail of the cut through the Dalitz plot for $\eta \to \pi^+ \pi^- \pi^0$ along the line
with $X = 0.125$. The solid lines represent the central values and the dotted lines the error band. The dashed line
marks the boundary of the physical region.}
\label{fig:resultsFitDalitzXYcDetail}

\vfill

\psset{xunit=5cm,yunit=32cm}
\begin{center}\begin{pspicture*}(-1.36,0.875)(1.05,1.081)

\psset{linewidth=\mylw,plotstyle=curve}

\readdata{\data}{data/ChPT/dalitzX01loop.dat}
\dataplot[linecolor=red]{\data}

\readdata{\data}{data/exp/dalitzX0PDGCentral.dat}
\dataplot[linecolor=black]{\data}
\readdata{\data}{data/exp/dalitzX0PDGUpper.dat}
\dataplot[linestyle=dotted,dotsep=.05,linecolor=black,linewidth=\mydotlw]{\data}
\readdata{\data}{data/exp/dalitzX0PDGLower.dat}
\dataplot[linestyle=dotted,dotsep=.05,linecolor=black,linewidth=\mydotlw]{\data}

\readdata{\data}{data/match/dalitzX0Central.dat}
\dataplot[linecolor=green]{\data}
\readdata{\data}{data/match/dalitzX0Lower.dat}
\dataplot[linestyle=dotted,dotsep=.05,linecolor=green,linewidth=\mydotlw]{\data}
\readdata{\data}{data/match/dalitzX0Upper.dat}
\dataplot[linestyle=dotted,dotsep=.05,linecolor=green,linewidth=\mydotlw]{\data}

\readdata{\data}{data/fit/dalitzX0Central.dat}
\dataplot[linecolor=blue]{\data}
\readdata{\data}{data/fit/dalitzX0Lower.dat}
\dataplot[linestyle=dotted,dotsep=.05,linecolor=blue,linewidth=\mydotlw]{\data}
\readdata{\data}{data/fit/dalitzX0Upper.dat}
\dataplot[linestyle=dotted,dotsep=.05,linecolor=blue,linewidth=\mydotlw]{\data}

\psframe(-1.002,0.905)(1.002,1.08)
\multips(-.5,0.905)(0.5,0){3}{\psline(0,5pt)}
\multips(-0.75,0.905)(0.5,0){4}{\psline(0,3pt)}
\multido{\n=-1.+0.5}{5}{\uput{.2}[270](\n,0.905){\n}}
\multips(-1,0.94)(0,0.04){4}{\psline(5pt,0)}
\multips(-1,0.92)(0,0.04){4}{\psline(3pt,0)}
\multido{\n=0.94+0.04}{4}{\uput{0.2}[180](-1,\n){\n}}
\uput{0.65}[270](0,0.905){\large $X$}
\uput{1.2}[180]{90}(-1,1.005){\large $\Gamma(X,0)$}

\psline[linestyle=dashed](-0.941,0.905)(-0.941,1.08)
\psline[linestyle=dashed](0.941,0.905)(0.941,1.08)

\psline[linecolor=red](-0.45,1.067)(-0.3,1.067) \rput[l](-0.25,1.067){\small one-loop \chpt}
\psline[linecolor=black](-0.45,1.053)(-0.3,1.053) \rput[l](-0.25,1.053){\small PDG}
\psline[linecolor=green](-0.45,1.039)(-0.3,1.039) \rput[l](-0.25,1.039){\small dispersive, matching}
\psline[linecolor=blue](-0.45,1.025)(-0.3,1.025) \rput[l](-0.25,1.025){\small dispersive, fit}

\end{pspicture*}\end{center}
\caption{The figure shows a cut along the line $Y = 0$ through the Dalitz plot for $\eta \to 3 \pi^0$. The solid lines
represent the central values and the dotted lines the error band. The vertical dashed lines mark the boundaries of the
physical region.}
\label{fig:resultsFitDalitzXY0}
\end{figure}

In Fig.~\ref{fig:resultsFitDalitz}, we show the Dalitz plot distribution and in Fig.~\ref{fig:resultsFitDalitzXYc}
two cuts through the Dalitz plot, both for the charged channel. In addition, we have plotted a detail
of the cut along $X=0.125$ in Fig.~\ref{fig:resultsFitDalitzXYcDetail}. The KLOE result is well approximated
by our curve. Figure~\ref{fig:resultsFitDalitzXY0} shows a cut through the Dalitz plot for the neutral channel. The
dispersive amplitude now indeed leads to the correct sign for the curvature.

For the Dalitz plot parameters in the charged channel, we find
\begin{align}\begin{aligned}
	a &= -1.077 \pm 0.025 \eolc \qquad 	&b &= 0.126 \pm 0.015 \eolc \\[1mm]
	d &= 0.062 \pm 0.008 \eolc \qquad 	&f &= 0.107 \pm 0.017 \eolc \\[1mm]
	h &= -0.037 \pm 0.008 \eolc
\end{aligned}\end{align}
with the correlation matrix
\begin{equation}
	\begin{array}{cccccc}
		  & a & b      & d      & f 		& h		\\[2mm]
		a & 1 & -0.371 & -0.335	& -0.179	& 0.363	\\[2mm]
		b &   & 1		& -0.546	& -0.802 & 0.425	\\[2mm]
		d &   & 			& 1		& 0.672  & -0.937	\\[2mm]
		f &   &			&			& 1		& -0.656	\\[2mm]
		h &	&			&			&			& 1
	\end{array}
\end{equation}
Again we have checked that parameters that vanish due to charge conjugation turn out to be consistent with zero if
included in the fit, while the other parameters are not affected. As opposed to the result by KLOE, we find the
parameter $h$ clearly different from zero and our value for $f$ is almost two standard deviations two small. The other
parameters agree with the KLOE result. Since $d$ and $h$ have relatively large negative correlation, it is to be
expected that setting $h$ to zero, will result in a smaller value for $d$, which is acceptable as our value for $d$ is
somewhat too large. Indeed, removing $h$ from the fit leads to
\begin{equation}
	a = -1.084 \eolc \quad b = 0.127 \eolc \quad d = 0.053 \eolc \quad f = 0.112 \eolc
\end{equation}
with similar uncertainties as before. The agreement with experiment has improved, but $f$ is still too small.
However, we have not fitted the subtraction constants to the original KLOE data set, but rather to a simulated data set
whose Dalitz plot parametrisation in Eq.~\eqref{eq:solDalitzSimData} agrees perfectly with what we have found. Given
that we have fitted to the simulated data set, this correspondence does not come as a surprise but is nevertheless
welcome as a consistency check.

In the neutral channel the situation is different, since we have not taken any experimental information on the decay
$\eta \to 3 \pi^0$ into account and can therefore make an independent prediction. We find
\begin{equation}
	\alpha = -0.045^{+0.008}_{-0.010}\eolc
\end{equation}
which differs from the PDG average by about 1.6 standard deviations. 

For the quark mass ratio $Q$ we get
\begin{equation}
	Q(\pi^+ \pi^- \pi^0) = 21.31^{+0.59}_{-0.50} \eolc \qquad Q(3 \pi^0) = 21.43^{+0.56}_{-0.50} \eolp
\end{equation}
Given that we have fitted our amplitude in the charged channel, we take the value for $Q$ derived from this channel as
our main result. For the branching ratio, we obtain
\begin{equation}
	r = 1.47^{+0.011}_{-0.015} \eolp
\end{equation}
Since the two results for $Q$ do not agree as well as before, the branching ratio lies further away from the
experimental result. The deviation is about 1.3 sigma.

\subsection{Error analysis}

\begin{table}[tb]
\renewcommand{\arraystretch}{1.6}
\begin{center}\begin{tabular}{lcccc}
								& $Q (\pi^+ \pi^- \pi^0)$	& $Q (3 \pi^0)$				& $r$				
								& $\alpha$			\\ \hline
$\Gamma$						& $\pm$ 0.29					& $\pm$ 0.29					& ---				
								& ---		\\
stat. KLOE					& $\pm$ 0.091					& $\pm$ 0.086					& $\pm$ 0.0068	
								& $\pm$ 0.0034			\\
syst. KLOE					& ${}^{+0.45}_{-0.30}$		& ${}^{+0.42}_{-0.28}$		& ${}^{+0.0078}_{-0.0125}$	
								& ${}^{+0.0067}_{-0.0094}$ 			\\
$\N$ KLOE					& ${}^{+0.030}_{-0.029}$	& ${}^{+0.030}_{-0.029}$	& ${}^{+0.0001}_{-0.0001}$		
								& ${}^{+0.0016}_{-0.0012}$			\\
$L_3$							& ${}^{+0.21}_{-0.25}$		& ${}^{+0.22}_{-0.26}$		& ${}^{+0.0020}_{-0.0021}$ 	
								& ${}^{+0.0018}_{-0.0015}$			\\
$\delta_I(s)$				& ${}^{+0.041}_{-0.053}$	& ${}^{+0.034}_{-0.048}$	& ${}^{+0.0014}_{-0.0018}$	
								& ${}^{+0.0020}_{-0.0017}$			\\
$W_A$							& ${}^{+0.000}_{-0.033}$	& ${}^{+0.000}_{-0.032}$	& ${}^{+0.0015}_{-0.0013}$		
								& ${}^{+0.0013}_{-0.0008}$			\\ \hline
total uncertainty			& ${}^{+0.59}_{-0.50}$		& ${}^{+0.56}_{-0.50}$		& ${}^{+0.011}_{-0.015}$
								& ${}^{+0.0083}_{-0.0104}$
\end{tabular}\end{center}
\caption{All the contributions to the uncertainties on $Q$, $r$, and $\alpha$.}
\label{tab:resultsFitErrorsQr}
\end{table}

The error analysis proceeds similarly as in the case of the matching but differs in a few points that we discuss in
detail in the following. The sources of errors on the amplitude are the statistical and systematic
uncertainties on the KLOE Dalitz plot parameters as well as the errors on $L_3$ and the phase shifts. In addition, the
result also depends on the weight factor $W_A$ by which the contribution to the $\chi^2$ from the region around the
Adler zero is multiplied and on the normalisation $\N$ that we apply to the simulated data set. As before, the
uncertainty on $Q$ gets an additional contribution due to the experimental error on the decay width. There is no
theoretical uncertainty coming from the integral cut-off in this case, since the effect from enlarging the cut-off can
be fully absorbed in the subtraction polynomial. Furthermore, we do not include an uncertainty due to inelastic
processes. Since the experimental data includes their contribution in its entirety, we believe that once we fit the
data, inelastic processes are adequately accounted for by the subtraction polynomial and their omission in the
dispersion relations does not lead to a noteworthy uncertainty.

The general procedure is again that we put a single input parameter to one of its extremal values, repeat the fit and
use the new results to estimate the uncertainty due to this parameter. The uncertainty owing to the
weight factor $W_A$ is estimated by varying it between 5 and 50. Also the experimental systematic
uncertainties are treated in this way. They are not included in the simulated data set, but we can use the
systematic errors on the Dalitz plot parameters to estimate their effect on our results. We shift one of the
Dalitz plot parameters by its systematic error and integrate the resulting change in the Dalitz distribution over each
bin. This additional contribution is added to the number of events in the bin and the modified data set is then used to
redo the fit and estimate the uncertainty due to the systematic error on that parameter.

The statistical uncertainties on the KLOE data must be treated in a special way. They are part of the simulated
data set and given for each bin, and they enter the fit through the definition of the $\chi^2$. In addition to the
central values for the subtraction constants, the fitting procedure also returns an estimate for the covariance matrix
from which follow the statistical uncertainties on the amplitude and all the quantities derived from it. Denoting the
subtraction constants by $\theta_i$ according to Eq.~\eqref{eq:solThetaSubConst}, the statistical uncertainty on some
function of the subtraction constants, $A(\vec \theta\,)$, is obtained from
\begin{equation}
	\sigma_A = \sum_{i,j = 1}^8 \frac{dA}{d \theta_i} ( \hat{\vec \theta}\,)\; 
					\frac{dA}{d \theta_j} ( \hat{\vec \theta}\,)\;
					\text{Cov}(\theta_i,\theta_j) \eolp
\end{equation}
Since the definition of $A$ is not known analytically, we estimate the derivatives numerically. The subtraction
constant $\hat{\theta}_i$ is shifted by one sigma, which is given by
\begin{equation}
	\sigma_{\theta_i} = \sqrt{ \text{Cov}(\theta_i,\theta_i) }\eolp
\end{equation}
Then, the iterations are performed with the new set of subtraction constants and the function $A$ is calculated. This
allows us to approximate the derivative by
\begin{equation}
	\frac{dA}{d \theta_i} ( \hat{\vec \theta}\,) \approx \frac{
				A(\hat{\theta}_1, \ldots, \hat{\theta}_i + \sigma_{\theta_i}, \ldots, \hat{\theta}_8)
				- A( \hat{\vec \theta}\,)}{\sigma_{\theta_i}} \eolp
\end{equation}

\section{Comparison with other results}

\begin{table}[tb]
	\renewcommand{\arraystretch}{1.6}
	\newcommand{\mc}[1]{\multicolumn{2}{c}{#1}}
	 { \small
	\begin{center}\begin{tabular}{lr@{\;}lc}
		
													& \mc{$Q$}								& \\ \hline
		dispersive (Walker)					& 22.8&$\pm 0.8$						& \cite{Anisovich+1996, Walker1998} \\ 
		dispersive (Kambor et al.)			& 22.4&$\pm 0.9$						& \cite{Kambor+1996} \\
		dispersive (Kampf et al.)			& 23.3&$\pm 0.8$						& \cite{Kampf+2011} \\
		\chpt, $\O(p^4)$						& 20.1&									& \cite{Bijnens+2007} \\
		\chpt, $\O(p^6)$						& 22.9&									& \cite{Bijnens+2007} \\ \hline
		no Dashen violation					& 24.3&									& \cite{Weinberg1977} \\
		with Dashen violation				& 20.7&$\pm1.2$						& \cite{Ananthanarayan+2004,Kastner+2008} \\
\hline
		lattice (FLAG average)				& 23.1&$\pm 1.5$						& \cite{Colangelo+2010a} \\ \hline
		dispersive, matching					& 22.74&${}^{+0.68}_{-0.67}$		& \\
		dispersive, fit						& 21.31&${}^{+0.59}_{-0.50}$		&
	\end{tabular} \end{center} }
	\caption{Various theoretical results for the quark mass double ratio $Q$. In order to simplify comparison, the same
values are also visualised in Fig.~\ref{fig:resultsQ}. The results are grouped according to their origin as follows:
from $\etapi$, from the kaon mass difference, from the lattice, and our own results. Note that the $\O(p^6)$ value in
the table does not agree with the number given in Ref.~\cite{Bijnens+2007}, which contains a misprint.}
	\label{tab:resultsQ}
\end{table}

\begin{figure}[tb]
	\begin{center}
	\psset{xunit=1.3cm, yunit=.60cm}
	\begin{pspicture*}(19.25,0)(29.05,-15.9)

		\psset{linewidth=\mylw}

		\setcounter{entrynum}{0}
		\renewcommand{\entryNum}{14}
		\setlength{\dotsize}{2.5pt}
		\renewcommand{\entryTextPos}{24.9}

		\psframe[fillstyle=solid,fillcolor=fillblue,linecolor=fillblue,linewidth=0](20.81,-\entryNum)(21.9,0)

		\psframe(19.3,0)(29.0,-\entryNum)
		\multips(20.0,-\entryNum)(1,0){5}{\psline(0,5pt)}
		\multips(19.5,-\entryNum)(1,0){6}{\psline(0,3pt)}
		\multido{\n=20+1}{5}{\uput{.2}[270](\n,-\entryNum){\n}}
		\uput{.7}[270](22,-\entryNum){\Large $Q$}

		\entry{dispersive (Walker) \cite{Anisovich+1996, Walker1998}}{22.8}{0.8}{0.8}
		\entry{dispersive (Kambor et al.) \cite{Kambor+1996}}{22.4}{0.9}{0.9}
		\entry{dispersive (Kampf et al.) \cite{Kampf+2011}}{23.3}{0.8}{0.8}
		\entry{\chpt\ $\O(p^4)$ \cite{Bijnens+2007}}{20.1}{}{}
		\entry{\chpt\ $\O(p^6)$ \cite{Bijnens+2007}}{22.9}{}{}
		\addtocounter{entrynum}{-1}
		\psline(19.3,\theentrynum)(29.0,\theentrynum)
		\entry{no Dashen violation \cite{Weinberg1977}}{24.3}{}{}
		\entry{with Dashen violation \cite{Ananthanarayan+2004,Kastner+2008}}{20.7}{1.2}{1.2}
		\addtocounter{entrynum}{-1}
		\psline(19.3,\theentrynum)(29.0,\theentrynum)
		\entry{lattice (FLAG average) \cite{Colangelo+2010a}}{23.1}{1.5}{1.5}
		\addtocounter{entrynum}{-1}
		\psline(19.3,\theentrynum)(29.0,\theentrynum)
		\psset{linecolor=green}
		\entry{dispersive, matching}{22.74}{0.67}{0.68}
		\psset{linecolor=blue}
		\entry{dispersive, fit}{21.31}{0.50}{0.59}

	\end{pspicture*}
\end{center}

	\caption{Visualisation of the results from Table~\ref{tab:resultsQ}. Our main result is marked by the blue band.}
	\label{fig:resultsQ}
\end{figure}
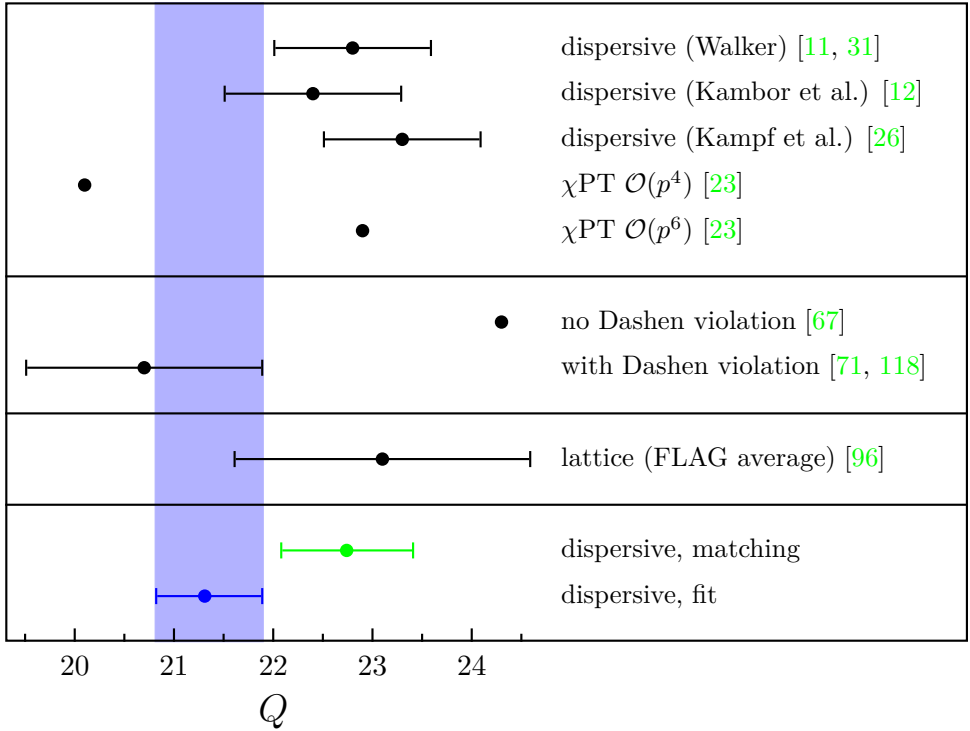

In this section we compare our results for the quark mass ratio $Q$ and the slope parameter $\alpha$ with theoretical
and experimental results from other authors.

Table~\ref{tab:resultsQ} lists a number of results for $Q$, which are visualised for easier comparison in
Fig.~\ref{fig:resultsQ}. The values in the first group have been obtained from $\etapi$ in the same way as our
results and differ only in the method that was used to calculate the decay amplitude. In particular, the list contains
the numbers from three other dispersive calculations that agree well with our matching, but are larger than what
we find from the fit. The results in the second group have been calculated from estimates of the kaon mass difference
according to Eq.~\eqref{eq:strongQmesonMasses}. In Fig.~\ref{fig:strongQFromKaon}, we have presented a few results of
this kind. Here, we only repeat the two extremal values obtained with no Dashen violation at all and with the
rather large deviation from Dashen's theorem that was found in Ref.~\cite{Ananthanarayan+2004}. The latter is the
result that agrees best with what we find from the fit. The list then ends with the lattice average from the FLAG and,
in a last group, our own results.

In Ref.~\cite{Amoros+2001}, $Q$ was determined from a fit of two-loop expressions for decay constants, meson masses,
and $K_{\ell 4}$ form factors to experimental results. The ratio $m_s/\hat{m} = 24$ was used as an input to the fit.
From Fig.~3(a) in Ref.~\cite{Amoros+2001}, one can however read off $Q$ also for other values of that ratio. Using the
average from the FLAG, $m_s/\hat{m} = 27.8$~\cite{Colangelo+2010a}, we find $Q \approx 21.4$, in good agreement with our
result from the fit. 	

The three dispersive results that rely on the one-loop result (Kambor et al., Walker, our matching) lead to
relatively large estimates for $Q$ that contain the two-loop result within their error bars. With the dispersion
relations one can include many more than two final-state rescattering processes. Nevertheless, it seems that if the
one-loop result is used to fix the subtraction constants one ends up with a result similar to \chpt\ at two loops.
Indeed, the Dalitz plot parameters that we obtain from the matching lie within the error bars of the
corresponding two-loop results (with the exception of $b$, which is slightly too large). This indicates that the bulk of
the final-state rescattering effects is already included in the two-loop result. However, since we are in clear
disagreement with the experimental Dalitz distribution, we have to conclude that the subtraction constants are not well
determined from one-loop \chpt. In particular, $\gamma_0$ and $\beta_1$ stand on not too firm theoretical grounds, since
for their determination, \chpt\ is used far away from the region of best convergence, where the other two constants are
matched.

\begin{table}[tb]
	\renewcommand{\arraystretch}{1.5}
	\newcommand{\mc}[1]{\multicolumn{2}{c}{#1}}
	\begin{center}\begin{tabular}{lcccc}
		
						& $|\alpha_0|$	& $|\beta_0|$	& $|\gamma_0|$	& $|\beta_1|$		\\ \hline
		matching		& 0.480			& 12.6			& 0				& 4.58 \\
		fit			& 0.542			& 15.6			& 41.4			& 17.8 \\ \hline
		change		& 13\%			& 24\%			& $\infty$		& 290\%
	\end{tabular} \end{center}
	\caption{Absolute values of the subtraction constants as obtained from the matching and the fit. We have omitted the
units. The change is given in per cent of the result from the matching.}
	\label{tab:resultsSubConst}
\end{table}

In order to examine the situation further, we compare the subtraction constants that we obtained from the two methods. 
Because the fit to the amplitude square does not constrain the phase of the amplitude, it makes more sense to actually
compare the absolute values. While $|\alpha_0|$ and $|\beta_0|$ are only affected at the level of less than 25 per cent,
the effect in the other two constants is indeed huge. There are two explanations for this circumstance. The first two
subtraction constants get their leading contribution already at tree level, while $\beta_1$ starts to contribute at
$\O(p^4)$ and $\gamma_0$ even only at $\O(p^6)$, since it is found to be zero in the matching at $\O(p^4)$. This renders
$\alpha_0$ and $\beta_0$ more robust against higher-order corrections. Furthermore, because $\gamma_0$ and $\beta_1$ are
strongly suppressed at small $s$, we expect that the fit around the Adler zero mainly determines the other two
constants, which must thus turn out similarly as in the matching. Even though the relative changes are of very
different size for each one of the two pairs of subtraction constants, their effects on $Q$ are very similar.

We have shown that the fit to the data does lead to a dispersive representation that is in good agreement with
experiment. At the same time, it lowers the value for $Q$ considerably. Taking only our results into account, this
seems to be a requirement of the data. The outcome of the analytical dispersive analysis in Ref.~\cite{Kampf+2011},
however, does not support this conclusion. Using the KLOE result to fit their subtraction polynomial, the authors found
a number for $Q$ that even surpasses two-loop \chpt, while being in agreement with the latter and the other dispersive
analyses. But the analysis by Kampf et al.\ includes only two final-state rescattering processes, much like the two-loop
result, and does not use the physical $\pi\pi$ phase shifts. Whether these two differences between our and their
approach can account for the discrepancy is yet to be seen.

Since lattice simulations are always performed in the isospin limit, they can not deliver a value for $Q$ without
external input. The FLAG average is obtained from lattice results for $\hat{m}$ and $m_s$ that are combined with
phenomenological input on on the kaon mass difference.

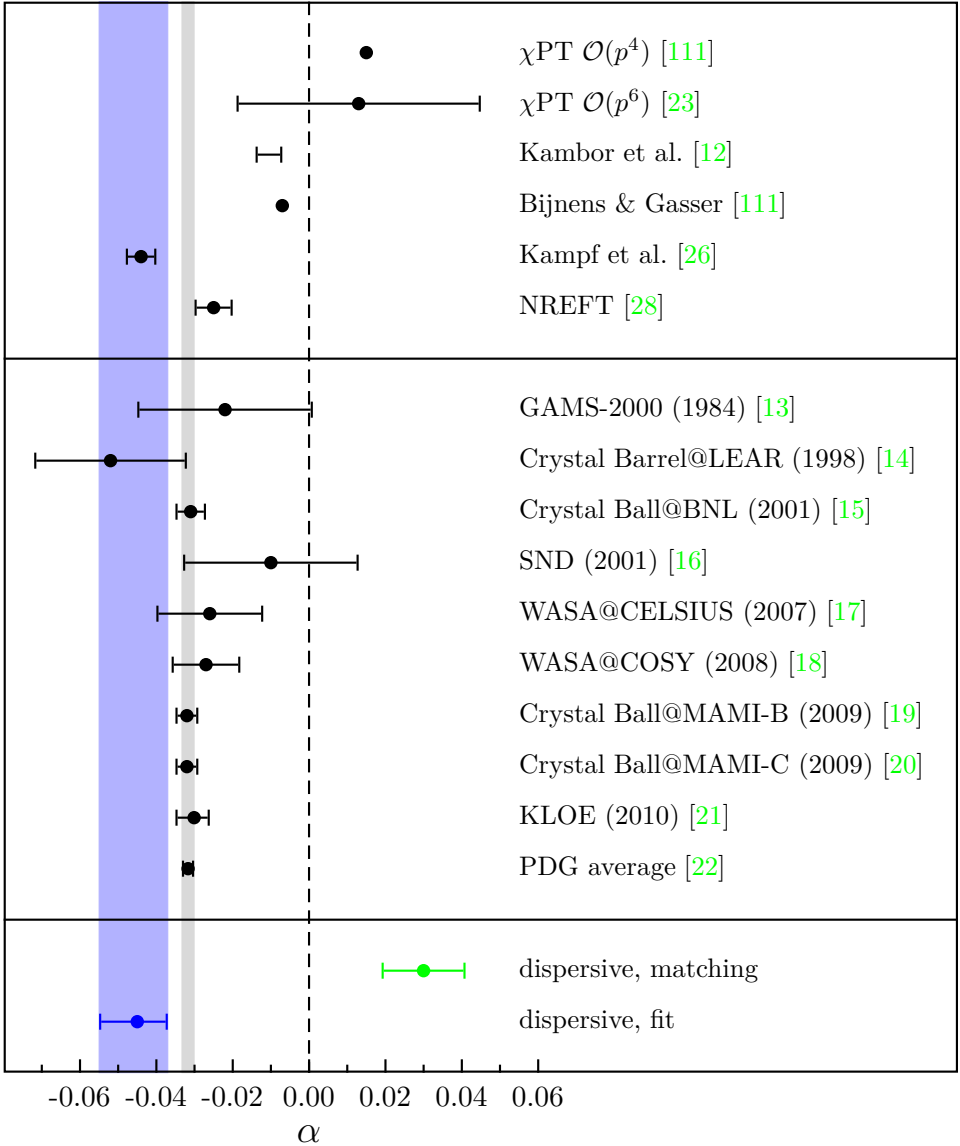
\begin{figure}[p]
\begin{center}
	\psset{xunit=50cm, yunit=.67cm}
	\begin{pspicture}(-0.08,0)(0.17,-22)

		\psset{linewidth=\mylw}

		\setcounter{entrynum}{0}
		\renewcommand{\entryNum}{21}
		\renewcommand{\entryTextPos}{0.055}
		\setlength{\dotsize}{2.5pt}

		\psline[linestyle=dashed](0,0)(0,-\entryNum)

		\psframe[fillstyle=solid,fillcolor=fillblue,linecolor=fillblue,linewidth=0](-0.037,-\entryNum)(-0.055,0)
		\psframe[fillstyle=solid,fillcolor=shadegray,linecolor=shadegray,linewidth=0](-0.0333,-\entryNum)(-0.0301,0)

		\psframe(-0.08,0)(0.17,-\entryNum)
		\multips(-0.06,-\entryNum)(0.02,0){7}{\psline(0,5pt)}
		\multips(-0.07,-\entryNum)(0.02,0){7}{\psline(0,3pt)}
		\multido{\n=-0.06+0.02}{7}{\uput{.2}[270](\n,-\entryNum){\n}}
		\uput{.7}[270](0,-\entryNum){\Large $\alpha$}

		\entry{\chpt\ $\O(p^4)$~\cite{Bijnens+2002}}{0.015}{}{}
		\entry{\chpt\ $\O(p^6)$~\cite{Bijnens+2007}}{0.013}{0.032}{0.032}
		\entry{Kambor et al.~\cite{Kambor+1996}}{}{-0.014}{-0.007}
		\entry{Bijnens \& Gasser~\cite{Bijnens+2002}}{-0.007}{}{}
		\entry{Kampf et al.~\cite{Kampf+2011}}{-0.044}{0.004}{0.004}
		\entry{NREFT~\cite{Schneider+2011}}{-0.025}{0.005}{0.005}
		
		\addtocounter{entrynum}{-1}
		\psline(-0.08,\theentrynum)(0.17,\theentrynum)

		\entry{GAMS-2000 (1984)~\cite{Alde+1984}}{-0.022}{0.023}{0.023}
		\entry{Crystal Barrel@LEAR (1998)~\cite{Abele+1998}}{-0.052}{0.020}{0.020}
		\entry{Crystal Ball@BNL (2001)~\cite{Tippens+2001}}{-0.031}{0.004}{0.004}
		\entry{SND (2001)~\cite{Achasov+2001}}{-0.010}{0.023}{0.023}
		\entry{WASA@CELSIUS (2007)~\cite{Bashkanov+2007}}{-0.026}{0.014}{0.014}
		\entry{WASA@COSY (2008)~\cite{Adolph+2009}}{-0.027}{0.009}{0.009}
		\entry{Crystal Ball@MAMI-B (2009)~\cite{Unverzagt+2009}}{-0.032}{0.003}{0.003}
		\entry{Crystal Ball@MAMI-C (2009)~\cite{Prakhov+2009}}{-0.032}{0.003}{0.003}
		\entry{KLOE (2010)~\cite{Ambrosino+2010}}{-0.0301}{0.0049}{0.0041}
		\entry{PDG average \cite{PDG2010}}{-0.0317}{0.0016}{0.0016}

		\addtocounter{entrynum}{-1}
		\psline(-0.08,\theentrynum)(0.17,\theentrynum)

		\psset{linecolor=green}
		\entry{dispersive, matching}{0.030}{0.011}{0.011}
		\psset{linecolor=blue}
		\entry{dispersive, fit}{-0.045}{0.01}{0.008}
	\end{pspicture}
\end{center}

\caption{Visualisation of the theoretical and experimental results that are listed in Table~\ref{tab:eta3piAlpha},
complemented with our own results. The blue band corresponds to our main result, the grey band to the PDG average.}
\label{fig:resultsAlpha}

\end{figure}

Let us now turn to the discussion of the slope parameter $\alpha$. We have visualised the results from
Table~\ref{tab:eta3piAlpha} together with our results in Fig.~\ref{fig:resultsAlpha}. The very accurate experimental
value for $\alpha$ offers a simple benchmark for the quality of theoretical results that is unfortunately absent in the
case of $Q$. From that point of view, the best prediction available comes from Ref.~\cite{Schneider+2011}. Our two
results support the statement we made in the discussion of $Q$: from the matching, we find a value that is rather close
to the results from \chpt\ at $\O(p^4)$ and $\O(p^6)$, but even further away from experiment than these. The fit to the
data then remedies the picture. Again the influence of the experimental input helps to detach the number for $\alpha$
from the perturbative results and brings it within 1.6 sigma from the experimental value. We stress again that our
calculation does not take any experimental input on the neutral channel into account and is thus an
independent prediction. Other than before, there is no discrepancy with the result from Ref.~\cite{Kampf+2011} and no
particular similarity can be observed between their result and \chpt\ at $\O(p^6)$. They have smaller error bars than
we do, however, these do not include the uncertainties coming from the experimental data, which are quite substantial
(see Table~\ref{tab:resultsFitErrorsQr}). Note that also from NREFT follows a value that is smaller than the PDG
average, if the KLOE data is taken into account. The authors of Ref.~\cite{Schneider+2011} found 
$\alpha = -0.062^{+0.006}_{-0.007}$ in rather strong disagreement with the PDG average and thus concluded ``that there
seems to be a significant tension between the available experimental results for charged and neutral Dalitz plot
parameters'' (p.~27). While we find a tension as well, it can not be called significant in our case.

Let us finally mention that only a few days ago, a new fit of the low-energy constants at $\O(p^4)$ and $\O(p^6)$ has
been published in Ref.~\cite{Bijnens+2011}. The main result (fit ``All'') for $L_3$ is
\begin{equation}
	L_3 = (-3.04 \pm 0.43)\ee{-3} \eolc
\end{equation}
which is considerably lower than the value that we have used: $L_3 = (-2.35 \pm 0.37)\ee{-3}$. We have not been able
to redo the computations with this new value, but from the error analysis we would expect $Q$ and $\alpha$ to
decrease, the latter only by a very small amount.

\section{Masses of the light quarks and the ratio \texorpdfstring{$R$}{R}}

\begin{table}[tb]
\renewcommand{\arraystretch}{1.6}
{ \small \newcommand{\mc}[1]{\multicolumn{1}{c}{#1}}
\begin{center}\begin{tabular}{cccllcc}
$Q$ from			& $\hat{m}$, $m_s$ from
										& $Q$		& \mc{$\hat{m}$}	& \mc{$m_s$}	& $m_u$				& $m_d$				\\ \hline
matching			& FLAG			& 22.74	& 3.4					& 95				& 2.12 $\pm$ 0.62	& 4.68 $\pm$ 0.38	\\
matching			& RBC/UKQCD		& 22.74	& 3.59				& 96.2			& 2.35 $\pm$ 0.30	& 4.83 $\pm$ 0.17	\\
matching			& BMW				& 22.74	& 3.469				& 95.5			& 2.20 $\pm$ 0.13	& 4.74 $\pm$ 0.10	\\ \hline
fit				& FLAG			& 21.31	& 3.4					& 95				& 1.94 $\pm$ 0.65	& 4.86 $\pm$ 0.39	\\
fit				& RBC/UKQCD		& 21.31	& 3.59				& 96.2			& 2.17 $\pm$ 0.31	& 5.01 $\pm$ 0.17	\\
fit				& BMW				& 21.31	& 3.469				& 95.5			& 2.02 $\pm$ 0.14 & 4.91 $\pm$ 0.11 
\end{tabular}\end{center}
}
\caption{Estimates for the light quark masses $m_u$ and $m_d$. They are calculated from both our results for 
	$Q (\pi^+ \pi^- \pi^0)$ and the lattice results from Refs.~\cite{Colangelo+2010a,Aoki+2010,Durr+2010}. All masses are
	in the $\msbar$ scheme at 2~GeV and given in MeV.}
\label{tab:resultsQuarkMasses}
\end{table}

Many lattice collaborations have published results for $m_s$ and for $\hat{m}$. Because the simulations are
performed in the isospin limit, the masses of the two lightest quarks can not be obtained from a pure lattice
calculation. The additional information that is
required can come, for example, from the kaon mass difference but also from the quark mass ratio $Q$:
\begin{equation}
	m_u = \hat{m} - \frac{m_s^2 - \hat{m}^2}{4 \hat{m} Q^2} \eolc \qquad
	m_d = \hat{m} + \frac{m_s^2 - \hat{m}^2}{4 \hat{m} Q^2} \eolp
\end{equation}
From these formul\ae\, we have calculated several estimates for $m_u$ and $m_d$, using both
our results for $Q (\pi^+ \pi^- \pi^0)$. As inputs for the quark masses $\hat{m}$ and $m_s$ we have used the averages
from FLAG \cite{Colangelo+2010a},
\begin{equation}
	\hat{m} = (3.4 \pm 0.4)~\MeV \eolc \qquad m_s = (95 \pm 10)~\MeV \eolc
\end{equation}
and two recent results that have been published after the FLAG average and are thus not yet included.
The BMW collaboration \cite{Durr+2010,Durr+2010a} has performed simulations at physical pion masses in very large
volumes and obtained the most precise results for the light quark masses that are currently available. They are 
\begin{equation}
	\hat{m} = (3.469 \pm 0.067)~\MeV \eolc \qquad m_s = (95.5 \pm 1.9)~\MeV \eolp
\end{equation}
Also the RBC/UKQCD collaboration has recently published a new result~\cite{Aoki+2010}:
\begin{equation}
	\hat{m} = (3.59 \pm 0.21)~\MeV \eolc \qquad m_s = (96.2 \pm 2.7)~\MeV \eolp
\end{equation}
All the masses are in the $\msbar$ scheme at $2~\GeV$. The three results are perfectly consistent.

Our results are summarised in Table~\ref{tab:resultsQuarkMasses}. As our main result we quote the number obtained with
$Q$ from the fit and the quark masses from the BMW collaboration, since these have the smallest uncertainties. For the
masses of the two lightest quarks we thus have
\begin{equation}
	m_u = (2.02 \pm 0.14)~\MeV \eolc \qquad m_d =  (4.91 \pm 0.11)~\MeV \eolp
\end{equation}
A vanishing up quark mass is clearly excluded.

By means of the lattice results for the quark masses, one can also obtain the quark mass ratio $R$ (see
Eq.~\eqref{eq:strongRDefinition}) from our result for $Q$. The two ratios are related by
\begin{equation}
	R = 2 Q^2 \left( 1 + \frac{m_s}{\hat{m}} \right)^{-1} \eolp
\end{equation}
With the FLAG average $m_s/\hat{m} = 27.8 \pm 1.0$ and our result for $Q$ from the fit, we find
\begin{equation}
	R = 31.5 \pm 2.1 \eolp
\end{equation}

%% file: figs/dalitz1.tex
\begingroup
  \makeatletter
  \providecommand\color[2][]{%
    \GenericError{(gnuplot) \space\space\space\@spaces}{%
      Package color not loaded in conjunction with
      terminal option `colourtext'%
    }{See the gnuplot documentation for explanation.%
    }{Either use 'blacktext' in gnuplot or load the package
      color.sty in LaTeX.}%
    \renewcommand\color[2][]{}%
  }%
  \providecommand\includegraphics[2][]{%
    \GenericError{(gnuplot) \space\space\space\@spaces}{%
      Package graphicx or graphics not loaded%
    }{See the gnuplot documentation for explanation.%
    }{The gnuplot epslatex terminal needs graphicx.sty or graphics.sty.}%
    \renewcommand\includegraphics[2][]{}%
  }%
  \providecommand\rotatebox[2]{#2}%
  \@ifundefined{ifGPcolor}{%
    \newif\ifGPcolor
    \GPcolortrue
  }{}%
  \@ifundefined{ifGPblacktext}{%
    \newif\ifGPblacktext
    \GPblacktexttrue
  }{}%
  \let\gplgaddtomacro\g@addto@macro
  \gdef\gplbacktext{}%
  \gdef\gplfronttext{}%
  \makeatother
  \ifGPblacktext
    \def\colorrgb#1{}%
    \def\colorgray#1{}%
  \else
    \ifGPcolor
      \def\colorrgb#1{\color[rgb]{#1}}%
      \def\colorgray#1{\color[gray]{#1}}%
      \expandafter\def\csname LTw\endcsname{\color{white}}%
      \expandafter\def\csname LTb\endcsname{\color{black}}%
      \expandafter\def\csname LTa\endcsname{\color{black}}%
      \expandafter\def\csname LT0\endcsname{\color[rgb]{1,0,0}}%
      \expandafter\def\csname LT1\endcsname{\color[rgb]{0,1,0}}%
      \expandafter\def\csname LT2\endcsname{\color[rgb]{0,0,1}}%
      \expandafter\def\csname LT3\endcsname{\color[rgb]{1,0,1}}%
      \expandafter\def\csname LT4\endcsname{\color[rgb]{0,1,1}}%
      \expandafter\def\csname LT5\endcsname{\color[rgb]{1,1,0}}%
      \expandafter\def\csname LT6\endcsname{\color[rgb]{0,0,0}}%
      \expandafter\def\csname LT7\endcsname{\color[rgb]{1,0.3,0}}%
      \expandafter\def\csname LT8\endcsname{\color[rgb]{0.5,0.5,0.5}}%
    \else
      \def\colorrgb#1{\color{black}}%
      \def\colorgray#1{\color[gray]{#1}}%
      \expandafter\def\csname LTw\endcsname{\color{white}}%
      \expandafter\def\csname LTb\endcsname{\color{black}}%
      \expandafter\def\csname LTa\endcsname{\color{black}}%
      \expandafter\def\csname LT0\endcsname{\color{black}}%
      \expandafter\def\csname LT1\endcsname{\color{black}}%
      \expandafter\def\csname LT2\endcsname{\color{black}}%
      \expandafter\def\csname LT3\endcsname{\color{black}}%
      \expandafter\def\csname LT4\endcsname{\color{black}}%
      \expandafter\def\csname LT5\endcsname{\color{black}}%
      \expandafter\def\csname LT6\endcsname{\color{black}}%
      \expandafter\def\csname LT7\endcsname{\color{black}}%
      \expandafter\def\csname LT8\endcsname{\color{black}}%
    \fi
  \fi
  \setlength{\unitlength}{0.0500bp}%
  \begin{picture}(7340.00,4284.00)%
    \gplgaddtomacro\gplbacktext{%
      \csname LTb\endcsname%
      \put(4059,497){\makebox(0,0)[l]{\strut{}-1}}%
      \put(4567,789){\makebox(0,0)[l]{\strut{}-0.5}}%
      \put(5075,1081){\makebox(0,0)[l]{\strut{} 0}}%
      \put(5583,1374){\makebox(0,0)[l]{\strut{} 0.5}}%
      \put(6091,1666){\makebox(0,0)[l]{\strut{} 1}}%
      \put(3713,410){\makebox(0,0){\strut{}-1}}%
      \put(3015,622){\makebox(0,0){\strut{}-0.5}}%
      \put(2316,834){\makebox(0,0){\strut{} 0}}%
      \put(1617,1047){\makebox(0,0){\strut{} 0.5}}%
      \put(918,1259){\makebox(0,0){\strut{} 1}}%
      \put(882,1442){\makebox(0,0)[r]{\strut{}0}}%
      \put(882,1675){\makebox(0,0)[r]{\strut{}}}%
      \put(882,1907){\makebox(0,0)[r]{\strut{}1}}%
      \put(882,2139){\makebox(0,0)[r]{\strut{}}}%
      \put(882,2370){\makebox(0,0)[r]{\strut{}2}}%
      \put(882,2603){\makebox(0,0)[r]{\strut{}}}%
      \put(302,2263){\makebox(0,0){\strut{}$\Gamma$(X,Y)}}%
    }%
    \gplgaddtomacro\gplfronttext{%
      \csname LTb\endcsname%
      \put(5584,878){\makebox(0,0){\strut{}X}}%
      \put(1964,507){\makebox(0,0){\strut{}Y}}%
      \put(302,2263){\makebox(0,0){\strut{}$\Gamma$(X,Y)}}%
      \put(7044,756){\makebox(0,0)[l]{\strut{} 0}}%
      \put(7044,1269){\makebox(0,0)[l]{\strut{} 0.5}}%
      \put(7044,1783){\makebox(0,0)[l]{\strut{} 1}}%
      \put(7044,2296){\makebox(0,0)[l]{\strut{} 1.5}}%
      \put(7044,2810){\makebox(0,0)[l]{\strut{} 2}}%
      \put(7044,3324){\makebox(0,0)[l]{\strut{} 2.5}}%
    }%
    \gplgaddtomacro\gplbacktext{%
    }%
    \gplgaddtomacro\gplfronttext{%
      \csname LTb\endcsname%
      \put(4059,497){\makebox(0,0)[l]{\strut{}-1}}%
      \put(4567,789){\makebox(0,0)[l]{\strut{}-0.5}}%
      \put(5075,1081){\makebox(0,0)[l]{\strut{} 0}}%
      \put(5583,1374){\makebox(0,0)[l]{\strut{} 0.5}}%
      \put(6091,1666){\makebox(0,0)[l]{\strut{} 1}}%
      \put(5584,878){\makebox(0,0){\strut{}X}}%
      \put(3713,410){\makebox(0,0){\strut{}-1}}%
      \put(3015,622){\makebox(0,0){\strut{}-0.5}}%
      \put(2316,834){\makebox(0,0){\strut{} 0}}%
      \put(1617,1047){\makebox(0,0){\strut{} 0.5}}%
      \put(918,1259){\makebox(0,0){\strut{} 1}}%
      \put(1964,507){\makebox(0,0){\strut{}Y}}%
      \put(882,1442){\makebox(0,0)[r]{\strut{}0}}%
      \put(882,1675){\makebox(0,0)[r]{\strut{}}}%
      \put(882,1907){\makebox(0,0)[r]{\strut{}1}}%
      \put(882,2139){\makebox(0,0)[r]{\strut{}}}%
      \put(882,2370){\makebox(0,0)[r]{\strut{}2}}%
      \put(882,2603){\makebox(0,0)[r]{\strut{}}}%
      \put(302,2263){\makebox(0,0){\strut{}$\Gamma$(X,Y)}}%
    }%
    \gplbacktext
    \put(0,0){\includegraphics{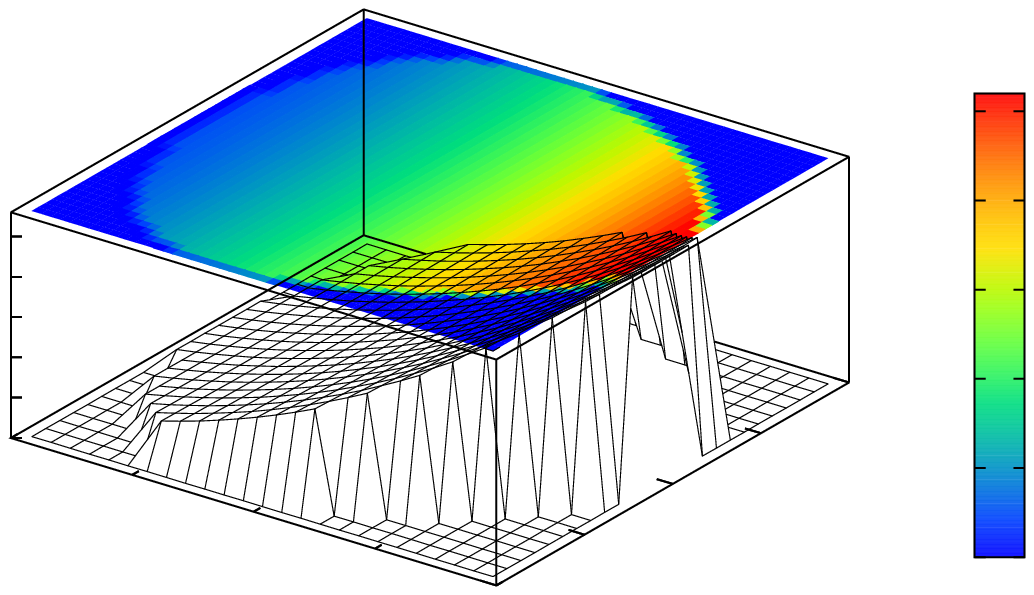}}%
    \gplfronttext
  \end{picture}%
\endgroup

%% file: figs/dalitz2.tex
\begingroup
  \makeatletter
  \providecommand\color[2][]{%
    \GenericError{(gnuplot) \space\space\space\@spaces}{%
      Package color not loaded in conjunction with
      terminal option `colourtext'%
    }{See the gnuplot documentation for explanation.%
    }{Either use 'blacktext' in gnuplot or load the package
      color.sty in LaTeX.}%
    \renewcommand\color[2][]{}%
  }%
  \providecommand\includegraphics[2][]{%
    \GenericError{(gnuplot) \space\space\space\@spaces}{%
      Package graphicx or graphics not loaded%
    }{See the gnuplot documentation for explanation.%
    }{The gnuplot epslatex terminal needs graphicx.sty or graphics.sty.}%
    \renewcommand\includegraphics[2][]{}%
  }%
  \providecommand\rotatebox[2]{#2}%
  \@ifundefined{ifGPcolor}{%
    \newif\ifGPcolor
    \GPcolortrue
  }{}%
  \@ifundefined{ifGPblacktext}{%
    \newif\ifGPblacktext
    \GPblacktexttrue
  }{}%
  \let\gplgaddtomacro\g@addto@macro
  \gdef\gplbacktext{}%
  \gdef\gplfronttext{}%
  \makeatother
  \ifGPblacktext
    \def\colorrgb#1{}%
    \def\colorgray#1{}%
  \else
    \ifGPcolor
      \def\colorrgb#1{\color[rgb]{#1}}%
      \def\colorgray#1{\color[gray]{#1}}%
      \expandafter\def\csname LTw\endcsname{\color{white}}%
      \expandafter\def\csname LTb\endcsname{\color{black}}%
      \expandafter\def\csname LTa\endcsname{\color{black}}%
      \expandafter\def\csname LT0\endcsname{\color[rgb]{1,0,0}}%
      \expandafter\def\csname LT1\endcsname{\color[rgb]{0,1,0}}%
      \expandafter\def\csname LT2\endcsname{\color[rgb]{0,0,1}}%
      \expandafter\def\csname LT3\endcsname{\color[rgb]{1,0,1}}%
      \expandafter\def\csname LT4\endcsname{\color[rgb]{0,1,1}}%
      \expandafter\def\csname LT5\endcsname{\color[rgb]{1,1,0}}%
      \expandafter\def\csname LT6\endcsname{\color[rgb]{0,0,0}}%
      \expandafter\def\csname LT7\endcsname{\color[rgb]{1,0.3,0}}%
      \expandafter\def\csname LT8\endcsname{\color[rgb]{0.5,0.5,0.5}}%
    \else
      \def\colorrgb#1{\color{black}}%
      \def\colorgray#1{\color[gray]{#1}}%
      \expandafter\def\csname LTw\endcsname{\color{white}}%
      \expandafter\def\csname LTb\endcsname{\color{black}}%
      \expandafter\def\csname LTa\endcsname{\color{black}}%
      \expandafter\def\csname LT0\endcsname{\color{black}}%
      \expandafter\def\csname LT1\endcsname{\color{black}}%
      \expandafter\def\csname LT2\endcsname{\color{black}}%
      \expandafter\def\csname LT3\endcsname{\color{black}}%
      \expandafter\def\csname LT4\endcsname{\color{black}}%
      \expandafter\def\csname LT5\endcsname{\color{black}}%
      \expandafter\def\csname LT6\endcsname{\color{black}}%
      \expandafter\def\csname LT7\endcsname{\color{black}}%
      \expandafter\def\csname LT8\endcsname{\color{black}}%
    \fi
  \fi
  \setlength{\unitlength}{0.0500bp}%
  \begin{picture}(7340.00,4284.00)%
    \gplgaddtomacro\gplbacktext{%
      \csname LTb\endcsname%
      \put(4059,497){\makebox(0,0)[l]{\strut{}-1}}%
      \put(4567,789){\makebox(0,0)[l]{\strut{}-0.5}}%
      \put(5075,1081){\makebox(0,0)[l]{\strut{} 0}}%
      \put(5583,1374){\makebox(0,0)[l]{\strut{} 0.5}}%
      \put(6091,1666){\makebox(0,0)[l]{\strut{} 1}}%
      \put(3713,410){\makebox(0,0){\strut{}-1}}%
      \put(3015,622){\makebox(0,0){\strut{}-0.5}}%
      \put(2316,834){\makebox(0,0){\strut{} 0}}%
      \put(1617,1047){\makebox(0,0){\strut{} 0.5}}%
      \put(918,1259){\makebox(0,0){\strut{} 1}}%
      \put(882,1442){\makebox(0,0)[r]{\strut{}0}}%
      \put(882,1675){\makebox(0,0)[r]{\strut{}}}%
      \put(882,1907){\makebox(0,0)[r]{\strut{}1}}%
      \put(882,2139){\makebox(0,0)[r]{\strut{}}}%
      \put(882,2370){\makebox(0,0)[r]{\strut{}2}}%
      \put(882,2603){\makebox(0,0)[r]{\strut{}}}%
      \put(302,2263){\makebox(0,0){\strut{}$\Gamma$(X,Y)}}%
    }%
    \gplgaddtomacro\gplfronttext{%
      \csname LTb\endcsname%
      \put(5584,878){\makebox(0,0){\strut{}X}}%
      \put(1964,507){\makebox(0,0){\strut{}Y}}%
      \put(302,2263){\makebox(0,0){\strut{}$\Gamma$(X,Y)}}%
      \put(7044,756){\makebox(0,0)[l]{\strut{} 0}}%
      \put(7044,1269){\makebox(0,0)[l]{\strut{} 0.5}}%
      \put(7044,1783){\makebox(0,0)[l]{\strut{} 1}}%
      \put(7044,2296){\makebox(0,0)[l]{\strut{} 1.5}}%
      \put(7044,2810){\makebox(0,0)[l]{\strut{} 2}}%
      \put(7044,3324){\makebox(0,0)[l]{\strut{} 2.5}}%
    }%
    \gplgaddtomacro\gplbacktext{%
    }%
    \gplgaddtomacro\gplfronttext{%
      \csname LTb\endcsname%
      \put(4059,497){\makebox(0,0)[l]{\strut{}-1}}%
      \put(4567,789){\makebox(0,0)[l]{\strut{}-0.5}}%
      \put(5075,1081){\makebox(0,0)[l]{\strut{} 0}}%
      \put(5583,1374){\makebox(0,0)[l]{\strut{} 0.5}}%
      \put(6091,1666){\makebox(0,0)[l]{\strut{} 1}}%
      \put(5584,878){\makebox(0,0){\strut{}X}}%
      \put(3713,410){\makebox(0,0){\strut{}-1}}%
      \put(3015,622){\makebox(0,0){\strut{}-0.5}}%
      \put(2316,834){\makebox(0,0){\strut{} 0}}%
      \put(1617,1047){\makebox(0,0){\strut{} 0.5}}%
      \put(918,1259){\makebox(0,0){\strut{} 1}}%
      \put(1964,507){\makebox(0,0){\strut{}Y}}%
      \put(882,1442){\makebox(0,0)[r]{\strut{}0}}%
      \put(882,1675){\makebox(0,0)[r]{\strut{}}}%
      \put(882,1907){\makebox(0,0)[r]{\strut{}1}}%
      \put(882,2139){\makebox(0,0)[r]{\strut{}}}%
      \put(882,2370){\makebox(0,0)[r]{\strut{}2}}%
      \put(882,2603){\makebox(0,0)[r]{\strut{}}}%
      \put(302,2263){\makebox(0,0){\strut{}$\Gamma$(X,Y)}}%
    }%
    \gplbacktext
    \put(0,0){\includegraphics{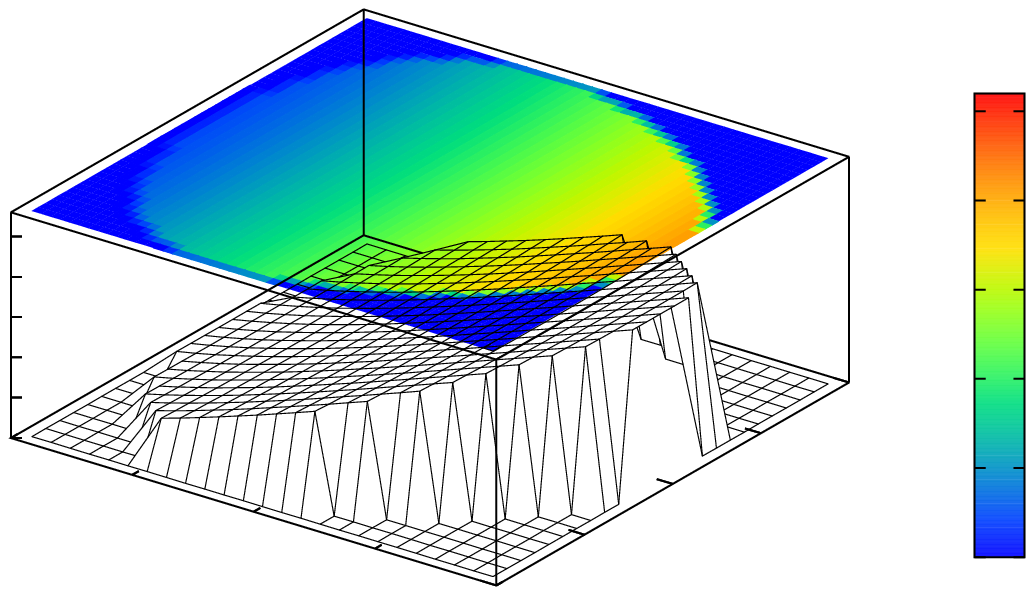}}%
    \gplfronttext
  \end{picture}%
\endgroup

%% file: main/conclusion.tex
\cleardoublepage
\phantomsection
\addcontentsline{toc}{chapter}{Conclusion and outlook}
\markboth{Conclusion and outlook}{Conclusion and outlook}
\chapter*{Conclusion and outlook}

We have discussed in great detail the derivation and solution of a set of dispersion relations for the decay $\etapi$.
The subtraction constants, which naturally appear in such equations, have been determined in two different ways. If we
rely exclusively on chiral perturbation theory, our result for the Dalitz plot parameters of the charged channel is
similar to the $O(p^6)$ result, but in clear disagreement with experiment. We find $Q = 22.74^{+0.68}_{-0.67}$ and
$\alpha = 0.030\pm0.011$, similarly to other results that rely on \chpt.

The disagreement between our amplitude and the measured Dalitz distribution in both channels is a strong motivation to
try and reduce the impact of \chpt\ and make use of the measurement by to KLOE collaboration instead. In this way, we
can successfully describe the Dalitz distribution in the charged channel. From the neutral channel, we obtain $\alpha =
-0.045^{+0.008}_{-0.010}$ in near-agreement with the PDG average. We stress again that this prediction is independent of
experimental information on the \emph{neutral} channel. For the quark mass ratio, we find $Q = 21.31^{+0.59}_{-0.50}$,
much lower than the result from the matching. This value also lies below most other results but is consistent with the
result from Ref.~\cite{Kastner+2008} that is based on a large deviation from Dashen's theorem in the kaon mass
difference.

Since $Q$ is a measure for the amount of isospin breaking in the strong interaction, it can be used to provide exactly
this information to lattice result, which are always obtained in the isospin limit. Combining our result for $Q$ from
the fit with the recent numbers for $\hat{m}$ and $m_s$ from the BMW collaboration~\cite{Durr+2010}, we find 
$m_u = (2.02 \pm 0.14)~\MeV$ and $m_d =  (4.91 \pm 0.11)~\MeV$ for the two lightest quark masses.

As we have outlined in the introduction, the rich body of work on $\etapi$ from different sources that appeared in
recent years still offers many possibilities to improve on the present analysis. Electromagnetic effects, for instance,
which influence the process through the pion mass difference as well as through additional interaction terms, have been
entirely neglected. The method we have chosen does not allow us to incorporate either one of these effects. The
analysis in the framework of NREFT~\cite{Schneider+2011}, however, can account for the pion mass difference and the
combination of our two approaches could lead to new insights. The computation of all electromagnetic effects up to 
$\O(e^2\M)$ in Ref.~\cite{Ditsche+2009} could possibly be used to gain information on the effects of electromagnetic
vertices on our analysis. 

We also hope to improve on our inputs. On the theoretical side, we plan to dispense entirely with the one-loop result
and make use of the two-loop result instead. On the experimental side, we are expecting new data from several
experiments soon, which will hopefully allow us to redo the fit with real data sets as input. It has also been announced
that these new experiments will considerably improve the statistics and the systematic uncertainties compared to the
KLOE measurement. In view of the fact that the systematic uncertainties on the data are the single largest error source
in our analysis, these are indeed bright prospects.

%% file: frontBack/acknowledgements.tex
\cleardoublepage
\phantomsection
\addcontentsline{toc}{chapter}{Acknowledgements}
\markboth{Acknowledgements}{Acknowledgements}

\chapter*{Acknowledgements}

Many others besides myself have played an important role in the genesis of this thesis. I consider myself lucky to be
surrounded by such wonderful people.

First of all, I am indebted to Gilberto Colangelo, who has ceaselessly supported and encouraged me during the last four
years. It has been a great pleasure to learn from his huge experience and deep knowledge. With his never-ending
optimism, he has many times managed to inspire me in moments of frustration.

I thank Emilie Passemar for the excellent collaboration, checks of parts of the numerical code, many interesting and
useful discussions, advice about survival in the world of physics, and proofreading parts of this thesis. 

Support from the Swiss National Science Foundation is gratefully acknowledged.

My thanks go also to all my friends at the institute, who have made life more pleasant but also very directly
contributed to this thesis through many fruitful discussions. Most particularly, I want to mention my office inmates
Christof Sch\"upbach and Peter Stoffer, as well as David Baumgartner, Lorenzo Mercolli (who also proofread parts of
the thesis), Matthias Nyfeler, Vidushi
Maillart, Daniel Arnold, and Markus Moser.

I also thank Heiri Leutwyler, J\"urg Gasser, Bastian Kubis, Stephan D\"urr, Andreas Fuhrer, Sebastian Schneider, and
Christoph Ditsche for enlightening discussions.

I am obliged to Andrzej Kup\'s\'c for creating the simulated data set from the KLOE Dalitz plot parametrisation and for
answering my questions regarding fits. Without his help, the fit of the subtraction constants would not have been
possible.

I thank Hans Bijnens for offering me the possibility to visit him in Lund, for many useful discussions, and for being
the coreferee of my thesis. My thanks go also to my office colleague Karol Kampf and all the others, who made my stay
enjoyable.

Peter Stoffer I thank for providing me with the source code of his beautiful Mandelstam diagrams and for help with
gnuplot, and Adrian W\"uthrich for advice on \LaTeX\ and for letting me use the master file of his PhD thesis.
The layout of the present work is based thereupon.

I thank the computer administrators---Markus Moser, Matthias Nyfeler, and Urs Gerber---for the luxury of a reliably
working computing environment and for their help with all kinds of computer related problems. To the secretaries of the
institute---Ottilia H\"anni, Ruth Bestgen, and Esther Fiechter---I am grateful for their help with administrative
matters and for their contribution to the pleasant atmosphere at the institute.

I thank the tireless developers of open source software, who make the results of their work available to the benefit of
all. In particular, my thanks go to the creators of Ubuntu, GNOME, \LaTeX, PSTricks, the Beamer-class, Kile, JabRef,
gnuplot, JaxoDraw and Grace. These programs have been indispensable in the preparation of this thesis.

I thank my parents, Helen and Andreas Lanz, who have for many years now supported me in all my ventures. I would not
have gotten far without them. Thanks go also to my parents-in-law, Regina and Toni Kunkler, for their ongoing support.
Especially in the last month of my work on this thesis, the four of them have helped to relieve me of many
non-physics-related duties.

My very special thanks go to those---mentioned above or not---that have been and are an encouragement to me
through their friendship.

From the bottom of my heart I thank my dear family, Mirjam, Ruben, and Liv, who have supported and encouraged me in
any possible way. You had to endure a much too often absent husband and father, especially in the last few months. 
This thesis is dedicated to you.

%% file: frontBack/appendix.tex
\begin{appendices}

\chapter{Fortran and NAG routines} \label{chp:appendixFortran}

The numerical programs used to solve the dispersion relations have been written in Fortran 95 and compiled with the
Intel Fortran Compiler IFORT, version 11.1. For the evaluation of numerical standard problems, such as for instance
integration or minimisation, we made use of the NAG Numerical Libraries for Fortran 77, Mark 21, by The Numerical
Algorithms Group Ltd. 

We give here a complete list of the routines from the NAG libraries that have been used in our programs together with a
short description of their purpose. A detailed documentation can be found online~\cite{NAG21}.

\begin{description}[leftmargin=1.8cm,style=sameline]
	\item[c05adf] zero of a continuous function in a given interval
	\item[d01ajf] one-dimensional integration over a finite interval
	\item[d01amf] one-dimensional integration over a semi-infinite or infinite interval
	\item[d01aqf] Cauchy principal value for a one-dimensional integral with weight function $1/(x-c)$
	\item[d01daf] two-dimensional integration over a finite region
	\item[d04aaf] first 14 numerical derivatives of a function of one real variable
	\item[e01baf] creation of a cubic spline interpolant for a function of one variable
	\item[e02bbf] evaluation of a cubic spline interpolant based on the output from \textsf{e01baf}
	\item[e02bcf] evaluation of the first three derivatives of a cubic spline interpolant based on the output
						from \textsf{e01baf}
	\item[e04fyf] unconstrained minimum of a sum of squares
	\item[e04jyf] minimum of a function of several variables
	\item[e04ycf] covariance matrix for nonlinear least-squares problem based on output from \textsf{e04fyf}
	\item[f01adf] inverse of a real, symmetric, positive-definite matrix
	\item[f04caf] solves a system of complex linear equations
	\item[s14baf] returns the regularised Gamma functions, used to calculate the cumulative
						$\chi^2$ distribution function
\end{description}

\chapter{\texorpdfstring{$\chi^2$-}{Chi-square }distribution} \label{chp:appendixChi2}

For a set of independent, standard normally-distributed random variables $X_1$, $\ldots$, $X_n$, the quantity
\begin{equation}
	Z = \sum_{i=1}^{n} X_i^2 
\end{equation}
is distributed according to the $\chi^2$-distribution with $n$ degrees of freedom. Its density and distribution
function are tabulated, but can also be written in terms of the Gamma function. The density is given by
\begin{equation}
	f_{\chi^2}(n,x) = \inv{2^{n/2}\, \Gamma(n/2)} x^{n/2-1} e^{-x/2} \eolc \qquad x \geq 0 \eolc
\end{equation}
with
\begin{equation}
	\Gamma(z) = \int\limits_0^\infty t^{z-1} e^{-t}\, dt \eolp
\end{equation}
The distribution function is defined as the integral over the density, leading to
\begin{equation}\begin{split}
	F_{\chi^2}(n,x) 	&= \int\limits_{-\infty}^x f_{\chi^2}(n,t)\, dt
				= \inv{2^{n/2} \Gamma(n/2)} \int\limits_0^x t^{n/2-1} e^{-t/2}\, dt\\[2mm]
				&= P(n/2,x/2) \eolc
	\label{eq:appChi2Dist}
\end{split}\end{equation}
where we have introduced the normalised incomplete Gamma function
\begin{equation}
	P(n,x) = \inv{\Gamma(n)} \int\limits_0^x t^{n-1} e^{-t} dt \eolp
\end{equation}

\begin{figure}[p]
	\psset{xunit=0.9cm,yunit=6.7cm}
	\begin{center}\begin{pspicture*}(-1,-.07)(10.5,1.04)
		\psset{linewidth=\mylw}
		\fileplot[linecolor=orange]{data/chi2Dens5.dat}
		\fileplot[linecolor=green]{data/chi2Dens4.dat}
		\fileplot[linecolor=red]{data/chi2Dens3.dat}
		\fileplot[linecolor=blue]{data/chi2Dens2.dat}
		\fileplot{data/chi2Dens1.dat}
		\psline(7,0.9)(7.5,0.9) \rput(8.2,0.9){\small $n=1$}
		\psline[linecolor=blue](7,0.82)(7.5,0.82) \rput(8.2,0.82){\small $n=2$}
		\psline[linecolor=red](7,0.74)(7.5,0.74) \rput(8.2,0.74){\small $n=3$}
		\psline[linecolor=green](7,0.66)(7.5,0.66) \rput(8.2,0.66){\small $n=4$}
		\psline[linecolor=orange](7,0.58)(7.5,0.58) \rput(8.2,0.58){\small $n=5$}
		\psaxes[Dx=2,dx=2,Ox=0,Dy=0.2,dy=0.2,Oy=0,axesstyle=frame,labels=all,ticks=all,showorigin=true,tickstyle=top,
ticksize=4pt](0,0)(10,1)
	\end{pspicture*}\end{center}
	\caption{Density function $f_{\chi^2}(n,x)$ of the $\chi^2$-distribution.}
	\label{fig:appendixDensity}
\end{figure}
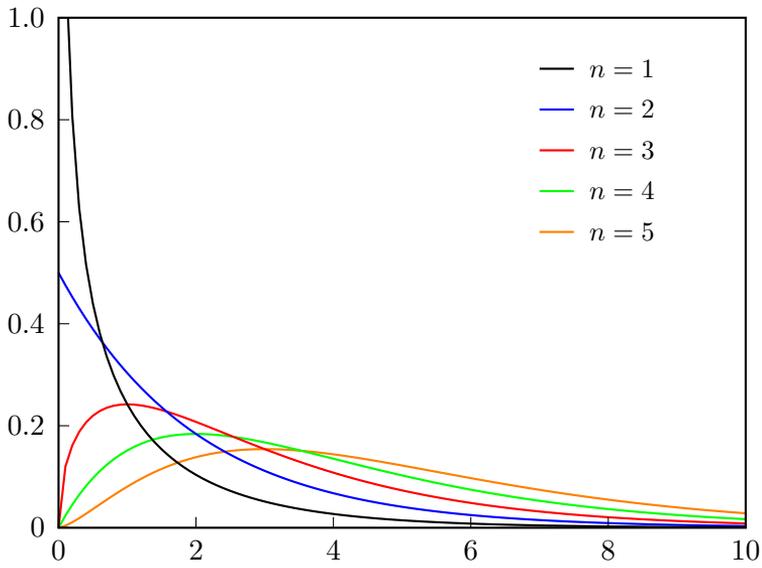

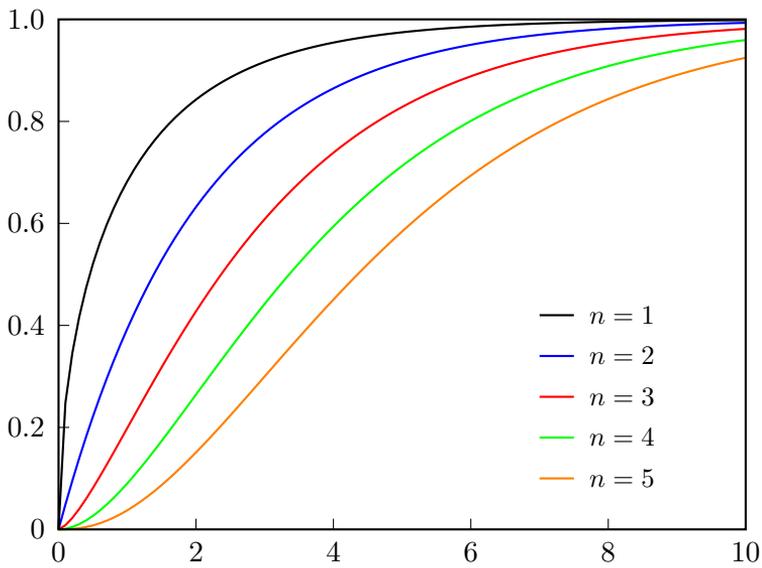
\begin{figure}[p]
	\psset{xunit=0.9cm,yunit=6.7cm}
	\begin{center}\begin{pspicture*}(-1,-.07)(10.5,1.04)
		\psset{linewidth=\mylw}
		\fileplot[linecolor=orange]{data/chi2Dist5.dat}
		\fileplot[linecolor=green]{data/chi2Dist4.dat}
		\fileplot[linecolor=red]{data/chi2Dist3.dat}
		\fileplot[linecolor=blue]{data/chi2Dist2.dat}
		\fileplot{data/chi2Dist1.dat}
		\psline(7,0.42)(7.5,0.42) \rput(8.2,0.42){\small $n=1$}
		\psline[linecolor=blue](7,0.34)(7.5,0.34) \rput(8.2,0.34){\small $n=2$}
		\psline[linecolor=red](7,0.26)(7.5,0.26) \rput(8.2,0.26){\small $n=3$}
		\psline[linecolor=green](7,0.18)(7.5,0.18) \rput(8.2,0.18){\small $n=4$}
		\psline[linecolor=orange](7,0.1)(7.5,0.1) \rput(8.2,0.1){\small $n=5$}
		\psaxes[Dx=2,dx=2,Ox=0,Dy=0.2,dy=0.2,Oy=0,axesstyle=frame,labels=all,ticks=all,showorigin=true,tickstyle=top,
ticksize=4pt](0,0)(10,1)
	\end{pspicture*}\end{center}
	\caption{Distribution function $F_{\chi^2}(n,x)$ of the $\chi^2$-distribution.}
	\label{fig:appendixDistribution}
\end{figure}

From the density one obtains for the moments about zero of the $\chi^2$-distribu\-tion with $n$ degrees of freedom
\begin{equation}
	E(X^m) = \int\limits_0^\infty x^m f_{\chi^2}(n,x)\, dx = 2^m \frac{\Gamma(m + n/2)}{\Gamma(n/2)} \eolp
\end{equation}
The expectation value and the variance follow immediately from the moments:
\begin{equation}
	E(X) = n \eolc \qquad \text{Var}(X) = E(X^2) - E(X)^2 = 2 n \eolp
\end{equation}
The densities and the distribution functions of the $\chi^2$-distribution with one to five degrees of freedom
are shown in Figs.~\ref{fig:appendixDensity} and \ref{fig:appendixDistribution}, respectively.

\markright{\leftmark}	

\chapter{Analytical integrals}

We derive the indefinite integrals that form the basis for the analytical evaluation of the definite integrals
$Q^{a/b}_{0/1}$ and $I^{a/b}_{0/1}$. First, we solve the indefinite integral
\begin{equation}
	\int dx\, \inv{\sqrt{-x} (x+\alpha)} \eolp
\end{equation}
For $x > 0$ and with the variable transformation $u^2 = x$ we find
\begin{equation}
	\int dx\, \inv{\sqrt{-x} (x+\alpha)} = \frac{2}{i} \int du\, \inv{u^2 + \alpha} 
				= \frac{2}{i \sqrt{\alpha}} \arctan \left( \frac{\sqrt{x}}{\sqrt{\alpha}} \, \right) \eolp
\end{equation}
Similarly, for $x < 0$,
\begin{equation}
	\int dx\, \inv{\sqrt{-x} (x+\alpha)} = \frac{2}{-i} \int du\, \inv{u^2 + \alpha} 
				= \frac{2i}{\sqrt{\alpha}} \arctan \left( \frac{\sqrt{x}}{\sqrt{\alpha}} \, \right) \eolp
\end{equation}
Since
\begin{equation}
	\frac{\sqrt{x}}{\sqrt{-x}} = \left\{
		\begin{array}{rl}
			-i \eolc \quad & x>0 \eolc \\[2mm]
			 i \eolc \quad & x<0 \eolc
		\end{array} \right.
\end{equation}
the two results can be summarised as
\begin{equation}
	\int dx\, \inv{\sqrt{-x} (x+\alpha)} 
		= \frac{2\sqrt{x}}{\sqrt{-x}}\, \frac{\arctan \left( \frac{\sqrt{x}}{\sqrt{\alpha}} \,\right)}{\sqrt{\alpha}}
		\eolp
		\label{eq:appendixIntegral1}
\end{equation}
The second integral we need is more complicated:
\begin{equation}
	\int dx\, \inv{(-x)^{3/2} (x+\alpha)} \eolp
\end{equation}
For $x > 0$ and with the variable transformation $u^2 = x$ we obtain
\begin{equation}
	\int dx\, \inv{(-x)^{3/2} (x+\alpha)} = \frac{2}{i} \int du\, \inv{u^2\,(u^2 + \alpha)} \eolc
\end{equation}
and, as before, the result for $x < 0$ differs only by the overall sign. The integrand can be decomposed as
\begin{equation}
	\inv{u^2\,(u^2 + \alpha)} = \inv{\alpha} \left( \inv{u^2} - \inv{u^2+\alpha} \right) \eolc
\end{equation}
leading to
\begin{equation}
	\int dx\, \inv{(-x)^{3/2} (x+\alpha)} = -\frac{2}{\alpha\, (-x)^{3/2}} 
				\left( x +  \frac{x^{3/2}\,\arctan \left( \frac{\sqrt{x}}{\sqrt{\alpha}} \,\right)}{\sqrt{\alpha}} \right)
	\eolc
	\label{eq:appendixIntegral2}
\end{equation}
where we made us of Eq.~\eqref{eq:appendixIntegral1} and
\begin{equation}
	\frac{x^{3/2}}{(-x)^{3/2}} = \left\{
		\begin{array}{rl}
			 i \eolc \quad & x>0 \eolc \\[2mm]
			-i \eolc \quad & x<0 \eolp
		\end{array} \right.
\end{equation}

\end{appendices}

%% file: frontBack/bibliography.tex
\cleardoublepage
\phantomsection
\addcontentsline{toc}{chapter}{Bibliography}

\input{thesis.bbl}

\cleardoublepage

%% file: thesis.bbl
\providecommand{\href}[2]{#2}\begingroup\raggedright\endgroup